\documentclass[11pt,a4paper]{article}
\pdfoutput=1

\usepackage{jheppub}

\usepackage{makecell} 
\usepackage{adjustbox}

\usepackage[T1]{fontenc}

\usepackage{amsmath}
\usepackage{amssymb}
\usepackage{amsthm}
\usepackage{amsfonts}
\usepackage{amscd}
\usepackage{bbm}
\usepackage{array}
\usepackage{blindtext}
\usepackage{booktabs}
\usepackage{enumerate}
\usepackage[shortlabels]{enumitem}
\usepackage{fancyhdr} 
\usepackage{float}
\usepackage{mdframed}
\usepackage{graphicx}
\usepackage{latexsym}
\usepackage{lmodern}
\usepackage{mathrsfs}
\usepackage{makeidx}
\usepackage{dsfont}
\usepackage{multirow}
\usepackage{tensor}
\usepackage{slashed}
\usepackage{url}
\usepackage{xspace}
\usepackage{tikz-cd}
\usetikzlibrary{arrows.meta, positioning,decorations.pathmorphing}

\usepackage{afterpage}
\usepackage{placeins}

\usepackage{caption}
\usepackage{ytableau}
\usepackage{bm}

\usepackage[normalem]{ulem}

\newcommand{\Z}{{\mathbb Z}}
\newcommand{\de}{{\partial}}
\renewcommand{\O}{{\mathcal O}}
\newcommand{\nn}{\nonumber}
\newcommand{\form}{\texttt{Form}\xspace}

\newcommand{\githubb}{\texttt{GitHub}\xspace}

\newcommand{\Rs}{$R^*$\xspace}

\newcommand{\ep}{\epsilon}
\newcommand{\EOM}{\text{EOM}}
\newcommand{\N}{n}

\usepackage{simplewick}
\usepackage{stmaryrd}

\usepackage{youngtab}

\title{EFT meets CFT: Multiloop renormalization of higher-dimensional operators in general $\boldsymbol{\phi^4}$ theories}

\author[1,2]{Johan Henriksson,}
\author[3,4]{Stefanos R. Kousvos,}
\author[5,6]{Jasper Roosmale Nepveu}
\affiliation[1]{Universit\'e Paris--Saclay, CEA, Institut de Physique Th\'eorique, 91191, Gif-sur-Yvette, France}
\affiliation[2]{Theoretical Physics Department, CERN, 1211, Geneva, Switzerland}
\affiliation[3]{Department of Physics, University of Pisa and INFN, section of Pisa,\\[-1pt] Largo Pontecorvo 3, I-56127 Pisa, Italy}
\affiliation[4]{Department of Physics, University of Torino and INFN, section of Torino,\\[-1pt] Via P.\ Giuria 1, 10125 Torino, Italy}
\affiliation[5]{Department of Physics and Center for Theoretical Physics, National Taiwan University, Taipei 10617, Taiwan}
\affiliation[6]{Leung Center for Cosmology and Particle Astrophysics, Taipei 10617, Taiwan}

\emailAdd{johan.henriksson@cern.ch}
\emailAdd{stefanosrobert.kousvos@unito.it}
\emailAdd{jasperrn@ntu.edu.tw}

\newcommand{\eps}{{\varepsilon}}

\numberwithin{equation}{section}

\abstract{
The renormalization of composite operators is a fundamental aspect of quantum field theory, relevant for the description of phase transitions and high energy phenomenology. We calculate the anomalous dimensions of a large set of operators in \emph{any} scalar $\phi^4$ theory in $d=4-\eps$ dimensions, up to five loops in most cases. The results have applications in both effective field theory (EFT) and conformal field theory (CFT). As  an EFT application, we extract the five-loop renormalization group (RG) equations of the Higgs sector of the Standard Model EFT at dimension six, and up to two loops at dimension eight, aligning our operator basis with custodial symmetry violation. Additionally, for CFT, by resumming the $\eps$-expanded results at the fixed-point, we determine the entire low-lying spectrum (i.e.\ up to dimension six and Lorentz rank two) of the Ising, $O(n)$ and hypercubic scalar CFTs. Our work enables future conformal bootstrap studies for numerous theories of interest. We include  introductions to EFT and CFT, and we illustrate our method and the structure in RG mixing matrices in several illuminating examples, which may also be of general interest. All results in the general theory are publicly available and we describe a systematic path towards applying them to more complicated CFTs.

}

\begin{document}

	\makeatletter
	\let\old@fpheader\@fpheader
	\renewcommand{\@fpheader}{  \vspace*{-0.1cm} \hfill CERN-TH-2025-258}
	\makeatother

\maketitle

\section{Introduction}
The renormalization group (RG) encodes the scale dependence of the parameters of a quantum field theory (QFT). It plays a key role in many branches of modern physics, such as the description of fundamental forces in particle physics and the characterization of critical phenomena in condensed-matter systems. 
For example, the renormalization of gauge theories~\cite{tHooft:1971akt,tHooft:1971qjg,tHooft:1972tcz}
and the discovery of asymptotic freedom~\cite{Politzer:1973fx,Gross:1973id, Gross:1973ju, Gross:1973zrg} have paved the way for a consistent treatment of the Standard Model. 
In the search for physics beyond the Standard Model using effective field theory (EFT), the RG also plays an essential role by relating the interactions across widely separated energy scales. 

The general RG framework by Wilson has provided an explanation of critical behavior and the universal features among different systems~\cite{Wilson:1971bg,Wilson:1971dh,Wilson:1973jj}. Over the years, numerous physical systems that exhibit critical behaviour have been studied~\cite{Pelissetto:2000ek}. Depending on the degrees of freedom at each lattice site (e.g.~the spins), and their type of interactions (ferromagnetic, antiferromagnetic, etc.), many different universality classes exist. 
From the theoretical point of view, these correspond to different conformal field theories (CFTs) with, in general, different global symmetries. Each universality class is characterized by a set of experimentally measurable critical exponents, which can be determined from the anomalous dimensions in the corresponding CFT.

In this paper, we consider the perturbative renormalization of composite operators. This is a central topic in QFT, with various applications; see Table~\ref{tab:applications}. 
Depending on the application, there is a need for anomalous dimension matrices at higher loop orders. For renormalizable and marginal interactions, the determination of $\beta$ functions is a classic topic, with multiloop results in QED \cite{Gorishnii:1990kd,Kataev:2012rf,Baikov:2012zm} and QCD \cite{Vanyashin:1965ple,Khriplovich:1969aa,Gross:1973id,Politzer:1973fx,Baikov:2016tgj,Herzog:2017ohr}. 
Seven-loop $\beta$ functions exist in the case of $O(n)$ symmetric scalar theory, with eight loops for the anomalous dimension of the field~\cite{Schnetz:2016fhy,Schnetz:2022nsc}. 
Operators with leading twist $\tau=\Delta-\ell$ (with scaling dimension~$\Delta$ and spin~$\ell$)
are important in the analysis of deep inelastic scattering~\cite{Gribov:1972ri,Dokshitzer:1977sg,Altarelli:1977zs}, for which results are derived for increasing spin at fixed twist $\tau=2$. 
Such results exist at three and four loops in QCD~\cite{Moch:2004pa,Vogt:2004mw,Velizhanin:2010ey,Moch:2017uml,Kniehl:2025jfs}
and at four loops in scalar theory~\cite{Derkachov:1997pf,Manashov:2017xtt,Henriksson:2018myn,Manashov:2025kgf}.
On the other hand, operators with zero Lorentz spin but higher scaling dimensions are relevant for EFT.
For instance, in the Standard Model EFT (SMEFT), full one-loop results exist at dimension six~\cite{Jenkins:2013zja,Jenkins:2013wua,Alonso:2013hga,Alonso:2014zka}, with partial results at two loops (e.g.~\cite{Jenkins:2023bls,Born:2024mgz,Haisch:2025vqj,DiNoi:2024ajj,Duhr:2025yor,Banik:2025wpi}), and partially at dimension eight at one loop (e.g.~\cite{Chala:2021pll,Helset:2022pde,DasBakshi:2022mwk,Bakshi:2024wzz}). 
In $O(n)$-symmetric scalar EFTs, calculations have been extended to even higher dimensions and loop orders~\cite{Cao:2021cdt,Cao:2023adc,RoosmaleNepveu:2024zlz}, laying some of the foundations for the work in the current paper.

\begin{table}
\centering
\caption{Applications of the multiloop renormalization of composite operators.}
\label{tab:applications}
{\small
\renewcommand{\arraystretch}{1.75}
\begin{tabular}{|l|p{6.5cm}|p{3.7cm}|}
\hline
&\textbf{Lorentz scalar operators} & \textbf{Spinning operators}
\\\hline
\textbf{Low twist} & Renormalizable theories, critical exponents & Deep inelastic scattering
\\\hline
\textbf{High twist} & Effective field theory and conformal field theory applications & Conformal field theory applications
\\\hline
\end{tabular}
}
\end{table}

\begin{figure}[!p]
    \centering
\includegraphics[width=0.48\textwidth]{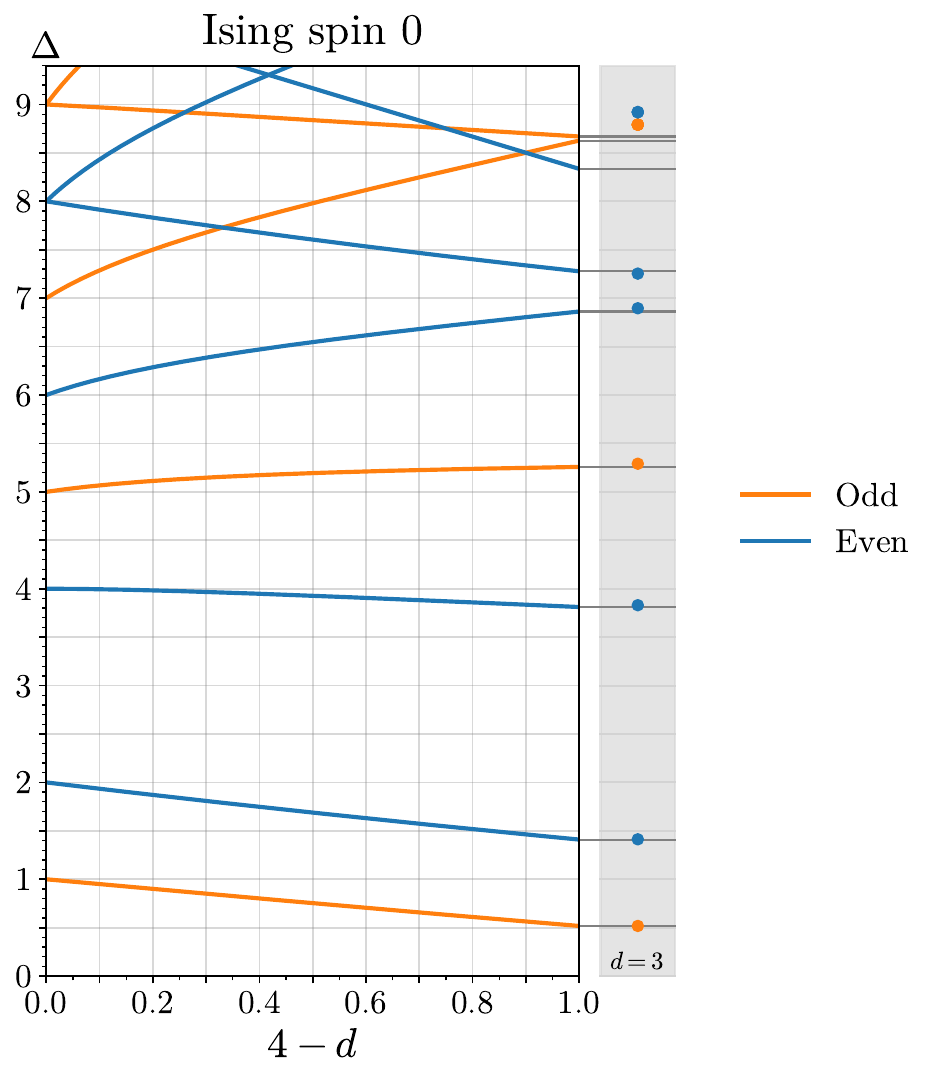}
\includegraphics[width=0.48\textwidth]{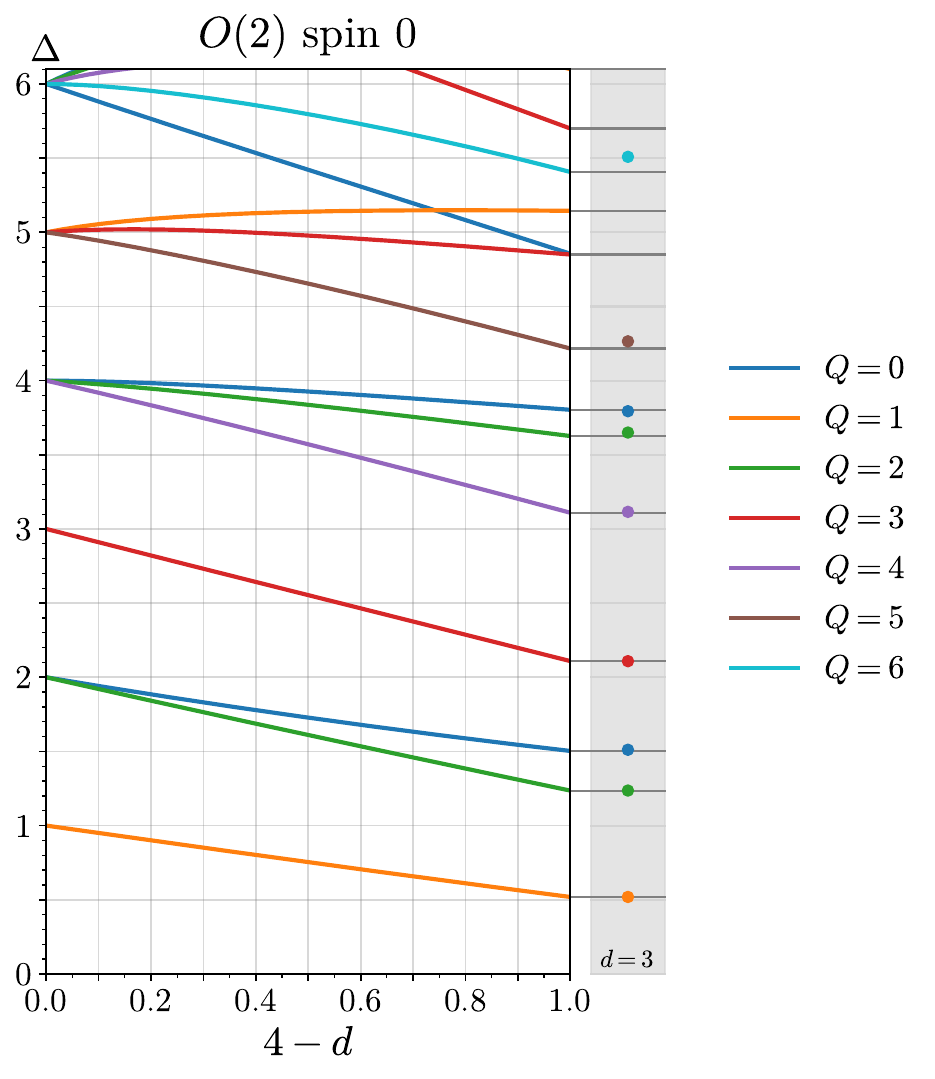}

\vspace{6mm}

\includegraphics[width=0.48\textwidth]{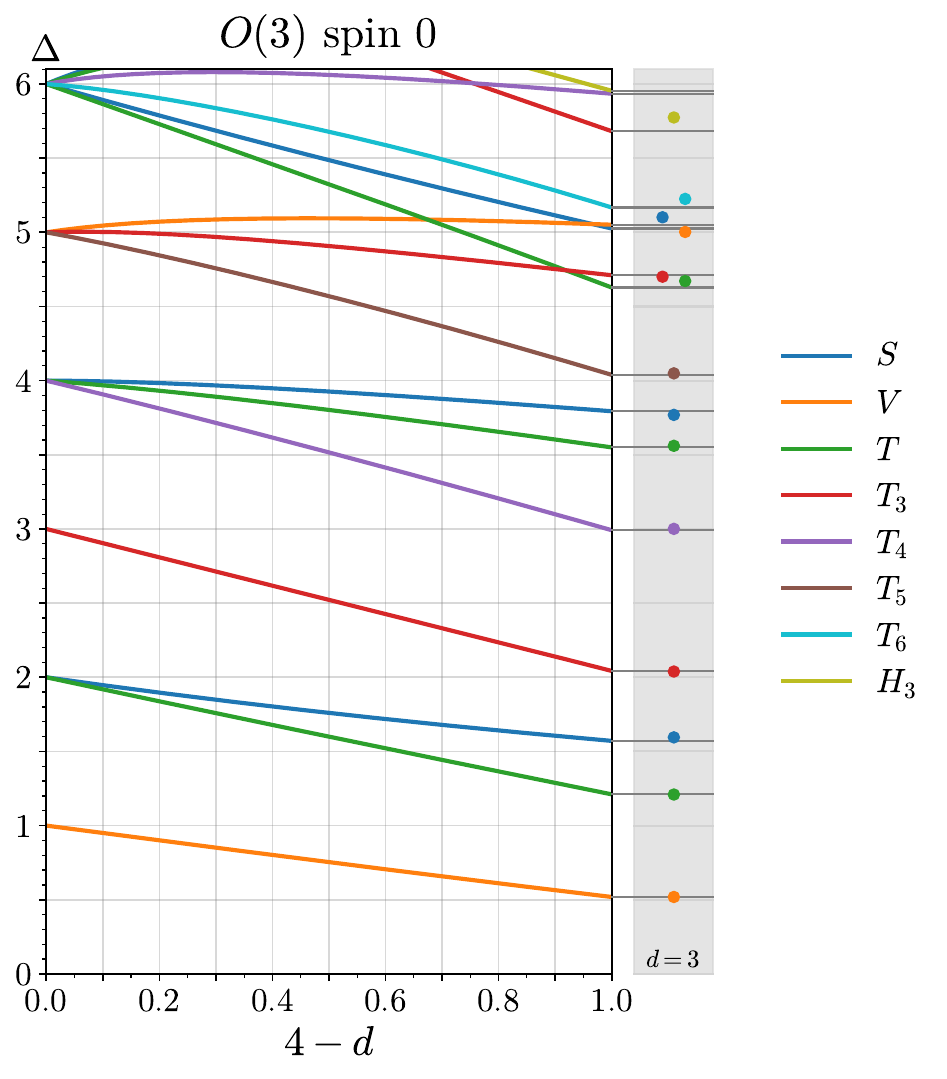}
\includegraphics[width=0.48\textwidth]{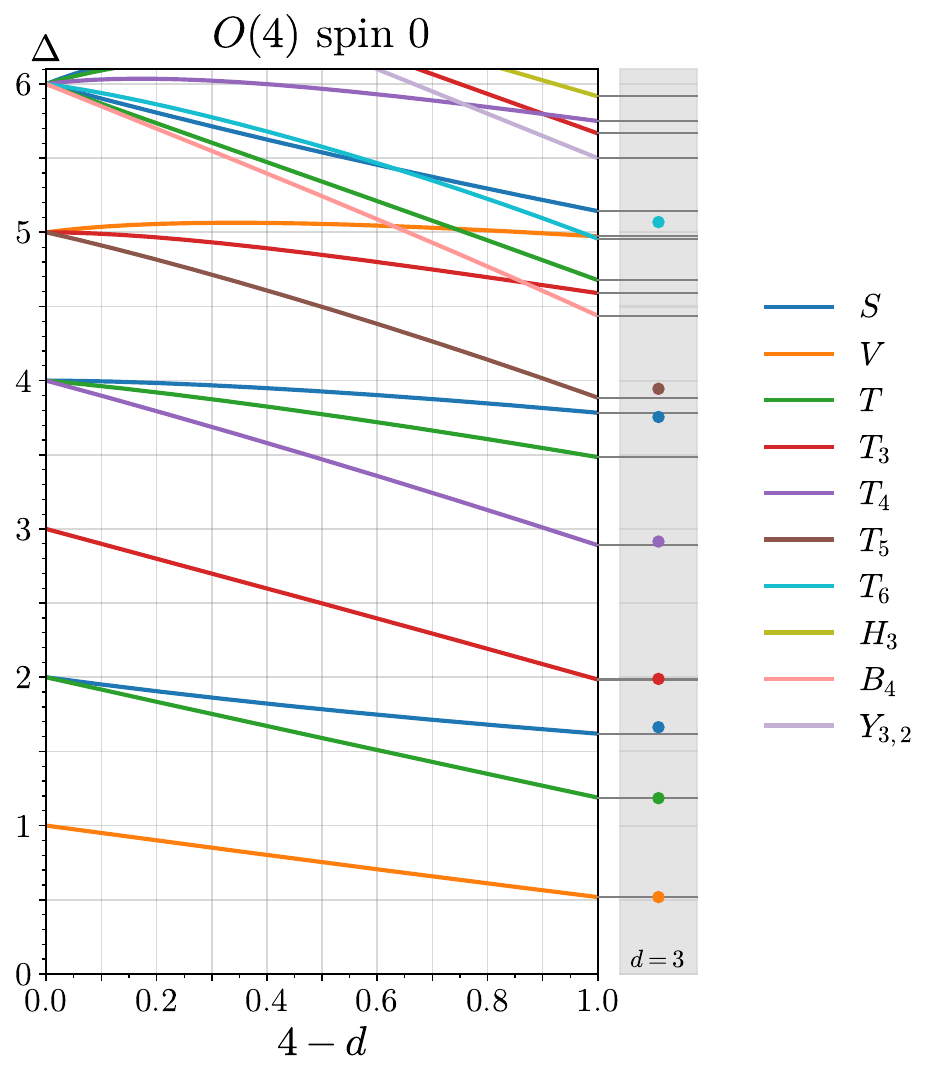}

\vspace{4mm}

    \caption{Scalar spectra for $O(n)$ symmetric theories with $n=1,2,3,4$. The energy levels shown on the very right are results in three dimensions: numerical bootstrap for $n=1$ \cite{Simmons-Duffin:2016wlq}, $n=2$ \cite{Chester:2019wfx}, $n=3$ \cite{Chester:2020iyt,Chester2020Unp,Rong:2023owx}, and Monte Carlo for $n=2$ (charge 5,6) \cite{Banerjee:2017fcx} and $n=4$ \cite{Calabrese:2002bm,Banerjee:2019jpw,Hasenbusch:2021rse}.
    Note that we have removed the representations and operators which vanish identically for low $n$; see \cite{Cao:2023psi}. The results of this paper concern five-loop anomalous dimensions for all operators with $\Delta(4d)\leqslant6$, together with additional results above this value in specific cases. We plot the spin-1 and spin-2 spectra in Figures~\ref{fig:vectors-all} and \ref{fig:tensors-all}, respectively.
    }
    \label{fig:scalar-all}
\end{figure}

In addition to the aforementioned applications, a major motivation for the renormalization of composite operators stems from CFT. 
Although precise results for many CFTs require non-perturbative methods, sometimes perturbation theory itself can provide precision data, for instance in the important class of $\lambda\phi^4$ theories using the famous $\eps$-expansion by Wilson and Fisher \cite{Wilson:1971dc,Wilson:1973jj}.
This expansion captures a large set of 3d CFTs with multicomponent scalar field content through a perturbative expansion in $d=4-\eps$ spacetime dimensions, where the coupling constant $\lambda=O(\eps)$ becomes infinitesimal. Remarkably, results from the $\eps$-expansion are in good agreement with 3d data when extrapolated to $\eps=1$ using various resummation techniques. This agreement has been observed for a long time for the leading operators in the simplest $O(n)$ symmetric theories~\cite{LeGuillou1987}. For larger parts of the spectrum, perturbative results have been more modest, chiefly at $O(\eps)$ in the $O(n)$ symmetric case \cite{Kehrein:1992fn,Kehrein:1994ff} with scattered results at subleading order, e.g.~\cite{Zhang1982,Kehrein:1995ia}; see \cite{Henriksson:2022rnm} for a summary of known results prior to this work.
The conformal bootstrap~\cite{Rattazzi:2008pe,Poland:2018epd,Chester:2019wfx,Rychkov:2023wsd,PerimeterCourse} allows one to study CFTs directly in $d=3$, but also across spacetime dimensions, making it possible to track parts of the spectrum across spacetime dimension from infinitesimal to finite $\eps$ \cite{El-Showk:2013nia,Cappelli:2018vir,Henriksson:2022gpa,Sirois:2022vth}. 
In these works, spinning, higher-dimensional and non-singlet operators feature at the same footing as the lowest scalar operators, and in our work we incorporate operators of all kind: both scalar and spinning operators in arbitrary global-symmetry representations. In Figure~\ref{fig:scalar-all} and later figures of this paper we demonstrate the wealth of operators for which we can compare (new) perturbative estimates with non-perturbative results in $d=3$. 

While one application of our perturbative study is the spectrum of $O(n)$ symmetric theories, for which precise non-perturbative data exist in 3d, an important motivation for our work is the extension to less symmetric $\phi^4$-type CFTs. These have been suggested to describe second-order phase transitions in various magnets as well as structural phase transitions~\cite{Pelissetto:2000ek}. 
It is generally a difficult task to derive rigorous precision results for these theories in 3d using the conformal bootstrap. Our results provide perturbative precision estimates for a large part of the spectrum, which will prove crucial in justifying assumptions used in future bootstrap studies. In fact, our estimates for theories with hypercubic symmetry were already important to separate the cubic from the very-nearby $O(3)$ CFT with the bootstrap \cite{Kousvos:2025ext}.

The approach taken in this paper, useful for both EFT and CFT applications, is to work in a general theory with an $n$-component field $\phi^a$ and arbitrary interaction $\lambda_{abcd}\phi^a\phi^b\phi^c\phi^d$. 
The results for particular theories with specific symmetries can subsequently be extracted from the general expressions. 
This not only covers a large set of theories at once, but it also captures the operators in all representations of the global symmetries.
The RG equations in the renormalizable sector of general QFTs with scalars, fermions and gauge bosons are completely known up to three loops~\cite{Mihaila:2013dta,Poole:2019kcm,Steudtner:2020tzo,Steudtner:2021fzs,Bednyakov:2021qxa,Davies:2021mnc,Steudtner:2024teg} (for Lorentz-scalar composite operators). 
For pure scalar theories, 
the most general operators of the form $\phi^a$, $\phi^a \phi^b$, $\phi^a \phi^b \phi^c$ and $\phi^a \phi^b \phi^c \phi^d$
were renormalized up to six loops in \cite{Bednyakov:2021ojn}. 
Recently, the same approach has been applied to higher-dimensional operators at one loop and up to dimension six in EFTs that include both scalar fields and gauge bosons~\cite{Fonseca:2025zjb,Aebischer:2025zxg,Misiak:2025xzq}. In the present work, we derive multiloop results beyond the renormalizable and spin-0 sector in the general scalar theory, paving the way for future work in more general theories.

\paragraph{Summary and main results.}

In this work, we renormalize composite operators in the most general scalar $\phi^4$ field theory in $d=4-2\epsilon$ dimensions,%
    \footnote{The CFT data that we derive will be presented in the more conventional $d=4-\varepsilon$ dimensions, where $\eps=2\ep$. This choice of notation 
    stems from different conventions within EFT and CFT literature.
    }
using minimal subtraction. 
In this way, we extend the results of \cite{Bednyakov:2021ojn} for scalar $\phi^4$ EFTs to the anomalous dimensions of all higher-dimensional operators in up to dimension six and Lorentz rank two, computed at five-loop level (except for a few operators), including operators with insertions of the fundamental field $\phi^a$ as well as insertions of derivatives (e.g.~$\phi^a \phi^b \partial_\mu \phi^c$ and $\phi^a\phi^b\phi^c\phi^d\phi^e$).
This constitutes a computational leap to multiloop calculations involving non-scalar operators (also of higher twist) as well as operators of higher dimension, both with general flavor index structure. 
In our calculation, we rely on  the computational tools that have already been developed for EFT. Our work demonstrates that, with suitable modifications, these can also be used for non-scalar (under Lorentz) and/or general non-singlet (under global symmetry) operators. 
A description of the computational method and the evaluation of the diagrams together with a first study of the results has previously been presented in~\cite{Henriksson:2025hwi}.

\begin{table}
\centering
\caption{The operators renormalized in this paper with generic global symmetry indices. With two derivatives, the Lorentz indices are either traceless-symmetric or antisymmetric.  We extended the four-loop entries to five loops for  hypercubic-symmetric (and hence also $O(n)$ symmetric) $\phi^4$ couplings.}
\label{tab:results-presented0}
{\small
\renewcommand{\arraystretch}{1.25}
\begin{tabular}{|lllll|}
\hline
$\Delta$ & scalars & spin-1 & spin-2 (symmetric) & spin-2 (antisymmetric)
\\\hline
$1$ & $\phi$ {\footnotesize (6-loop \cite{Bednyakov:2021ojn})} & --- & ---
&---
\\
$2$ & $\phi^2$ {\footnotesize (6-loop \cite{Bednyakov:2021ojn})} & --- & ---
&---
\\
$3$ & $\phi^3$ {\footnotesize (6-loop \cite{Bednyakov:2021ojn})} & $\phi^2\partial^\mu$ 5-loop & --- &---
\\
$4$ & $\phi^4$ {\footnotesize (6-loop \cite{Bednyakov:2021ojn})} & $\phi^3\partial^\mu$ 5-loop & $\phi^2\partial^{(\mu}\partial^{\nu)}$ 5-loop
&---
\\
$5$ & $\phi^5$ 5-loop & $\phi^4\partial^\mu$ 5-loop & $\phi^3\partial^{(\mu}\partial^{\nu)}$ 5-loop & $\phi^3\partial^{[\mu}\partial^{\nu]}$ 5-loop
\\
$6$ & $\{\phi^6,\square\phi^4\}$ 4-loop & $\phi^5\partial^\mu$ 4-loop & $\phi^4\partial^{(\mu}\partial^{\nu)}$ 5-loop & $\phi^4\partial^{[\mu}\partial^{\nu]}$ 5-loop
\\
\hline
\end{tabular}
}
\end{table}

A schematic overview of our method is as follows:
\begin{equation*}
    \begin{matrix} \text{operator} \\[-1.5mm]  \text{basis}\end{matrix}
    \ \to \ 
        \begin{matrix} \text{diagram generation} \\[-1.5mm]  \text{and evaluation}\end{matrix}
        \ \to \
    \begin{matrix} 
    \text{general}\\[-1.5mm]\text{results}
\end{matrix}
\ \to \ 
      \begin{matrix} \text{tensor structures,} \\[-1.5mm]  \text{critical coupling}\end{matrix}
\ \to \ 
     \begin{matrix} \text{specific} \\[-1.5mm]  \text{results}\end{matrix}
\end{equation*}
We summarize the determination of operator bases in the general EFT below, together with our treatment of redundant operators, which appear because we perform the renormalization with off-shell external kinematics.
As described in \cite{Henriksson:2025hwi}, we use an implementation of the \Rs method for the diagram evaluation, relying in part on the \form~\cite{Vermaseren:2000nd,Ruijl:2017dtg} 
programs \texttt{Forcer}~\cite{Ruijl:2017cxj,Baikov:2010hf,Lee:2011jt} and \texttt{Opiter}~\cite{Goode:2024cfy}, and the diagram generator of Refs.~\cite{Kaneko:1994fd,Kaneko:2017wzd}.
We have summarized the scope of our calculation in the general EFT in Table~\ref{tab:results-presented0}.
The computed anomalous dimensions are publicly available in the \githubb repository 
\begin{center}
    \href{https://github.com/jasperrn/EFT-RGE}{https://github.com/jasperrn/EFT-RGE}\,,
\end{center}
where we also list our conventions.\footnote{
The results extracted for $O(n)$ symmetry are available at \href{https://github.com/johhen1/ON-model}{https://github.com/johhen1/ON-model} while those for hypercubic symmetry can be obtained from the authors upon request.}

From the general results, the spectra of 
different EFTs and CFTs can be extracted by imposing a global symmetry.
In this paper, we consider theories with
$O(n)$ and hypercubic symmetries,
which have been of interest in the conformal bootstrap literature \cite{El-Showk:2012cjh,El-Showk:2014dwa,Kos:2014bka,Kos:2016ysd,Chang:2024whx,Kos:2013tga,Kos:2015mba,Chester:2019ifh,Chester:2020iyt,Rong:2023owx,Rong:2017cow,Stergiou:2018gjj,Kousvos:2018rhl,Kousvos:2019hgc,Kousvos:2025ext}. 
To illustrate the use and validity of these estimates, we present plots such as those in Figure~\ref{fig:scalar-all} for the scalar spectra in $O(n)$-symmetric theories across spacetime dimensions, making comparison with non-perturbative results in three dimensions from the conformal bootstrap and Monte Carlo simulations. Additional plots including spin-one and spin-two operators are presented in Section~\ref{sec:newData}.
Our perturbative results agree well with the non-perturbative data, even for large $\eps$. 
Beyond the theories considered in this paper, we emphasize that the spectrum of any $\phi^4$ theory can be extracted from our results, leading to a plethora of new operator dimensions for many theories of interest. 
These provide important estimates for future studies, giving a precise indication of the approximate spectrum. 
We explain the procedure for doing such extractions in Section~\ref{s:Specific}. Their exhaustive implementation is left for future work.

\paragraph{Operator bases.}
An important aspect of our methodology is the systematic construction of operator bases in the general theory, suitable for both EFT and CFT, and the identification of redundant operators. The \emph{Green's basis} is spanned by the operators that remain after removing all total derivatives (descendants). 
A convenient choice of Green's basis (for both EFT and CFT purposes) consists of operators proportional to $\partial^2\phi^a$ and operators  
that are annihilated by the generator of special conformal transformations in the free theory. The latter are primary operators of the free theory; see Section~\ref{s:prim} for more details. 
When interactions are included, these operators mix under RG with a mixing matrix of the form
\begin{equation}
\label{eq:GreensBasisIntro}
    \Gamma_{\text{Green's}}=\begin{pmatrix}
        \gamma_{pp}(\lambda) & \gamma_{pr}(\lambda)\\\gamma_{rp}(\lambda) & \gamma_{rr}(\lambda)
    \end{pmatrix}\,,
\end{equation}
where we have labeled the submatrices by $p$ (primary) and $r$ (proportional to $\partial^2\phi^a$). 
The eigenvectors of this matrix correspond to actual primary and redundant operators of the interacting theory. 
At leading order in $\lambda$, the eigenvalues of the primary operators depend only on $\gamma_{pp}$, which can be used to determine the one-loop spectrum~\cite{Kehrein:1992fn,Kehrein:1994ff}. However, at higher orders in $\lambda$, the mixing between primary and redundant operators is important. We will treat this mixing systematically.

The generic mixing matrix~\eqref{eq:GreensBasisIntro} can be simplified by trading the operators proportional to $\partial^2\phi^a$ for operators proportional to the interacting equations of motion (EOM), denoted by $\bar r$. In this case, the mixing matrix becomes triangular,
\begin{equation}
\label{eq:MinimalBasisIntro}
    \Gamma_{\text{minimal+EOM}}=\begin{pmatrix}
        \bar \gamma_{pp}(\lambda) & 0\\\bar\gamma_{\bar rp}(\lambda) & \bar\gamma_{\bar r\bar r}(\lambda)
    \end{pmatrix}\,,
\end{equation}
and the eigenvalues of the primary operators can be directly extracted from $\bar \gamma_{pp}$ at all orders. 
The results in our work are presented only for $\bar \gamma_{pp}$, which is determined from only 
$\gamma_{pp}$ and $\gamma_{ rp}$ in~\eqref{eq:GreensBasisIntro}.
This allows for an efficient calculation, because diagrams with insertions of redundant operators need not be computed.
We discuss this in more detail in Section \ref{s:ADcalc} and we will work out in detail a pedagogical example in Section \ref{s:Redex}.

\FloatBarrier

\paragraph{Structure of this paper.}
The paper is organised as follows. In Section~\ref{sec:EFTmeetsCFT}, we provide a brief introduction to EFT basics for CFT experts, and vice versa. Then, in Section~\ref{sec:method} we review important aspects of our computation, with a special emphasis on the construction of a consistent operator basis in the general theory. Throughout Section~\ref{s:examples} we work out examples of operator mixing relevant to this work, showing how to treat redundant operators systematically.
Section~\ref{s:SpecificEFT} contains applications of our general results to EFT, deriving new results for the scalar sector of the SMEFT. Moving to CFT in Section~\ref{s:Specific}, we detail how to extract data from our general results for several Wilson--Fisher theories relevant in the condensed matter and high-energy theory literature. Section~\ref{sec:newData} collects the results from our exhaustive extraction of the spectra for Ising, $O(n)$ and hypercubic CFTs. Finally, we conclude and present future directions in Section~\ref{sec:outlook}. Our work is also supplemented by four appendices. 
In Appendix~\ref{app:Tables} we tabulate results for the $O(n)$ and hypercubic CFTs at small values of $n$. 
In Appendix~\ref{app:scalarPotentials} we provide further examples of operator mixing relevant to the literature. Appendix~\ref{app:example1App} constitutes a calculation in an alternative operator basis, with perhaps a counter-intuitive way to compute scaling dimensions, which is nevertheless still equally correct. Finally, in Appendix~\ref{s:Operators} we provide information on all operators studied and their symmetry properties in order to facilitate future data extraction by the reader.

\section{EFT meets CFT}
\label{sec:EFTmeetsCFT}

A main motivation of our work is to harness the dramatic progress within EFT to make predictions for CFT. To bridge across the subjects, we will now give some fundamentals on the two topics. The omniscient reader can skip this section and proceed to Section~\ref{sec:method}.%

\subsection{EFT introduction for CFT experts}

EFTs are widespread across many branches of physics. 
They are useful when there is a separation of scales (e.g.~in energy or length)
between regimes in which different degrees of freedom are important. 
In this section, we review the terminology and technology developed in the context of (relativistic) effective field theories of particle physics, which has the SMEFT, Low-energy EFT (LEFT) 
and chiral perturbation theory as important examples.\footnote{Since we will work with scalar EFTs in the rest of this work, we will omit commonly discussed aspects related to spin-$\tfrac{1}{2}$ and spin-$1$ particles, such as Fierz identities and evanescent operators; see e.g.~\cite{Brivio:2017vri,Manohar:2018aog,Isidori:2023pyp,Falkowski:2023hsg} for more complete introductions. Evanescent operators in pure-scalar theories do however also exist, starting at high mass dimensions~\cite{Hogervorst:2015akt}.}

For the purpose of presentation, consider the case where we have a single UV scale $M_\text{UV}$, and an effective field theory defined below that scale. We assume that the UV theory is renormalizable with a light and a heavy field,
\begin{equation}
    \mathcal L_\text{UV}=\mathcal L^{\text{ren.}}_\text{UV}[\phi,\Phi]\,,
\end{equation}
where the mass of $\Phi$ is $M_\text{UV}$, and $\phi$ has mass $m\ll M_\text{UV}$. In this theory, observables such as S-matrix elements are computed with both $\phi$ and $\Phi$ as propagating degrees of freedom, with renormalizable interaction vertices.

Observables at low energies $E\ll M_\text{UV}$ can instead be described by an EFT in which the relevant degree of freedom is only $\phi$, while the imprint of the UV dynamics is given by higher-dimensional point interactions,
\begin{equation}
\label{eq:L-EFT}
    \mathcal L_\text{EFT}=\mathcal L_\text{EFT}^{\text{ren.}}+\sum_{\Delta>4,\,i} \frac{c^{(\Delta)}_i}{
    M_\text{UV}^{\Delta-4}}
    \mathcal{O}^{(\Delta)}_i\,,
\end{equation}
where $\mathcal{L}^\text{ren}_\text{EFT}$ is the renormalizable part of the theory. 
The operators $\mathcal{O}^{(\Delta)}_i$ are constructed out of $\phi$ and derivatives, and have engineering dimension $\Delta$, which is also called mass dimension.
The label $i$ runs over the set of independent operators at each mass dimension.
In this way, the UV degrees of freedom have been traded for an infinite set of higher-dimensional operators, parametrized by $c_i^{(\Delta)}$.
When the operators are normalized by appropriate powers of $4\pi$, the EFT parameters are expected to be of order one~\cite{Manohar:1983md,Gavela:2016bzc}. 
At low energies, the separation of scales $E\ll M_\text{UV}$ then allows to truncate the EFT expansion at a finite order. 
Computations in the EFT are thus performed by keeping a finite number of $\mathcal{O}^{(\Delta)}_i$, typically up to a given dimension.

\begin{figure}
    \centering
\includegraphics[width=0.8\textwidth]{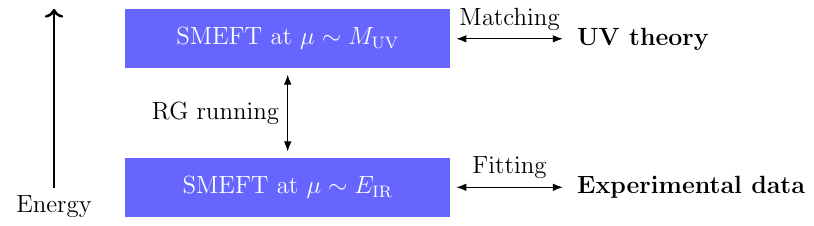} 
    \caption{
    The parameters of an EFT can be determined by matching to a UV theory and by fitting to experimental data. In the regime of validity of the EFT, the anomalous dimensions determine the running of the couplings with renormalization scale $\mu$. 
    }
    \label{fig:EFTsimple}
\end{figure}

EFTs facilitate the comparison between a large class of UV theories and experiments, as illustrated in Figure~\ref{fig:EFTsimple}.
The values of the EFT coefficients $c^{(\Delta)}_i$ can be determined either
by \emph{matching} with a specific UV theory, in which case the values are obtained by “integrating out” the heavy degrees of freedom; or by \emph{fitting} the coefficients to experimental data while remaining agnostic about the UV theory.
In the latter case, the theory becomes predictive since only a finite number of experiments are needed to fit the coefficients at low orders. 
Regardless of the approach, the coefficients $c_i^{(\Delta)}$ depend on the renormalization scale $\mu$ (in dimensional regularization). Due to the assumed scale separation between the UV matching scale and experiments at low energies, the computation of the $\mu$ dependence of the EFT parameters is crucial in EFT studies. 
Schematically, EFT allows to separate large logarithms $\log(E_\text{UV}/E_\text{IR})$, which may result in a breakdown of perturbation theory, into two parts: $\log(E_\text{UV}/\mu_1)-\log(\mu_2/E_\text{IR})$. 
The logarithm of each contribution can be made small by appropriate choices of the renormalization scale, provided the renormalization group equations are applied to run between $\mu_1$ and $\mu_2$.

The RG running is determined by the beta function of the couplings 
\begin{align}
\label{eq:definitionCouplingBeta}
    \frac{d \, c_i}{d\log \mu} \equiv \beta_{c_i} = \gamma_{ij}\,c_j + \gamma_{ijk} \, c_j \, c_k + ...\,,
\end{align} 
where $\gamma_{ij}$ and $\gamma_{ijk}$ depend on the couplings of the renormalizable Lagrangian and we have omitted terms at higher orders in the EFT parameters.
In this paper, we will work in the massless limit and at leading order in the EFT parameters, restricting to the anomalous dimension matrix $\gamma_{ij}$.%
    \footnote{\label{footnoteTensors1}
    The anomalous dimension tensors, such as $\gamma_{ijk}$, encode the mixing between operators with different mass dimensions. 
    For instance, the running of dimension-eight parameters in the SMEFT receives important contributions from double insertions of dimension-six operators~\cite{Chala:2021pll,Bakshi:2024wzz}. See also~\cite{Cao:2023adc} for multiloop results for $\gamma_{ijk}$ in scalar EFTs.
    }

Besides the determination of the RG running and the matching conditions, an important aspect in the systematic study of EFTs is the enumeration of independent operators using Hilbert series~\cite{Henning:2015alf,Henning:2017fpj},
and the formulation of standard operator bases, such as the conventional Warsaw basis for the SMEFT at dimension six \cite{Grzadkowski:2010es}. 
In the rest of this section, we will
discuss different types of operators and how they contribute to observables. 

The independent operators of an EFT are those that contribute independently to the S-matrix.
EFT operators are thus in one-to-one correspondence to the contact-interaction contributions to tree-level amplitudes. 
As an example, let us consider the tree-level S-matrix of
an EFT of massless scalar particles, before discussing operators at the Lagrangian level. Assuming the absence of three-point interactions for simplicity, any $2\to2$ amplitude at low energies can be written as
\begin{equation}\label{eq:EFTSmatrix}
    \mathcal{A}(\phi_1\phi_2\to\phi_3\phi_4) = c_0 
    + c^{(6)}_1\frac{s}{M^2} + c^{(6)}_2\frac{t}{M^2} 
    + c_1^{(8)} \frac{s^2}{M^4}
    + c_2^{(8)} \frac{s\,t}{M^4} 
    + c_3^{(8)} \frac{t^2}{M^4}
        + O(M^{-6})\,.
\end{equation}
Here we used that the external particles are on shell, $p_i^2=0$, and we defined the Mandelstam variables $s=(p_1+p_2)^2$ and $t=(p_1-p_3)^2$. We applied momentum conservation and the on-shell condition to write the third Mandelstam variable in terms of these two: ${u=(p_1-p_4)^2=-s-t}$. Finally, the mass scale $M$ is included using dimensional analysis.

Symmetries of the low-energy theory may impose additional relations between the parameters. For instance in a single-scalar theory, Bose symmetry implies $c_1^{(6)}=c_2^{(6)}=0$ and 
$c_1^{(8)} = c_2^{(8)} = c_3^{(8)}$.
Furthermore, dispersive arguments with minimal assumptions on the physics at high energies (e.g.~unitarity and analyticity) place bounds on the parameters of the EFT, such as $c^{(8)}_1>0$, called positivity bounds~\cite{Adams:2006sv}, and bounds on ratios of such parameters \cite{Caron-Huot:2020cmc,Bellazzini:2020cot,Tolley:2020gtv,Arkani-Hamed:2020blm}. 
Ultimately, the $c_i^{(\Delta)}$ should be fixed by experiments or matched to a known UV theory. We will now exemplify the latter.

When the UV theory of interest is known, 
the parameters in \eqref{eq:EFTSmatrix} can be determined through matching.%
    \footnote{There exist a variety of methods and tools to compute the matching conditions, e.g.~\cite{Fuentes-Martin:2022jrf,DeAngelis:2023bmd}. 
    Moreover, there exist dictionaries for the matching of any beyond the SM theory onto the SMEFT~\cite{deBlas:2017xtg,Guedes:2024vuf}, as well as the matching of the SMEFT onto the LEFT~\cite{Dekens:2019ept}.}  
As a simple example, consider a UV theory with a global $O(n)$ symmetry given by
\begin{equation}
    \mathcal{L}_\text{UV} = \frac{1}{2}\partial_\mu\phi_a\partial^\mu \phi_a + 
    \frac{1}{2}\partial_\mu \Phi \partial^\mu\Phi - \frac12 M^2 \Phi^2
    -\frac{\lambda}{8}(\phi_a\phi_a)^2 - \frac{\kappa\, M}{2}\, \phi_a\phi_a\Phi\,,
\end{equation}
where $\phi$ is a massless
particle in the fundamental representation and 
$\Phi$ is an $O(n)$ singlet with a large mass $M$.%
\footnote{Restricting to tree level for simplicity, we ignore loop corrections to the mass of $\phi$, and we suppress other interactions that are allowed by the symmetry. 
} 
We have also normalized the three-point coupling by $M$, such that $\kappa$ is dimensionless.
The four-point amplitude of light scalars is then
\begin{align}
    \mathcal{A}_\text{UV}(\phi_1(p_1)\phi_1(p_2)\to\phi_2(p_3)\phi_2(p_4)) 
    &= -\lambda - \frac{\kappa^2M^2}{s-M^2}
     \nonumber\\&
    \approx -\lambda + \kappa^2 + \kappa^2\frac{s}{M^2} + \kappa^2 \frac{s^2}{M^4} + O(s^3/M^6)\,,
\end{align}
where we fixed the choice of particle flavors for simplicity, and we expanded the second line at low energy ($s\ll M^2$). 
Matching with the general EFT expansion in~\eqref{eq:EFTSmatrix}, we thus find (for the chosen flavor structure)
$c_0 = -\lambda + \kappa^2$, 
$c_1^{(6)} = \kappa^2$, 
$c_1^{(8)} = \kappa^2$,
and $c_2^{(6)} = c_2^{(8)} = c_3^{(8)} = 0$, 
where $\kappa^2 \ll 16\pi^2$ for the tree-level approximation to be accurate.

Now let us discuss the possible operators that we can add to \eqref{eq:L-EFT}, and how they relate to the parameters in the S-matrix.
Firstly, we only consider operators $\mathcal{O}_i^{(\Delta)}$ that are Lorentz singlet operators and invariant (singlet) under the gauge and global symmetries of the theory.%
\footnote{In theories with symmetry group $H$, it may happen that the theory is invariant under a larger symmetry group $G$ in some approximation (i.e.~setting some couplings to zero), with $H\subset G$.
It is then useful to organize the operators, which are invariant under $H$, in terms of various representations of $G$. We exemplify this in the context of RG in Section~\ref{s:SpecificEFT}.}
Next, not all operators that are allowed by the symmetries contribute independently to physical observables. For instance, in a translation-invariant theory, total derivatives vanish in the action,
\begin{align}
    \int d^4x \,\partial_\mu (\bullet) = 0\,.
\end{align}
Such operators have vanishing contributions to the S-matrix due to momentum conservation. If instead of S-matrix elements, one computes form factors, which involve operator insertions that are not integrated over spacetime, it is necessary to keep total derivatives in the basis of operators. 

Besides momentum conservation, another constraint on the kinematics of the S-matrix is the on-shell condition, $p_i^2=m_i^2$. 
This condition is enforced by the LSZ reduction formula~\cite{Lehmann:1954rq,Lehmann:1957zz}. As a consequence of the LSZ reduction,
S-matrix elements are invariant under redefinitions of the fields in the Lagrangian~\cite{Chisholm:1961tha,Kamefuchi:1961sb,Politzer:1980me,Arzt:1993gz}; see also the more recent works~\cite{Passarino:2016saj,Criado:2018sdb,Cohen:2024fak,Criado:2024mpx}.
To illustrate, consider field redefinitions of the form
\begin{equation}\label{eq:FieldRedef}
    \phi\to\phi'[\phi] = \phi + F(\phi)\,,
\end{equation}
where $F$ is a polynomial function of the fields and derivatives; see \cite{Cohen:2024fak} for the discussion of more general field redefinitions. 
In this case, both $\phi$ and $\phi'$ are interpolating 
fields for the same physical state. This means that both $\phi$ and $\phi'$ can produce the same single-particle state by acting on the vacuum, and
\begin{align}
    \int d^4x \, e^{ipx} \langle\phi(x)\phi(0)\rangle = \frac{i\,Z}{p^2 - m^2 + i0} + ...\,,\hspace{1mm}
    &&
    \int d^4x \, e^{ipx} \langle\phi'(x)\phi'(0)\rangle = \frac{i\,Z'}{p^2 - m^2 + i0} + ...\,,
\end{align}
where the omitted terms are finite at the physical mass, $p^2=m^2$. The LSZ reduction formula then relates 
the $n$-particle pole of the Green's function of $n$ fields $\phi$ (or $\phi'$) to the corresponding S-matrix element, up to wavefunction normalization factors $Z$ (or $Z'$).

The field redefinition~\eqref{eq:FieldRedef} leads to a change in the action of the schematic form
\begin{align}\label{eq:FieldRedefEffect}
    I[\phi] = \int d^4x \, \mathcal{L}_\text{EFT}[\phi] 
    \to I[\phi'] = I[\phi] &+ 
    \int d^4 x\,\frac{\delta I[\phi]}{\delta \phi} F[\phi] 
    + \int d^4x\,d^4y\, \frac{\delta^2 I[\phi]}{\delta \phi^2} F[\phi] F[\phi]\nn\\
    &\hspace{1cm}+ O(F^3)\,.
\end{align}
Here, the first correction term is proportional to the classical equations of motion,
\begin{equation}
    (\text{EOM})^a = \frac{\delta I[\phi]}{\delta \phi^a} 
    = -\partial^2 \phi^a - m^2 \phi^a + O(\lambda, c_i^{(\Delta)})\,.
\end{equation}
For the purpose of constructing an operator basis, this means that operators proportional to $\partial^2\phi^a$ can be replaced by other operators.
Such operators are considered to be redundant in the free theory, which is the relevant starting point for determining an operator basis. On the other hand, in the interacting theory, operators proportional to the full dimension-four EOM are redundant. This is valid at leading order in the EFT couplings, which is what we restrict to in the rest of this work.%
    \footnote{We emphasize that it is generally important to keep higher-order terms in the EOM, as well as non-linear contributions from $\delta^2 I/\delta\phi^2$, when relating the action of the same theory in two different field bases; see e.g.~\cite{Scherer:1994wi} or~\cite{Criado:2018sdb} for a recent pedagogical discussion. 
    } 
We summarize the different types of operators in Table~\ref{tab:Operators}.

Upon selecting an operator basis, the EFT has to be renormalized. EFTs are formally non-renormalizable, because infinitely many operators can be generated through the renormalization group equations. In practice, however, the EFT is truncated at fixed mass dimension and thus becomes predictive. In this work, we will renormalize scalar effective field theories using dimensional regularization in $4-2\epsilon$ dimensions. We will compute the corresponding $\beta$ functions \eqref{eq:definitionCouplingBeta}, to leading order in $c_i^{(\Delta)}$ (thus not determining e.g.~$\gamma_{ijk}$), and to a given loop order in the renormalizable couplings. 
We will only consider contributions from the marginal $\phi^4$ interaction
and not the super-renormalizable couplings like 
those of $\phi^3$ interactions. This is motivated by our CFT applications, in which point only the $\phi^4$ interaction survives at the fixed-point.

\begin{table}
\centering
\caption{Types of operators in EFT and CFT. 
We consider special cases in which this classification is more subtle in Section~\ref{s:Bases}.}
\label{tab:Operators}
{\small
\renewcommand{\arraystretch}{1.75}
\begin{tabular}{|l|p{3.1cm}|p{2.3cm}|p{4.5cm}|}
\hline
&Redundant & Descendants & Primary operators
\\\hline
Form in the free theory 
    & $(\bullet)\partial^2\phi$ 
    & $\partial_\mu (\bullet)$
    & $[K_\mu,\mathcal{O}(0)]=0$ \\\hline
    Relevant object in EFT 
    & Off-shell correlation functions 
    & Form factors
    & Off-shell corr.\ functions, form factors, S-matrix elements
    \\\hline
Role in CFT
    & Not in spectrum
    & Determined from primaries
    & Fundamental building blocks of spectrum and observables\\\hline
\end{tabular}
}
\end{table}

Instead of renormalizing the S-matrix directly (e.g.~as in~\cite{Caron-Huot:2016cwu}), we choose to renormalize off-shell correlation functions, also called Green's functions.
This reduces the number of diagrams, because it allows for the restriction to one-particle-irreducible (1PI) diagrams. Furthermore, the off-shell external momenta serve as an infrared regulator. However, working off shell does require the use of a larger operator basis, since correlation functions depend on the choice of fields. That is, they are not invariant under field redefinitions.
The set of operators necessary to absorb all counterterms is thus spanned by the minimal basis for the S-matrix supplemented also by all redundant operators. 
We call such an operator basis a Green's basis. 
Nevertheless, it is possible to leverage the fact that redundant operators do not mix into non-redundant operators under RG; see \eqref{eq:MinimalBasisIntro}, allowing for a more economical computation and presentation of the results. We will discuss this in more detail in Section~\ref{sec:method} and through examples in Section~\ref{s:examples}.

Before closing this section, let us briefly remark on some of the factors that have driven recent progress in EFT, focussing on the SMEFT.
The SMEFT provides an ideal framework to parametrize new physics in a model-independent way. However, its generality, which translates to the large number of operators even at low mass dimensions, comes with computational challenges. 
These have spurred a lot of recent work on the SMEFT in particular and EFTs in general. 
For instance, it is crucial to understand the interplay between assumptions on the type of physics beyond the SM and the importance of particular (classes of) operators. 
Moreover, it is necessary to assess the scenarios under which operators of higher mass dimensions need to be included and when calculations need to be extended to higher loops. 
An important indicator on the relevance of such higher-order corrections is the quantitative impact of the renormalization group equations on physical observables.
For phenomenology at the LHC and future colliders, a common first target comprises one-loop matching at dimension six,
because there exist observables that receive their leading contributions at loop level. Since the one-loop matching conditions depend on the regularization scheme (in particular the prescription of dealing with $\gamma_5$ in $d$ dimensions), they should be accompanied by two-loop renormalization for scheme-independent results~\cite{Buras:1989xd,Ciuchini:1993ks,Ciuchini:1993fk}.

While the scalar EFT considered in this work is not of immediate phenomenological relevance, our work contributes to the recent developments by increasing the understanding of general EFTs, with future extensions in mind towards theories with gauge symmetries and fermionic matter. 
In addition, as we will turn to now, anomalous dimensions of EFT operators ($\gamma_{ij}$) are directly related to important data in CFT, namely the scaling dimensions of primary operators.

\subsection{CFT introduction for EFT experts}

A CFT is a quantum field theory with conformal invariance; 
see e.g.~\cite{Simmons-Duffin:2016gjk,Rychkov:2016iqz} for recent introductions.
They commonly arise at fixed-points of the RG flow, however non-conformally invariant fixed-points also exist \cite{Polchinski:1987dy,Nakayama:2013is}. %
A remarkable feature of the RG flow is that CFTs can also appear as IR fixed-points in systems whose microscopic description is not a QFT. Examples are famous statistical physics models such as the Ising model, and real-world systems undergoing continuous phase transitions. The observables, e.g.~critical exponents, at these phase transitions are related to (anomalous) dimensions of operators in the corresponding CFT. In the same basin of attraction of the RG flow as these systems, one can often formulate a UV quantum field theory, providing a way to study critical phenomena using quantum field theory methods. 

An important aspect in the study of CFT is that it can be formulated axiomatically, without reference to a Lagrangian. This means that CFTs can be studied using both conventional QFT methods such as Feynman diagrams, and with axiomatic methods. 
In either case, the notion of operators is central. Operators are endowed with quantum numbers: a scaling dimension $\Delta$ and a spin $\ell$ (more generally a Lorentz representation). These determine the transformation properties of operators under the conformal group. Importantly, under a rescaling they transform as 
$\O(\alpha \, x)=\alpha^{-\Delta}\O(x)$, fixing already the form of the two-point function. 
Additionally, there are three-point constants $\lambda_{ijk}$ (OPE coefficients) that are proportional to the three-point functions. All higher point functions can be expressed in terms of the aforementioned data by making use of the operator product expansion.
The data that describes a CFT is thus 
\begin{equation}
\text{Data of a CFT }= \big\{{(\Delta_i,\ell_i)}, \ \lambda_{ijk}\big\},
\end{equation}
where $(\Delta_i,\ell_i)$ are quantum numbers of operators $\O_i$. 

If the CFT is described by a (weakly-coupled) Lagrangian, this data can, in principle, be computed (perturbatively). It can also be seen as an abstract collection of numbers. In many interacting CFTs, the precise values of these numbers are unknown. If one can match the abstract CFT to a perturbative description, for instance via a weak-coupling limit, the scaling dimensions can be written as a classical plus an anomalous part:
\begin{equation}
    \Delta_i^{\mathrm{CFT}}=[\O_i]+\gamma_i\,,
\end{equation}
where $[\O_i]$ denotes the classical operator dimension (in $d$ dimensions), %
and $\gamma_i$ can be obtained from the eigenvalues of the anomalous dimension matrix $\gamma_{ij}$ of the EFT parameters 
introduced in~\eqref{eq:definitionCouplingBeta}. 
The precise relation involves a convention-dependent constant shift; 
we define this within our conventions in Section~\ref{s:basics}.
Typically, the anomalous dimensions of interest are $O(1)$ numbers in the CFT.\footnote{Broadly speaking, three-dimensional CFTs have couplings of $O(1)$, but often admit a weak-coupling expansion, such as the $\eps$-expansion in $d=4-\eps$ dimensions. These are the theories we consider here. Four-dimensional CFTs of interest, for instance supersymmetric gauge theories, often have a marginal coupling and can be considered across the range of couplings including a weakly-coupled point.}

The canonical example is the Ising CFT, which is the IR fixed-point of the theory of a single massless scalar field perturbed a $\lambda\phi^4$ interaction.  
It describes phase transitions in fluids and in uniaxial magnets including the Ising model. 
Within Lagrangian field theory one can write
\begin{equation}
\label{eq:IsingPlusHigher}
    \mathcal L=\frac12(\partial\phi)^2-\frac12m^2\phi^2-\frac\lambda{24}\phi^4+\sum_i c_i \O_i\,.
\end{equation}
Here we distinguish different types of couplings in $d=4-\eps$\,: a) \emph{Relevant} couplings, like the renormalized mass $m^2$, must be tuned to zero to reach a controlled fixed-point. This renders the CFT gapless. b) \emph{Marginally relevant} couplings (here $\lambda$) which will flow under the RG flow and may attain non-zero values at the fixed-point.\footnote{Under this flow, the operator $\phi^4$
goes from being marginally relevant at the start of the flow to irrelevant at the fixed-point. 
For other flows, potentially also with more than one marginally relevant operator, whether an operator becomes irrelevant at the fixed-point depends on the specific fixed-point and operator.} c) \emph{Irrelevant} couplings like the $c_i$ flow to zero in the IR, hence the notion irrelevant. The irrelevant couplings are normally not written out, however in this work we display them to make a connection with EFT and the computational procedure. The couplings $c_i$ in \eqref{eq:IsingPlusHigher} should be seen as probe couplings, since they are not needed to reach the fixed-point \emph{per se}, but their beta functions encode information about the anomalous dimensions of operators they multiply.

\begin{figure}
    \centering
   \includegraphics{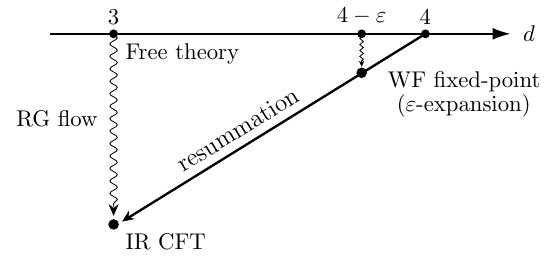}\qquad
    \caption{Long RG flow in three dimensions versus short RG flow in $4-\eps$ dimensions and resummation to reach the 3d IR CFT.
    }
    \label{fig:eps-expansion}
\end{figure}

Here we consider a class of CFTs that can be studied using perturbation theory, including \eqref{eq:IsingPlusHigher} and its generalizations to $n$ fields and generic tensor couplings $\lambda_{abcd}$. The perturbative computation works in $d=4-\eps$ dimensions, where crucially $\eps$ will be kept finite in order to give results for the theory in $d<4$. This leads to the famous $\eps$-expansion originated by Wilson and Fisher \cite{Wilson:1971dc,Wilson:1973jj} to approach three-dimensional CFTs; see Figure~\ref{fig:eps-expansion}. The renormalized mass $m^2$ and all irrelevant couplings $c_i$ are set to zero at the fixed-points we will consider. However, we will still insert them (at linear order) as sources for the corresponding operators needed in the computation of anomalous dimensions later. All orders of $\lambda_{abcd}$ are kept.

The leading-order $\beta$ function in $d=4-\eps$ is\footnote{Here, as in the rest of the paper, we absorb factors of $16\pi^2$ in the definition of the couplings.}
\begin{equation}
    \beta_{abcd}(\lambda)
    =-\eps\,
    \lambda_{abcd}
    +(\lambda_{abef}\lambda_{efcd}
    +{\text{2 perms.}})
    +O(\lambda^3).
\end{equation}
The zero(s) of the beta function can be solved for order by order in $\eps$, leading to a set of fixed-points, called Wilson--Fisher fixed-points. For instance, in the case of a single scalar case
\begin{equation}
\label{eq:critCouplingIsing}
    \beta(\lambda_*)=0\quad \Rightarrow \quad \lambda_*=\frac\eps3+\frac{17}{81}\eps^2+\left(\frac{709}{17496}-\frac{4 \zeta _3}{27}\right)\eps^3+\ldots
\end{equation}
Then, we can write the physical scaling dimension of an operator $\O_i$ as
\begin{equation}
    \Delta_i = [\O_i]+\gamma_i(\lambda_*)\,,
\end{equation}
where the anomalous dimension ($\gamma_i$ eigenvalue of the matrix $\gamma_{ij}$) is re-expanded in $\eps$:
\begin{equation}
\label{eq:gammaieps}
    \gamma_i(\eps)=\gamma_i(\lambda_*(\eps))=\gamma_{\O}^{(1)}\eps+\gamma_{\O}^{(2)}\eps^2+\ldots.
\end{equation}
The anomalous dimensions at the fixed-point are observables quantities, and therefore scheme independent.  

The computation of a set of anomalous dimensions gives the (low-lying) spectrum of the CFT as a series expansion in $\eps$. For comparing with experiments and Monte Carlo simulations, we would like to consider the spectrum at $\eps=1$. At this value, the theory is beyond the weakly-coupled limit and the anomalous dimensions become $O(1)$ numbers. The series expansions~\eqref{eq:gammaieps} are in general asymptotic and direct truncations give bad results. Instead typically one employs resummations, for instance Pad\'e approximants. With a set of Pad\'e approximants computed from multiloop anomalous dimensions, one arrives at figures such as the one in Figure~\ref{fig:scalar-all}.

The CFTs approached by the $\eps$-expansion can also be studied with methods not referring to a Lagrangian. In such approaches, the operators are often given alternative names not referring to their UV field content, e.g.~$\sigma$ for $\phi$, $\epsilon$ for $\phi^2$, $\epsilon'$ for $\phi^4$ etc.\footnote{This notion is commonplace for 2d CFTs but employed also for 3d CFTs in the context of the conformal bootstrap. Operators with primes transform in the same global-symmetry and Lorentz representation as the corresponding un-primed operator, but subleading scaling dimension.} Results from the bootstrap are quoted in this fashion, \emph{e.g.} $\Delta_\sigma=0.518148806(24)$, $\Delta_\epsilon = 1.41262528(29)$  for the 3d Ising CFT \cite{Chang:2024whx}. Note that these do not appear to take any simple rational expressions. 

Historically, there has been a focus on relevant operators or marginally irrelevant operators. The dimensions of these operators are related to (leading) critical exponents, which are traditional observables associated to the scaling of certain quantities near the critical point. For instance, $\eta=2\Delta_\phi+2-d$ determines the two-point function of the order parameter, $\nu=(d-\Delta_{\phi^2})^{-1}$ the scaling of the correlation length, and $\omega=\Delta_{\phi^4}-d$ the leading correction to scaling. 
The operators of interest also include those that are not invariant under global and/or spacetime symmetries. They describe, for instance, crossover exponents in the case of global symmetries, and exponents measuring rotational symmetry breaking on lattices in the case of spacetime symmetries. 

The modern, axiomatic, perspective on CFT emphasizes the whole spectrum of operators, including RG-irrelevant ones.
Before discussing this field of research in more detail below, 
let us consider the different type of operators.
Local operators in a non-perturbative formulation of CFT are classified as primary operators, which are annihilated by special conformal transformations (see section~\ref{s:prim} below) when inserted at the origin; and descendant operators, which are produced by acting on primaries with partial derivatives. 
Conformal symmetry completely dictates the form of two- and three-point functions of primary operators,\footnote{Analogous expressions for spinning operators also exist. Note that three-point functions involving more than one spinning operator contain more than one independent OPE coefficient; see e.g.~\cite{Kravchuk:2016qvl}.} 
\begin{equation}
    \langle \O_i(x_1)\O_j(x_2)\rangle = \frac{\delta_{ij}}{x_{12}^{2\Delta_{i}}}, \quad \langle \O_i(x_1)\O_j(x_2)\O_k(x_3)\rangle =\frac{\lambda_{ijk}}{x_{12}^{\Delta_i+\Delta_j-\Delta_k}x_{13}^{\Delta_i-\Delta_j+\Delta_k}x_{23}^{-\Delta_i+\Delta_j+\Delta_k}}\,,
\end{equation}
where $x_{ij}^2=(x_i-x_j)^2$.
Descendant correlation functions are given by acting with derivatives on these expressions. Both primary and descendant operator insertions give rise to observable quantities. 

In addition to primaries and descendants, the perturbative renormalization group, which we will use to describe CFTs in this work, also contains redundant operators, which are proportional to the equation of motion itself, or are composites of it. An interesting example is the divergence of the global symmetry current in the presence of a continuous global symmetry $\partial_\mu J^{\mu} = \EOM^\prime =0$, where $\EOM^\prime$ is a composite $\EOM$ operator; see e.g.~\cite{Kousvos:2025ext}. 
Redundant operators have correlators that are non-zero only at coincident points (contact terms) and they are not considered part of the CFT spectrum.\footnote{For example, they are not part of the operator--state correspondence. Moreover, in the conformal bootstrap program, one only ever considers correlators at non-coincident points. That being said, operators proportional to the $\EOM$ cannot be forgotten entirely, since they lead to Ward identities. One example is Ward identities involving the stress-tensor which fix ratios of OPE coefficients, a condition that is usually imposed in the bootstrap.} 
All redundant operators have scaling dimensions at the fixed-point that are related to other scaling dimensions in the theory (the simplest example is $\Delta_\EOM = d-\Delta_\phi$). 

An interesting observable in CFT is the four-point correlation function, which depends non-trivially on kinematic variables and can be determined in terms of an operator product expansion in which an infinite number of operators appear:
\begin{equation}
\label{eq:CBexp}
    \langle \O_i(x_1)\O_j(x_2)\O_k(x_3)\O_l(x_4)\rangle =\sum_\O \lambda_{ij\O}\lambda_{kl\O} G_{\Delta_i-\Delta_j,\Delta_k-\Delta_l;\Delta,\ell}[x_i]
\end{equation}
where $(\Delta,\ell)$ are the dimension and spin of primary operators $\O$. Here $G$ are kinematic functions called conformal blocks which sum up all the descendant contributions to the respective primaries in the sum. Eq.~\eqref{eq:CBexp} contains an infinite number of operators and an infinite number of constraints. It is thus sensitive to the spectrum of the CFT -- in particular the anomalous dimensions. This is the starting point for the numerical bootstrap method.

\paragraph{The space of CFTs and the bootstrap.}
\label{sec:futureBootstrap}

An important notion in the study of CFT is the idea of ``theory space,'' see e.g.~\cite{Poland:2022zhe}, the putative space of all mathematically consistent theories for physics. In this space, a QFT can be thought of as an RG flow between two fixed-points (which will often be CFTs). The CFTs themselves satisfy the axioms of a conformal field theory. Following this idea, it is therefore crucial to classify and characterize the set of possible CFTs. This programme bridges between classical critical phenomena in the form of 3d CFTs \cite{Pelissetto:2000ek}, exactly solvable models in two dimensions \cite{DiFrancesco:1997nk}, and supersymmetric CFTs via geometric/string constructions in higher dimensions \cite{Gaiotto:2009we,Heckman:2018jxk}. 
It also connects to the conformal bootstrap \cite{Rattazzi:2008pe}, reviewed in~\cite{Poland:2018epd,Rychkov:2023wsd}, which is a non-perturbative numerical method to study CFTs based on self-consistency. 

The bootstrap aims to isolate CFTs in increasingly small isolated islands in the parameter space spanned by the first few operator dimensions. The meaning of such islands is that the points surrounding them have been rigorously excluded as incompatible with conformal symmetry (including crossing and unitarity) and additional assumptions used. This has been achieved for the $O(n)$ symmetric scalar theories where the state of the art includes \cite{Chester:2019ifh,Chester:2020iyt}, and the Ising CFT at $n=1$ \cite{Chang:2024whx}, and some scalar-fermion theories in three dimensions \cite{Atanasov:2022bpi,Erramilli:2022kgp}.
Other $\phi^4$ theories have not been isolated in such small islands (to be discussed in detail below).

Let us focus on 3d scalar theories. A coarse classification, following Landau--Ginzburg--Wilson, is to organize CFTs by global symmetry. For several choices of symmetry groups ($\Z_2$, $O(n)$, hypercubic symmetry, \ldots), there is a scalar CFT in $d=3$ dimensions with that symmetry and which inherits the name: ``$O(n)$ CFT'', ``hypercubic CFT'', etc of the symmetry group. However, in general there could be many different CFTs with the same symmetry group. For instance, for the symmetry group $O(m)\times O(n)/Z_2$ there are two non-trivial coupled unitary scalar CFTs that can be found in the $\varepsilon$-expansion for appropriate values of $m,n$ \cite{Kawamura1998}. For a broad study aimed at finding such CFTs; see \cite{Osborn:2017ucf,Osborn:2020cnf} and references therein. 

An important motivation for this work is that perturbative estimates play an important role in the bootstrap, even though the the method ultimately derives rigorous \emph{non-perturbative} results. The reason is that the bootstrap algorithm needs to be fed in a set of assumptions, typically spectral gaps in various representations of Lorentz and global symmetry, which aim to exclude other potentially nearby CFTs in theory space, and eventually isolate the theories of interest in islands. Perturbative estimates is therefore used to provide a well-founded set of such assumptions. 
In particular, new perturbative estimates are especially important for bootstrapping theories beyond $O(n)$ symmetry,\footnote{Even the aforementioned scalar-fermion theories isolated to small bootstrap islands have $O(n)$ global symmetry.} see~\cite{Kousvos:2025ext}. 
For scalar field theories the group $O(n)$ is the maximal global symmetry a theory can have. Consequently, the spectrum of the $O(n)$-symmetric theory is particularly sparse compared to all other Wilson--Fisher CFTs, making it much easier to study. To understand this, we note that apart from gap assumptions, the bootstrap makes use of crossing equations which implement the global symmetry. When considering groups $G$ of a smaller (subgroup) symmetry, the crossing equations of that system will also satisfied by $O(n)$ global symmetry. Said otherwise, the parameter space of bootstrap solutions for a group $G$, when $G$ is not maximally symmetric, is rather dense, and to ``zoom in'' on a particular theory of interest is made particularly difficult. For example, one may end up with islands in parameter space which include more than one theory. In order to disentangle the different theories in such a setup, 
precise perturbative results are crucial for providing sharp assumptions.
In summary:
\begin{equation*}
    \text{Estimates}\to \text{gap assumptions}\to \text{bootstrap} \to \text{rigorous bounds}\,.
\end{equation*}
In addition to bootstrapping new theories, deriving perturbative precision estimates will also facilitate a cross-check of earlier gap assumptions made on spectra of theories in the bootstrap.

Finally, we note that it is also possible to compute perturbative data directly in $d=3$ spacetime dimensions~\cite{Parisi:1980gya}. 
However, since the CFT axioms, except for unitarity, hold in continuous $d$,\footnote{Some discussion about making continuation in group parameters rigorous was given in~\cite{Binder:2019zqc}; see also~\cite{Grans-Samuelsson:2021uor,Jacobsen:2022nxs,Cao:2023psi}} 
it is suggestive to consider families of CFTs parametrized by some $d<4$ \cite{El-Showk:2013nia,Hogervorst:2015akt,Cappelli:2018vir,Henriksson:2022gpa,Bonanno:2022ztf,Sirois:2022vth,Reehorst:2024vyq,Henriksson:2025kws}.
An advantage of perturbation theory in $4-\eps$ dimensions is therefore that the $\eps$-expansion continually connects a family of CFTs across dimensions; see Figure~\ref{fig:eps-expansion}. 
A precision study of the perturbative spectrum allows for a comparison to non-perturbative results at non-integer $d$, providing for instance insight into genuinely non-perturbative physics such as level repulsion \cite{Henriksson:2022gpa}. 
This could also  be a way to better understand the space of CFTs by identifying which perturbative fixed-points in $d=4-\eps$ dimensions correspond to which bootstrap islands in $d=3$, especially since various things can happen across dimensions, such as merger and annihilation of fixed-points \cite{Chester:2022hzt,Reehorst:2024vyq}. It may also provide a route to derive bootstrap islands at all, where the perturbative estimates can be taken to be particularly reliable at small $\eps$, allowing for aggressive assumptions, which can then be changed in a controlled way as $\eps$ increases.

\section{Method}
\label{sec:method}

We compute the anomalous dimensions of \emph{all} composite operators in \emph{any} scalar $\phi^4$ theory up to dimension six and Lorentz rank up to two in $4-2\ep$ spacetime dimensions,
using the minimal subtraction scheme. In this section, we list our conventions and provide additional background details on the calculation. Details on the diagram evaluation using the small-momentum asymptotic expansion of the $R^*$ method were previously outlined in \cite{Henriksson:2025hwi}.

\subsection{Lagrangian}

The full Lagrangian we will be concerned with is~\cite{Henriksson:2025hwi}
\begin{equation}\label{eq:gentheory}
    \mathcal{L} = \frac12 \partial_\mu \phi^a \partial^\mu \phi^a 
    - \frac{1}{4!} \lambda^{abcd} \phi^a \phi^b \phi^c \phi^d 
    + \mathcal{L}^{(\Delta<4)}
    + \mathcal{L}^{(\ell=0)}
    + u^{\mu}\mathcal{L}^{(\ell=1)}_\mu 
    + v^{(\mu\nu)} \mathcal{L}^{(\ell=2)}_{(\mu\nu)}
    + w^{[\mu\nu]} \mathcal{L}^{(\ell=\{1,1\})}_{[\mu\nu]}
    ,
\end{equation}
where $u^\mu$, $v^{(\mu\nu)}$ and $w^{[\mu\nu]}$ are reference tensors in different representations of the Lorentz group: vector, traceless symmetric and antisymmetric. 
This includes the composite operators for the superrenormalizable part at zero spin 
\begin{align}
    \mathcal{L}^{(\Delta<4)} 
    &= 
        - t_a \phi^a 
        - \frac{m^2_{ab}}{2}\,\phi^a\phi^b
        - \frac{h_{abc}}{3!}\,\phi^a\phi^b\phi^c\,,
    \end{align}
    and
    {\allowdisplaybreaks
    \begin{align}
    \mathcal{L}^{(\ell=0)} &=
    \frac{c_{abcde}^{(5,0)}}{5!} \, 
        \phi^a \phi^b \phi^c \phi^d \phi^e
    +\frac{c_{abcdef}^{(6,0)}}{6!}\,
        \phi^a\phi^b\phi^c\phi^d\phi^e \phi^f
    -\frac{c_{abcd}^{(6,0)}}{4}\,
        \phi^a\phi^b\partial_\mu\phi^c\partial^\mu \phi^d\,,\nonumber
        \displaybreak[2]\\[4mm]
    \mathcal{L}^{(\ell=1)}_\mu &=
    c_{ab}^{(3,1)}\,\phi^a \partial_\mu \phi^b
    +\frac{c_{abc}^{(4,1)}}{2} \phi^a \phi^b \partial_\mu \phi^c
    +\frac{c_{abcd}^{(5,1)}}{3!}  \phi^a \phi^b \phi^c \partial_\mu \phi^d
    \nonumber
    \\&\qquad 
    +c_{ab}^{(5,1)} \left(
      3\,\partial_\mu \phi^{a} \partial^2 \phi^{b}
      -\phi^{a} \partial_\mu \partial^2 \phi^{b}
   \right)
    +\frac{c_{abcde}^{(6,1)}}{4!} \phi^a\phi^b\phi^c\phi^d \partial_\mu\phi^e\,,
    \displaybreak[2]\nonumber\\[4mm]
    \mathcal{L}^{(\ell=2)}_{(\mu\nu)} &=
    c^{(4,2)}_{ab}
    \left(\phi^a\partial_\mu\partial_\nu \phi^b -2\,\partial_\mu\phi^a\partial_\nu\phi^b\right)
    + 2 \, c_{abc}^{(5,2)} \left( 
    \phi^a\phi^b\partial_\mu\partial_\nu\phi^c 
    -2 \, \phi^a \partial_\mu\phi^b \partial_\nu \phi^c 
    \right)\nonumber\\&\qquad 
    + c_{abcd}^{(6,2)}
    \left(
    \phi^a\phi^b\phi^c
    \partial_\mu\partial_\nu\phi^d
    -2 \, \phi^a\phi^b
    \partial_\mu\phi^c\partial_\nu\phi^d
    \right)
    +c_{ab}^{(6,2)}
      \partial_\mu \partial_\nu \phi^{a} 
      \partial^2 \phi^{b}\,,
      \nonumber
      \displaybreak[2]\\[4mm]
    \mathcal{L}^{(\ell=\{1,1\})}_{[\mu\nu]} &=
    -c_{abc}^{(5,\{1,1\})} \, \phi^a \partial_\mu \phi^b \partial_\nu \phi^c 
    -c_{abcd}^{(6,\{1,1\})}\,
    \phi^a\phi^b
    \partial_\mu\phi^c\partial_\nu\phi^{d}\,,
    \label{eq:LagrsGenTheory}
\end{align}
for} the remaining operators. 
The permutation symmetries and other constraints on the coupling constant tensors are defined in Appendix~\ref{s:Operators}.
Throughout this work, we will freely raise and lower the scalar (flavor) indices. 

In our calculations, we consider single insertions of the composite operators, but with an arbitrary number of insertions of the dimension-four operator $\lambda^{abcd}\phi^a\phi^b\phi^c\phi^d$ 
up to five loops in most cases.%
     \footnote{In the general theory, we refer to e.g.~$\lambda^{abcd}\phi^a\phi^b\phi^c\phi^d$ as a single operator, even though it involves multiple independent components. When choosing the parameters appropriately to restrict to a specific theory, this general operator then splits into multiple operators.}    
The only exceptions are the anomalous dimensions of $c_{abcdef}^{(6,0)}$, $c_{abcd}^{(6,0)}$ and $c_{abcde}^{(5,1)}$, which we calculate up to four loops.
The terms in $\mathcal{L}^{(\Delta<4)}$ (i.e.\ $\phi^a$, $\phi^a\phi^b$ and $\phi^a\phi^b\phi^c$) were previously renormalized up to six loops in \cite{Bednyakov:2021ojn}. We recomputed these results up to five loops (with single insertions) and find full agreement.%
\footnote{We used the \texttt{SimTeEx} package~\cite{Fonseca:2024rcg} to facilitate the comparison between the tensor expressions.} 
For specific scalar theories, namely the single-scalar $\Z_2$ theory, the $O(n)$ model and the hypercubic theory, we also compute anomalous dimensions beyond mass dimension six. For instance, we renormalized dimension eight $O(n)$ invariant Lorentz scalar operators up to four loops. In addition to this, for these specific theories, all dimensions for operators up to mass dimension six, and Lorentz rank two, are computed at five loops (compared to some four loop results given for generic theories).

To keep the presentation simple, we have suppressed dimensionful parameters and factors of $16\pi^2$ in 
Eqs.~\eqref{eq:gentheory}--\eqref{eq:LagrsGenTheory}.
These can be restored by dimensional analysis~\cite{Manohar:1983md}. For example,
\begin{align}
    \lambda^{abcd} &\to 16\pi^2\,\mu^{2\ep} \,\lambda^{abcd}\,, &&
    c^{(6,0)}_{abcdef} \to \frac{\left(16\pi^2\,\mu^{2\ep}\right)^2}{\mu^2}\,c_{abcdef}^{(6,0)}\,,
\end{align}
where $\mu$ is the renormalization scale. 
More generally, for the coupling of an $m$-point operator with engineering dimension $[\mathcal{O}]$ (in $d=4-2\ep$ dimensions),
\begin{align}\label{eq:dimAna}
    c_m\to
    (4\pi)^{m-2} \, \mu^{d-[\mathcal{O}]} \, c_m\,.
\end{align}
The factors of $4\pi$ are included in such a way that the expressions of the anomalous dimensions of the couplings are free of such loop factors. 
To facilitate the relation to scaling dimensions in CFT, we have chosen to normalize dimension-$n$ operators by a factor of $1/\mu^{n-4}$.
In EFT, it is more standard to normalize operators by a $\mu$-independent EFT scale $\Lambda$. 
The two conventions differ only in how the classical $\mu$-dependence of the couplings is defined; the full scaling dimension of the operators is of course unaffected by this.

The Lagrangian \eqref{eq:LagrsGenTheory} is (almost) minimal in the sense that we have removed redundant operators associated to equations of motion and total derivatives. This imposes constraints on the tensors, which we systematically derive, as we will describe in Section~\ref{s:prim}.
However, we do
keep the tensors $c_{ab}^{(5,1)}$ and $c_{ab}^{(6,2)}$, which could in principle be removed by a field redefinition. 
This amounts to the inclusion of redundant operators in the mixing matrices for these cases. 
We chose to keep these operators in our basis because there are additional components of $c_{abcd}^{(5,1)}$ and $c_{abcd}^{(6,2)}$ that are also redundant, which are more subtle to systematically eliminate; see Section~\ref{s:tHooft} and the example in Section~\ref{s:Redundant} for more details.

\subsection{Basic definitions}\label{s:basics}

Before giving more details on our basis choice and the calculation, let us briefly list the basic conventions used in the rest of this work. 
We provide more technical details on the calculation of the anomalous dimensions in Section~\ref{s:ADcalc}.

In general, the Lagrangian of a QFT contains a collection of operators $\O_i$ (built of powers of the renormalized field $\phi$ and derivatives) with their coupling constants (leaving flavor indices implicit here),
\begin{equation}
    \mathcal L=\sum_i c_i \, \O_i(x)\,.
\end{equation}
These depend on the renormalization scale as dictated by the renormalization group,
\begin{equation}
\label{eq:generalRG}
    \mu\frac{d}{d\mu} \sum_i c_i \,\O_i = \sum_{i,j}  c_i\,\gamma_{ji}\,\O_j
    -\sum_i n_i \, \gamma_\phi \, c_i \, \O_i
    \,,
\end{equation}
where $n_i$ is the number of $\phi$'s in $\mathcal{O}_i$,
$\gamma_\phi$ is the anomalous dimension of the field (defined below) and 
$\gamma_{ij}$ is the anomalous dimension mixing matrix. We have ignored contributions from multiple operator insertions on the RHS.%
    \footnote{At non-linear order, the evolution of coupling constants (beta functions) also involve anomalous dimension tensors, $\gamma_{ijk}^{A B_j B_k} c_j^{B_j}c_k^{B_k}$, etc. (We use capital indices to denote groups of additional flavor indices.) These play an important role in EFT, but they do not contribute to the scaling dimensions in CFT. We leave their computation in the general theory to future work.} 
From here, two equivalent routes can be taken: the scale dependence can either be carried by the operators, or by the couplings. Here we take the latter, and in anticipation of this we have defined the anomalous dimension matrix with transposed indices as in \eqref{eq:generalRG}. 

We will compute $\gamma_{ij}$ by demanding that correlation functions with $n$ insertions of $\phi$, and single insertions of the operators $\O_i$ are finite. This will introduce a set of counterterms, $Z_{ij}$, from which the anomalous dimension is extracted,
\begin{equation}
    \gamma_{ij}=-\frac {d\,(\log Z)_{ij}}{d\log\mu}
    \,.
\end{equation}
To be more precise, in the general theory, we define beta functions corresponding to the local operators as
\begin{align}
    \beta_{c_i}^A = \frac{d\,c_i^A}{d\log\mu} \,.
\end{align}
where $A$ represents a collection of indices. 
Working at linear order in the composite operators,
the beta function can be written as
\begin{align}
    \beta_{c_i}^A = \sum_j \sum_{B_j}\gamma_{ij}^{AB_j} c_j^{B_j} 
                  = \left([\mathcal{O}_i]-d\right)c_i^A + O(\lambda)\,,
\end{align}
for the coupling $c_i$ of an operator with engineering dimension $[\mathcal{O}_i]$
(recall the normalization of~\eqref{eq:dimAna}). 
The sum runs over all coupling constants at the same engineering dimension. Since there may be mixing between operators with different number of fields, we use $B_j$ to indicate the set of indices of coupling $c_j$. 
We call $\gamma_{ij}$ the anomalous dimension (mixing) matrix.

For the kinetic term, we write
\begin{equation}\label{eq:bareKin}
    \mathcal{L}_0 \supset
     \frac12 \partial_\mu\phi^a_0 \, \partial^\mu\phi_0^a
    =\frac12 Z_2^{ab} 
    \partial_\mu \phi^a \partial^\mu \phi^b\,,
\end{equation}
where $\phi_0^a$ is the bare field and the counterterm matrix $Z_2^{ab}$ is symmetric.
 
The bare kinetic term is independent of the renormalization scale, such that we can derive the anomalous dimension of the field $\gamma_\phi$ and its scaling dimension $\Delta_\phi$ in terms of $Z_2^{ab}$.
Our conventions are
\begin{align}\label{eq:gammaPhiDef}
    \frac{d\,\phi^a}{d\log\mu} = -\gamma^{ab}_\phi \phi^b\,,
    &&
    \Delta_\phi^{ab} = 
    (1-\ep)\,\delta^{ab} +\gamma_\phi^{ab}\,,
\end{align}
in $4-2\ep$ dimensions, such that 
\begin{equation}
    \gamma^{ab}_\phi = \frac12 (Z_2^{-1})^{ac}
    \frac{d\,Z_2^{cb}}{d\log\mu}\,.
\end{equation}
This definition is not unique, due to ambiguities related to the antisymmetric part of $\gamma_\phi$~\cite{Bednyakov:2014pia,Herren:2017uxn}. We take the anomalous dimension of the field to be a symmetric matrix. 

Similarly, the beta function of $\lambda$ can be derived from the bare interaction term in the Lagrangian
\begin{align}\label{eq:bareph4}
    \mathcal{L}_0 \supset -\frac{16\pi^2}{4!} \,
    \mu^{2\ep} \lambda^{abcd}_0 \phi_0^a\phi_0^b\phi_0^c\phi_0^d
    = 
    -\frac{16\pi^2}{4!} \,
    \mu^{2\ep} \sum_{A,B}Z_4^{AB}\lambda^{B} (\phi^4)^A\,,
\end{align}
where we use capital indices to denote collections of indices. For example, $\lambda^B = \lambda^{b_1b_2b_3b_4}$ and $(\phi^4)^A=\phi^{a_1}\phi^{a_2}\phi^{a_3}\phi^{a_4}$.
We have defined the counterterm tensor $Z_4$,
which cancels the UV divergences of 
four-point 1PI correlations functions. 
$Z_4$ becomes the renormalization factor of $\lambda$ after dividing out factors of $\sqrt{Z_2}$.
The beta function of~$\lambda$,
\begin{equation}
    \beta_\lambda^A = \frac{d\,\lambda^A}{d\log\mu}\,,
\end{equation}
can be computed from the fact that the bare term in \eqref{eq:bareph4} does not depend on the renormalization scale $\mu$.

\paragraph{Scaling dimension and dilatation operator.} In the context of CFT, it is more conventional to report the dimension of operators instead of couplings. The RG mixing of composite operators is given by ($D$ being the dilatation generator)
\begin{align}\label{3.16}
    [D, \mathcal{O}_i^A(0)] = (4-2\ep) \, \mathcal{O}_i^{A} +  \sum_j \sum_{B_j}\mathcal{O}_j^{B_j}\gamma_{ji}^{B_j A}
    = [\mathcal{O}_i] \, \mathcal{O}_i^A + O(\lambda)\,,
\end{align}
where we note that the mixing matrix $\gamma_{ij}$ appears transposed with respect to \eqref{3.16}, and the definition of engineering dimensions differs by an additional $d=4-2\ep$.\footnote{Note that in this way, our definition of anomalous dimension is not the usual one of the CFT literature, namely $[D,\O(0)]= \Delta^{\mathrm{CFT}}_\O\O(0)$, with $\Delta^{\mathrm{CFT}}_\O=[\O]+\gamma_{\O}^{\mathrm{CFT}}$. The difference is just a constant shift of $d$ for any operator. We again emphasise that the end result is of course the same.}
In a specific theory (after flavor indices have been collected in overall tensors which can be divided out; see Section~\ref{s:Specific}), the scaling dimensions can be obtained from the eigenvalues of 
$\gamma_{ij}$,
\begin{equation}\label{3.15}
    \Delta_{\mathcal{O}_i} = 4-2\ep+\text{eig}_i(\gamma)\,.
\end{equation}

For future reference, the running of the mass parameter is given by
\begin{align}
    \frac{d\,m^2_{ab}}{d\log\mu} = \gamma_{m^2}^{abcd}\,m^2_{cd}\,,
\end{align}
where $\gamma_{m^2}$ is a function of $\lambda^{abcd}$. In most cases of interest discussed below, there is a single mass term and the above equation simplifies to 
\begin{align}
    \frac{d\,m^2}{d\log\mu} = \gamma_{m^2} \, m^2 = -2\,m^2+O(\lambda)\,, 
    && \Delta_{\phi^2} = (4-2\ep)+\gamma_{m^2} \,.
\end{align}
We have fixed our conventions in accordance with \eqref{3.15}.
This is to be contrasted with the usual convention in CFT, $\Delta_{\phi^2}=2-\eps+\gamma^{\mathrm{CFT}}_{\phi^2}$ (where $\eps=2\epsilon$) so that
$    \gamma^{\mathrm{CFT}}_{\phi^2}= \gamma_{m^2}+2.$

\subsection{The basis of conformal primary operators}\label{s:prim}

Upon fixing the field content of an effective field theory, the spectrum of operators needs to be determined. Let us point out that for both EFT as well as CFT applications, it is possible to focus on primary operators. 
The enumeration of a minimal basis of operators in generic EFTs can therefore be accomplished through the computation of the so-called Hilbert series~\cite{Henning:2015alf,Henning:2017fpj}, which counts the number of primary operators of a free CFT~\cite{Cardy:1991kr,Dolan:2005wy}. Even in a pure scalar theory, this is non-trivial due to redundancies in the action stemming from the fact that total derivatives vanish and field redefinitions leave the S-matrix invariant. 
Although the choice of basis is not unique, Refs.~\cite{Henning:2015alf,Henning:2017fpj} isolated one choice of basis of EFT operators as the set of conformal primary operators.

We use the basis of primary operators throughout this work (except in specific examples in Section~\ref{s:examples} that illustrate the aforementioned redundancies). This uniquely determines the operators and the constraints on the coupling constant tensors in~\eqref{eq:gentheory}, up to normalizations. In the basis construction we avoid linear combinations of operators with different field and derivative content. In the following, we define the basis of primary operators and consider simple examples. We list the full operator basis in Appendix~\ref{s:Operators}.

\paragraph{Definition.}
We define the basis of primary operators in $d=4$ spacetime dimensions in the free theory limit.
That is, operators are constructed from fields $\phi^a$ and derivatives $\partial_\mu$. The complete set of operators is spanned by linear combinations of monomials in the fields and their derivatives. This includes descendants (total derivatives) and operators proportional to the free equations of motion (${\partial^2\phi^a=0}$), both of which we drop. 
We define the basis of primary operators to be spanned by linear combinations of monomials that are not proportional to $\partial^2\phi^a$, with the additional condition
\begin{equation}\label{eq:primCondition}
    i\,[K_\mu,\mathcal{O}(0)] = 0\,,
\end{equation}
where $K_\mu$ is the generator of special conformal transformations defined to act in the free theory. 

The action of $K_\mu$ on operators can be determined through standard procedure applying the conformal algebra and using that $\phi^a(x)$ is a spin-0 primary operator of dimension one, thus satisfying 
\begin{align}
    i\,[K_\mu, \phi^a(0)] = 0\,,
\qquad
    i\,[M_{\mu\nu},\phi^a(0)] = 0\,,\qquad
    i\,[D,\phi^a(0)] = \phi^a(0)\,,
\end{align}
where $D$ and $M_{\mu\nu}$ are the generators of dilatation and rotations respectively. 
Derivatives result from the action of the generator of translations, $P_\mu$,
\begin{equation}
    i\,[P_\mu,\mathcal{O}(x)] = \partial_\mu\mathcal{O}(x)\,.
\end{equation}
The action of $K_\mu$ on a generic operator then follows from the conformal algebra. 
In particular, the following commutation relations are useful:\footnote{The complete conformal algebra can be found in standard references such as \cite{DiFrancesco:1997nk,Rychkov:2016iqz,Simmons-Duffin:2016gjk}.}
\begin{align}
    [K_\mu, P_\nu] &= -2i\,(\eta_{\mu\nu} D+M_{\mu\nu}) \,,
    \nn\\
    [M_{\mu\nu},P_\rho]&=i\,(\delta_{\nu\rho}P_\mu - \delta_{\mu \rho}P_\nu)\,,
    \nn\\
    [D,P_\mu] &= -i\,P_\mu\,.
\end{align}
One of the advantages of defining the operator basis through \eqref{eq:primCondition} is that this condition can be applied in theories with general flavor structure and in theories with specific global symmetries alike. 
This therefore facilitates the reduction from the former to the latter, without the need for additional field redefinitions. (When multiple primary operators with the same field content exist, the condition \eqref{eq:primCondition} does not fully specify the basis, in which case 
only coupling redefinitions will be needed in the specialization.)
Mapping between generic (non-primary) operator bases requires the use of field redefinitions.
We exemplify this in Section~\ref{s:reducingEFT}.

\paragraph{Examples.}
Let us consider some simple examples. It directly follows that any product of fields without derivatives, such as the dimension six operator $\mathcal{O}_{6,1}^{(6,0)} = c^{abcdef}\phi^a\phi^b\phi^c\phi^d\phi^e\phi^f$, is a primary operator in the free theory. These are therefore included in the EFT basis.
The coupling $c^{abcdef}$ is automatically fully permutation invariant, but no additional constraints are imposed by the condition \eqref{eq:primCondition}.
Excluding operators proportional to $\partial^2\phi^a$ means that any dimension-six operator with two contracted derivatives takes the form
\begin{equation}
    \mathcal{O}_{4,1}^{(6,0)} = -\frac14c^{abcd}\phi^a\phi^b\partial_\mu\phi^c\partial^\mu \phi^d\,,
\end{equation}
where $c^{abcd}$ is symmetric in the first two indices and symmetric in the last two indices. 
Using that $i\,[K_\nu,\partial_\mu\phi^a] = 2\,\eta_{\mu\nu}\phi^a$, the condition \eqref{eq:primCondition} requires
\begin{equation}
    -\,c^{abcd}\phi^a\phi^b\phi^c\partial_\nu \phi^d=0\,.
\end{equation}
This imposes the constraints 
\begin{equation}\label{primdim6}
c^{abcd}+c^{bcad}+c^{cabd}=0\,,
\end{equation}
and $c^{abcd}=c^{cdab}$ on the coupling constant tensor.
In the results which we make available on \githubb\!\!, the expressions have been simplified using the permutation symmetries. We also imposed the condition~\eqref{primdim6} via the replacement rule
\begin{equation}
    c^{abcd} \to \frac23 c^{abcd} - \frac13 c^{bcad} - \frac13 c^{cabd}\,,
\end{equation}
and similarly for all other operators.

It is illustrative to identify the redundancies which are present if \eqref{eq:primCondition} is not satisfied. For instance, take $\tilde c^{abcd} = c^{abcd} + \alpha^{abcd}$, where $\alpha^{abcd}$ is permutation invariant in its first three indices. In this case, the operator
\begin{equation}
    -\frac14 \alpha^{abcd} \phi^a\phi^b\partial_\mu\phi^c\partial^\mu\phi^d = 
    -\frac1{12} \alpha^{abcd} \partial_\mu \! \left(\phi^a\phi^b\phi^c\partial^\mu\phi^d \right)
    + \frac14 \alpha^{abcd}\phi^a\phi^b\phi^c\partial^2\phi^d
\end{equation}
is a linear combination of a total derivative and an operator proportional to $\partial^2\phi^a$. 
The former has vanishing contributions to the S-matrix, while the contributions of the latter can be captured by other operators after a field redefinition.

\subsection{Diagram evaluation}\label{s:R*}

We compute the UV divergences of off-shell 1PI correlation functions in dimensional regularization. Our framework relies on infrared rearrangement~\cite{Vladimirov:1979zm}, the \Rs method~\cite{Chetyrkin:1982nn,Chetyrkin:1984xa,Smirnov:1985yck,Larin:2002sc,Kleinert:2001hn,Batkovich:2014rka,Chetyrkin:2017ppe,Herzog:2017bjx,deVries:2019nsu,Beekveldt:2020kzk} (see also \cite{Collins:1984xc}) and the small-momentum asymptotic expansion~\cite{Chetyrkin:1982zq, Chetyrkin:1983qlc, Gorishnii:1983su, Gorishnii:1986mv, LlewellynSmith:1987jx, Chetyrkin:1988zz, Chetyrkin:1988cu, Gorishnii:1989dd, Smirnov:1990rz, Smirnov:1994tg,Smirnov:2002pj,Chakraborty2023,Chakraborty:2024uzz}, with an implementation in \texttt{Maple} \cite{maple} and \form~\cite{Vermaseren:2000nd,Ruijl:2017dtg}.
The \Rs method allows for the factorisation of an $L$-loop multiscale integral into a product of massless propagator-type integrals of at most $L-1$ loops.
Tensor reduction is performed with \texttt{Opiter}~\cite{Goode:2024cfy} and the integrals are evaluated using 
\texttt{Forcer}~\cite{Ruijl:2017cxj}, which relies in part on the results of~\cite{Baikov:2010hf,Lee:2011jt}.
We previously described our method in~\cite{Henriksson:2025hwi}. %
Since the \Rs operation automatically subtracts sub-divergences, the polynomial form of the resulting UV counterterms (i.e.~the absence of logarithms) is a non-trivial validation of the implementation.
In the rest of this section 
we give a brief overview of the computational resources used.

\begin{table}
\caption{Number of diagram topologies for the calculation of $m$-point correlators at mass dimension six and spin $\ell=0,1,2$.
}\label{tab:diagnumbers}
\centering{ 
\renewcommand{\arraystretch}{1.2}
\begin{tabular}{|c|c|c|c|c|c|}
\hline
Loop order & 1 &2&3&4&5          
\\\hline
$\ell=0, \ m=6 $&2&11&77&561&4461\\
$ \ell=0, \ m=4$&1&5&22&117&697 \\
$ \ell=0, \ m=2$&0&1&2&10&46 \\
$ \ell=1, \ m=5$&1&4&20&121&807\\
$ \ell=1, \  m=3$&0&1&4&18&97\\
$\ell=2, \ m=4$&1&4&18&96&567\\
$\ell=2, \ m=2$&0&1&2&8&39
\\\hline
\end{tabular}
}
\end{table}

We employed the Feynman diagram generator of~\cite{Kaneko:1994fd,Kaneko:2017wzd} which is implemented in version 5.0 of \form. 
In particular, we benefited from the option to symmetrize over the external legs. This avoids separate calculation of diagrams that are related by permutations. 
For reference, we have listed the number of diagrams at dimension six for the calculations up to spin two and five loops in Table~\ref{tab:diagnumbers}.

The implementation of operators with arbitrary flavor structure and living in non-singlet Lorentz representations was systematically worked out in~\cite{Henriksson:2025hwi}.
The evaluation of the loop integrals is insensitive to the dressing by flavor factors in the general scalar theory. 
However, the resulting tensors form large expressions, especially after taking $m!$ permutations of the external indices and momenta of the $m$-point diagrams.
We identified terms that are equivalent under relabeling of dummy indices by mapping each tensor expression to a canonical one defined by diagrams. This is necessary to reduce the size of the expressions, and to cross-check the consistency of the counterterms at different loop orders; see Section~\ref{s:tHooft}.

All calculations were performed on single desktop computers (28 cores, 60GB RAM, 1TB disk space). 
The majority of the calculations required less than one hour of running time, even at five loops. The most computationally expensive diagram evaluations were the six- and four-point correlator at mass dimension six and five loops. 
On 24 cores, these calculations took 42 and 9 hours, respectively. 
The resulting six-point counterterm takes up to 4MB of disk space before taking the $6!$ permutations of the external legs, which increases the size of intermediate expressions by multiple orders of magnitude. 
The manipulations required for large expressions in the general theory when performing field redefinitions, and especially the  consistency checks, are time consuming due to the large size of the expressions. 
This is in fact one of the obstacles for extensions to higher mass dimensions at multiloop order in the general theory.

\subsection{Anomalous dimensions and field redefinitions}\label{s:ADcalc}

The evaluation of 1PI correlation functions $\Gamma_m[\mathcal{O}_i]$ of $m$ fields $\phi$ and an operator $\O_i$, 
results in a set of counterterms for the UV divergences, which can be mapped onto an operator basis, 
\begin{align}
    \mathcal{Z}(\Gamma_m[\mathcal{O}_i]) &= 
     \sum_j \mathcal{O}_j \, \delta Z_{ji}
     + \sum_j \mathcal{R}_j \, \delta \tilde Z_{ji}\,, 
     \nn\\ 
     \label{eq:GammaR}
     \mathcal{Z}(\Gamma_m[\mathcal{R}_i]) &= 
     \sum_j \mathcal{O}_j \, \delta Z'_{ji}
     + \sum_j \mathcal{R}_j \, \delta \tilde Z'_{ji}\,,
\end{align}
where we suppressed the flavor indices for simplicity, and we have split the operator basis into operators $\O_i$ and $\mathcal{R}_i$, on which we comment below.\footnote{These operators in fact correspond to the submatrices with labels $p$ and $r$ in the introduction.}
The BPHZ counterterm operator $\mathcal{Z}$~\cite{Bogoliubov:1957gp,Hepp:1966eg,Zimmermann:1969jj}, which extracts (minus) the overall UV divergence, is taken in the minimal subtraction scheme.
The interpretation of the right-hand side is the linear combination of $m$-point operators needed to reproduce the divergence on the left-hand side through their Feynman rules. 
In other words, the $\mathcal Z$ operator makes sure that the correlators
\begin{equation}
    \Gamma_m [\O_i +\O_j \delta Z_{ji}+ \mathcal R_j \delta \tilde{Z} _{ji}]\,, \qquad
    \Gamma_m [\mathcal R_i +\O_j \delta Z^\prime_{ji}+ \mathcal R_j \delta \tilde{Z}^\prime _{ji}]
\end{equation}
are finite. Note that we have taken the contributions from the wavefunction renormalization, $(Z_2)^{m/2}$, as part of the definition of the counterterms.

We have split the operator basis into the basis of primary operators, spanned by the $\mathcal{O}_i$,%
    \footnote{We take the basis of primary operators for concreteness, but the following does not depend on the choice of basis for the S-matrix.}
and an additional set of independent operators $\mathcal{R}_i$.
The latter are necessary to span the operators proportional to the equations of motion, which are needed to renormalize off-shell correlation functions. 
The complete basis for correlation functions, spanned by all $\mathcal{O}_i$ and $\mathcal{R}_i$ is called a Green's basis. 
One choice of operators $\mathcal{R}_i$ consists of those proportional to $\partial^2\phi^a$, which we have discarded in the construction of the basis of primary operators.
We remind the reader that we have defined 
this basis in the free theory limit, meaning that the operator mixing is not yet diagonalized in the interacting theory. 

The counterterms in \eqref{eq:GammaR} are included in the Lagrangian through the bare couplings
\begin{align}
    (Z_2)^{m/2} \mu^{[\mathcal{O}_i]-d}\,c^b_i &= \sum_j (\delta_{ij}+\delta Z_{ij}) c_j 
            + \sum_j \delta Z'_{ij} d_j\,,\nn\\
            \label{3.35}
    (Z_2)^{m/2}\mu^{[\mathcal{R}_i]-d}\,d^b_i &= \sum_j \delta \tilde Z_{ij}c_j 
            + \sum_j (\delta_{ij}+\delta\tilde Z'_{ij}) d_j\,,
\end{align}   
where $c_i$ and $d_i$ are the coefficients of $\mathcal{O}_i$ and $\mathcal{R}_i$, respectively.
In the following, we will argue that only $\delta Z$ and $\delta\tilde Z$ (and $\gamma_\phi$ and $\beta(\lambda)$) are needed to compute the anomalous dimensions of the primary operators.
That is, the renormalized couplings $d_i$ can effectively be set to zero.

At leading order in the EFT parameters, field redefinitions result in the addition of operators proportional to the equations of motion of the dimension-four Lagrangian, 
$\text{EOM}^a = \partial^2\phi^a + \frac{1}{6}\lambda^{abcd}\phi^b\phi^c\phi^d$. 
In a generic basis, field redefinitions can be used to replace operators $\mathcal{R}_i$ by linear combinations of $\mathcal{O}_j$. 
A particular basis in which the operators 
$\mathcal{R}_i$ are proportional to $\text{EOM}$
can be obtained through a coupling redefinition of the form
\begin{align}
    c^b_i = \bar c^b_i + f_i(\bar d^b_j)\,,&&
    d^b_i=\bar d^b_i \,,
\end{align}
where $f_i$ is a function of the couplings $\bar d_j$ and $\lambda$. 
In such a basis, the coefficients $\bar d_i$ 
do not contribute to the S-matrix, while the coefficients $\bar c_i$ do.
This holds for any choice of the renormalization scale $\mu$, which implies that the operator mixing is simplified in this basis: the $\bar d_i$ do not mix into $\bar c_i$~\cite{Politzer:1980me}.
Said otherwise, the bare couplings
\begin{equation}
   \mu^{[\mathcal{O}_i]-d}\, \bar c^b_i = \bar Z_{ij} \bar c_j = 
(\delta_{ij}+\delta \bar Z_{ij})\, \bar c_j
\end{equation} 
depend only on $\bar c_j$, not on $\bar d_j$. 
Noting that $\bar d_i = d_i$, we conclude that the couplings $d_i$ can effectively be set to zero, even when working in a basis with a generic choice of 
$\mathcal{R}_i$.%
    \footnote{
We remark that the couplings $\bar d_i$ do correct the anomalous dimensions of the fields, such that prematurely setting them to zero may lead to inconsistencies~\cite{Manohar:2024xbh}. We briefly review this in Section~\ref{s:tHooft}.
    }
However, we do emphasize that both $\delta Z$ and $\delta \tilde Z$ are necessary to determine $\delta \bar Z$ in this case. We have also divided out $(Z_2)^{m/2}$. 
We exemplify this in Section~\ref{s:Redex}.
In conclusion, correlation functions with insertions of the operators $\mathcal{R}_i$~\eqref{eq:GammaR} need not be computed to determine the anomalous dimensions of primary operators.%
    \footnote{
    In principle, correlation functions with insertions of $\mathcal{R}_i$ are necessary to subtract sub-divergences of an off-shell correlator. These are subtracted automatically in the \Rs method we employ in this work.}

The beta functions can now be extracted from the counterterms $\bar Z$, using
\begin{equation}
    \frac{d\,\bar c_i}{d\log\mu}
    = ([\mathcal{O}_i]-d)\,\bar c_i - (\bar Z^{-1})_{ij} \frac{d \, \bar Z_{jk}}{d\log\mu} \bar c_k\,,
\end{equation}
where $d\,\bar Z_{jk}/d\log\mu = \beta_\lambda \, d\bar Z_{jk}/d\log\lambda$.
However, this equation becomes cumbersome upon reinstating the flavor indices. Note, for instance, the need for the inverse of higher-rank tensors $\bar Z$ and the square root of $Z_2$ in \eqref{3.35}, which is generally not unique when $Z_2$ is a matrix.
In practice, we instead solved for the anomalous dimensions recursively order by order for increasing loops. For example, consider the single operator at spin one and mass dimension four.
The bare Lagrangian term does not depend on $\mu$,
\begin{equation}\label{3.38}
    \frac{d}{d\log\mu}\left( \frac{\mu^{\epsilon}}{2} (c^{(4,1)}_{abc}+K^{(4,1)}_{abc}) \phi^a \phi^b \partial_\nu \phi^c
    \right) = 0\,,
\end{equation}
where $K_{abc}^{(4,1)}$ is the counterterm for the 1PI correlation function, including dependence on EFT couplings and before including the wavefunction renormalization. 
That is, the fields in \eqref{3.38} are renormalized (as opposed to bare) fields and thus depend on the scale $\mu$.
The anomalous dimension of the coupling can then be determined recursively from 
\begin{align}\label{3.39}
    &
    \left( \epsilon + \frac{d}{d\log\mu}\right)  
    (c^{(4,1)}_{ a b c} + K^{(4,1)}_{ a b c})-
    (c^{(4,1)}_{\bar a\bar b\bar c} + K^{(4,1)}_{\bar a\bar b\bar c})
    (\gamma_\phi^{\bar a a} \delta^{\bar b b} \delta^{\bar c c}
    +
    \delta^{\bar a a} \gamma_\phi^{\bar b b} \delta^{\bar c c}
    +
    \delta^{\bar a a} \delta^{\bar b b} \gamma_\phi^{\bar c c}
    )
    =0\,,
\end{align}
where $dK_{abc}^{(4,1)}/d\log\mu = \beta_\lambda\, dK_{abc}^{(4,1)}/d\lambda + \sum_i \beta_{c_i}\, dK_{abc}^{(4,1)}/dc_i$, and we recall the definition of $\gamma_\phi$ in \eqref{eq:gammaPhiDef}.
Strictly speaking, the part of $c_{abc}^{(4,1)}$ that is antisymmetric in $a\leftrightarrow b$ remains unconstrained by this equation. To avoid ambiguities, we choose to fix the symmetries of the coupling constant tensors as implied by the operator it multiplies, including the primary condition~\eqref{eq:primCondition}.
This approach generalizes straightforwardly to other operators.

\subsection{Consistency conditions}
\label{s:tHooft}

Expanding \eqref{3.39} around $\ep=0$, the Laurent series gives 
\begin{equation}\label{3.40}
    \frac{d\,c_{abc}^{(4,1)}}{d\log\mu} = 
    (\ldots)
    + 2\,\sum_{L} L\,K_{abc}^{(4,1)}\bigg|_{L,\ep^{-1}}
    +c^{(4,1)}_{\bar a b c}
    \gamma_\phi^{\bar a a} 
    +c^{(4,1)}_{ a\bar b c}
    \gamma_\phi^{\bar b b} 
    +c^{(4,1)}_{ a b\bar c}
    \gamma_\phi^{\bar c c} 
    -\ep\,c_{abc}^{(4,1)} 
    \,,
\end{equation}
where $ K_{abc}^{(4,1)}\big|_{L,\ep^{-1}}$ is the coefficient of the $1/\ep$ pole at $L$ loops; see for instance~\cite{Jenkins:2023rtg,Fonseca:2025zjb} for recent derivations.
We suppressed all terms with $1/\ep^n$ poles. 
In fact, these pole terms must cancel to ensure that the anomalous dimensions (of observable quantities) are finite as 
$\ep\to0$~\cite{tHooft:1973mfk}.
In particular, this fully determines the coefficients of $1/\ep^{n>1}$ in terms of the coefficient of $1/\ep$ in
the counterterms. The latter are sufficient to determine the anomalous dimensions. 
In our calculation, we have not used the shortcut~\eqref{3.40}. 
Instead we determined the anomalous dimensions using the full equations of the form~\eqref{3.39}. The subsequent cancellation of the $1/\ep$ poles constitutes a strong internal consistency check on the counterterms and the applied field redefinitions.
In this section, we will describe how a violation of these consistency conditions can be encountered in two specific cases of our work, 
and how it originates from a subtle way in which the basis of primary operators, as defined in Section~\ref{s:prim}, is over-complete.

Typically, the failure of the cancellation of pole terms in \eqref{3.40} indicates a mistake in the calculation. However, there are various consistent ways in which some anomalous dimensions may diverge as 
$\ep\to0$. 
Importantly, the RG running of physical observables remains finite in such cases.
For instance, infinite anomalous dimensions may be the result of ambiguities in the antisymmetric part of $\gamma_\phi^{ab}$ or in taking the square root to obtain wavefunction renormalization factors~\cite{Jack:1990eb,Fortin:2012hn,Bednyakov:2014pia,Herren:2017uxn,Herren:2021yur,Pannell:2024sia}. 
Another violation of the consistency conditions was encountered in~\cite{Jenkins:2023rtg, Jenkins:2023bls, Manohar:2024xbh}, in which
infinite field anomalous dimensions were the result of prematurely setting redundant operators to zero.

In Section~\ref{s:ADcalc}, we argued that the couplings of EOM-proportional operators $d_i$ can be set to zero without affecting the RG running of parameters that contribute to the S-matrix. 
Crucially, the S-matrix does not depend on the choice of fields, due to the LSZ reduction formula~\cite{Lehmann:1954rq, Lehmann:1957zz}. 
As a result, in theories with a mass, the field anomalous dimensions retain dependence on the couplings of EOM-proportional operators~\cite{Manohar:2024xbh}.
Indeed, the kinetic term can itself be seen as part of such an operator, namely
$-\frac12 \phi^a \left(\partial^2\phi^a +m^2 \phi^a + \frac{1}{6}\lambda^{abcd}\phi^b\phi^c\phi^d\right)$. 
It was thus found in Refs.~\cite{Jenkins:2023rtg, Jenkins:2023bls, Manohar:2024xbh} that ignoring the contributions of $d_i$ leads to a violation of the consistency conditions and thus to infinite anomalous dimensions of the field~$\phi^a$.

A similar phenomenon occurs in our calculations at dimension five spin one and at dimension six spin two. 
These instances can be generalized to all mass dimensions, which we do in Section~\ref{s:Bases}.
Let us consider dimension five spin one for concreteness. Upon removing total derivatives, the relevant operators are
\begin{align}\label{Lagr51}
    \mathcal{L}^{(5,1)} = 
    \frac{1}{6}c_{abcd}^{(5,1)} \,
    \phi^a\phi^b\phi^c\partial_\mu\phi^d
    + c_{ab}^{(5,1)} \left( 
        3\,\partial_\mu \phi^a\partial^2\phi^b
        -\phi^a\partial_\mu\partial^2\phi^b
    \right) ,
\end{align}
where $c_{ab}^{(5,1)}$ is antisymmetric, while $c_{abcd}^{(5,1)}$ is symmetric in the first three indices and satisfies the condition
\begin{equation}\label{3.44}
    c_{abcd}^{(5,1)} + c_{abdc}^{(5,1)} + c_{acdb}^{(5,1)} + c_{bcda}^{(5,1)} = 0\,,
\end{equation}
which follows from \eqref{eq:primCondition}.
In the free theory ($\lambda=0$), $c_{abcd}^{(5,1)}$ and $c_{ab}^{(5,1)}$ parametrize primary and EOM-proportional operators, respectively. 
Contrary to the arguments outlined in Section~\ref{s:ADcalc}, however, setting $c_{ab}^{(5,1)}$ to zero leads to infinite anomalous dimensions of some components of $c_{abcd}^{(5,1)}$.

The consistency conditions fail in these components of $c_{abcd}^{(5,1)}$ because they parameterize EOM-proportional operators in the interacting theory ($\lambda\neq0$). 
The operator basis is thus overcomplete, even after removing operators proportional to $\partial^2\phi^a$. 
Indeed, for symmetric $r_{ab}$ we can write
\begin{equation}\label{4.22}
    \mathcal{L}^{(5,1)}\supset 
    \frac{1}{6} r_{ab} 
    \, \lambda_{axyz}
    \, \phi^x\phi^y\phi^z \partial_\mu \phi^b
    \stackrel{\textsc{ibp}}{=}
    r_{ab} 
    \, (\partial^2\phi^a + 
    \frac16 \lambda_{axyz}
    \phi^x\phi^y\phi^z) \, \partial_\mu \phi^b
    = r_{ab} (\text{EOM})^a \partial_\mu \phi^b
    \,,
\end{equation}  
where we used integration by parts (at the action level) to relate
\begin{equation}\label{eq:d2phTotalDer}
    r_{ab}\int\!d^dx \,\partial^2\phi^a \, \partial_\mu \phi^b
= r_{ab}\int\! d^dx\,  \partial^\nu
\left( 
    \partial_\nu \phi^a \partial_\mu \phi^b
    -\frac12 \eta_{\mu\nu} \partial_\rho \phi^a \partial_\rho \phi^b
\right)
\stackrel{\textsc{ibp}}{=} 0\,.
\end{equation}
For this reason, the couplings $c_{ab}^{(5,1)}$ do mix into $r_{ab}$ and therefore $c^{(5,1)}_{ab}$ cannot be set to zero for a consistent calculation of the anomalous dimensions of $r_{ab}$. Ignoring the couplings $c_{ab}^{(5,1)}$ results in infinite anomalous dimensions starting at four loops.
We will study this mixing structure explicitly in Section~\ref{s:Redundant}.

Faced with the fact that some components in $c_{abcd}^{(5,1)}$ are redundant in the interacting theory, we could remove them together with the $c_{ab}^{(5,1)}$ parameters through a field redefinition 
\begin{equation}\label{eq:redef} 
    \phi^a \to  
    \phi^a + r_{ab} \, \partial^\mu \phi^b
    + 4\,c_{ab}^{(5,1)} \, \partial^\mu \phi^b
    \,,
\end{equation}
where we recall that $r_{ab}$ is symmetric and $c_{ab}^{(5,1)}$ is antisymmetric. After this field redefinition, the $c_{ab}^{(5,1)}$ parameters in the Lagrangian can be absorbed into $c_{abcd}^{(5,1)}$, while the $r_{ab}$ drop out because they multiply a total derivative.
We instead decided to renormalize the full set of primary and EOM-proportional operators,
because the field redefinition \eqref{eq:redef} alters the two-point function $\langle \phi(x) \phi(y) \rangle$. 
Such field redefinitions affect the wavefunction normalization factor in the LSZ reduction formula~\cite{Lehmann:1954rq,Lehmann:1957zz}.

We close this section by noting that, in a similar vein as above, operators of the form $\frac{1}{6}c^a\lambda^{abcd}\phi^b\phi^c\phi^d$ are primary operators in the free theory. But they
can be combined into the EOM operator
$
c^a\!\left(\partial^2\phi^a + \frac{1}{6}
\lambda^{abcd}\phi^b\phi^c\phi^d\right)$ in the interacting theory, where $\partial^2\phi^a$ is a total derivative.
For example, in a single-scalar theory, the Lagrangian term $\kappa \, \phi^3$ can be removed by field redefinitions in favour of 
$\frac{\kappa}{\lambda}\partial^2\phi$; see Ref.~\cite{Cohen:2024fak} for a recent discussion.  
This implies that the RG equation of $\kappa$ can be written in terms of $\gamma_\phi$ and $\beta_\lambda$:
\begin{equation}
    \gamma_\kappa = \frac{1}{\kappa}\frac{d\,\kappa}{d\log\mu}   
    =-1+\ep-\gamma_\phi + \frac{\beta_\lambda}{\lambda} = 
    -\Delta_\phi+\frac{\beta_\lambda}{\lambda}
    \,.
\end{equation}
Following our convention~\eqref{3.15}, the final expression is consistent with the EOM operator's scaling dimension $\Delta_\text{EOM} = 
d-\Delta_\phi$ at the fixed-point;
see Appendix~\ref{app:scalarPotentials} for more details.

\section{Examples of operator mixing}\label{s:examples}

In this section, we give a few examples to illustrate the discussion in the previous section. In the first two examples, we renormalize a redundant set of operators that in addition to primary operators also includes operators proportional to the EOM or total derivatives. The resulting zeros in the mixing matrices illustrate why it is possible to restrict to a minimal basis in practical calculations, as we have done to produce the results in the rest of this work.
For simplicity, we present these examples in a theory of a single scalar field, but the same conclusions follow in the scalar theory with general flavor structure.
In Section~\ref{s:Redundant}, we consider an example in the general theory to illustrate the RG mixing structure in the case that some total derivative operators are also proportional to $\partial^2\phi$.

\subsection{Dimension-five scalars and redundant operators}\label{s:Redex}

We first consider mass-dimension five at spin zero. At this order, there is a single primary operator in the single-scalar theory, namely $\phi^5$. 
We will show that its anomalous dimension can be obtained either as one of the eigenvalues of a larger (redundant) system, or through a calculation that involves only $\phi^5$. The latter requires the use of field redefinitions to remove redundant operators in intermediate steps. 
The dimension-five operators at spin zero have previously been discussed in \cite{Nicoll1981,Zhang1982} to three-loop order. We repeat the discussion and extend it to four-loop order. 
We provide the anomalous dimension of $\phi^5$ to five loops in \eqref{eq:fiveloopPhi5}.

We define
\begin{equation}
    \mathcal L=\frac12\de_\mu\phi\de^\mu\phi
    -\frac{\lambda}{24}\phi^4+\frac{1}{\mu}\mathcal L'\,,
\end{equation}
and we will write down different choices for $\mathcal L'$. Without any restrictions, there are four operators that we can write down here
\begin{equation}
\label{eq:allOps}
    \mathcal L_{\text{all}}' = a_1\,\phi^5
    +a_2\,\phi^2\de^2\phi
    +a_3\,\phi\de_\mu\phi\de^\mu\phi
    +a_4\,(\de^2)^2\phi\,.
\end{equation}
We shall now discuss choices that lead to reduction in size.

\paragraph{Removing descendant operators.}

The first step is to work in a framework that allows us to exclude the effect of descendant operators (total derivatives). Their anomalous dimensions will be related to anomalous dimensions of operators with lower dimension in the same theory. For illustration, we consider a redundant system including descendants in Section~\ref{s:descExample}. Total derivatives can be avoided in two equivalent ways:
1) by working with integrated operators (position space), or 2) by imposing momentum conservation (momentum space).

A simple change of basis from \eqref{eq:allOps} gives
\begin{equation}
    \mathcal L_{\text{all}}' =  
    \frac{b_1}{5!} \, \phi^5
    +\frac{b_2}{2}\, \phi^2\de^2\phi
    +b_3(2\,\phi\de_\mu\phi\de^\mu\phi
    +\phi^2\de^2\phi)+b_4(\de^2)^2\phi\,,
\end{equation}
where now the operators proportional to $b_3$ and $b_4$ are total derivatives, which we drop. We call the resulting terms the Green's basis,
\begin{equation}
\label{eq:naive}
    \mathcal L_{\text{Green's}}' =  
    \frac{b_1}{5!} \phi^5
    -\frac{b_2}{2}\phi^2\de^2\phi\,,
\end{equation}
where the numerical prefactors are conventional.
Note that we could alternatively have chosen the operator $\phi\de_\mu\phi\de^\mu\phi$ to multiply $b_2$. The operator $\phi^2\partial^2\phi$ is preferred with the foresight that the EOM-proportional operator $\phi^2(\partial^2\phi+\frac{\lambda}{6}\phi^3)$ is equivalent to zero at the fixed-point, or more generally equivalent to terms with mass dimension other than five in a more general EFT.

Having set total derivatives to zero, we will impose momentum conservation in the rest of this subsection.

\paragraph{Results in the Green's basis.}
Away from the fixed-point in $d=4-2\epsilon$ spacetime dimensions, we write the bare couplings as $b_1 = (4\pi)^3 \mu^{3\epsilon}b_1(\mu)+O(\lambda)$
and 
$b_2 = 4\pi\,\mu^{\epsilon}b_2(\mu)+O(\lambda)$.
The scale dependence of the renormalized couplings $b_1(\mu)$ and $b_2(\mu)$ is given by 
\begin{equation}
     \frac{d}{d\log\mu} \begin{pmatrix}
        b_1(\mu)\\b_2(\mu)
    \end{pmatrix}
    =
\begin{pmatrix}
    \gamma_{11}&\gamma_{12}\\
    \gamma_{21}&\gamma_{22}
\end{pmatrix}
\begin{pmatrix}
        b_1(\mu)\\b_2(\mu)
    \end{pmatrix},
\end{equation} 
where the matrix entries are given by
{\allowdisplaybreaks
\begin{align}    
\gamma_{11} &= 
1-3 \epsilon
+10 \lambda
-\frac{475 \lambda ^2}{12}
 +\left(120 \zeta _3+\frac{3515}{16}\right) \lambda ^3
 \nonumber\\ \nonumber &\qquad 
    +\left(-1185 \zeta _3+240 \zeta _4-1800 \zeta _5-\frac{292655}{192}\right) \lambda ^4\,,\\[2mm]
  \nonumber
 \gamma_{12}&= 
 60 \lambda ^2 -330 \lambda ^3
+\left(1080 \zeta _3+1980\right) \lambda ^4
\nonumber\\&\qquad 
     - \left(11055 \zeta_3-2250 \zeta _4+16800 \zeta _5+\frac{344665}{24}\right) \lambda ^5 \,, \nonumber\\[2mm]
\gamma_{21} &=
\frac{\lambda }{6}
-\frac{5 \lambda ^2}{12}
+ \frac{169 \lambda ^3}{48}\,,  \nonumber\\[2mm]
 \gamma_{22} &= 1-\epsilon 
 +\lambda
 +\frac{3 \lambda ^2}{4}
 -\frac{29 \lambda ^3}{48}
 +\left(-\frac{3 \zeta _3}{2}-3 \zeta _4+\frac{3833}{192}\right) \lambda ^4
 \,.
 \label{eq:gammaPhi5example}
\end{align}
We calculated these anomalous dimensions by renormalizing all off-shell 1PI Green's functions.
The expressions agree with \cite{Nicoll1981,Zhang1982} up to $O(\lambda^2)$, and up to $O(\lambda^3)$ when it comes to eigenvalues on the critical coupling \eqref{eq:critCouplingIsing}. Recall that our conventions are such that the full scaling dimension is given by adding $d$ to the eigenvalues of the anomalous dimension matrix; see \eqref{3.15}.

\paragraph{Removing the redundant operator.}
The change of basis from the Green's basis choice in~\eqref{eq:naive} to one with an EOM-proportional operator,\footnote{We consider yet another alternative choice of basis in Appendix~\ref{app:example1App}.}
\begin{equation}\label{eq:Lmin}
    \mathcal{L}'_\text{minimal} =
        \frac{c_1}{5!} \phi^5 
        - \frac{c_2}{2} \, 
            \phi^2 \left(\partial^2\phi + 
            \frac{\lambda}{6}\phi^3 \right) \,,
\end{equation}
is achieved by 
\begin{equation}\label{changeofBasisToMinimal0}
    \begin{split}
        b_1 &= c_1
        - 10\lambda \, c_2\,,\\
        b_2 &= c_2\,.
    \end{split}
\end{equation}
This means that the anomalous dimensions in the new basis are given by
\begin{equation}
\label{eq:basisChange}
\begin{pmatrix}
    \bar \gamma_{11}& \bar \gamma_{12}\\
    \bar \gamma_{21}&\bar \gamma_{22}
\end{pmatrix}
\begin{pmatrix}
        c_1(\mu)\\c_2(\mu)
\end{pmatrix} 
    =
\left[
\begin{pmatrix}
    1&10\lambda \\
    0&1
\end{pmatrix}
\begin{pmatrix}
     \gamma_{11}&  \gamma_{12}\\
     \gamma_{21}& \gamma_{22}
\end{pmatrix}
\begin{pmatrix}
    1&-10\lambda \\
    0&1
\end{pmatrix}
+
\begin{pmatrix}
     0&10 \, \beta_\lambda\\
     0&0
\end{pmatrix}
\right]
\begin{pmatrix}
        c_1(\mu)\\c_2(\mu)
\end{pmatrix} \,,
\end{equation} 
where
{\allowdisplaybreaks
\begin{align}    
\nonumber
\bar \gamma_{11} &= 
\gamma_{11}+10\lambda \gamma_{21}
\nonumber\\&=
1-3 \epsilon
+10 \lambda
-\frac{455 \lambda ^2}{12}
 +\left(120 \zeta _3+\frac{10345}{48}\right) \lambda ^3
 \nonumber\\ &\qquad 
    -\left(1185 \zeta _3-240 \zeta _4+1800 \zeta _5+\frac{285895}{192}\right) \lambda ^4\,,\nonumber\\[2mm]
\nonumber
 \bar \gamma_{12}&= 
 \gamma_{12} + 10\lambda \gamma_{22} -10 \lambda \gamma_{11} -100 \lambda^2 \gamma_{21} + 10 \beta_\lambda=0 \,,\\[2mm]
\bar \gamma_{21} &=
\gamma_{21}=
\frac{\lambda }{6}
-\frac{5 \lambda ^2}{12}
+ \frac{169 \lambda ^3}{48}\,,\nonumber \\[2mm]
\nonumber
 \bar \gamma_{22} &= 
 \gamma_{22}-10\lambda\gamma_{21}
\\&=
 1-\epsilon 
 +\lambda
 -\frac{11 \lambda ^2}{12}
 +\frac{57 \lambda ^3}{16}
 -\left(\frac{3 \zeta _3}{2}+3 \zeta _4+\frac{2927}{192}\right) \lambda ^4
 \,.
 \label{eq:gammaHatPhi5example}
\end{align}
It} is a general (and crucial) feature that primaries do not mix into EOM-proportional operators, $\bar \gamma_{12} = 0$~\cite{Politzer:1980me}.\footnote{The cancellations in $\bar\gamma_{12}$ are not obvious until the involved quantities are explicitly inserted. It thus constitutes a non-trivial check of the present computation.} Due to this zero, the scaling dimension of $\phi^5$ in the minimal basis (i.e.\ when the EOM-proportional operator is dropped) can therefore be read off as $\bar \gamma_{11}$, which depends on $\gamma_{11}$, $\gamma_{21}$ and the specific basis change, but not on $\gamma_{12}$ nor $\gamma_{22}$. This follows from the reasoning of Section~\ref{s:ADcalc}.

The anomalous dimension of the EOM-proportional operator can be expressed in terms of other anomalous dimensions,
\begin{equation}
    \bar \gamma_{22} = 
    1-\epsilon
    -\gamma_\phi 
    +(2+\gamma_{m^2})
    = 
     \Delta_{\phi^2} - \Delta_\phi
    \,;
\end{equation}
see Section~\ref{s:basics} for the definitions.
In terms of scaling dimensions of the operator, this becomes
\begin{equation}
    \Delta_{\phi^2\text{EOM}} =\Delta_{\phi^2}+ (d-\Delta_\phi) \,,
\end{equation}
as expected~\cite{Brezin:1974zr,Brown:1979pq}. More generally, we have $\Delta_{\mathcal{O}\,\text{(EOM)}} = \Delta_\mathcal{O}+\Delta_\text{EOM}$, under the assumption that $\mathcal{O}$ is a scaling operator. One way to understand this is by inserting these composite operators into $n$-point correlation functions, and deriving Ward-like identities, such as in~\cite[Eq.~2.9]{Kousvos:2025ext}, using the fact that the EOM is related to the variation of the action.

We note that the eigenvalues of $\gamma_{ij}$ and $\bar \gamma_{ij}$ are different due to the $\lambda$-dependence in the change of basis Eq.~\eqref{changeofBasisToMinimal0}. 
The physical interpretation of this is that since the matrix implementing the change of basis depends on $\lambda$, it is also affected by running. In practice, this effect shows up in as the second term inside the square-brackets of \eqref{eq:basisChange}. At the critical coupling, where $\beta_\lambda=0$, the eigenvalues agree. 
In an alternative approach, it is possible to first rescale operators by appropriate factors of $\lambda$, after which a change of basis becomes a simple rotation; see e.g.\ Ref.~\cite{Cao:2021cdt}.

As explained in Section~\ref{s:ADcalc}, in the calculation in the rest of this work, we compute diagrams from a Lagrangian in the minimal basis, so in this example only $\phi^5$. 
However, the counterterms to render these diagrams finite may require redundant operators. 
Said otherwise, we compute only entries $\gamma_{11}$ and $\gamma_{21}$, and we do not consider diagrams with insertions of the operator $\phi^2\partial^2\phi$. 
Since we know the change of basis towards the minimal basis, this allows us to determine the scaling dimension of $\phi^5$ in the minimal basis, $\bar \gamma_{11}$, which we did up to five loops,
\begin{align}
\resizebox{\textwidth}{!}{$\begin{aligned}
    \gamma_5 &= 
    1-3\,\ep
    +10 \lambda 
    -\frac{455 \lambda^2}{12}
    +\left(120 \zeta_3+\frac{10345}{48}\right) \lambda ^3
    -\left(1185 \zeta _3-240 \zeta _4+1800 \zeta
   _5+\frac{285895}{192}\right) \lambda ^4
   \\&\quad
   +\lambda ^5 \left(810 \zeta
   _3^2+\frac{545275 \zeta _3}{48}-\frac{5605 \zeta
   _4}{2}+20950 \zeta _5-6375 \zeta_6+28140
   \zeta_7+\frac{8853395}{768}\right) +O(\lambda^6)\,.
   \end{aligned}$}
   \label{eq:fiveloopPhi5}
\end{align}
and $\Delta_{\phi^5}=4-2\ep+\gamma_5$. 
To obtain this result, we evaluated 
$1+4+20+121+807$ five-point diagrams (separated by loop order) and 
$0+1+4+18+97$ three-point diagrams.

We evaluate this result at the critical coupling to find
\begin{align}
 \gamma_{* \phi^5}&=
    1-\frac{3}{2}\varepsilon+
\frac{10 }{3}\varepsilon
-\frac{685 }{324}\varepsilon ^2+\left(\frac{107855}{34992}+\frac{80}{27} \zeta_3\right) \varepsilon^3
  \nonumber
   \\
   & \quad  -
   \left(\frac{22238765}{3779136}+\frac{2875 \zeta _3}{729}-\frac{20 \zeta _4}{9}+\frac{1400 \zeta
   _5}{81}\right)\varepsilon^4
   \nonumber
   \\
   & \quad +\left(
   \frac{5143144325}{408146688}+\frac{1168565 \zeta _3}{104976}-\frac{2875 \zeta
   _4}{972}+\frac{5560 \zeta _5}{243}
   -\frac{1750 \zeta _6}{81}-\frac{100 \zeta _3^2}{81}+\frac{7910 \zeta_7}{81}
   \right)\eps^5\nonumber
   \\
   & \quad +O(\eps^6)\,.
\end{align}
We will return to this operator in Section~\ref{sec:Isingextractions}, where we compare different Pad\'e approximants with conformal bootstrap results; see Figure~\ref{fig:Deltaphi5} below.

\paragraph{Remark.}
The purpose of this section was to illustrate how a suitable basis helps structuring the computation. After removing descendants we evaluated the $2\times2$ anomalous dimension matrix using a naive $(b_1,b_2)$ and a minimal $(c_1,c_2)$ Green's basis. Even though the operator multiplying $b_1$ and $c_1$ is the same, the result of the computations differ, even in the $(1,1)$ entry! This serves as a reminder that a consistent computation always requires a complete Green's basis at any dimension. The change with RG scale of the coefficient of $\phi^5$ is different if $b_2$ is held fixed compared to if $c_2$ is held fixed. Likewise, if at a particular scale the expressions agree by the virtue of $c_2=b_2=0$, at any other scale these coefficients would be non-zero by running effects.

\subsection{Dimension-six Lorentz tensors and total derivatives}\label{s:descExample}

Next we consider operators in a non-trivial representation of the Lorentz group. At mass dimension six and spin two, there are the following seven operators,
\begin{equation}
\resizebox{\textwidth}{!}{$\begin{aligned}
    \Big\{
    \phi^3 \partial_\mu \partial_\nu \phi \,,\
    \phi^2 \partial_\mu \phi \partial_\nu \phi \,,\ 
    \phi \partial_\mu\partial_\nu \partial^2 \phi \,, \ 
    \partial_\mu \phi \partial_\nu \partial^2\phi  
    +(\mu\leftrightarrow\nu), \ 
    \partial_\mu \partial_\nu \phi \partial^2\phi\,, \ 
    \partial_\rho \phi \partial_\mu \partial_\nu \partial^\rho \phi \,, \ 
    \partial_\mu \partial_\rho \phi \partial_\nu \partial^\rho \phi 
    \Big\} \,,
\end{aligned}$}
\end{equation}
where we restrict to the traceless symmetric Lorentz representation. In this system of operators there are three linear combinations that are proportional to the equations of motion. We deal with these as described in Section~\ref{s:Redex}, removing any operator proportional to the EOM. 
Note that this also holds in the interacting theory.%
    \footnote{
This is in contrast to the general theory case at dimension six and spin two, in which we did choose to keep operators proportional to the EOM; 
see Section~\ref{s:Redundant}.
The single-scalar theory case considered in this section is simpler, because the ``problematic'' operator exactly vanishes: 
\begin{equation*}
    \partial^2\phi^a \de_\mu \de_\nu\phi^b
    - \de_\mu \de_\nu\phi^a \partial^2\phi^b
    = \partial_\alpha \left( 
    2 \,\de_\mu\de_\nu\phi^a\de^\alpha\phi^b
    +\eta_{\alpha\mu}\de_\nu\de_\rho\phi^a
        \de^\rho\phi^b
    +\eta_{\alpha\nu}\de_\mu\de_\rho\phi^a
        \de^\rho\phi^b
    -(a\leftrightarrow b)
    \right)
    \stackrel{n=1}{=}0\,.
\end{equation*}
}

We will thus consider the mixing matrix that includes primary operators, as well as descendants.
For this purpose, we choose the following basis of operators:
\begin{align}
    \mathcal{O}_{4,1} &= \frac{1}{2}\left( \phi^3 \partial_\mu \partial_\nu \phi 
    - 2 \, \phi^2 \partial_\mu \phi \partial_\nu \phi \right) \,, \nonumber\\
    \mathcal{O}_{4,2} &= -\frac{1}{4!} \, \partial_\mu \partial_\nu (\phi^4)\,, \nonumber\\
    \mathcal{O}_{2,1} &= 
    \frac{1}{2}\partial_\mu \partial_\nu (\partial_\rho\phi \partial^\rho\phi) \,,\nonumber\\
    \mathcal{O}_{2,2} &= 
    \partial_\rho (\partial_\mu\partial_\nu \phi \partial^\rho\phi)\,,
\end{align}
where $\mathcal{O}_{4,1}$ is a conformal primary operator in $d=4$ dimensions and the remaining operators are total derivatives. 
We find the four-loop anomalous dimensions of the couplings of these operators to be 
{\allowdisplaybreaks
\begin{align}
    \mu \frac{d\,c_{4,1}}{d\mu} &=
        \bigg(
        2-2\epsilon
        +\tfrac{13}{3}\lambda 
        -\tfrac{941}{108}\lambda^2
        +\left(\tfrac{241963}{7776}
                +\tfrac{64}{3}\zeta_3\right)\lambda^3
                \nonumber\\&\hspace{1cm}
        -\left(
            \tfrac{71926849}{559872}
            +\tfrac{2200}{9}\zeta_5
            -35\zeta_4
            +\tfrac{11123}{81}\zeta_3
        \right)\lambda^4
        \bigg)c_{4,1}\,, \nonumber\\[2mm]
    \mu \frac{d\,c_{4,2}}{d\mu} &=
    \left(
            -\tfrac{151}{144}\lambda^3 
            +\left(
                \tfrac{40123}{3888}
                +\tfrac{14}{9}\zeta_3
            \right)\lambda^4
        \right) c_{4,1} 
        \nonumber\\& \hspace{-0.7cm}
        +\left(2-2\epsilon 
        +6\lambda
        -17\lambda^2
        +\left(\tfrac{431}{6}
                +48\zeta_3\right)\lambda^3
        -\left(
            \tfrac{51781}{144}
            +600\zeta_5
            -90\zeta_4
            +390\zeta_3
        \right)\lambda^4
        \right)c_{4,2} \nonumber\\&
        \hspace{-0.7cm}
        +\left(-4\lambda^2+21\lambda^3
        - \left(\tfrac{1205}{12}+72\zeta_3\right)\lambda^4
            +\left( 
                \tfrac{78275}{144}
                +960\zeta_5
                -150\zeta_4
                +621\zeta_3
            \right)\lambda^5 
        \right)c_{2,1} \nonumber\\&
        \hspace{-0.7cm}
        +\left(
            -\tfrac53\lambda^2
            +\tfrac{94}{9}\lambda^3
            -\left(\tfrac{397}{8}
                    +36\zeta_3\right)\lambda^4
            +\left(
                \tfrac{58109}{216}
                +480\zeta_5
                -76\zeta_4
                +\tfrac{934}{3}\zeta_3
            \right)\lambda^5
        \right)c_{2,2}\,, \nonumber\\[2mm]
    \mu \frac{d\,c_{2,1}}{d\mu} &= 
        \left(
            2
            +\lambda
            -\tfrac56 \lambda^2
            +\tfrac{25}{6}\lambda^3
            -\left(
                \tfrac{5701}{288}
                +3\zeta_4
                +\tfrac32\zeta_3
            \right)\lambda^4
        \right)c_{2,1} \nonumber\\&\hspace{0.6cm}
        +\left(-\tfrac{13}{72}\lambda^2
                +\tfrac{3511}{7776}\lambda^3
        \right)c_{4,1}
        +\left(
        -\tfrac{1}{3}\lambda^2
        +\tfrac{22}{9}\lambda^3
        \right)c_{4,2} \nonumber\\&\hspace{0.6cm}
        +\left(
            \tfrac{2}{3}\lambda
            -\tfrac{4}{9}\lambda^2
            +\tfrac{19}{8}\lambda^3
            -\left(
                \tfrac{4873}{432}
                +2\zeta_4
                +\tfrac13\zeta_3
            \right)\lambda^4
        \right)c_{2,2}\,, \nonumber\\[2mm]
    \mu \frac{d\,c_{2,2}}{d\mu} &=     
        2\,c_{2,2}+
        \left(\tfrac{5}{18}\lambda^2
        -\tfrac{1955}{2592}\lambda^3\right)
        c_{4,1}\,,
\end{align}
where} we note that we avoided imposing momentum conservation to determine these values. 
In this basis, the anomalous dimension matrix therefore takes the form
\begin{equation}
    \Gamma=\begin{pmatrix}
        \times & 0 & 0 & 0
        \\ \times &\times &\times &\times 
        \\ \times &\times &\times &\times 
        \\ \times &\times &\times &\times 
    \end{pmatrix},
\end{equation}
where the zeros are the entries that vanish due to the division of operators into primaries and descendants. We can see that the couplings of descendant operators do not mix into those of primary operator. Indeed, any correlator with the insertion of a total derivative is proportional to the sum of all external momenta. Their UV divergences can therefore be cancelled by operators that are total derivatives as well. 
This is also easy to see from the operator point of view: once a primary is renormalized, its total derivative is also renormalized, hence it does not necessitate any further additive renormalization (i.e.~mixing) with primary operators. 
To determine the running of $c_{4,1}$, one can therefore simply drop descendant operators throughout the calculation, which is what we do for the determination of the anomalous dimensions in this work, by imposing momentum conservation.

Recalling the definition \eqref{3.15}, we recognize the scaling dimensions of the three total derivative operators in this system to be
\begin{align}
    \Delta_{\mathcal{O}_2} &= 2+d\nn\,,\\
    \Delta_{\mathcal{O}_3} &= 4+\Delta_{\phi^2}\,,\nn\\
    \Delta_{\mathcal{O}_4} &= 2+\Delta_{\phi^4}\,,
\end{align}
which we checked at the critical coupling.
The operator with dimension $2+d$ is a total derivative of the stress tensor, $\square T_{\mu \nu}$.
We have also computed a five-loop result for the primary operator. Evaluated at the fixed-point in $d=4-\eps$ dimensions, this becomes
\begin{align}
&\Delta_{\mathcal{O}^{(6,2)}_{n=1}} = 
6
-\frac{5 \eps}{9}
-\frac{19 \eps^2}{324}
+\eps^3 \left(\frac{4 \zeta_3}{27}
+\frac{22879}{209952}\right)
+\eps^4 \left(\frac{181 \zeta_3}{6561}
+\frac{\zeta_4}{9}-\frac{640 \zeta_5}{729}
+\frac{983687}{45349632}\right) \nn\\& \ 
+\eps^5 \left(-\frac{452 \zeta_3^2}{2187}-\frac{568505 \zeta_3}{2834352}
+\frac{181 \zeta_4}{8748}-\frac{7697 \zeta_5}{19683}-\frac{800 \zeta_6}{729}
+\frac{1127 \zeta_7}{243}
+\frac{115637399}{3265173504}\right).
\label{eq:DeltaTprim}
\end{align}
Only the $O(\eps)$ result had been determined previously \cite{Kehrein:1994ff}. 
In Section~\ref{sec:Isingextractions}, Figure~\ref{fig:Tprim}, we compare different Pad\'e approximants with conformal bootstrap results.

\subsection[Dimension five spin one and total derivatives proportional to 
\texorpdfstring{$\partial^2\phi$}{d\textasciicircum2 phi}]{Dimension five spin one and total derivatives proportional to 
\texorpdfstring{\boldmath$\partial^2\phi$\unboldmath}{d\textasciicircum2 phi}\unboldmath}\label{s:Redundant}

In this section, we discuss the RG mixing structure at mass dimension five and spin one. 
At this order, there is an operator proportional to $\partial^2\phi^a$ which is also a total derivative, as previously identified in~\eqref{eq:d2phTotalDer}.
This obscures the identification of EOM operators. 
Here we illustrate that this identification is necessary for the correct removal of redundant operators, since generally all EOM operators mix with each other under RG.\footnote{We remind the reader that a consistent calculation either removes either all, or non of them. An inconsistent calculation leads to remaining $1/\ep$ poles in the anomalous dimensions.}
We will work in the general theory to avoid simplifications that hide important details, at the cost of more complicated expressions.

In the general theory, the Green's basis is spanned by two operators, with coupling constant tensors $c^{(5,1)}_{abcd}$ and $c_{ab}^{(5,1)}$; see~\eqref{Lagr51}.
The RG equations in this basis are
\begin{align}
\frac{d\,c^{(5,1)}_{ab}}{d\log\mu} &= 
      c^{(5,1)}_{ab}
      + l^2\bigg[
\frac{1}{6} 
c^{(5,1)}_{xw} \lambda_{axyz} \lambda_{bwyz}
+\frac{1}{12} \left(
c^{(5,1)}_{aw} \lambda_{bxyz} \lambda_{wxyz}
- \
c^{(5,1)}_{bw} \lambda_{axyz} \lambda_{wxyz}
\right) \nn\\&\hspace{1.34cm}
+\frac{1}{48} \left(
\lambda_{axyz} c^{(5,1)}_{xyzb}
- \lambda_{bxyz}c^{(5,1)}_{xyza}
- \lambda_{axyz} c^{(5,1)}_{bxyz}
+ \lambda_{bxyz} c^{(5,1)}_{axyz}
\right)
      \bigg] 
      +O(l^3)\,,
      \nn\\
\frac{d\,c^{(5,1)}_{abcd}}{d\log\mu} &= 
      (1-2\,\ep)c^{(5,1)}_{abcd}
      + T_{abcd}
      \,,      
      \label{4.17}
\end{align}
where the tensor $T_{abcd}$ is fixed by symmetries to be of the form
\begin{align}
T_{abcd} = 3\,\tilde T_{abcd} - \tilde T_{abdc}
      - \tilde T_{acdb}
      - \tilde T_{bcda}\,,
    \end{align}
with
\begin{align} 
      \tilde T_{abcd} = 
        l\left[\frac{1}{8}\lambda^{abxy}c_{xycd}^{(5,1)}
        -\frac{1}{8}\lambda^{abxy}c_{xcdy}^{(5,1)}
        +\frac12 \lambda^{abxz}\lambda^{cdyz}c_{xy}^{(5,1)}
        + \text{perms of } (a,b,c)\right]
        +O(l^2)
        \,,\label{4.16}
\end{align}
$l=1$ being a loop counting parameter. 
The mixing of $c_{ab}^{(5,1)}$ into $c_{abcd}^{(5,1)}$ thus starts at one loop and $c_{abcd}^{(5,1)}$ mixes into $c_{ab}^{(5,1)}$ at two loops.

As described in Section \ref{s:ADcalc}, this mixing can be simplified by choosing the redundant operators to be proportional to the EOM.
Couplings of EOM-proportional operators mix only amongst themselves.
We therefore consider a second choice of basis,
\begin{align}
    \mathcal{L}^{(5,1)} &= 
    \frac{1}{6}\tilde c_{abcd}^{(5,1)} \,
    \phi^a\phi^b\phi^c\partial_\mu\phi^d
    + \tilde c_{ab}^{(5,1)} 
    \bigg( 
        3\,\partial_\mu \phi^a\Big(
        \partial^2\phi^b+\frac16\lambda^{bxyz}\phi^x\phi^y\phi^z\Big)
        \nn\\&\hspace{5.4cm}
        -\phi^a\partial_\mu\Big(
        \partial^2\phi^b
        +\frac16\lambda^{bxyz}\phi^x\phi^y\phi^z\Big)
    \bigg) \,,
\end{align}
which can be obtained by the coupling redefinition
\begin{align}
    c_{abcd}^{(5,1)} &= \tilde c_{abcd}^{(5,1)}
        +\left(3\lambda^{abcx}\tilde c_{xd}^{(5,1)}
        - \lambda^{abdx}\tilde c_{xc}^{(5,1)}
        - \lambda^{acdx}\tilde c_{xb}^{(5,1)}
        - \lambda^{bcdx}\tilde c_{xa}^{(5,1)}
        \right)
        ,
    \nn\\
    c_{ab}^{(5,1)} &= \tilde c_{ab}^{(5,1)}\,.
\end{align}
In this basis, the RG equation of $\tilde c_{abcd}^{(5,1)}$ takes the form of \eqref{4.17} with
\begin{align}
    \tilde T_{abcd} &= 
-l^2\left[\frac{1}{12} \tilde c^{(5,1)}_{xy} \lambda_{vwyz}
\lambda_{abcx} \lambda_{dvwz}
+\frac{1}{12} \tilde c^{(5,1)}_{dx} \lambda_{vwxz}\lambda_{vwyz}
\lambda_{abcy}   + O(\tilde c_{abcd}^{(5,1)})\right] + O(l^3)
        \,,
\end{align}
where we suppress the terms that involve $\tilde c_{abcd}^{(5,1)}$ for brevity.
We thus find that $\tilde c_{ab}^{(5,1)}$ does not mix into $\tilde c_{abcd}^{(5,1)}$ at one loop, but it  \emph{does} mix into 
(some components of) 
$\tilde c_{abcd}^{(5,1)}$ at two loops.
The associated term in the Lagrangian that receives these RG mixing contributions can be written as
\begin{equation}\label{4.27}
    \mathcal{L}^{(5,1)}\supset 
    \frac{1}{6} r_{ab} 
    \, \lambda_{axyz}
    \, \phi^x\phi^y\phi^z \partial_\mu \phi^b\,, 
\end{equation}  
with a symmetric coupling constant matrix $r_{ab}$ that satisfies
\begin{equation}\label{4.20}
    \frac{d\,r_{ab}}{d\log\mu} = 
    -\frac{1}{3} 
    \tilde c^{(5,1)}_{ax} \lambda_{xyzw}\lambda_{byzw}
    - \frac{1}{3} 
    \tilde c^{(5,1)}_{bx} \lambda_{xyzw}\lambda_{ayzw}\,.
\end{equation}
In conclusion, we have identified the components of 
$\tilde c_{abcd}^{(5,1)}$ that receive RG mixing contributions from the couplings of EOM operators. 
Even though \eqref{4.27} does not look like it, this operator is itself also proportional to the EOM up to total derivatives; see \eqref{4.22}.

Importantly, it is not consistent to ignore this contribution to the RG of $r_{ab}$. In particular, $\tilde c_{abcd}^{(5,1)}$ mixes into $\tilde c_{ab}^{(5,1)}$ at two loops as well, so it is not possible to set $\tilde c_{ab}^{(5,1)}$ to zero at all scales. 
We find that ignoring contributions from $\tilde c_{ab}^{(5,1)}$ mixing into $r_{ab}$ (or $\tilde c_{abcd}^{(5,1)}$) leads to
a violation of the consistency conditions on the coefficients of the $1/\ep$ poles (see Section~\ref{s:tHooft})
at four loops.
The consistency conditions can be restored either by ignoring all contributions from both $\tilde c_{ab}^{(5,1)}$ and $r_{ab}$, or by keeping all contributions. We opted for the latter in our calculation. 

\paragraph{Remarks.}
The above situation occurs in our calculations when we remove a term due to it being a total derivative, but this same term is also proportional to $\partial^2 \phi^a$. This  term can in turn be written as $\partial^{\mu_i} \mathcal J_{\mu_1\ldots \mu_\ell}$, where $\mathcal J_{\mu_1\ldots \mu_\ell}$ is some higher spin current that is conserved specifically in the free theory. In the interacting theory these conservation equations break \cite{Skvortsov:2015pea}, where the linear combination $\partial^{\mu_1} \mathcal J_{\mu_1\ldots \mu_\ell} - \O_{\mu_2 \ldots \mu_\ell}$, with $\O_{\mu_2 \ldots \mu_\ell}$ some primary operator in the free theory, is now an EOM-proportional operator. In other words, prematurely removing the operator proportional to $\partial^2 \phi^a$ hampers our ability to identify the EOM operator related to the non-conservation equation. We discuss this for arbitrary mass dimension in the next subsection. Examples where primary operators cease to exist due to equation of motion effects are often referred to as ``multiplet recombination'' in CFT literature  \cite{Rychkov:2015naa}. We also note that \eqref{4.20} is zero in any single-mass theory, because then $\lambda^{axyz}\lambda^{bxyz}\propto \delta^{ab}$, while $\tilde c_{ab}^{(5,1)} = -\tilde c_{ba}^{(5,1)}$.

\subsection{General lessons and comparison with one-loop dilatation}
\label{s:Bases}

We have given three examples in this section, illustrating the reduction of a basis of operators from overcomplete to minimal. 
We showed that considering a larger system reproduces the same eigenvalues of primary operators, supplemented by the set of expected eigenvalues associated to redundant and descendant operators. In summary, the basis reduction is performed through the following steps:
\begin{enumerate}
    \item \textbf{Removal of descendants.} This corresponds to imposing momentum conservation. In position space, the same reduction would be achieved by considering integrated operators. At the level of operator construction, the removal of descendants can be achieved in the free theory by imposing $[K_\mu,\O(0)]=0$; see \eqref{eq:primCondition}.
    
    \item \textbf{Removal of redundant operators.} 
    Contrary to descendants, redundant operators do contribute to off-shell correlation functions. It is essential to take them into account to obtain the correct dimensions of the primary operators at higher loop order.
    By working in a suitable Green's basis, these contributions can be accounted for in a systematic way.
    Generically, the mixing matrix between (free-theory) primary operators ($p$) and operators proportional to $\partial^2\phi$ ($r$) takes the form
    \begin{equation}
\label{eq:GreensMinimalBasesMain0}
    \Gamma_{\text{Green's}}=\begin{pmatrix}
        \gamma_{pp}(\lambda) & \gamma_{pr}(\lambda)\\\gamma_{rp}(\lambda) & \gamma_{rr}(\lambda)
    \end{pmatrix}.
\end{equation}    
    At the level of operator construction (in the free theory limit), the Green's basis operators are primaries if they have no factor $\partial^2\phi^a$ and additionally satisfy $[K_\mu,\O(0)]=0$.
    After a change of basis to one in which the non-primary operators are proportional to the EOM of the interacting theory, this matrix becomes
    \begin{equation}
\label{eq:GreensMinimalBasesMain}
   \Gamma_{\text{minimal+EOM}}=\begin{pmatrix}
        \bar \gamma_{pp}(\lambda) & 0\\\bar\gamma_{\bar rp}(\lambda) & \bar\gamma_{\bar r\bar r}(\lambda)
    \end{pmatrix},
\end{equation}
where $\bar r$ denotes operators proportional to the EOM.
This makes it possible to present the results for the minimal basis only, given by $\bar \gamma_{pp}$. 
This sub-matrix depends on the entries $\gamma_{pp}$ and $\gamma_{rp}$, as well as on the change of basis. Importantly, the change of basis from
\eqref{eq:GreensMinimalBasesMain0} to \eqref{eq:GreensMinimalBasesMain}
can be determined without explicitly computing the entries $\gamma_{pr}$ and $\gamma_{rr}$, which are thus unnecessary for the determination of the dimensions of the primary operators.

\item
\textbf{Remaining redundant operators.}
In some cases, the set of operators still contains some redundant operators, even when all operators proportional to $\partial^2\phi^a$ have been removed,
as discussed in Section~\ref{s:tHooft} and exemplified in Section~\ref{s:Redundant}. 
This may happen when a total derivative of an operator is also proportional to $\partial^2 \phi^a$. 
In particular, this is the case for the divergence of higher-spin currents in the free theory limit, $\partial^{\mu_i}\mathcal J_{\mu_1\ldots \mu_\ell} \propto \partial^2 \phi$.
In our setup of scalar $\phi^4$ theories there is a minimal classification of the orders at which these operators can appear,
\begin{equation}\label{eq:whereRecombRed}
     (\Delta=4+\ell,\ell=0,1,2,3,\ldots)\,.
\end{equation}
Since we have treated the scalar cases separately (see Appendix~\ref{app:scalarPotentials}), this effect only shows up in our results at dimension-five spin-1 and dimension-six spin-2. 
Consistent computations either keep all possible redundant operators or remove all such operators. We opted for the former because identifying all EOM-proportional operators is not an automatic task to implement (as we saw, simply removing factors of $\partial^2 \phi$ does not suffice).
That is, we renormalized the full Green's basis in these particular cases.
The presence of redundant operators then results in additional eigenvalues which need to be removed by hand. These additional dimensions corresponding to redundant operators take a simple form, namely $\Delta=d+1$ at dim.-5, spin-1 and $\Delta=d+2$ at dim.-6, spin-2.\footnote{These values are easily derived from $\Delta_{(\partial_\mu\phi)  (\EOM)}=(\Delta_\phi+1)+(d-\Delta_\phi)=d+1$ and $\Delta_{(\partial_\mu \partial_\nu\phi ) (\EOM)} = (\Delta_\phi +2) + (d-\Delta_\phi)=d+2$, where the forms of the operators in the subscripts are to be understood as schematic. These can be explicitly checked order by order in perturbation theory.} 
\end{enumerate}
Even though we focus on scalar theories in this work, we remark that the same operator classification can be made in more general theories, with corresponding mixing structure.
In a gauge theory, the minimal basis also requires the restriction to gauge-invariant operators, in addition to the treatment of redundant operators.
The role that gauge-variant (so-called alien/nuisance) operators play is analogous to that of redundant operators considered here. In general, there is mixing between gauge-invariant and gauge-variant operators. In addition, they are necessary to renormalize sub-divergences of higher-loop correlation functions, even in the background-field method~\cite{Abbott:1980hw,Abbott:1983zw}; see e.g.~\cite{Falcioni:2024xav,Naterop:2024cfx} for more details. Using the \Rs method, 
these sub-divergences are automatically subtracted. This means that counterterms to gauge-variant operators need not be computed explicitly in the background-field gauge.

\paragraph{Comparison with one-loop dilatation operator.}
In this section, we have emphasized the necessity to keep track of corrections due to the mixing with redundant operators. 
However, in scalar field theories, one can restrict to the minimal basis when applying one-loop dilatation operators, thus completely ignoring terms proportional to $\partial^2\phi^a$.
This is a peculiarity of $\lambda\phi^4$ theory\footnote{
This is for instance not the case in gauge theories, where Feynman graphs have three-point vertices proportional to $g$, but the one-loop term represents order $g^2=4\pi\alpha$.} 
and is due to the fact that the one-loop dilatation operator does not mix operators with different number of fields \cite{Kehrein:1992fn,Kehrein:1994ff}.\footnote{The full dilatation operator is defined by $
[D_{\text{full}},\O(0)]=\Delta_{\text{full}}\O(0)
$ on its eigenoperators, and the one-loop counterpart is defined by $[D_{\text{1-loop}},\O(0)]=\gamma^{(1)}\O(0)$ where $\gamma^{(1)}$ is the $O(\lambda)$ part. For the case of operators with no derivatives, it takes the simple form $D_{\text{1-loop}}=\frac1{2}\phi^a\phi^b\lambda_{abcd}\check\phi^c\check\phi^d$ \cite{Jack:1983sk,Kehrein:1992fn,Kehrein:1994ff}, where $\check\phi^a$ annihilates a field $\phi^a(0)$.} 
Moreover, a set of operators, namely the $\O_{\mu_2\cdots\mu_\ell}$ mentioned in the remark at the end of section~\ref{s:Redundant}, also need to be removed by hand in this approach, a fact that it related to the ``remaining redundant operators'' discussed above.

When the full basis contains no redundant part (e.g.~with all derivatives and field indices in a traceless-symmetric configuration), a two-loop dilatation operator may be formed \cite{Kehrein:1995ia}. For literature on (one-loop) dilatation operators; see \cite{Kehrein:1992fn,Kehrein:1994ff,Kehrein:1995ia,Derkachov:1997qv,Beisert:2003tq,Beisert:2003jj,Hogervorst:2015akt,Hogervorst:2015tka,Liendo:2017wsn,Osborn:2017ucf,Antipin:2019vdg,Bednyakov:2023lfj}. Cases with no mixing of derivatives can also be formulated in terms of scalar potentials; see Appendix~\ref{app:scalarPotentials}.

\section{Applications to EFT}\label{s:SpecificEFT}

In this section, we exemplify the extraction of anomalous dimensions from the results in the general theory with two applications in EFT (i.e.~away from any fixed-point). 
The EFTs considered have global symmetries given by a subgroup of $O(n)$, which is the maximal symmetry group of a theory with $n$ scalar particles.
Nevertheless, in both examples, it is useful to decompose the set of EFT operators into irreducible representations of $O(n)$. 
The reason is that the $O(n)$ symmetry can be recovered in particular limits of the theory, which leaves an imprint in the RG mixing structure. We start with some general considerations on how to specify the general theory results to an EFT of choice.

\subsection{Reducing to specific EFTs}\label{s:reducingEFT}

One of the main advantages of the primary basis is that it streamlines the reduction from general to specific theories.
Let us compare, for example, the dimension-six operators with zero spin in the primary basis 
of the general theory, the $O(n)$ model and the single scalar ($n=1$) theory,
\begin{align}
    \mathcal{L}^{(6,0)}_\text{general} &= \frac{c_{abcdef}}{6!} \phi^a\phi^b\phi^c\phi^d\phi^e\phi^f 
    - \frac{c_{abcd}}{4} \phi^a\phi^b\partial_\mu\phi^c \partial^\mu \partial^d\,,\nonumber\\[2mm]
    \mathcal{L}^{(6,0)}_{O(n)} &= \frac{c_{O(n),6}}{48} (\phi\cdot\phi)^3 
    - \frac{c_{O(n),4}}{4} \Big((\phi\cdot\phi)(\partial_\mu\phi\cdot\partial^\mu\phi) - 
    (\phi\cdot\partial_\mu\phi)^2\Big)\,,\nonumber\\[2mm]
    \mathcal{L}^{(6,0)}_{n=1} &= \frac{c_{n=1}}{6!} \phi^6 \,.
\end{align}
Here we restrict to $O(n)$ singlet operators. The reduction to other irreps works in the same way, however, the tensor structures may become more involved in more complicated cases. We discuss the systematics of tensor structures in Section~\ref{s:Specific}.

The operators in $\mathcal{L}_{O(n)}^{(6,0)}$ and 
$\mathcal{L}_{n=1}^{(6,0)}$ are obtained from 
$\mathcal{L}_\text{general}^{(6,0)}$ through 
\begin{align}
    &c_{abcdef}\to c_{O(n),6}(\delta_{ab}\delta_{cd}\delta_{ef} + 14 \text{ perms})\,,
    &&
    c_{abcd}\to c_{O(n),4} \left(\delta_{ab}\delta_{cd}-\frac12 \delta_{ac}\delta_{bd} -\frac12 \delta_{ad}\delta_{bc}\right)\,;
    \label{eq:ONsubstitutions}
    \\
    &c_{abcdef}\to c_{n=1}\,,
    &&
    c_{abcd}\to 0\,,
\end{align}
respectively. Up to normalizations, these are the only possible expressions consistent with the primary condition \eqref{eq:primCondition}. 
In addition, the operator in the single-scalar theory can be obtained from the $O(n)$ model by setting $n\to1$ and normalizing $c_{O(n),6}\to \frac{1}{15} c_{n=1}$. Had we chosen a different operator basis for any of these theories, the map between them would require field redefinitions in addition to similar maps for the couplings.
Note, however, that in some cases, even if there is a single structure in the general theory, there can be several structures in a specific theory. For instance, there are three $\phi^6$-type singlet operators under hypercubic symmetry, all captured by the general $c_{abcdef}$, these correspond to the existence of three invariant tensors of six indices: $\delta_{abcdef}$, $\delta_{abcd}\delta_{ef}$ and $\delta_{ab}\delta_{cd}\delta_{ef}$.

\paragraph{Representations of \bm{$O(n)$}.}
In the following examples, we will specify to EFTs with global symmetry groups that are subgroups of $O(n)$. 
That means that the theory includes operators that are in non-singlet representations of $O(n)$.
For this reason, we recall the scalar operators with an even number of fields in the $O(n)$ model:\footnote{We use names for the $O(n)$ irreps corresponding to those used in \cite{Henriksson:2022rnm}.}
\begin{align}
\label{eq:ONscalarContent}
    \phi^2 &:\ S,T\,, \nonumber\\
    \phi^4&:\ S, T,T_4\,, \nonumber\\
    \phi^6 &:\ S,T,T_4,T_6 \,,&
    \phi^4\partial^2&: S,T,B_4\,, \nonumber
    \\
    \phi^8 &:\ S,T,T_4,T_6,T_8\,,
    & \phi^6 \partial^2 &:\ S,2\,T,T_4,B_4,H_4,Y_{4,2}\,, &
    \phi^4\partial^4 &:\ 2\,S, 2\,T, T_4, B_4\,,
\end{align}
which will be used in the following. 

\subsection{The Higgs sector of the SMEFT and accidental symmetries}\label{s:Higgs}

In this section, we relate the results of the general scalar theory to the Higgs sector of the SMEFT. 
This sector has an \emph{accidental} $O(4)$ symmetry. 
That is, the dimension-four Lagrangian is invariant under $O(4)$ transformations, but this symmetry is explicitly broken by higher-dimensional operators. We find (up to dimension eight) that the operators that break the $O(4)$ symmetry transform in the $B_4$ representation, with Young tableau $ \,\raisebox{1mm}{\scalebox{0.35}{$\ydiagram{2,2}$}}$\,.
Choosing the operators in the singlet and $B_4$ representation thus separates the operator basis into custodial-preserving and violating operators.
In addition, this perspective allows to extract all anomalous dimensions from the results in the $O(n)$ model, which we do up to five-loop order at dimension six and two-loop order (four loops for the $O(n)$-singlet operators) at dimension eight.

\paragraph{Dimension-4.}

The Higgs sector of the SMEFT is given by
\begin{align}
    \mathcal{L}_H = \partial_\mu H^\dagger \partial^\mu H 
    -\lambda_H \, (H^\dagger H)^2
    + \sum_i \mathcal{O}_i
    \,,
\end{align}
where we will consider the higher-dimensional operators $\mathcal{O}_i$ below. 
The Higgs doublet can be written in terms of four real scalars via $H = \frac{1}{\sqrt{2}}(\phi_1 + i\phi_2, \, \phi_3+i\phi_4)\,.$ 
In this basis, the Higgs Lagrangian becomes
\begin{align}\label{eq:HiggsLagr}
    \mathcal{L}_H = \frac{1}{2}
    \partial_\mu \phi^a \partial^\mu \phi^a
    -\frac{\lambda_H}{4} \, \phi^a \phi^a \phi^b \phi^b
    + \sum_i \mathcal{O}_i
    \,,
\end{align}
which is given by \eqref{eq:gentheory} with 
\begin{equation}\label{eq:Lamdim4SMEFT}
    \lambda^{abcd} = 2\, \lambda_H
    \left(\delta^{ab}\delta^{cd} + \delta^{ac}\delta^{bd} + \delta^{ad}\delta^{bc}\right).
\end{equation}
The dimension-four part of the Lagrangian \eqref{eq:HiggsLagr} is thus invariant under $O(4)$ transformations, which we call custodial symmetry and  which contains the Standard Model 
$SU(2)_L\times U(1)_Y$ as subgroup. 
Custodial symmetry is explicitly broken in the SM by gauge and Yukawa interactions. Importantly, even with pure-Higgs operators, higher-dimensional operators in the SMEFT may break custodial symmetry as well. 
In this sense, it is an accidental symmetry of the Higgs sector, which is preserved at lowest order in the EFT. 

Due to the custodial $O(4)$ symmetry, the $\lambda_H$ dependence in the beta function of $\lambda_H$ can be extracted from the beta function of the $O(4)$ model,
\begin{equation}
    \mu \frac{d}{d\mu} \lambda_H = -2 \ep \lambda_H
    + 24\,\lambda_H^2
    - 312\,\lambda_H^3 
    + 24 \,\lambda_H^4(299+168\zeta_3)
    + O(\lambda_H^5)
    \,,
\end{equation}
which was computed up to seven loops in~\cite{Schnetz:2016fhy}. 
We write $\lambda_H$ with subscript $H$ to emphasize that the normalization differs from our $O(n)$ model conventions elsewhere in this work. Nevertheless, for brevity we will omit the subscript in the rest of this subsection.

\paragraph{Dimension six.}
The Higgs operators in the SMEFT at dimension six are conventionally taken to be~\cite{Grzadkowski:2010es}
\begin{align}\label{eq:HiggsDim6}
    &\mathcal{O}_{H} = (H^\dagger H)^3\,, &&
    \mathcal{O}_{H\square} = (H^\dagger H)\partial^2 (H^\dagger H)\,, &&
    \mathcal{O}_{HD} = (H^\dagger \partial_\mu H)(\partial^\mu H^\dagger H)\,,
\end{align}
where the operators are linear combinations of primary operators (in $d=4$ dimensions) and redundant (EOM) operators.
Mapping our results (which were computed in the basis of primary operators) to these operators thus requires the use of field redefinitions. To illustrate the extraction from our data files, we will first determine the anomalous dimensions in the basis of Eq.~\eqref{eq:HiggsDim6}. We subsequently compute the anomalous dimensions of the Higgs operators in the basis of primary operators from the anomalous dimensions in the $O(n)$ model by leveraging custodial symmetry.
Ref.~\cite{Elias-Miro:2013mua} 
previously studied the imprint of custodial symmetry on the one-loop RG of the SMEFT at dimension six.

The operators in \eqref{eq:HiggsDim6} can be obtained from the general Lagrangian \eqref{eq:gentheory} by replacing
\begin{align}
    c^{abcdef} &\to \left(6\,c_H-16 c_{H\square}\lambda +4 c_{HD}\lambda \right)
    \left(\delta^{ab}\delta^{cd}\delta^{ef} + \text{ 14 permutations} \right) ,\nn\\[2mm]
    c^{abcd} &\to 
    -\frac43c_{H\square}\left(\delta^{ab}\delta^{cd} 
    -\frac12 \delta^{ac}\delta^{bd} - \frac12\delta^{ad}\delta^{bc}\right)
    \nonumber\\&\qquad \label{eq:Repl2}
    -\frac12c_{HD} \left(\Omega^{ad}\Omega^{bc} - \Omega^{ac}\Omega^{db}
    -\frac23\delta^{ab}\delta^{cd} 
    +\frac13 \delta^{ac}\delta^{bd} + \frac13\delta^{ad}\delta^{bc}
    \right).
\end{align}
Here we note that the replacement of the six-point tensor $c^{abcdef}$ involves the four-point couplings $c_{H\square}$ and $c_{HD}$, due to the fact that \eqref{eq:HiggsDim6} is not in the basis of primary operators. A field redefinition would be required to relate $(H^\dagger H)^3$ to $(H^\dagger H)(\partial^2 H^\dagger H + H^\dagger \partial^2 H)$, but this is not explicitly needed for the extraction of the results. We also defined the matrix
\begin{equation}
    \Omega = \begin{pmatrix}
        0 &\mathbbm{1}\\-\mathbbm{1}&0
    \end{pmatrix},
\end{equation}
which is an $SU(2)$-invariant tensor, and we included Kronecker delta functions in the second term in \eqref{eq:Repl2} to render the combination traceless.
With these replacement rules, the anomalous dimensions can be extracted from the general RGEs in our results,
{\allowdisplaybreaks
\begin{align}
    \mu \frac{d}{d\mu} c_H &= 
        (2-4\,\ep)\,c_H 
        + \left\{ 
        108\,c_H\,\lambda + (-160\,c_{H\square} + 48\,c_{HD})\,\lambda^2
        \right\}_1 \nonumber\\&\hspace{-1.2cm}
        + \left\{
-3444\,c_H\,\lambda^2 + (7968\,c_{H\square} - 1992\,c_{HD})\,\lambda^3
        \right\}_2 
        \nonumber\\&\hspace{-1.2cm} 
        + \left\{
36\,c_H\,(4653 + 1984\,\zeta_3)\,\lambda^3 + (528\,c_{HD}\,(199 + 
84\,\zeta_3) - 816\,c_{H\square}\,(521 + 224\,\zeta_3))\,\lambda^4
        \right\}_3 \nonumber\\&\hspace{-1.2cm}
        + \Big\{
-12\,c_H\,(849787 + 505488\,\zeta_3 - 98064\,\zeta_4 + 
692160\,\zeta_5)\,\lambda^4 
\nonumber\\&\hspace{-0.4cm}
+ \Big(-8\,c_{HD}\,(826273 + 485160\,\zeta_3 
- 93744\,\zeta_4 + 664320\,\zeta_5) 
\nonumber\\&\hspace{0.4cm}
+ 16\,c_{H\square}\,(1685449 + 
975344\,\zeta_3 - 189792\,\zeta_4 + 1387840\,\zeta_5)\Big)\,\lambda^5
        \Big\}_4\,,
    \nn\\[5mm]
    \mu \frac{d}{d\mu} c_{H\square} &= 
   c_{H\square} \,( 2-2\,\ep)
    + \left\{ 
        24\,c_{H\square}\,\lambda
    \right\}_1
    + \left\{ 
        -204\,c_{H\square}\,\lambda^2
    \right\}_2 
    + \left\{ 
        144\,c_{H\square}\,(39 + 8\,\zeta_3)\,\lambda^3
    \right\}_3
    \nonumber\\&\hspace{-1.2cm}
    + \left\{ 
-2880\,c_H\,\lambda^3 + \Big(-1920\,c_{HD} - 24\,c_{H\square}\,(5003 + \
2852\,\zeta_3 + 144\,\zeta_4 + 4000\,\zeta_5)\Big)\,\lambda^4
    \right\}_4\,,
     \nn\\[5mm]
    \mu \frac{d}{d\mu} c_{HD} &=
   c_{HD} \,( 2-2\,\ep)
    + \left\{ 
12\,c_{HD}\,\lambda
    \right\}_1
    + \left\{ 
-144\,c_{HD}\,\lambda^2
    \right\}_2 
    + \left\{ 
12\,c_{HD}\,(227 + 72\,\zeta_3)\,\lambda^3
    \right\}_3
    \nonumber\\&\hspace{-1.2cm}
    + \left\{ 
-96\,c_{HD}\,(789 + 409\,\zeta_3 - 36\,\zeta_4 + \
560\,\zeta_5)\,\lambda^4
    \right\}_4\,,
\end{align}
which} agree with the two-loop results of~\cite{Jenkins:2023bls} and extend them to four loops. 

The mixing between $c_{H\square}$ and $c_{HD}$ is diagonal up to three loops.
In fact, $c_{H\square}$ and $c_H$ will not mix into $c_{HD}$ at any loop order in this basis.
However, $c_{HD}$ does mix into $c_{H\square}$ starting at four loops. 
This follows from the fact that the custodial violating operator $\mathcal{O}_{HD}$ can be seen as a linear combination of a four-field
operator in the $B_4$ irreducible representation of $O(4)$ and a six-field $O(4)$-singlet operator, as becomes clear from the replacement rules in \eqref{eq:Repl2}. Operators in different irreducible representations do not mix under RG, while the mixing of the six-field singlet into the four-field singlet starts at four loops.%
    \footnote{The Feynman diagram for the mixing of $\phi^6$ into $\phi^4\partial^2$ trivially vanishes in dimensional regularization at one loop. The two-loop zero is predicted by the non-renormalization theorem of~\cite{Bern:2019wie}. It was first observed in \cite{Cao:2021cdt} that this mixing entry is also zero at three loops.}

To better expose the mixing structure at all loops, let us therefore consider the basis of $SU(2)\times U(1)$ singlet operators that is aligned with the branching from $O(4)$ irreducible representations according to
\begin{equation}
    \phi^6:\ S, \qquad \phi^4\de^2:\ S,B_4.
\end{equation}
Such a basis can be obtained from the general theory~\eqref{eq:gentheory} by the replacements
\begin{align}\label{eq:ruleBetter1}
    c^{abcdef} &\to c_6
    \left(\delta^{ab}\delta^{cd}\delta^{ef} + \text{ 14 permutations} \right) ,\\[3mm]
    c^{abcd} &\to 
    c_{4,1}\left(\delta^{ab}\delta^{cd} 
    -\frac12 \delta^{ac}\delta^{bd} - \frac12\delta^{ad}\delta^{bc}\right)
    \nonumber\\&\qquad 
    +c_{4,2} \left(\Omega^{ad}\Omega^{bc} - \Omega^{ac}\Omega^{db}
    -\frac23\delta^{ab}\delta^{cd} 
    +\frac13 \delta^{ac}\delta^{bd} + \frac13\delta^{ad}\delta^{bc}
    \right).
    \label{eq:ruleBetter2}
\end{align}
Here the structures are the same as in \eqref{eq:Repl2}, but with single coefficients multiplying them, not requiring any field redefinitions to reach this form. 
In this basis, the mixing matrix of $(c_6, c_{4,1}, c_{4,2})$ is block diagonal,
\begin{equation}
\Gamma=  \begin{pmatrix}
2-4 \ep + 108 \lambda -3444 \lambda^2+\ldots\!\! & 
-720\lambda^3+\ldots & 0 \\
 640\lambda^3+\ldots &\!\! 2-2 \ep +24 \lambda -204\lambda^2+\ldots\!\! & 0 \\
 0 & 0 & \!\!2-2 \ep + 12 \lambda - 144 \lambda^2+\ldots
    \end{pmatrix},
\end{equation}
where the zeros extend to all loop orders. These zeros follow from the fact that our renormalizable interaction respects the $O(4)$ symmetry exactly. They are non-zero if gauge and Yukawa interactions from the SMEFT are considered. 
Some of these zeros are effectively hidden if the results are given in the conventional basis described earlier.

Our complete expressions up to five loops are
{\allowdisplaybreaks
\begin{align}
    \Gamma_{11} &= 
    2-4\ep 
   +108 \lambda
   -3444 \lambda
   ^2
   +\left(71424 \zeta
   _3+167508\right) \lambda ^3 
   \nn\\&\quad
   -\left(6065856 \zeta _3-1176768
   \zeta _4+8305920 \zeta _5+10189764\right)
   \lambda ^4 \nn\\&\quad
    +\big(11816064 \zeta
   _3^2+486064224 \zeta _3-114810912 \zeta
   _4+829944576 \zeta _5 \nn\\&\hspace{3.5cm} 
    -239716800 \zeta
   _6+1001879424 \zeta _7+707542968\big)
   \lambda ^5\,,\nn\\[2mm] 
\Gamma_{12} &= 
   -720 \lambda ^3
   +\left(27648 \zeta _3+56520\right)
   \lambda ^4\nn\\&\quad
   -\left(711936 \zeta _3-82944 \zeta
   _4+4769280 \zeta _5+2993112\right) \lambda
   ^5 \nn\\&\quad
+\big(-1748736 \zeta _3^2-29273184 \zeta
   _3-1508544 \zeta _4+179316864 \zeta
   _5\nn\\&\hspace{3.5cm}
   -53654400 \zeta _6+679335552 \zeta
   _7+145659474\big) \lambda
   ^6\,,\nn\\[2mm] 
\Gamma_{21} &= 
    640 \lambda ^3
    -\left(21504 \zeta _3+59584\right) \lambda
   ^4 \,,\nn\\[2mm] 
\Gamma_{22} &= 
2-2\ep 
+24 \lambda
-204 \lambda^2
+\left(1152
   \zeta _3+5616\right) \lambda ^3 
    \nn\\&\quad
-\left(68448
   \zeta _3+3456 \zeta _4+96000 \zeta
   _5+127752\right) \lambda ^4 \nn\\&\quad
   +\big(-449280 \zeta _3^2+3914688
   \zeta _3-169344 \zeta _4+7139328 \zeta
   _5\nn\\&\hspace{3.5cm}
   -777600 \zeta _6+4318272 \zeta
   _7+3896904\big) \lambda ^5
\,,\nn\\[2mm] 
\Gamma_{33} & = 
2-2\ep 
   +12
   \lambda
   -144 \lambda ^2
   +\left(864 \zeta
   _3+2724\right) \lambda ^3\nn\\&\quad 
   -\left(39264
   \zeta _3-3456 \zeta _4+53760 \zeta
   _5+75744\right) \lambda ^4\nn\\&\quad
+\big(-203904 \zeta _3^2+2407488
   \zeta _3-158112 \zeta _4+3399168 \zeta
   _5\nn\\&\hspace{3.5cm} 
   -561600 \zeta _6+2413152 \zeta
   _7+2313624\big) \lambda ^5\,.
\end{align}
These} are specializations to $n=4$ from results which we derived in the $O(n)$ model for general $n$. 
Note that in these expressions $\lambda=\lambda_H$ as defined by \eqref{eq:Lamdim4SMEFT}.

In the complex Higgs basis, the operators that are obtained by the replacement rules Eqs.~(\ref{eq:ruleBetter1}, \ref{eq:ruleBetter2}) are
\begin{align}
    \mathcal{O}_6 &= \frac{1}{6} (H^\dagger H)^3\,, \nonumber\\
    \mathcal{O}_{4,1} &= -(H^\dagger H)(\partial_\mu H^\dagger \partial^\mu H)
    +\frac{1}{4}\partial_\mu(H^\dagger H)\partial^\mu(H^\dagger H)\,,\nonumber\\
    \mathcal{O}_{4,2} &= \frac{1}{2}(\partial_\mu H^\dagger H - H^\dagger \partial_\mu H)^2 -\frac{2}{3} \mathcal{O}_{4,1}\,,
\end{align}
which is a choice of primary operators for the Higgs sector of SMEFT at dimension six that is aligned with custodial symmetry breaking.
We note that both terms of $\mathcal{O}_{4,1}$ preserve custodial symmetry. Their relative coefficient is fixed by the primary condition~\eqref{eq:primCondition}. 
The antisymmetric combination in the first term of $\mathcal{O}_{4,2}$ violates custodial symmetry, following a general pattern of the breaking of accidental symmetries identified recently in~\cite{Grinstein:2024jqt}.
We include the second term in $\mathcal{O}_{4,2}$ to render the operator traceless in flavor space.

The block-diagonal form of the mixing in this basis exposes the structure imposed by custodial symmetry breaking $O(4)\to SU(2)\times U(1)$ by higher-dimensional operators. The blocks correspond to two singlets and one operator in the $B_4$ representation of $O(4)$. 
It would be interesting to algorithmically determine which $O(4)$ irreps correspond to $SU(2)\times U(1)$ singlets in general.

\paragraph{Dimension eight.}

In the SMEFT at dimension eight, there are six pure-scalar operators; see e.g.~\cite{Murphy:2020rsh} for one choice of basis for these operators. 
Instead of using this basis,
we align our basis with patterns of custodial symmetry breaking using $O(4)$ representation theory, as we did at dimension six.
We need to consider operators in the following irreps of $O(n)$, which are singlets under $SU(2)\times U(1)$:
\begin{equation}
 \phi^8:\ S, \qquad\phi^6\partial^2 :\ S,B_4 ,\qquad\phi^4\partial^4 :\ 2S,B_4.
\end{equation}
The operators in these representations can be obtained from operators in the general scalar theory, defined in \eqref{eq:dim8GeneralOperators} in the appendix, by the replacement rules
\begin{align}
    c^{abcdefgh} &\to 
    c_{S,8} \left( \delta^{ab}\delta^{cd}\delta^{ef}\delta^{gh} + \text{104 permutations}
        \right) , \nn\\[2mm] \label{5.22}
    c^{abcdef} &\to 
    c_{S,6}\,D_{S,6}^{abcdef} + c_{B_4,6}\Big( 
        \delta^{ab}(\Omega^{ce}\Omega^{df} + \Omega^{cf}\Omega^{de})+ \text{5 permutations} - \frac{1}{3}D_{S,6}^{abcdef}
    \Big) \,,\nn\\[2mm] 
    c_1^{abcd} &\to 
        c_{S,4,1}\left(\delta^{ab}\delta^{cd} + \delta^{ac}\delta^{bd} + \delta^{ad}\delta^{bc}\right),\nn\\[2mm]
    c_2^{abcd} &\to 
        c_{S,4,2}\left(\delta^{ab}\delta^{cd} -\frac{1}{2}\delta^{ac}\delta^{bd} -
        \frac{1}{2}\delta^{ad}\delta^{bc}\right) \nn\\&\hspace{1cm}
        +c_{B4,4}\left(\Omega^{ad}\Omega^{bc} - \Omega^{ac}\Omega^{db}
        -\frac{2}{3}\delta^{ab}\delta^{cd} 
        +\frac{1}{3}\delta^{ac}\delta^{bd} +
        \frac{1}{3}\delta^{ad}\delta^{bc}
        \right)        ,
\end{align}
where the five permutations that are left implicit are permutations of the indices $a,b,c,d$ that result in new terms. We also defined 
\begin{align}
    D_{S,6}^{abcdef} &= 4(\delta^{ab}\delta^{cd} + \delta^{ac}\delta^{bd}+\delta^{ad}\delta^{bc})\delta^{ef}
        -\delta^{ab}(\delta^{ce}\delta^{df}+\delta^{cf}\delta^{de})
        -\delta^{ac}(\delta^{be}\delta^{df}+\delta^{bf}\delta^{de})
        \nonumber\\&\hspace{-1.3cm} 
        -\delta^{ad}(\delta^{ce}\delta^{bf}+\delta^{cf}\delta^{be})
        -\delta^{bc}(\delta^{ae}\delta^{df}+\delta^{af}\delta^{de})
        -\delta^{bd}(\delta^{ae}\delta^{bf}+\delta^{cf}\delta^{ae})
        -\delta^{cd}(\delta^{ae}\delta^{bf}+\delta^{af}\delta^{be})\,,
\end{align}
and we note that this convention differs by a factor $6$ with the operator definition in 
our auxiliary files on \githubb. 

The mixing matrix that defines the RG running of 
$(c_{S,8},c_{S,6},c_{S,4,1},c_{S,4,2},c_{B_4,6},c_{B_4,4})$ is
\begin{align}\label{eq:MixingDim8}
    \Gamma=\text{diag}(4-6\ep,4-4\ep,4-2\ep,4-2\ep,4-4\ep,4-2\ep)+
    \begin{pmatrix}
        \bm A &\boldsymbol{0}\\ \boldsymbol{0} &\bm B
    \end{pmatrix}, 
\end{align} 
with
\begin{align} \nn
\bm A = 
    \begin{pmatrix}
        192\lambda-8400\lambda^2
        & -11520\lambda^3
        & \frac{76032}{5}\lambda^3-1170048\lambda^4
        & 1728 \lambda^3 + 4992 \lambda^4 \\[2mm]
        0
        & 68\lambda-\frac{4384}{3}\lambda^2
        & -\frac{1624}{15}\lambda^2+\frac{155728}{27}\lambda^3
        & \frac{424}{3}\lambda^2 - \frac{129320}{27}\lambda^3\\[2mm]
        0
        &0
        & \frac{80}{3}\lambda-\frac{5752}{27}\lambda^2
        & -\frac{40}{3}\lambda+\frac{1160}{27}\lambda^2\\[2mm]
        %
        0
        & 0 
        & -\frac{16}{3}\lambda +\frac{608}{27}\lambda^2
        & \frac{32}{3}\lambda - \frac{2776}{27}\lambda^2
    \end{pmatrix} ,
\end{align}
and 
\begin{align}\nn
    \bm B=\begin{pmatrix}
        & 56\lambda -\frac{3856}{3}\lambda^2
        & \frac{1040}{3}\lambda^2 - \frac{232768}{27}\lambda^3
        \\[2mm]
        &0
        & \frac{16}{3}\lambda -\frac{2144}{27}\lambda^2
    \end{pmatrix}.
\end{align}
Here the zero matrices 
$\boldsymbol{0}$ vanish for all loops, while the zeros inside $ \bm A$ and $ \bm B$ become non-zero at higher loops. 
This basis thereby exposes additional structure
which is not visible in other operator bases. 

We confirmed that the eigenvalues of \eqref{eq:MixingDim8} agree with the one-loop results of~\cite{DasBakshi:2022mwk}, where the renormalization was computed in the basis of~\cite{Murphy:2020rsh}, and we provide four-loop results for the mixing among the $O(n)$ singlets in the \githubb repository.\footnote{We remind the reader that the conventions of this subsection differ slightly with those of Appendix~\ref{s:Operators} and on \githubb.}

\subsection[Approximate 
\texorpdfstring{$O(n)$}{O(n)} symmetry in a hypercubic EFT]{Approximate 
\texorpdfstring{\boldmath$O(n)$\unboldmath}{O(n)}\ symmetry in a hypercubic EFT}
\label{s:approx}

As a second example, we consider an EFT in which 
the $O(n)$ symmetry is explicitly broken to the hypercubic symmetry group $S_n\ltimes (\Z_2)^n\subset O(n)$, but we will consider this breaking to be small. 
The $O(n)$ symmetry is recovered as a small symmetry breaking parameter (called $y$ below) goes to zero.
In this case, $O(n)$ is an \emph{approximate} symmetry of the theory. It is then again useful to organize the operators in terms of $O(n)$ irreps to expose flavor selection rules in the RG mixing. In addition, such an organization of the operator basis helps identifying directions in parameter space associated to additional sources of flavor symmetry breaking in the UV.
An analogous type of flavor selection rules was considered in the SMEFT and LEFT in  Refs.~\cite{Machado:2022ozb,Renner:2025cmd}.

The hypercubic EFT of $n$ scalars is defined by the dimension-four Lagrangian
\begin{equation}
\label{eq:lambdaCubicShifted}
    \lambda_{abcd}=\lambda \frac{\delta_{ab}\delta_{cd}+\delta_{ac}\delta_{bd}+\delta_{ad}\delta_{bc}}3+y\left(\delta_{abcd}- \frac{\delta_{ab}\delta_{cd}+\delta_{ac}\delta_{bd}+\delta_{ad}\delta_{bc}}{n+2}\right),
\end{equation}
where we will consider $y$ to be ``phenomenologically'' small. 
Here, $\delta^{abcd}$ is the generalized Kronecker delta function, which is 1 if all indices are equal and 0 otherwise.
We consider an alternative choice of basis in which the symmetry breaking operator is not traceless further below.

The theory with \eqref{eq:lambdaCubicShifted} has an approximate $O(n)$ symmetry, which becomes exact in the $y\to0$ limit. In fact, the small-$y$ limit can be motivated by working at/near the fixed-point -- at the hypercubic fixed-point for $n=4$ scalars in $d=3.9$ spacetime dimensions, one has $\lambda_*/y_* \approx 15$. For this reason, and to simplify the exposition, we will restrict to $n=4$ in what follows. Notice that in \eqref{eq:lambdaCubicShifted} we have written the Lagrangian in terms of operators that remain irreducible under $O(n)$: the first term transforms as a singlet $S$, and the second as a four-index traceless symmetric tensor $T_4$. We study the conformal data at the cubic fixed-point in Section~\ref{sec:cubicRes}.

The theory defined by \eqref{eq:lambdaCubicShifted} has an exact hypercubic symmetry $C_4 = S_4\ltimes(\mathbb{Z}_2)^4 \subset O(4)$. 
The EFT is therefore spanned by all singlet operators of the hypercubic symmetry. In complete generality, we will decompose this set of operators into irreducible representations of $O(4)$. For this purpose, we observe the branching rules from irreps of $C_4$ to $O(4)$ (see e.g.~\cite{Antipin:2019vdg,Bednyakov:2023lfj} as well as equation~\eqref{eq:branching-cubic} below),
\begin{align}
\label{eq:cubic-branchings}
    &S\to S, 
    &&T_4\to S\oplus\ldots, 
    &&T_6\to S\oplus \ldots, 
    &&T_8\to 2S\oplus \ldots,
\end{align}
which imply that the EFT will include all traceless symmetric $O(4)$ irreps of even rank $\neq2$,
in addition to the $O(4)$ singlet operators. 
With foresight, we have chosen the interaction parametrized by $y$ in \eqref{eq:lambdaCubicShifted} to be an operator in the $T_4$ irrep.

The fact that irreducible representations do not mix under RG shows in the absence of the $O(y)$ term in the beta function of $\lambda$,
\begin{align}
    \mu \frac{d}{d\mu} \lambda &= 
    -2\ep \lambda 
    +4 \lambda ^2
    -\frac{26 \lambda ^3}{3}
    +\frac{1}{216} \lambda ^4 (4032 \zeta_3+7176)
    +O(y^2\lambda,\lambda^5)\,.
\end{align} 
Furthermore, in the $y\to0$ limit, the 
beta function of $y$ is nothing but $y$ times the scaling dimension of the first $T_4$ operator in the $O(n)$ model,
\begin{align} 
    \mu \frac{d}{d\mu} y &= 
    y\left(
        -2\ep 
        +4 \lambda
        -\frac{34 \lambda ^2}{3}
        +\frac{1}{108} \left(3456 \zeta _3+4548\right) \lambda ^3
    \right)+O(y^2,y\,\lambda^4)\,.
\end{align} 
Expressions for general $n$, or at higher orders in $y$ and $\lambda$ can straightforwardly be extracted from our data files or from the results of~\cite{Bednyakov:2021ojn}.

\paragraph{Dimension six.}

By the branching rules, the invariant (singlet) operators in the hypercubic theory are captured by operators in the $O(n)$ model in the following irreps 
\begin{equation}
    \phi^6: \ S,\,T_4,\,T_6\,, \qquad 
    \phi^4\de^2:\, S.
\end{equation}
These operators can be obtained from the general theory by taking the tensors to be
\begin{align}
    c^{abcdef} &\to c_{6,S} \,D_{6,S}^{abcdef} 
    + c_{6,T_4} \, D_{6,T_4}^{abcdef} 
    + c_{6,T_6}\left(\delta^{abcdef}-\frac{1}{48}D_{6,1}^{abcdef}-\frac{1}{12}D_{6,2}^{abcdef}\right),\nn\\\label{eq:dim6CubicApprox2}
    c^{abcd} &\to c_{4,S}\left(\delta^{ab}\delta^{cd}
        -\frac{1}{2}\delta^{ac}\delta^{bd} 
        -\frac{1}{2}\delta^{ad}\delta^{bc}\right),
\end{align}
where we defined
\begin{align}
    D_{6,S}^{abcdef} &= \delta^{ab}\delta^{cd}\delta^{ef} + \text{14 permutations}\,,\nn\\
    D_{6,T_4}^{abcdef}&=\left(\delta^{abcd}-\frac{1}{6}\left(\delta^{ab}\delta^{cd}
            +\delta^{ac}\delta^{bd}
            +\delta^{ad}\delta^{bc}\right)\right) + \text{14 permutations}\,.
\end{align}
In this way, the mixing matrix of $(c_{6,T_4},c_{6,T_6},c_{6,S},c_{4,S})$ can be straightforwardly extracted from the data files (up to five loops for generic $y$ and any number of scalars $n$). At two loops, the mixing matrix is\\[1mm] $\Gamma = \text{diag}(2-4\ep,2-4\ep,2-4\ep,2-2\ep)+\phantom{.}$
\begin{align}  
\left(
\begin{array}{cccc}
14 \lambda  -69 \lambda ^2 &  0& 0 & 0 \\
0 & 10 \lambda -\frac{127 \lambda ^2}{3} & 0 & 0 \\
 0 & 0 & 18 \lambda-\frac{287 \lambda ^2}{3}  & -\frac{10 \lambda ^3}{3} \\
  0&0 &\frac{80}{27}\lambda^3 & 4 \lambda-\frac{17 \lambda ^2}{3} \\
\end{array}
\right)+O(y)\,,
\end{align}
where we display the four-loop (i.e.~leading order) entry 
$\frac{80}{27}\lambda^3$, while all other entries are written up to two loops.
We emphasize that the block-diagonal structure becomes exact in the $y\to0$ limit, in which case the eigenvalues are nothing but the scaling dimensions of the operators in the $O(4)$ model. 
However, all entries are generically non-zero for finite $y$.

\paragraph{Dimension eight.}

At mass dimension eight, the singlet operators under the hypercubic EFT descend from the following $O(4)$ irreps denoted with their respective multiplicities
\begin{equation}
    \phi^8:\ S,\,T_4,\,T_6,\, 2T_8 ,\qquad\phi^6\partial^2:\ S,\,T_4, \qquad \phi^4\de^4:\ 2S,\,T_4,
\end{equation}
where the 2 in front of $T_8$ is because this irrep contains the cubic singlet twice (see \eqref{eq:cubic-branchings}), while the 2 in front of $S$ indicates that there are two distinct $O(n)$ singlets of the form $\phi^4\partial^4$ (see \eqref{eq:ONscalarContent}). 
The mixing matrix of these operators can be obtained from the anomalous dimensions in the general theory using replacement rules of the form of \eqref{eq:dim6CubicApprox2}. 

The mixing matrix of $(c_{8,S},c_{6,S},c_{4,S,1},c_{4,S,2},
                        c_{8,T_4},c_{6,T_4},c_{4,T_4},
                        c_{8,T_6},c_{8,T_8,1},c_{8,T_8,2})$ is
\begin{equation}
    \Gamma= 
    \begin{pmatrix}
        \bm{A}_S&&&\\
        &\bm{B}_{T_4}&&\\
        &&\bm C_{T_6}&\\
        &&&\bm D_{T_{8}}
    \end{pmatrix}\,,
\end{equation}
where all off-diagonal entries are $O(y)$ (or higher) and 
{\allowdisplaybreaks
\begin{align}  
\bm A_S &= 
\scalebox{0.9}{$
\left(
\begin{array}{cccccccccc}
4-6\ep+ 32 \lambda -\frac{700 \lambda ^2}{3}\!\! & -\frac{80 \lambda ^3}{9} &
   \frac{352 \lambda ^3}{5}-\frac{24376 \lambda ^4}{27} & \frac{104
   \lambda ^4}{27}+8 \lambda ^3  \\
 0 &\!\! 4 - 4 \ep+\frac{34 \lambda }{3}-\frac{1096 \lambda ^2}{27} & \frac{38932
   \lambda ^3}{243}-\frac{812 \lambda ^2}{45} & \frac{212 \lambda
   ^2}{9}-\frac{32330 \lambda ^3}{243}  \\
 0 & 0 & 4 - 2 \ep+\frac{40 \lambda }{9}-\frac{1438 \lambda ^2}{243} & \frac{290
   \lambda ^2}{243}-\frac{20 \lambda }{9}  \\
 0 & 0 & \frac{152 \lambda ^2}{243}-\frac{8 \lambda }{9} & 4 - 2 \ep+\frac{16
   \lambda }{9}-\frac{694 \lambda ^2}{243}  \\
\end{array}
\right)$}\,,\nn\\[2mm]
\bm B_{T_4} &= 
 \left(
\begin{array}{cccccccccc}
4 - 6 \ep+28 \lambda -\frac{584 \lambda ^2}{3} &
   -\frac{1120 \lambda ^3}{9} & \frac{32 \lambda ^3}{3}-\frac{47816
   \lambda ^4}{405}  \\
0&4 - 4 \ep+\frac{22 \lambda }{3}-\frac{607 \lambda
   ^2}{27} & \frac{46981 \lambda ^3}{102060}-\frac{31 \lambda ^2}{378}  \\
0 & 0 &4 - 2 \ep+\frac{20 \lambda }{9}-\frac{866 \lambda^2}{243}
\end{array}
\right)\,,\nn\\[3mm]
 \bm C_{T_6} &= 
 4 - 6 \ep+24 \lambda -156 \lambda ^2  \,,\nn\\[3mm]
 \bm D_{T_8} &= \left(
\begin{array}{cccccccccc}
  4 - 6 \ep+\frac{56 \lambda }{3}-\frac{940 \lambda ^2}{9} &
   0 \\
 0 & 4 - 6 \ep+\frac{56 \lambda }{3}-\frac{940
   \lambda ^2}{9} \\
\end{array}
\right)\,.
\end{align}
In} this case, the eigenvalues of this matrix in the $y\to0$ limit are the scaling dimensions of the scalar $O(n)$ operators in the singlet and traceless symmetric representations.
Since they originate from the same operator in $O(n)$, the anomalous dimensions of $c_{8,T_8,1}$ and $c_{8,T_8,2}$ are exactly the same in the $y\to0$ limit. This degeneracy is lifted by small $y$ corrections.

\paragraph{Remark.}

Another choice of couplings would be 
\begin{equation}
\label{eq:lambdaCubicNaive}
    \lambda_{abcd}=\lambda' \frac{\delta_{ab}\delta_{cd}+\delta_{ac}\delta_{bd}+\delta_{ad}\delta_{bc}}3+y \, \delta_{abcd}\,.
\end{equation}
Here $\lambda'=\lambda$ as $y=0$. In this basis, the perturbation proportional to $y$ is not traceless, and the matrix
\begin{align}
\label{eq:betaCubicExample1}
   &\begin{pmatrix}
        \frac{\partial\beta_{\lambda'}}{\partial\lambda'} &\frac{\partial\beta_{\lambda'}}{\partial y} 
        \\
        \frac{\partial\beta_y}{\partial\lambda'} &\frac{\partial\beta_y}{\partial y} 
    \end{pmatrix}
    \nonumber\\&= 
    \resizebox{0.9\textwidth}{!}{$\begin{pmatrix}
    -2 \epsilon+\frac{2}{3} \lambda'(n+8)-\lambda^{\prime 2} (3 n+14)+ \left(2-\frac{44 \lambda' }{3}\right)y & 2 \lambda'
    -\frac{22 \lambda^{\prime 2}}{3}+ \left(-\frac{10 \lambda'
   }{3}\right)y \\
 \left(4+\left(-\frac{10 n}{9}-\frac{164}{9}\right) \lambda' \right) y & -2 \epsilon+4 \lambda'
   +\lambda^{\prime 2} \left(-\frac{5 n}{9}-\frac{82}{9}\right)+ \left(6-\frac{92 \lambda' }{3}\right)y
   \end{pmatrix}$}
\end{align}
is not diagonal in the limit $y=0$. The eigenvalues however are unchanged, noting that $\lambda'=\lambda-\frac{3y}{n+2}$. In fact,
\begin{equation}
     \begin{pmatrix}
        \frac{\partial\beta_\lambda}{\partial\lambda} &\frac{\partial\beta_\lambda}{\partial y} 
        \\
        \frac{\partial\beta_y}{\partial\lambda} &\frac{\partial\beta_y}{\partial y} 
    \end{pmatrix}
    = M\begin{pmatrix}
        \frac{\partial\beta_{\lambda'}}{\partial\lambda'} &\frac{\partial\beta_{\lambda'}}{\partial y} 
        \\
        \frac{\partial\beta_y}{\partial\lambda'} &\frac{\partial\beta_y}{\partial y} 
    \end{pmatrix}
     M^{-1}, \qquad M=\begin{pmatrix}
        1&\frac3{n+2}\\0&1
    \end{pmatrix}\,.
\end{equation}

\section{Applications to CFT: Tensor structures}\label{s:Specific}

In the previous section, we discussed how to
extract results in two specific EFTs from our general expressions.
In this section, we will systematise this procedure with CFT applications in mind, outlining the extraction of scaling dimensions in various theories.
We do not aim for a completely general algorithm, because the exact auxiliary tensors that need to be implemented, and their corresponding properties, are group-theory dependent. Instead, we introduce a collection of techniques through a broad set of examples, which can also be applied in cases that we do not cover.
We start with dimension-six operators in $O(n)$ vector models. 
This case is of particular interest, since it involves the mixing between free-theory (i.e.~bare) operators with different number of fundamental fields. 
After establishing some basics, we then consider more complicated symmetry groups encountered in condensed matter and/or CFT applications.

The groups we will consider and their corresponding CFTs are summarised in  Table~\ref{tab:CFTs}. These correspond to various $\lambda\phi^4$ theories that have been proposed to have experimental relevance for critical phenomena in 3d; see \cite{Pelissetto:2000ek}. The reader purely interested in the results can proceed to Section~\ref{sec:newData}.

\begin{table}
    \centering
    \caption{Fixed-points considered organised by symmetry group. The third column lists the number of non-zero couplings required to reach each fixed-point.}
    \label{tab:CFTs}
    \begin{tabular}{|llcc|}
        \hline
        $G$ & CFT name & \# $\lambda's$ &  Section \\
        \hline
         $O(n)$ & $O(n)$ &  $1$ &  
         \ref{s:Higgs}, \ref{generalconsiderations}, \ref{sec:ONextractions}
        \\
        $S_n\times \Z_2$ &
        Hypertetrahedral &
        2&
        \ref{sec:hypertetrahedral}
        \\
        $\Z_2^{\ n}\rtimes S_n$ & Hypercubic &  $2$ & 
        \ref{s:approx}, \ref{sec:HypercubicGT},\ \ref{sec:cubicRes}
        \\
        $O(m)^n\rtimes S_n$ & MN &  $2$ &  \ref{sec:MN}
        \\ 
        $O(m)\times O(n)/\Z_2$ & $O(m)\times O(n)$ &  $2$ & \ref{sec:OmOn}
        \\ 
        $O(m)\times O(n)$ & Biconical & $3$ &  \ref{sec:MMtheories}
        \\ 
        $U(m)\times U(n)/U(1)$ & $U(m)\times U(n)$ &   $2$ &  \ref{sec:UmUn}
        \\
        \hline
    \end{tabular}
\end{table}

\subsection{General considerations and simple examples}
\label{generalconsiderations}

The general expressions in our results take the form
\begin{equation}\label{genform}
    \frac{d\,D_A^{(i)}}{d\log\mu} = \beta_A(\lambda, \{D^{(j)}\})\,,
\end{equation}
where we collectively denote sequences of indices by a capital letter index $A={abcd\cdots}$\,.
There are potentially multiple coupling constant tensors $D^{(i)}_{A_i}$ that mix.
The basic procedure to extract data for specific theories is to organize the indices in the general expressions into a discrete set of tensor structures $T^{(i)}_A$, which are compatible with the index structure and the global-symmetry representation under consideration. This can be achieved through the following steps:
\begin{enumerate}
    \item Substitute the dimension-four coupling constant by tensor structures that are compatible with the global symmetries: $\lambda_{abcd}=\sum_i \lambda_iL^{(i)}_{abcd}$. The coupling constants $\lambda_i$ will take non-zero values at the fixed-point. 
    
    \item Substitute the general tensors of the composite operators (on both sides of \eqref{genform})
    by a set of tensor structures compatible 
    with the symmetries of $D_A$,
    \begin{equation}
        D_A = \sum_i d_iT^{(i)}_A\,.
    \end{equation}
    Here $d_i$ are coupling constants without flavor indices.
    However, the resulting expression still has a set of open indices.
   
    \item Extract index-free results by acting on the expression with ``projectors'' $P^{(i)}_A $ such that\footnote{The term ``projectors'' is meant in a broad sense. There are different ways to implement this step, where the ``projectors'' do not necessarily satisfy the orthonormality condition. In the simplest cases, this step can be implemented as just ``taking the coefficient of'' various tensor structures, taking into account any overall factors that arise by this method. The classical part of the scaling dimension provides a useful cross-check. We give examples below.}
    \begin{equation}\label{eq:proj1}
        \sum_A P^{(i)}_A T^{(j)}_A=\delta^{ij}.
    \end{equation}
    \item Take coefficients of the couplings $d_i$. This gives the anomalous dimension matrix $\gamma_{ij}$ where $i,j$ runs over the all operators in a given representation with compatible tensor structures. The anomalous dimensions of scaling operators are the eigenvalues of this matrix. 
\end{enumerate}
Below we will first consider some straightforward examples of this procedure for dimension-six scalar operators in a theory with $O(n)$ symmetry. We subsequently discuss different strategies for more involved symmetry groups of experimentally relevant CFTs.

\paragraph{Simple examples.}

Let us consider the results for scalars at dimension six with $O(n)$ symmetry. The coupling substitution is 
\begin{equation}
\label{eq:ONcouplingSub}
\lambda_{abcd} = \frac\lambda3 (\delta_{ab}\delta_{cd} + \delta_{ac}\delta_{bd} +\delta_{ad}\delta_{bc})\,.  
\end{equation}
We give the fixed-point value for $\lambda$ in \eqref{eq:critCouplingON} below. 
The data files of the \githubb repository stores two expressions, which we denote $R^{(6)}_{abcdef}$ and $R^{(4)}_{abcd}$.\footnote{In the data file, they are denoted \texttt{res[6,0,6,1]} and \texttt{res[6,0,4,1]} and the indices are \texttt{a1},\ldots,\texttt{a6} and \texttt{a1},\ldots,\texttt{a4}, respectively. The results in the data files also contain factors of \texttt l, which is a formal loop counting parameter and which we will set to 1 here.} These depend on two coupling constant tensors that satisfy the following symmetries: $D^{(6,6)}_{abcdef}$ is completely symmetric, while $D^{(6,4)}_{abcd}$ is symmetric in the first two and the last two indices and in addition satisfies
(c.f.~\eqref{primdim6})
\begin{equation}
    D^{(6,4)}_{abcd}=D^{(6,4)}_{cdab}, \qquad D^{(6,4)}_{abcd}+D^{(6,4)}_{cabd}+D^{(6,4)}_{bcad}=0\,,
\end{equation}
corresponding to the primary condition.

\vspace{2mm}\noindent
Let us go through all irreps of $O(n)$ that have scalar operators at dimension six.

\begin{itemize}
    \item 
The simplest case is the rank-six traceless symmetric irrep called $T_6$, with tensor structures
\begin{equation}
\label{eq:substT6}
    D^{(6,6)}_{abcdef}= C_1 \,t_at_bt_ct_dt_et_f\,,\qquad D^{(6,4)}_{abcd}=0\,,
\end{equation}
where $t_a$ is an ancillary polarization vector satisfying $t_at_a=0$. 
There is no four-index operator in this representation and hence also $R^{(4)}_{abcd}$ is set to zero by the replacements. 
The substitutions \eqref{eq:substT6} together with the coupling replacement \eqref{eq:ONcouplingSub} give
\begin{align}
    R^{(6)}_{abcdef}&= 2\,C_1\,t_at_bt_ct_dt_et_f
    -4\epsilon \, C_1\,t_at_bt_ct_dt_et_f
    +10 \lambda \, C_1\,t_at_bt_ct_dt_et_f\nonumber\\
&\quad 
    -\left(\frac{109}3+\frac{3n}2\right)\lambda^2\,C_1\,t_at_bt_ct_dt_et_f+\ldots.
\end{align}
Taking the coefficient of $t_at_bt_ct_dt_et_f$ (this step corresponds to acting with one of the aforementioned ``projectors'') and of $C_1$ (giving a $1\times1$ ``matrix''), and adding $d=4-\eps=4-2\epsilon$, we find
\begin{align}
\nonumber
    \Delta_{\phi^6_{T_6}}&=6-3\eps+10\lambda-\left(\frac{109}3+\frac{3n}2\right)\lambda^2+\ldots
    \\
    &=6-3\eps+\frac{30}{n+8}\eps-\frac{3  (9 n^2+110 n+904)}{2 (n+8)^3}\eps^2+\ldots,
\end{align}
where in the last line we substituted the critical coupling \eqref{eq:critCouplingON}. This result agrees with \cite{Bednyakov:2022guj} to the five-loop order computed here.

\item Next, we consider the $T_4$ irrep. Again, only $D^{(6,6)}_{abcdef}$ is compatible with the symmetry (four traceless-symmetric indices), but our substitution requires a sum over permutations,
\begin{equation}
\label{eq:substT4}
    D^{(6,6)}_{abcdef}= C_1 (\delta_{ab}t_ct_dt_et_f+\text{14 perms})\,,\qquad D^{(6,4)}_{abcd}= 0 \,.
\end{equation}
The final result is
\begin{equation}
    \Delta_{\phi^6_{T_4}}=6 - 3 \eps + \frac{n+38}{n+8}\eps - \frac{ 49 n^2+742n+4600}{2(n+8)^3}\eps^2+\ldots ,
\end{equation}
which is a new result at $O(\eps^5)$, improving on $O(\eps)$ from \cite{Wegner:1972zz}.

\item For the $T_2$ irrep, there now exist compatible structures for both $D^{(6,6)}$ and $D^{(6,4)}$,
\begin{align}
    D^{(6,6)}_{abcdef}&= C_1 (\delta_{ab}\delta_{cd}t_et_f+\text{44 perms})\,,\nonumber
    \\    
    D^{(6,4)}_{abcd}&=C_2(2\delta_{ab}t_ct_d+2\delta_{cd}t_at_b-\delta_{ac}t_bt_d-\delta_{ad}t_bt_c-\delta_{bc}t_at_d-\delta_{bd}t_at_c) \,.
\end{align}
The step of taking coefficients to generate the $2\times2$ anomalous dimension matrix now proceeds as follows. In $R^{(6)}_{abcdef}$ take coefficient of $\delta_{ab}\delta_{cd}t_et_f$ and in $R^{(4)}_{abcd}$ take half the coefficient of $\delta_{ab}t_ct_d$. 
The resulting two expressions then correspond to ${dC_1/d\log\mu=\beta_{C_1}(C_1,C_2)}$ and ${dC_2/d\log\mu=\beta_{C_2}(C_1,C_2)}$.
These expressions can be combined in matrix form by taking the coefficients of $C_1$ and $C_2$ on the RHS, resulting in\footnote{Keeping track of the loop order, the top-right corner is two-loop while the bottom-left corner is four-loop.}
\begin{equation}
    \Gamma= \begin{pmatrix}
        2-2\eps+\left(14+\frac{2n}3\right)\lambda+\ldots & \frac{20(2-n)}{27}\lambda^3
        \\
       \frac{5 (n+4) (n+6)}{81}  \lambda ^3+\ldots& 2-\eps+\frac{4+n}3\lambda+\ldots
    \end{pmatrix}.
\end{equation}
The eigenvalues can then be studied at the fixed-point:
\begin{equation}
    \Delta_{\square\phi^4_T}=6-2\eps+\frac{n+4}{n+8}\eps+\ldots, \quad 
\Delta_{\phi^6_T}=6-3\eps+\frac{2n+42}{n+8}\eps+\ldots,
\end{equation}
where we indicate that the first of these eigenvalues corresponds to an operator with the leading-order field content $\phi^4\de^2$, and the second with leading-order field content $\phi^6$. 
\item The extractions of singlets $S$ proceeds similarly, with the appropriate substitutions given in \eqref{eq:ONsubstitutions}. The results have already been extracted to five-loop order in \cite{RoosmaleNepveu:2024zlz}, and at the fixed-point they take the form
\begin{equation}
    \Delta_{\square\phi^4_S}=6-2\eps+\frac{2(n+2)}{n+8}\eps+\ldots, \quad 
\Delta_{\phi^6_S}=6-3\eps+\frac{3(n+14)}{n+8}\eps+\ldots.
\end{equation}
\item The $B_4$ irrep (with Young tableau $ \,\raisebox{1mm}{\scalebox{0.35}{$\ydiagram{2,2}$}}$\,) only admits a $\phi^4\de^2$ structure. We use the substitutions $D^{(6,6)}_{abcdef}=0$ and\footnote{We impose $t_at_a=t_as_a=s_as_a=0$. Note that the tensor in \eqref{eq:substB4} does not fully capture all transformation properties of the $B_4$ irrep, but for the case at hand, it is enough to correctly extract the operator.}
\begin{equation}
\label{eq:substB4}
   D^{(6,4)}_{abcd} = C(2t_at_bs_cs_d+2t_ct_ds_as_b-t_at_cs_bs_d-t_at_ds_bs_c-t_bt_cs_as_d-t_bt_ds_as_c)\,.
\end{equation}
This leads to
\begin{equation}
  \Delta_{\square\phi^4_{B_4}}=6-2\eps+\frac{6}{n+8}\eps -\frac{7 n^2-8 n-152}{2 (n+8)^3}\eps^2+\ldots,
\end{equation}
finishing the set of dimension-6 scalar operators in the $O(n)$ CFT. 
\end{itemize}

\vspace{2mm}\noindent 
A few remarks on these examples and their generalizations are in order. 

\paragraph{Remark 1 - Multiple compatible tensor structures.}
There are scenarios in which a given irrep may support more than one tensor structure for a given number of fundamental fields $\phi$ and derivative insertions $\partial$. 
This was encountered already in Section~\ref{s:Higgs} above, see \eqref{eq:ruleBetter2}, and is discussed in detail in \cite[Sec.~3.1]{Bednyakov:2023lfj}.

\paragraph{Remark 2 - Contraction of indices.}
It is often convenient to contract a beta function with auxiliary tensors that respect its symmetries. For instance, fully permutation invariant expressions can be contracted with a set of commuting auxiliary vectors $A_a$, such that the result has no open indices left, e.g. 
\begin{align}
    &\beta_{abcd} = \frac{d\lambda_{abcd}}{d\log\mu} \quad
    \rightarrow \quad 
    A_a A_b A_c A_d \, \beta_{abcd} = \beta (\lambda)\,.
\end{align}
Upon additionally replacing $\lambda_{abcd}$ with tensor structures, this expression does not have any indices at all.
In the previous example with the $\phi^6$ operator in the $T_6$ representation of $O(n)$, with coupling $D_{abcdef}^{(6,6)} = C_1\,t_a t_b t_c t_d t_e t_f$, the beta function similarly becomes
\begin{equation}
     A_a A_b A_c A_d A_e A_f\,
    \frac{d\,D_{abcdef}^{(6,6)}}
    {d\log\mu}
    = \beta_{C_1} \, T_6 \,,
\end{equation}
where $T_6 \equiv (A_a t_a)^6$ 
and $\beta_{C_1} = dC_1/d\log\mu$, now an index-free expression (a power series in the critical coupling) from which we can easily read off the anomalous dimension.

As the number of indices considered increases, for example by considering heavier composite operators, multiplying by the $A$ vectors provides a significant speed up, without affecting the generality of the computation. For example, in the case of $\phi^6$ operators it equates up to $6!=720$ terms. 

If we had a coupling that was not totally symmetric in all its indices, the above procedure should be modified by multiplying by some auxiliary tensor with the same symmetry properties as the coupling. For example, if the coupling was totally antisymmetric, we would then multiply by a totally antisymmetric auxiliary tensor $A_{ijk...}$.

\paragraph{Remark 3 - Change of index labels.}

Not all fixed-points are conveniently expressed in terms of a fundamental field\footnote{By which we mean the field from which the Lagrangian is built.} with a single index. For example, in cases where the fundamental operator transforms in the defining representation of some group that can parameterized as a replica theory $S_n \ltimes G^n$ \cite{Kousvos:2021rar}, where $G$ is some group, the field is more naturally parameterized as $\phi^a_r$. 
Here $a=1,...,n$ labels the different replicas, while $r=1,...,m$ runs through the components of the vector representation of $G$.  
We will see examples of this below, where we discuss how to relate multi-index fields to single-index fields, starting from Section~\ref{sec:MN}.

\subsection{Hypercubic Theories}
\label{sec:HypercubicGT}
Next we consider hypercubic theories with global symmetry $S_n \ltimes (\Z_2)^n$. This is one of the main applications of our paper, and in Section~\ref{sec:cubicRes} we perform a comprehensive extraction of the low-lying operator spectrum at five loops for $n=3$ and $n=4$. 
The main practical difference with the above examples concerns the existence of additional invariant tensors compared to $O(n)$-symmetric theories. These are multi-index Kronecker deltas such as $\delta_{abcd}$. We thus need to introduce new identities between auxiliary tensors that take these into account. For example, in the case of the auxiliary vectors $t_a$ discussed in the preceding subsections for $O(n)$, which satisfy $\delta_{ab} t_a t_b =0$, we now need to supplement these identities with new ones, such as $\delta_{abcd} t_a t_b =0$. Such identities have been laid out in \cite{Bednyakov:2023lfj}.

The Lagrangian needed to reach the fixed-point is 
\begin{equation}
    \mathcal{L} = \mathcal{L}_{kin} - \frac{\lambda_{4,1}}{4!}\frac{\delta_{ab}\delta_{cd}+\delta_{ac}\delta_{bd}+\delta_{ad}\delta_{bc}}{3}\,\phi^a \phi^b \phi^c \phi^d -\frac{\lambda_{4,2}}{4!}\,\delta_{abcd}\,\phi^a \phi^b \phi^c \phi^d\,,
    \label{hypercubicLag}
\end{equation}
where one recovers the $O(n)$ model as $\lambda_{4,2} \rightarrow 0$. The procedure, then, for extracting a given operator in a given representation of the global symmetry proceeds in the same way as the $O(n)$ models of the preceding section.

\subsection{MN and General Replica Theories}
\label{sec:MN}

We now proceed to a slightly more complicated example that will involve transforming notations between multi-indexed fundamental fields (which are often the most natural way to express group action) to single-indexed fields (the notation in which the results of this paper are made public). A first example is to generalize $S_n \ltimes (\Z_2)^n $ to $S_n \ltimes O(m)^n$, after which the generalization to $S_n \ltimes G^n$ for any $G$ should be straightforward. Theories with $S_n \ltimes O(m)^n$ global symmetry have been of active interest in the bootstrap; see e.g. \cite{Stergiou:2019dcv,Henriksson:2021lwn,Kousvos:2021rar} and references therein for applications and earlier perturbative studies.

Since the action of the symmetry group on $\phi$ can no longer be expressed naturally using a single index, we need to introduce new tensors which we will call $T$.\footnote{Several examples involving the use of such tensors can be found in \cite{Bednyakov:2021ojn}.} The relations that these tensors obey will depend on the specific group in question, but their practical utility is to transform a field carrying a single index, into a field with more indices. Invariant tensors will then be expressible solely in terms of generalized Kronecker delta functions (e.g.~$\delta_{abcdef}$, $\delta_{abcdefgh}$, ...), which are easy to implement and manipulate. 
The fact that everything can be expressed solely in terms of Kronecker deltas is the reason we call this choice ``natural''. The tensors $T$ are defined through the relation $\phi^a_r = (T^a_r)^i \phi^i$, which maps the field between notations. This map is in fact a trivial relabeling of the fields, as we will illustrate.

The Lagrangian with global symmetry $S_n \ltimes O(m)^n$ in terms of the two-index fundamental fields $\phi^a_r$, where the upper index $a$ transforms under the permutations $S_n$, and the lower index $r$ transforms under $O(m)$, can be written as 
\begin{align}
    \mathcal{L} = \mathcal{L}_{kin} &+ \lambda^1_{4} (\delta_{ab}\delta_{cd} \, \delta_{rs}\delta_{tu}+\delta_{ac}\delta_{bd} \, \delta_{rt}\delta_{su}+\delta_{ad}\delta_{bc} \, \delta_{ru}\delta_{ts}) \phi^a_r \phi^b_s \phi^c_t \phi^d_u \nn\\
    &+ \lambda^2_4 \delta_{abcd} (\delta_{rs}\delta_{tu}+\delta_{rt}\delta_{su}+\delta_{ru}\delta_{ts})\phi^a_r \phi^b_s \phi^c_t \phi^d_u\,.
\end{align}
In order to extract results from our general formulas we need to bring this into the form 
\begin{equation}
    \mathcal{L} = \mathcal{L}_{kin} + \lambda_{abcd} \, \phi^a \phi^b \phi^c \phi^d\,.
\end{equation}
We can do this using the $(T^a_r)^i$ tensors, to read off 
\begin{align}
    \lambda_{ijkl} & =\lambda^1_{4} (\delta_{ab}\delta_{cd}\delta_{rs}\delta_{tu}+\delta_{ac}\delta_{bd}\delta_{rt}\delta_{su}+\delta_{ad}\delta_{bc}\delta_{ru}\delta_{ts}) (T^a_r)^i (T^a_s)^j (T^a_t)^k (T^a_u)^l \nn\\
    & \quad+ \lambda^2_4 \delta_{abcd} (\delta_{rs}\delta_{tu}+\delta_{rt}\delta_{su}+\delta_{ru}\delta_{ts}) (T^a_r)^i (T^a_s)^j (T^a_t)^k (T^a_u)^l\,.
\end{align}
In order to perform practical calculations we will also need the identities
\begin{align}
    (T^a_r)^i (T^b_s)^j \delta_{ij} &= \delta_{ab} \delta_{rs}\,,\\
    (T^a_r)^i (T^a_r)^j &= \delta_{ij} \,,
\end{align}
which follow directly, e.g. in the case of $n=2$ and $m=2$ for illustration purposes, if we make the labeling choice\footnote{They can also be obtained by calculating the two point function in the two notations: $\langle \phi_i \phi_j \rangle \sim \delta_{ij} \sim (T^a_r)^i (T^b_s)^j \langle \phi^a_r \phi^b_s \rangle\sim (T^a_r)^i (T^b_s)^j \delta_{ab} \delta_{rs} $. Similar statements will hold for all subsequent symmetry groups we consider below, which necessitate $T$ tensors to change the number of indices the fundamental field $\phi$ carries.}
\begin{align}
    (\phi^1,\phi^2,\phi^3,\phi^4)=(\phi^1_1,\phi^1_2,\phi^2_1,\phi^2_2)\,.
\end{align}
We reiterate that the above is simply a relabeling of the fields which makes the transformation properties of the global symmetry more transparent. Its practical utility is that all fields on the right hand side have correlators that can be expressed solely in terms of Kronecker deltas, which are particularly easy to deal with systematically. 
\subsection{Direct-product theories}

Direct-product theories (as the name implies) have group theory that is determined by the product of the involved groups. However, there is an important subtlety, which we will now explain. 
When referring to direct-product theories, we will mod out some subgroup from the product of groups we are considering. For example, if we consider a fundamental field $\phi_{ar}$, where the first index $a$ transforms in the vector representation of some group $O(m)$ and the second index $r$ transforms in the vector of some group $O(n)$, the correct symmetry of the resulting fixed-point is $G=O(m)\times O(n)/\Z_2$. This is because the $\Z_2$ subgroups in both factors act (identically) as $\phi_{ar} \rightarrow -\phi_{ar}$, resulting in an overcounting. This is different from the multi-mass theories which will be discussed later, where $G=O(m)\times O(n)$ (without the modded out $\Z_2$). In these the $O(m)$ index is carried by a field $\phi_a$, and the $O(n)$ index is carried by a separate field $\tilde{\phi}_i$.

Direct product theories have received a large amount of attention in the conformal bootstrap; see e.g. \cite{Nakayama:2014lva,Nakayama:2014sba,Henriksson:2020fqi,Dowens:2020cua,Reehorst:2024vyq,Kousvos:2022ewl} and references therein.

\paragraph{$\bm{O(m)\times O(n)/Z_2}$.}
\label{sec:OmOn}

Let us start by expressing the Lagrangian in its most convenient and intuitive form,
\begin{align}
    \mathcal{L} = \mathcal{L}_{kin} &+ \lambda^1_{4} (\delta_{ab}\delta_{cd}\delta_{rs}\delta_{tu}+\delta_{ac}\delta_{bd}\delta_{rt}\delta_{su}+\delta_{ad}\delta_{bc}\delta_{ru}\delta_{ts}) \phi_{ar} \phi_{bs} \phi_{ct} \phi_{du} \nn\\
    &+ \lambda^2_4 \delta_{ab}\delta_{cd}(\delta_{rt}\delta_{su}+\delta_{ru}\delta_{ts}) \phi_{ar} \phi_{bs} \phi_{ct} \phi_{du}\nn\\
    &+ \lambda^2_4 \delta_{ac}\delta_{bd}(\delta_{rs}\delta_{tu}+\delta_{ru}\delta_{ts})\phi_{ar} \phi_{bs} \phi_{ct} \phi_{du} \nn\\
    &+ \lambda^2_4 \delta_{ad}\delta_{bc}(\delta_{rs}\delta_{tu}+\delta_{rt}\delta_{su})\phi_{ar} \phi_{bs} \phi_{ct} \phi_{du}\,.
\label{OmOnLag}
\end{align}
The first line corresponds to the usual $O(N)$ symmetric term, with $N=mn$, whereas the consequent lines break the symmetry from $O(mn)$ to $O(m)\times O(n) /\Z_2$. In order to rewrite this in terms of the original interaction term notation $\lambda_{ijkl} \phi_i \phi_j \phi_k \phi_l$, we employ the $T$ introduced above. In particular, we substitute $\phi_{ar}=T_{ar}^i \phi_i$, which as discussed is related to a simple relabeling of the fields. We thus obtain
\begin{align}
    \lambda_{ijkl} =  & \lambda^1_{4} (\delta_{ab}\delta_{cd}\delta_{rs}\delta_{tu}+\delta_{ac}\delta_{bd}\delta_{rt}\delta_{su}+\delta_{ad}\delta_{bc}\delta_{ru}\delta_{ts}) T_{ar}^i T_{bs}^j T_{ct}^k T_{du}^l \nn\\
    +& \lambda^2_4 \delta_{ab}\delta_{cd}(\delta_{rt}\delta_{su}+\delta_{ru}\delta_{ts}) T_{ar}^i T_{bs}^j T_{ct}^k T_{du}^l \nn\\
    +& \lambda^2_4 \delta_{ac}\delta_{bd}(\delta_{rs}\delta_{tu}+\delta_{ru}\delta_{ts})T_{ar}^i T_{bs}^j T_{ct}^k T_{du}^l \nn\\
    +& \lambda^2_4 \delta_{ad}\delta_{bc}(\delta_{rs}\delta_{tu}+\delta_{rt}\delta_{su})T_{ar}^i T_{bs}^j T_{ct}^k T_{du}^l \,,
\end{align}
and using the relation $T_{ar}^i T_{bs}^i = \delta_{ab}\delta_{rs}$ as well as the relations between Kronecker deltas, one can extract the necessary CFT data. 

\paragraph{$\bm{U(m)\times U(n)/U(1)}$.}
\label{sec:UmUn}
We now proceed to a generalization of $O(m)\times O(n)/\Z_2$ (see \cite{Kousvos:2022ewl}) by making the fundamental field of the Lagrangian complex,
\begin{equation}
    \mathcal{L} = \mathcal{L}_{kin} + \lambda^1_4 (\Phi^\dagger_{ar} \Phi^\dagger_{bs} \Phi_{ar} \Phi_{bs}) + \lambda^2_4 (\Phi^\dagger_{ar} \Phi^\dagger_{bs} \Phi_{as} \Phi_{br})\,,
\end{equation}
where, similarly to before, the operator multiplying $\lambda^1_4$ is $O(N)$ symmetric, with $N=2mn$ now, and the operator multiplying $\lambda^2_4$ breaks the symmetry down to $U(m)\times U(n) /U(1)$. The additional complication compared to $O(m)\times O(n)/\Z_2$ is that we need to reduce the original complex fields $\Phi_{ar}$ onto real-field components $\phi_i$. This can still be done using $T$ tensors, but we will now use two of them; $T$ and ${\tilde T}$; one for the real part and one for the imaginary part,
\begin{equation}
    \Phi_{ar} = T_{ar}^i \phi_i + i  {\tilde T}_{ar}^i \phi_i\,.
\end{equation}
The logic of the preceding subsections now follows through verbatim. We obtain
\begin{align}
    \lambda_{ijkl}&=\lambda^1_2 (T^i_{ar}-i {\tilde T}^i_{ar}) (T^i_{bs}-i {\tilde T}^i_{bs}) (T^i_{ar}+i {\tilde T}^i_{ar}) (T^i_{bs}+i {\tilde T}^i_{bs})\nonumber\\
    &\quad + \lambda^1_2 (T^i_{ar}-i {\tilde T}^i_{ar}) (T^i_{bs}-i {\tilde T}^i_{bs}) (T^i_{as}+i {\tilde T}^i_{as}) (T^i_{br}+i {\tilde T}^i_{br})\,.
    \label{UmUncoupling}
\end{align}
While not explicitly done in \eqref{UmUncoupling}, one should symmetrize the expression over the indices $i,j,k,l$. It now suffices to use the following identities in order to extract data: $T^i_{ar} T^i_{bs} =\delta_{ar} \delta_{bs}$, ${\tilde T}^i_{ar} {\tilde T}^i_{bs} =\delta_{ar}\delta_{bs} $ and $T^i_{ar} {\tilde T}^i_{bs}=0$, in addition to the usual identities between Kronecker deltas. Various relations between the possible notations for these theories are given in the appendices of \cite{Kousvos:2022ewl}. 

\subsection{Multi-mass theories}
\label{sec:MMtheories}

Let us now proceed to a theory that has more than one mass term. 
This means a case where the global symmetry allows multiple invariant operators built from two fields and no derivatives.
In terms of critical phenomena, this means that multiple masses would need to be tuned to zero simultaneously to reach a critical point.\footnote{
A rather amusing exotic possibility, however, is that in a theory with $N$ distinct masses the operators multiplying all but one of them obtain a large enough anomalous dimension to become irrelevant. Thus leading the theory to become critical instead of multicritical. We are however not aware of any such examples.} 

Let us start by writing down the Lagrangian 
\begin{equation}
    \mathcal{L}(\phi,\tilde \phi) = \mathcal{L}_{O(m)}(\phi) 
    + \mathcal{L}_{O(n)}(\tilde \phi) 
    + \lambda_3 \, \phi_i \phi_i {\tilde \phi}_a {\tilde \phi}_a\,,
    \label{biconicalLag}
\end{equation}
where both $\phi$ and ${\tilde \phi}$ are real fields. 
Additionally, we have 
\begin{align}
    \mathcal{L}_{O(m)}(\phi)&=\mathcal{L}_{kin}+ \lambda_1 (\delta_{ij}\delta_{kl}+\delta_{ik}\delta_{jl}+\delta_{il}\delta_{jk}) \phi_i \phi_j \phi_k \phi_l \,, \nn\\
    \mathcal{L}_{O(n)}(\tilde \phi)&=\mathcal{L}_{kin}+ \lambda_2 (\delta_{ab}\delta_{cd}+\delta_{ac}\delta_{bd}+\delta_{ad}\delta_{bc}){\tilde \phi_a} {\tilde \phi_b} {\tilde \phi_c} {\tilde \phi_d}\,.
\end{align}
Where we have not explicitly written out the masses for $\mathcal{L}_{O(m)}$ and for $\mathcal{L}_{O(n)}$. This means that in $d=4-\varepsilon$ there are at least two relevant operators at the interacting fixed-point.
Furthermore, we observe that if we set $\lambda_3=0$ we recover two decoupled Lagrangians with $O(m)$ and $O(n)$ symmetry,  respectively. 

To match with our results, we need to repackage the Lagrangian \eqref{biconicalLag} in terms of single index fields $\chi_r$, where the index $r$ runs from $1$ to $m+n$.\footnote{We use $\chi$ to avoid any notational confusion with $\phi$ and ${\tilde \phi}$.} First, we define
\begin{equation}
    \chi_r = T_r^i \phi_i + {\tilde T}_r^a {\tilde \phi}_a\,.
\end{equation}
Or, conversely
\begin{equation}
    \phi_i = T_r^i \chi_r\,,\qquad 
    {\tilde \phi}_a = {\tilde T}_r^a \chi_r\,,
\end{equation}
which can be derived using $T_r^i {\tilde T}_r^a=0$, $T_r^i T_r^j = \delta_{ij}$ and ${\tilde T}_r^a {\tilde T}_r^b = \delta_{ab}$. As a vector, we have
\begin{gather}
 \chi_r
 =
  \begin{bmatrix} \phi_i \\ {\tilde \phi_a}\end{bmatrix}.
\end{gather}
Finally, if we set
\begin{align}
    \lambda_{rstu}&=\lambda_1 T^i_r T^i_s T^j_t T^j_u
                  +\lambda_2 {\tilde T}^a_r {\tilde T}^a_s {\tilde T}^b_t {\tilde T^b}_u
                  +\lambda_3  T^i_r T^j_s {\tilde T}^b_t {\tilde T^b}_u\,,
\end{align}
where the RHS should be symmetrized over $r,s,t,u$ (but we didn't, for the sake of compactness), and plug this into 
\begin{equation}
    \mathcal{L}_{O(m) \times O(n)} =\mathcal{L}_{kin}+ \lambda_{rstu} \chi_r \chi_s \chi_t \chi_u\,,
\end{equation}
one recovers \eqref{biconicalLag}. Once again, everything can now be computed by applying identities of Kronecker delta functions and the $T$ tensors.

\subsection{Hypertetrahedral}
\label{sec:hypertetrahedral}
As a last example let us briefly mention hypertetrahedral theories \cite{Zia:1975ha}. These have global symmetry $G=S_n \times \Z_2$, where the purpose of the $\Z_2$ is to forbid $\phi^3$ type invariants that are otherwise allowed by $S_n$. These theories have been of interest in the bootstrap in \cite{Rong:2017cow,Stergiou:2018gjj}. In particular, the fundamental field of the Lagrangian transforms in the $(n-1)$-dimensional representation of $G$, since the $n$-dimensional defining irrep is in fact reducible:
\begin{gather}
    \phi_i =    \begin{bmatrix} \phi_1 \\ \phi_2 \\ \ldots\\
    \phi_n\end{bmatrix}=   \begin{bmatrix} \phi_1 -\frac{1}{n} \left(\phi_1+\phi_2\ldots + \phi_n\right) \\ \phi_2 -\frac{1}{n} \left(\phi_1+\phi_2\ldots + \phi_n\right) \\ \ldots\\
    \phi_n -\frac{1}{n} \left(\phi_1+\phi_2\ldots + \phi_n\right)\end{bmatrix} + \frac{1}{n}\left(\phi_1+\phi_2+\ldots + \phi_n\right) = \tilde{\phi}_i+\frac{1}{n}R\,,
\end{gather}
where the $\phi_i$ represents a field in the defining representation, $\tilde{\phi}_i$ is the fundamental field with which the Lagrangian will be built, and $R$ is simply a one-dimensional irrep that is odd under the $\Z_2$ in $G$ (which acts as $\phi_i\to-\phi_i$). Hence, $\sum_i \tilde{\phi}_i =0$. The Lagrangian can then be written as
\begin{equation}
    \mathcal{L}=\mathcal{L}_{kin} + \lambda_1 \delta_{abcd} \tilde{\phi}^a \tilde{\phi}^b \tilde{\phi}^c \tilde{\phi}^d + \lambda_2 (\delta_{ab}\delta_{cd}+\delta_{ad}\delta_{bc}+\delta_{ac}\delta_{bd}) \tilde{\phi}^a \tilde{\phi}^b \tilde{\phi}^c \tilde{\phi}^d\,.
\end{equation}
Given that the $n$ fields $\tilde{\phi}^a$ can be expressed in terms of $n-1$ independent fields, it is customary to introduce vielbeins: $\tilde{\phi}^a ={e^a}_i \chi_i$  (see \cite{Zia:1975ha} for relations between them).\footnote{In fact, one can think of them as a special case of the $T$ tensors introduced previously.} Here the $\chi_i$ are a collection of $n-1$ real fields.
The upper index on the vielbeins runs from $a=1$ to $n$, whereas the lower runs from $i=1$ to $n-1$. It thus becomes clear, then, that to consider hypertetrahedral theories we need to make the following replacement in our theory-independent results
\begin{equation}
    \lambda_{ijkl} = \left(\lambda_1  \delta_{abcd} +\lambda_2 \left(\delta_{ab}\delta_{cd}+\delta_{ad}\delta_{bc}+\delta_{ac}\delta_{bd} \right) \right){e^a}_i {e^b}_j {e^c}_k {e^d}_l\,.
\end{equation}
At this point, a couple of esoteric comments are at hand: firstly, notice that since the invariant tensors of the hypertetrahedral theory approach those of the hypercubic theory at large $n$, the two should have the same large-$n$ limit. For example, the two-point function has the following tensor structure $\langle \tilde{\phi}_i \tilde{\phi}_j \rangle \sim (\delta_{ij}-\frac{1}{n}\omega_i \omega_j)$, where $\omega_i$ is equal to one for any value of $i$. The hypertetrahedral and hypercubic tensors differ from each other in factors of $\omega_i$ that are $1/n$ suppressed. Secondly, if one wishes to express everything solely in terms  of Kronecker deltas, it suffices to express the fundamental field in terms of two indices $X_{ij}$, where $\delta_{ij}X_{ij}=0$ and $X_{ij}=0$ if $i  \neq j$. Then, for example, the two-point function becomes $\langle X_{ij} X_{kl} \rangle \sim (\delta_{ijkl}-\frac{1}{n}\delta_{ij}\delta_{kl})$.

\section[\texorpdfstring{Applications to CFT: New data for $\Z_2$, $\boldsymbol{O(n)}$ and hypercubic symmetry}{Applications to CFT: New data for Z_2, O(n) and hypercubic symmetry}]{Applications to CFT: New data for $\Z_2$, $\boldsymbol{O(n)}$ and hypercubic symmetry}
\label{sec:newData}

In Subsections \ref{sec:Isingextractions}, \ref{sec:ONextractions} and \ref{sec:cubicRes} we provide an exhaustive tabulation of all extracted data from our general results, for the Ising, $O(n)$ and hypercubic theories respectively.
For each theory we extract the scaling dimensions of all possible operators with engineering dimension up to $\Delta=6$
and that transform in Lorentz irreps up to rank two (i.e.~given by Young tableaux with at most two boxes). In particular, for $O(n)$ symmetric theories we provide five-loop data for more than $30$ operators, and in the hypercubic case for more than $100$. 
In each respective subsection, we outline the impact these results have had, or are expected to have, when paired with the numerical conformal bootstrap and other non-perturbative methods.

To make the new results for Ising and $O(n)$ available in computer-readable format, the data file of the Arxiv submission of \cite{Henriksson:2022rnm} has been updated. A version of this data file, alongside a \texttt{pdf} summary of all new results since the publication of \cite{Henriksson:2022rnm}, can also be found at
\begin{center}
    \href{https://github.com/johhen1/ON-model}{https://github.com/johhen1/ON-model}\,.
\end{center}

\subsection{New data for the Ising CFT}
\label{sec:Isingextractions}

The Ising CFT is the interacting IR fixed-point of a single-field $\lambda\phi^4$ theory defined in $d<4$ dimensions. It serves as a prototype for the class of $\phi^4$ theories and more generally for an interacting IR CFT. It describes real-world phase transitions such as the critical liquid--gas transition, transitions in binary fluid mixtures, and in uniaxial magnets, and is easy to simulate on a lattice as well as with the fuzzy sphere approach. These simulations and in particular the conformal bootstrap have led to a relatively large number of operators for which we have numerical data to compare with.

In \cite{Henriksson:2025hwi} we presented our new results for the Ising CFT, and they can be extracted in computer-readable format from the updated data file of \cite{Henriksson:2022rnm}. The previous perturbative state-of-the-art includes many operators computed at leading order \cite{Kehrein:1992fn,Kehrein:1994ff,Henriksson:2022rnm}, and a small sample of previous results beyond leading order.\footnote{
 The results were computed in \cite{Derkachov:1997pf} for twist-2 operators, in \cite{Bertucci:2022ptt} for twist-3 operators, in \cite{Zhang1982} for $\phi^5$ and in \cite{Schnetz:2016fhy,Schnetz:2022nsc} for the renormalizable part} 
The new state-of-the-art for operators with $\Delta\leqslant8$ (scalars up to $\Delta\leqslant 10$) is summarized in Table~\ref{tab:Isingdata-pres}. Our new predictions for scalars and spin-2 operators are presented in Figures~\ref{fig:scalar-all} and \ref{fig:tensors-all}. We do not show any results at spin-1 since the first spin-1 operator is at $\Delta=10$, known to one loop \cite{Kehrein:1994ff}.

\begin{table}
\caption{Perturbative state of the art for all low-lying operators in the Ising CFT with $\Delta\leqslant8$ (and scalars up to $\Delta=10$). In addition to the spin-$\ell$ operators shown, there are operators in the $(2,2)$ ($\Delta=7$ and $\Delta=8$) and $(4,1)$ ($\Delta=8$) Lorentz representations, known to $O(\eps)$. Results in bold are new.}\label{tab:Isingdata-pres}
\centering{ \small
\renewcommand{\arraystretch}{1.25}
\begin{tabular}{|ll|lp{0.5cm}|ll|ll|ll|ll|ll|}
\cline{1-3}
\multicolumn{3}{|l|}{$\phi^{10},\square^2\phi^6,\square^3\phi^4$   $\boldsymbol{\eps^2}$}
\\\cline{1-3}
$\phi^9$,$\square^2\phi^5\!\!\!\!$ &  $\boldsymbol{\eps^3}$
\\\cline{1-2}\cline{5-14}
$\phi^8$,$\square^2\phi^4\!\!\!\!$ &  $\boldsymbol{\eps^4}$ && & $\de^2\phi^6,\de^2\square\phi^4\!\!\!\!$ &  $\eps^1$ & $\de^3\phi^5$ &  $\eps^1$ & $2\!\times\!\de^4\phi^4\!\!\!\!$ &  $\eps^1$  & $\de^5\phi^3$ &  $\eps^2$ & $\de^6\phi^2$ &  $\eps^4$
\\\cline{1-2}\cline{5-14}
  $\phi^7$ &  $\boldsymbol{\eps^5}$&
  \multicolumn{2}{l|}{}
  & $\de^2\phi^5$ &  $\eps^1$ & & & $\de^4\phi^3$ &  $\eps^2$ 
  \\\cline{1-2}\cline{5-10}
    $\phi^6$ & $\boldsymbol{\eps^5}$ &
    \multicolumn{2}{l|}{}
    &   $\de^2\phi^4$ & $\boldsymbol{\eps^5}$ & $\de^3\phi^3$ & $\eps^2$ & $\de^4\phi^2$ & $\eps^4$
\\\cline{1-2}\cline{5-10}
    $\phi^5$ & $\boldsymbol{\eps^5}$ &\multicolumn{2}{l|}{} &   $\de^2\phi^3$ & $\boldsymbol{\eps^5}$ & 
\\\cline{1-2}\cline{5-6}
    $\phi^4$ & $\eps^7$ & \multicolumn{2}{l|}{} &   $T^{\mu\nu}$ & {\!\!\!\!\!\!\!\!\!\! exact}  & 
\\\cline{1-2}\cline{5-6}
\\\cline{1-2}
$\phi^2$ & $\eps^7$
\\\cline{1-2}
$\phi$ & $\eps^8$
\\\hline
\multicolumn{2}{c}{$\ell=0$}& \multicolumn{2}{c}{$\ell=1$}&\multicolumn{2}{c}{$\ell=2$}& \multicolumn{2}{c}{$\ell=3$}& \multicolumn{2}{c}{$\ell=4$}& \multicolumn{2}{c}{$\ell=5$}&  \multicolumn{2}{c}{$\ell=6$}
\end{tabular}
}
\end{table}

We now describe a few interesting aspects of our new predictions, complementing the examples given in \cite{Henriksson:2025hwi}.

\paragraph{The \boldsymbol{$\phi^5$} operator across $\boldsymbol{d}$.}

Already in section~\ref{s:Redex} we gave details on the extraction of the anomalous dimension of $\phi^5$, allowing to estimate the dimension $\Delta_{\phi^5}$ in three dimensions. Here we compare the previous available and our new estimate with results from the boostrap and MC simulations (fuzzy sphere)
\begin{align}
    \text{Pad\'e$_{[1,2]}$:} && 5.325845\,, && \text{Bootstrap (rigorous error):} &&& 5.262(89)\text{ \cite{Reehorst:2021hmp}}\,,
    \nonumber
    \\\nonumber
    \text{Pad\'e$_{[2,3]}$:} &&5.257395\,, && \text{Bootstrap (statistical error):} &&& 5.2906(11) \text{ \cite{Simmons-Duffin:2016wlq}}\,,
    \\
  &&&&  \text{Fuzzy sphere:} &&&5.30346  \text{ \cite{Zhu:2022gjc}}\,,
\end{align}
where the result quoted as ``Bootstrap (rigorous error)'' implies that the true operator dimension must lie in the interval $[5.173,\,3.351]$. 
Our new Pad\'e$_{[2,3]}$ approximant for $d=3$ moves us somewhat closer the center of this interval compared to the Pad\'e$_{[1,2]}$ constructed from the $O(\eps^3)$ results available from \cite{Zhang1982}. 

In Figure~\ref{fig:Deltaphi5}, we display the two Pad\'e approximants alongside bootstrap data from \cite{Henriksson:2022gpa} at many values of $\eps$. We see that the bootstrap data starts deviating from our estimates in the region $\eps > 1$, however we can find a curve that nicely interpolates all bootstrap data if we construct a Pad\'e$_{[3,4]}$ approximant forcing it to go through the points $(d,\Delta)=(3,5.29068)$ from \cite{Henriksson:2022gpa} and the exact 2d value $(d,\Delta)=(2,\tfrac{49}{8})$.\footnote{Although it is not generically clear how to continue operators across $d$, for low-lying operators it is possible to deduce plausible guesses based on the low-lying spectrum of the 2d theory (which is exactly known). Both $\phi^5$ and the operator with dimension $\frac{49}8$ (a Virasoro descendant of $\sigma$) are the second-lowest (quasi)primaries in the scalar $\Z_2$-odd sector, so it is natural to identify the two.}
\begin{figure}
    \centering
\includegraphics[width=\textwidth]{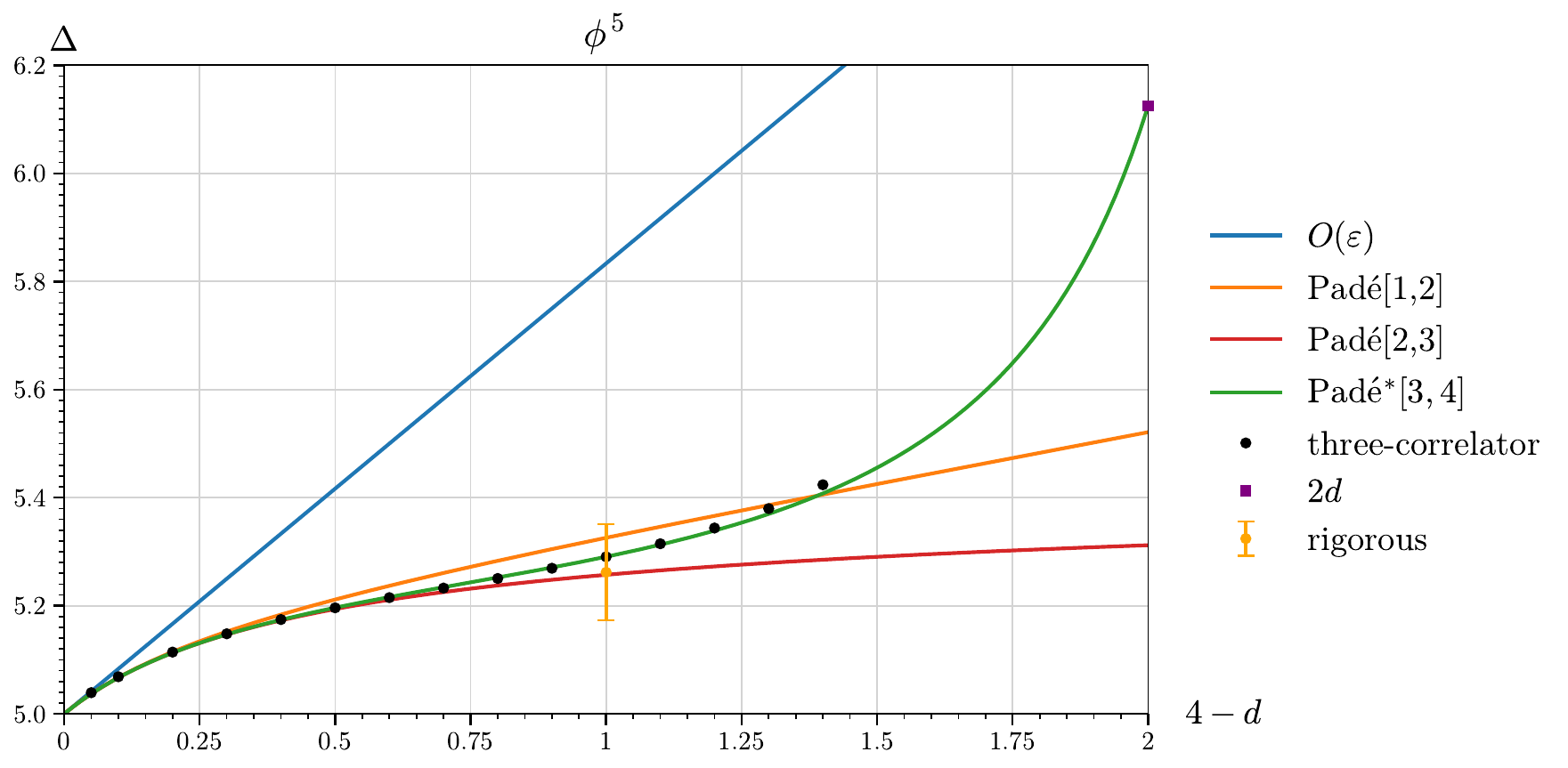}
    \caption{
Comparing perturbative and non-perturbative results for the operator $\phi^5$. Prior to our work, three-loop results enabled the Pad\'e$_{[1,2]}$ estimate. 
Pad\'e$^*_{[3,4]}$ was obtained from our five-loop results, constrained to go through $5.29068$ for $d=3$~\cite{Henriksson:2022gpa} and the exact 2d value $6.125$.}
    \label{fig:Deltaphi5}
\end{figure}

\paragraph{The subleading tensor operator across \bm{$d$}.}
Next we consider the operator $\phi^4\de^2$, which in the bootstrap literature is denoted $T'$, since it is the next operator after the stress-energy-momentum tensor in the spin-2 $\Z_2$-even sector. We gave its five-loop anomalous dimension in \eqref{eq:DeltaTprim} above, improving on the previous determination at $O(\eps)$ from \cite{Kehrein:1994ff}. 
We study this operator across dimensions by presenting its scaling dimension in the range $d\in[2,4]$ in Figure~\ref{fig:Tprim}. 

In $d=3$, we can compare across different results 
\begin{align}
    \text{$O(\eps)$ truncation:} && 5.444444\,, && \text{Bootstrap (rigorous error):} &&& 5.499(17)\text{ \cite{Reehorst:2021hmp}}\,,
    \nonumber
    \\\nonumber
    \text{Pad\'e$_{[2,3]}$:} && 5.465026\,, && \text{Bootstrap (statistical error):} &&& 5.50915(44)\text{ \cite{Simmons-Duffin:2016wlq}}\,,
    \\
  &&&&  \text{Fuzzy sphere:} &&& 5.5827144 \text{ \cite{Zhu:2022gjc}}\,,
\end{align}
where the rigorous interval is $[5.482,\,5.516]$. 
Our new Pad\'e$_{[2,3]}$ approximant moves us closer to, but is not inside, this interval. As seen from the figure, there is good evidence from the bootstrap \cite{Cappelli:2018vir,Henriksson:2022gpa} that this operator connects to an operator with dimension $\Delta=6$ in 2d. Following the example with $\phi^5$, we find a more consistent picture if we include Pad\'e approximants going through the point $(d,\Delta)=(2,6)$, one of which moves us inside the rigorous error interval.

\begin{figure}
    \centering
\includegraphics[width=\textwidth]{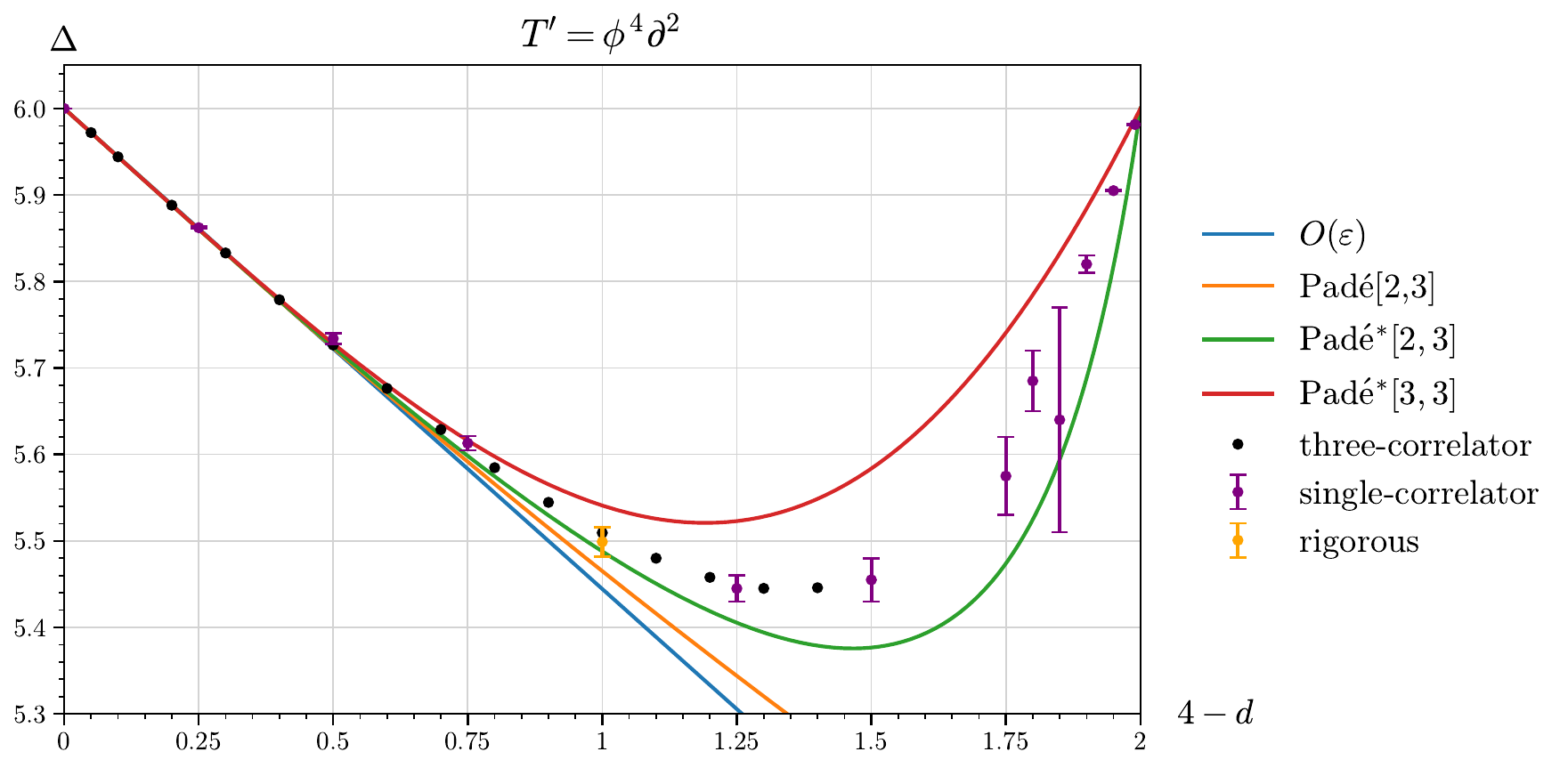}
    \caption{Pad\'e approximants and conformal bootstrap results for the operator dimension $T'=\phi^4\de^2$ in the Ising CFT. The three-correlator data are from \cite{Henriksson:2022gpa} which does not provide error-bars, and the single-correlator data are from \cite{Cappelli:2018vir}. The star denotes Pad\'e approximants fixed to go through the point $(d,\Delta)=(2,6)$. 
    }
    \label{fig:Tprim}
\end{figure}

\paragraph{Pure-power scalar operators.}
By the result of \cite{Antipin:2024ekk}, the anomalous dimension of the tower of operators of the form $\phi^k$, $k\neq3$, at $O(\eps^l)$ is given by $k$ times a degree-$l$ polynomial in $k$. Moreover, \cite{Antipin:2024ekk} determined the leading power $\eps^lk^{l+1}$ to be given by 
\begin{equation}
\label{eq:predictionGammaK}
    1,\frac{1}{6},-\frac{17}{324},\frac{125}{3888},-\frac{3563}{139968},\frac{29183}{1259712},-\frac{1044133}{45349632},\ldots,
\end{equation}
 \emph{c.f.} their Equation~19.
We now have five-loop results for all the operators $\phi$, $\phi^2$, $\phi^4$, $\phi^5$, $\phi^6$, $\phi^7$, providing six data points. This means we can fix enough coefficients at five-loop order to derive the following general expression
\begin{align}
    \Delta_{\phi^k}&=k\left[1-\tfrac\eps2\right]+\frac{k(k-1) }{6}  \eps-\frac{k}{324}  \left(17 k^2-67 k+47\right) \eps^2\nonumber\\&\quad +k\left(\frac{ 1125 k^3-7433 k^2+15034 k-8399}{34992}+\frac{2(k-1)(k-3)}{27} \zeta_3 \right) \eps^3  
    \nonumber\\
    &  \quad +\gamma^{(4)}\eps^4+\gamma^{(5)}\eps^5+O(\eps^6)\,,
    \label{eq:phik}
\end{align}
where
\begin{align}
\nonumber
    \gamma^{(4)}&=\frac{k(-96201 k^4+875313 k^3-2854576 k^2+3882758 k-1785643)}{3779136}+\frac{k(k-1) (k-3)}{18} \zeta _4 \\ &\quad -\frac{ k(261 k^3-1772 k^2+3679 k-2120}{5832}\zeta _3-\frac{5k(k+2) (k-1) (k-3) }{81} \zeta _5\,,
\end{align} and \begin{align}
\nonumber
   \gamma^{(5)}&=k\left(\frac{29183 k^4}{1259712}-\frac{4045207
   k^3}{15116544}+\frac{6020699 k^2}{5038848}-\frac{522964753
   k}{204073344}-\frac{409364045}{408146688
   k}+\frac{534173381}{204073344}\right) 
   \nonumber\\
   &\quad +\frac{k (3996 k^4-35124 k^3+113521 k^2-156983
   k+73603)}{104976} \zeta_3\nonumber\\
   &\quad +-\frac{k (261 k^3-1772 k^2+3679 k-2120)
  }{7776}\zeta_4\nonumber\\
   &\quad +\frac{k (180 k^4-995 k^3+1022 k^2+1607 k-1694)}{4374} \zeta_5-\frac{25 k (k^3-2 k^2-5 k+6)}{324} \zeta_6
   \nonumber\\
   &\quad +\frac{7k
   (8 k^4-17 k^3+28 k^2-211 k+192)}{972} \zeta_7+\frac{k(k^3-14 k^2+43 k-30)}{162} \zeta_3^2 \,.
\end{align}
The coefficients of $\eps^3k^4$, $\eps^4k^5$, and $\eps^5k^6$ agree with \eqref{eq:predictionGammaK}, providing a non-trivial check. 

\paragraph{Other towers of operators.}
Recently also the subleading coefficients at large $k$, i.e.~those of $\eps^lk^l$ at any loop order $l$, have been derived \cite{Antipin:2025ilv}. For the tower of $\phi^k$ operators, these agree with \eqref{eq:phik}, as also confirmed in \cite{Antipin:2025ilv}. In fact, the results of \cite{Antipin:2025ilv} for the leading and subleading coefficients make it possible to revisit some of the towers of primary operators introduced already in \cite[Table~4]{Kehrein:1994ff}. 
These towers have fixed number of derivatives and $k\geqslant k_0$ fields, where $k_0$ may be different for different towers. Reference \cite{Antipin:2025ilv} showed that in these towers, the leading coefficients of $\eps^lk^{l+1}$ are equal to \eqref{eq:phik} for all towers, and provided a way to compute the subleading coefficients of $\eps^lk^k$, which differ tower by tower. 

By combining the form predicted by \cite{Antipin:2025ilv} with some of our new results, we can derive two-loop anomalous dimensions for two additional towers:
\begin{itemize}
    \item Spin-2 operators of the form $\de^2\phi^k$ for $k\geqslant2$. Their dimensions are
    \begin{equation}
       \Delta_k= 2+k\left[1-\tfrac \eps2\right]+\frac{(k-2)(3k+1)}{18}\eps-\frac{(k-2)(102 k^2-335 k-235)}{1944}\eps^2+O(\eps^3),
    \end{equation}
    where $O(\eps)$ agrees with \cite{Kehrein:1994ff}, and the four coefficients at $O(\eps^2)$ were fixed from three operators at $k=2,3,4$ and two conditions from \cite{Antipin:2025ilv} (this gives one overconstrained parameter providing a consistency check). These operators are visible in Figure~\ref{fig:tensors-all} below.
    \item Scalar operators of the form $\square^2\phi^k$ for $k\geqslant4$. Their dimensions are
    \begin{equation}
        \Delta_k=4+k\left[1-\tfrac \eps2\right]+\frac{k(3k-7)}{18}\eps-\frac{51 k^3-338 k^2+443 k+108}{972}\eps^2 +O(\eps^3),
    \end{equation}
  where again the $O(\eps)$ term agrees with \cite{Kehrein:1994ff}, and the four coefficients at $O(\eps^2)$ were fixed from three operators at $k=4,5,6$ (see \cite{Henriksson:2025hwi}) and two conditions from \cite{Antipin:2025ilv}. 
\end{itemize}

\subsection[New data for the critical 
\texorpdfstring{$O(n)$}{O(n)} CFT]{
New data for the critical 
\texorpdfstring{$\bm{O(n)}$}{O(n)} CFT}
\label{sec:ONextractions}

The simplest generalization of the single-scalar case is the $O(n)$ CFT for $n=1,2,3,\ldots$. This theory has been studied with various methods, including precision bootstrap studies for $n=2,3$~\cite{Chester:2019ifh,Chester:2020iyt} and a sequence of bootstrap islands for larger $n$ \cite{Kos:2015mba}.\footnote{The cases $n=-2$ and $n=0$ are relevant for loop-erased random walks and self-avoiding random walks \cite{deGennes:1972zz,Duplantier1992,Peled2017,Wiese:2018dow}, and there are even results for non-integer $n$ \cite{Liu:2012ca,Sirois:2022vth}.} There are also results from Monte Carlo simulations \cite{Hasenbusch2011,Banerjee:2017fcx,Hasenbusch:2019jkj,Banerjee:2021bbw,Hasenbusch:2020pwj,Hasenbusch:2021rse,Hasenbusch:2025yrl} and the fuzzy sphere \cite{Han:2023lky,Dey:2025zgn}.

For $O(n)$ symmetry, the coupling is $\lambda_{abcd}=\frac13(\delta_{ab}\delta_{cd}+\delta_{ac}\delta_{bd}+\delta_{ad}\delta_{bc})$ leading to a unique interacting fixed-point with
\begin{equation}
   \lambda_*=\!\frac{3}{n+8}\eps\!+\!\frac{9(3 n+14)}{(n+8)^3}\eps^2\!+\!\left(\frac{99 n^3-330 n^2-5280 n-13632}{8 (n+8)^5}\!+\!\frac{36 (5 n+22)}{(n+8)^4} \zeta _3\!\right)\eps^3\!+\!\ldots.
\label{eq:critCouplingON}
\end{equation}
The full expression is stored in our data file on \githubb, as well as in the data file of \cite{Henriksson:2022rnm}.

\begin{table}
\caption{The $O(n)$ operators for which there are now five-loop results. Below $\Delta=6$ there are also spin-3 and spin-4 operators, for which we do not provide new results. 
}\label{tab:ON-operators-pres}
\centering{ \small
\renewcommand{\arraystretch}{1.25}
\begin{tabular}{|ll|ll|ll|ll|}
\hline
  $\phi^6$ & $S,T,T_2,T_4,T_6$ & \multirow{2}{*}{ $\de\phi^5$ }& $V,T_3,$&\multirow{2}{*}{ $\de^2\phi^4$} & $2\!\times\!S,3\!\times\!T,$
  &\multirow{2}{*}{ $\de^{[1,1]}\phi^4$} & \multirow{2}{*}{$A,Y_{2,1,1}$}  
  \\
    $\square\phi^4\!\!$ & $S,T,B_4$ & & $H_3,H_5$ & &  $T_4,H_4,B_4$& &  
\\\hline
       $\phi^5$ & $V,T_3,T_5$  & $\de\phi^4$  & $A,H_4$ &    $\de^2\phi^3$ & $2\!\times\!V,T_3,H_3$ & $\de^{[1,1]}\phi^3$ & $A_3$ 
\\\hline
   $\phi^4$ & $S,T,T_4$ & $\de\phi^3$ & $V,H_3$ & $\de^2\phi^2$ & $S,T$
\\\cline{1-6}
$\phi^3$ & $T_3$ & $\de\phi^2$ & $A$
\\\cline{1-4}
$\phi^2$ & $S,T$
\\\cline{1-2}
$\phi$ & $V$
\\\hline
\multicolumn{2}{c}{$\ell=0$}& \multicolumn{2}{c}{$\ell=1$}&\multicolumn{2}{c}{$\ell=2$}& \multicolumn{2}{c}{$\ell=[1,1]$} 
\end{tabular}
}
\end{table}

\begin{figure}
\hspace{7.5cm}
\includegraphics[width=0.48\textwidth]{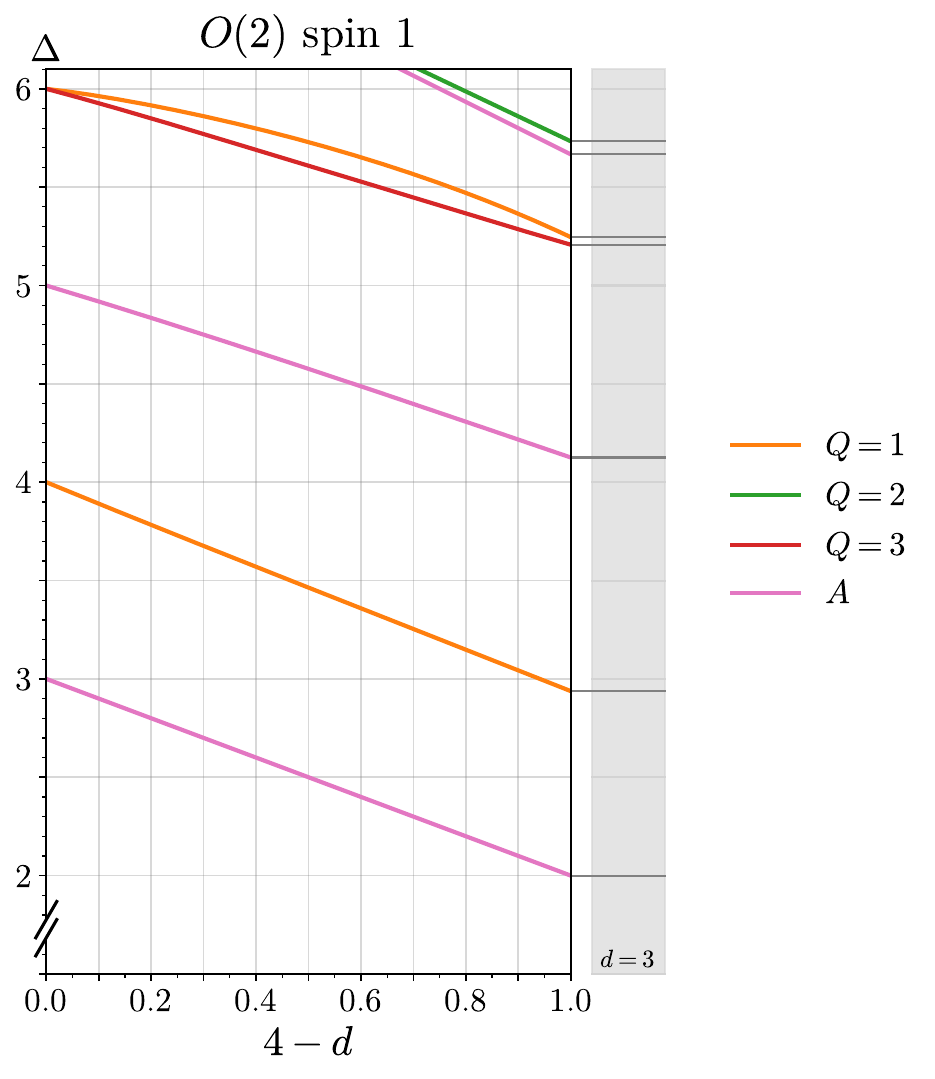}

\vspace{5mm}

\centering
\includegraphics[width=0.48\textwidth]{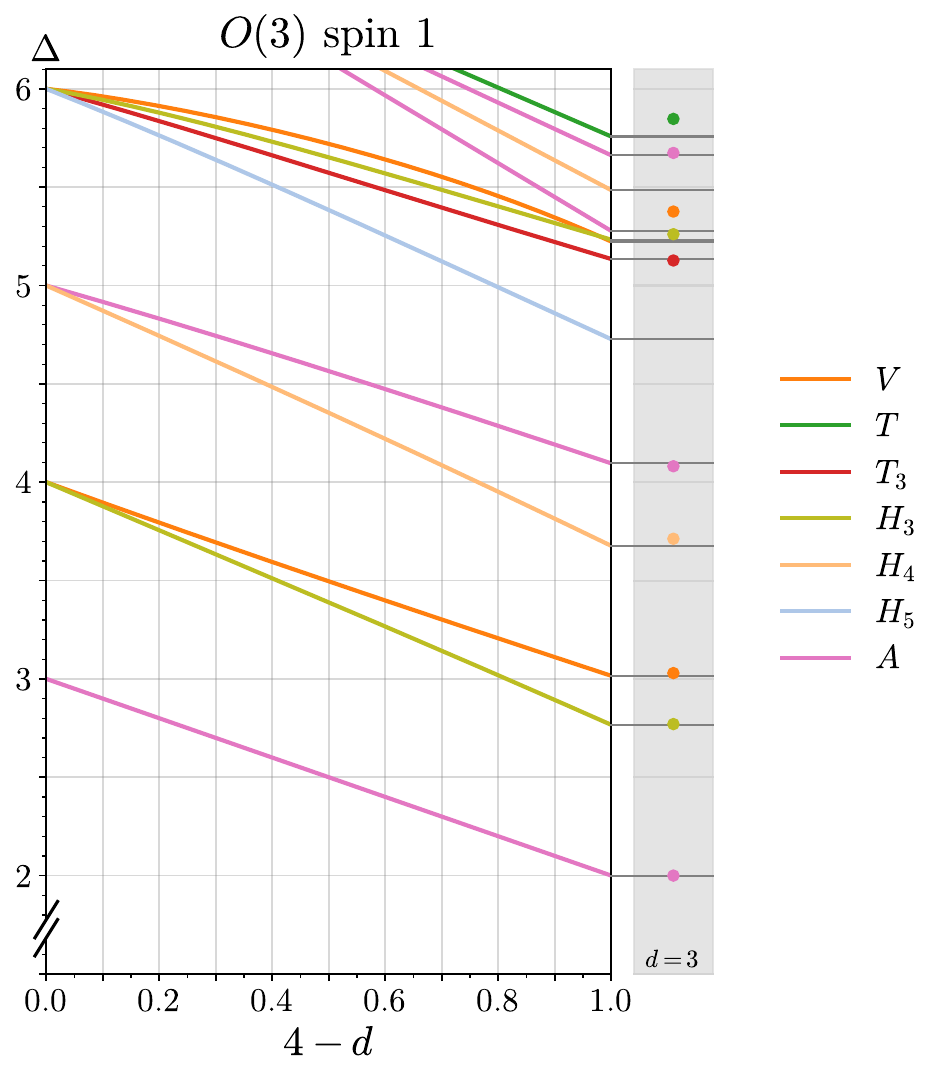}
\includegraphics[width=0.48\textwidth]{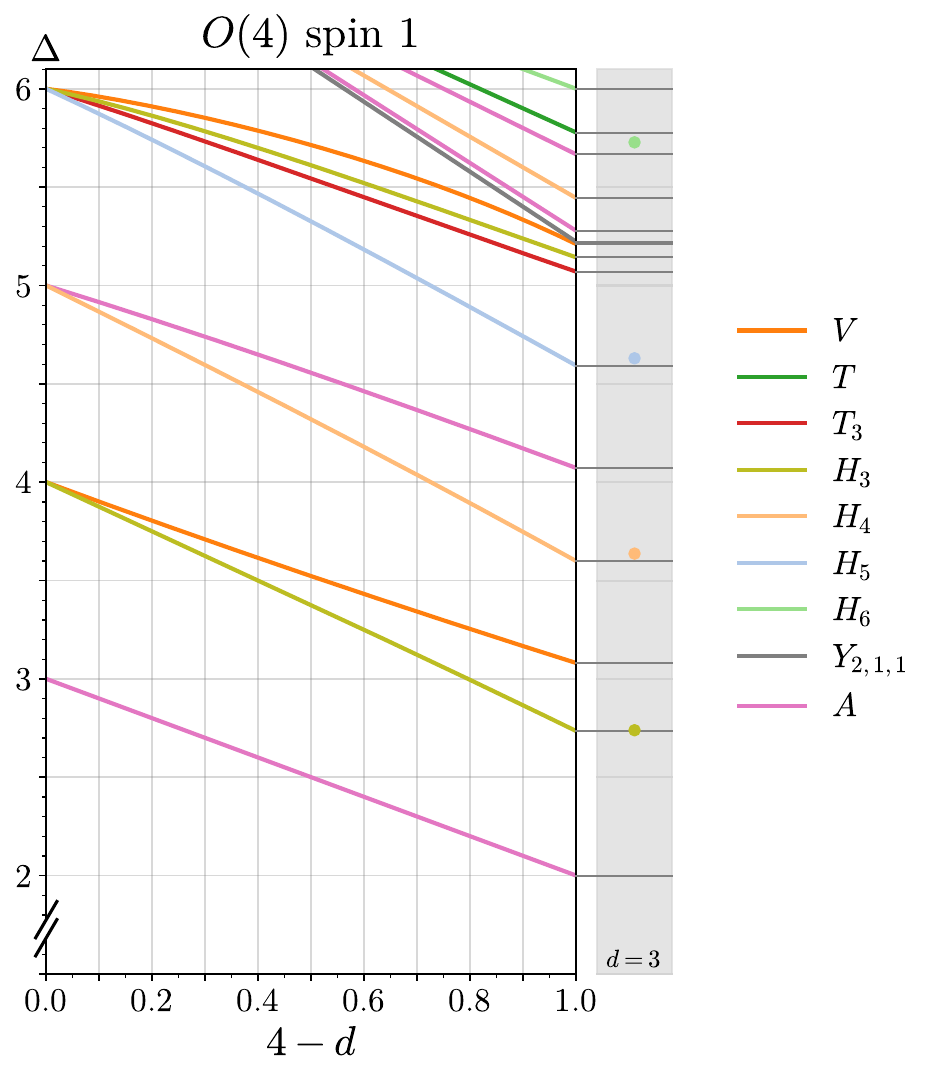}

\vspace{2mm}

    \caption{Spin-1 spectra for $O(n)$ symmetric theories with $n=2,3,4$ (there are no spin-1 operators in the $n=1$ theory for which we have new results). The energy levels shown on the very right are results in three dimensions: conformal bootstrap, $n=3$ \cite{Chester:2020iyt} (and unpublished by the same authors) and Monte Carlo $n=4$ \cite{Banerjee:2021bbw}. We do not know of any numerical data for $O(2)$ at spin 1.
    We plot the spin-0 and spin-2 spectra in Figures~\ref{fig:scalar-all} and \ref{fig:tensors-all}, respectively.
    }
    \label{fig:vectors-all}
\end{figure}

In Table~\ref{tab:ON-operators-pres} we display the operators for which there are now results up to $O(\eps^5)$, where $\phi^k$ with $k\leqslant 4$ were computed earlier \cite{Schnetz:2016fhy,Bednyakov:2021ojn}.
Here $Y_{\lambda_1,\ldots,\lambda_k}$ denotes a representation of $O(n)$ labeled by a Young tableau with $k$ rows of lengths $\lambda_i$. We also use the special names, following the conventions of \cite{Henriksson:2022rnm},
\begin{align}
    S&=Y_{\emptyset}\,, &  V&=Y_1\,, & T&=Y_2\,,  & T_m&=Y_m\,,  \nonumber \\
     A&=Y_{1,1}\,, & H_m&=Y_{m-1,1}\,,  & B_4&=Y_{2,2}\,,  & A_m&=Y_{1^m}\,. 
\end{align}
For instance, in the data file of \cite{Henriksson:2022rnm}, the five-loop result of the $\phi^6$-type scalar operator in the rank-2 traceless symmetric $T$ irrep is stored as \texttt{DeltaE[Op[T,0,4]]}.\footnote{In the data file of \cite{Henriksson:2022rnm}, the notation used to denote operators is \texttt{Op[$\langle R\rangle$,$l$,$i$]} for the operator in global symmetry irrep $\langle R\rangle$ of spin $l$ and $i$th lowest scaling dimension. For Ising, the representations are $\langle R\rangle=\texttt{E},\texttt O$ for $\Z_2$-even and $\Z_2$-odd. For $O(n)$, the irreps $T_m$, $A_m$ and $H_m$ are denoted \texttt{Tm[$m$]}, \texttt{Am[$m$]}, \texttt{Hm[$m$]} respectively, and lowlying irreps are given by symbols \texttt S, \texttt V, \texttt T, \texttt A, \texttt {A3}, \texttt{B4}.} This gives the dimension at the critical coupling \eqref{eq:critCouplingON}. The dimension at generic coupling can be reverse-engineered by inverting that relation.\footnote{For example: we have $\Delta_{\phi^6_T}=6-3\eps+\frac{2n+42}{n+8}\eps-\frac{71 n^2+1066 n+5544}{2 (n+8)^3}\eps^2+\ldots$, where $6-3\eps$ is the free-theory part which should be separated before substituting $\eps=\frac{n+8}3\lambda-\frac{3n+14}{3}\lambda^2+\ldots$ (this is the order-by-order solution for $\eps$ in terms of $\lambda$ in \eqref{eq:critCouplingON}). This gives the expression $\Delta_{\phi^6_T}=6-3\eps+\frac{2n+42}{3}\lambda-\frac{107n+1134}{18}\lambda^2+\ldots$ at general $\lambda$.}

\begin{figure}
    \centering
\includegraphics[width=0.48\textwidth]{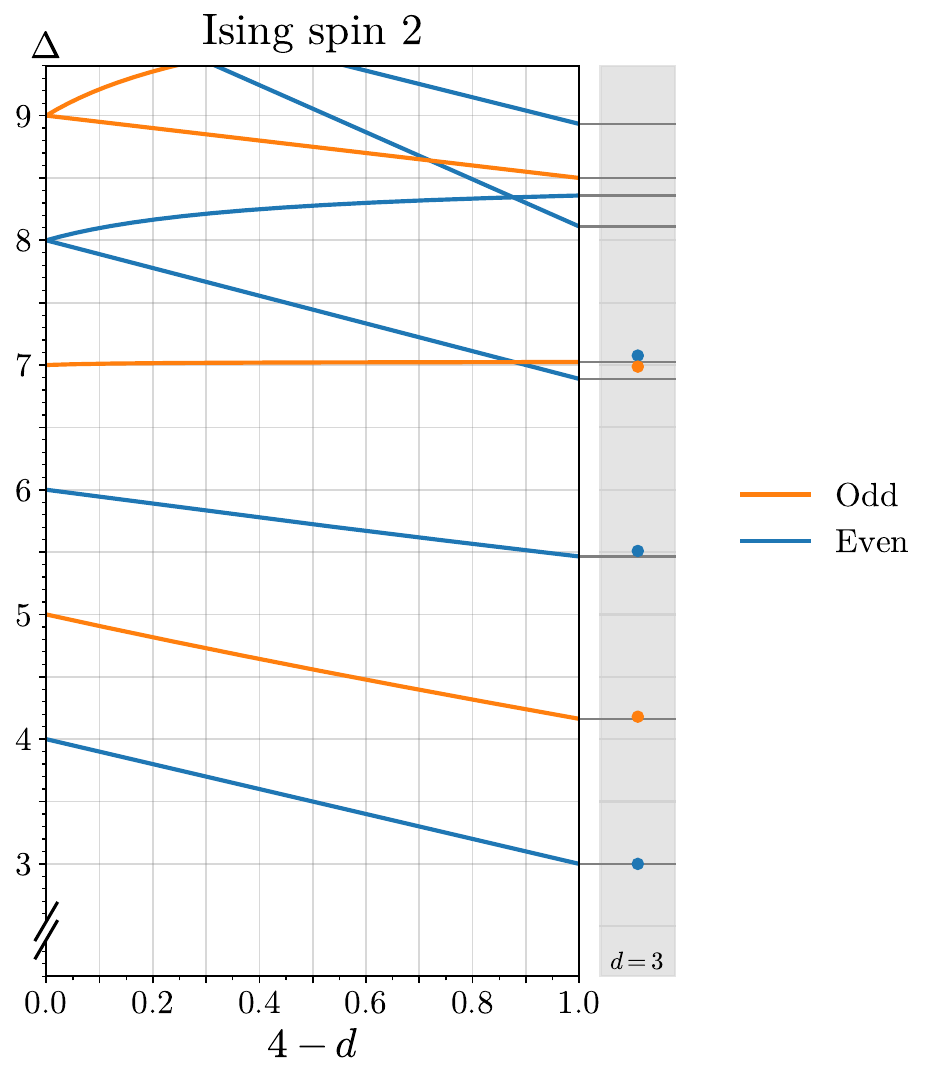}
\includegraphics[width=0.48\textwidth]{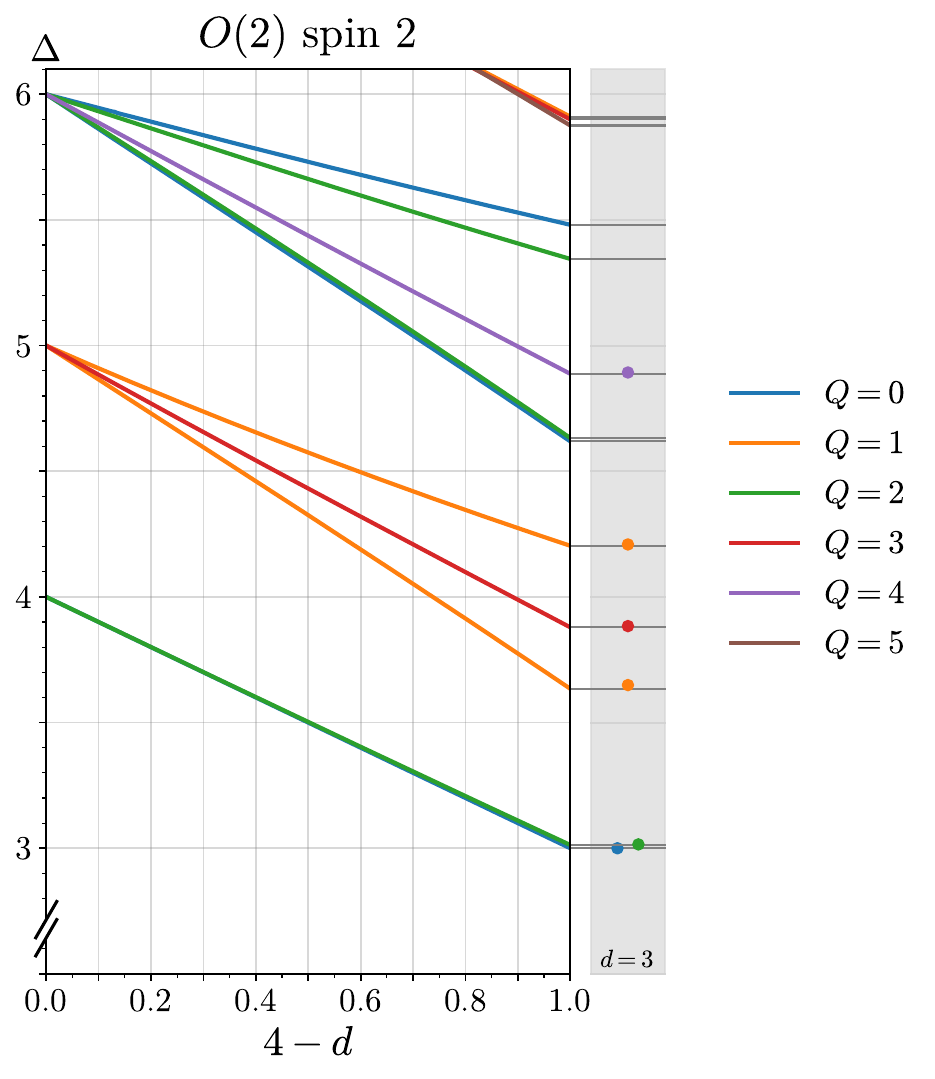}

\vspace{8mm}

\includegraphics[width=0.48\textwidth]{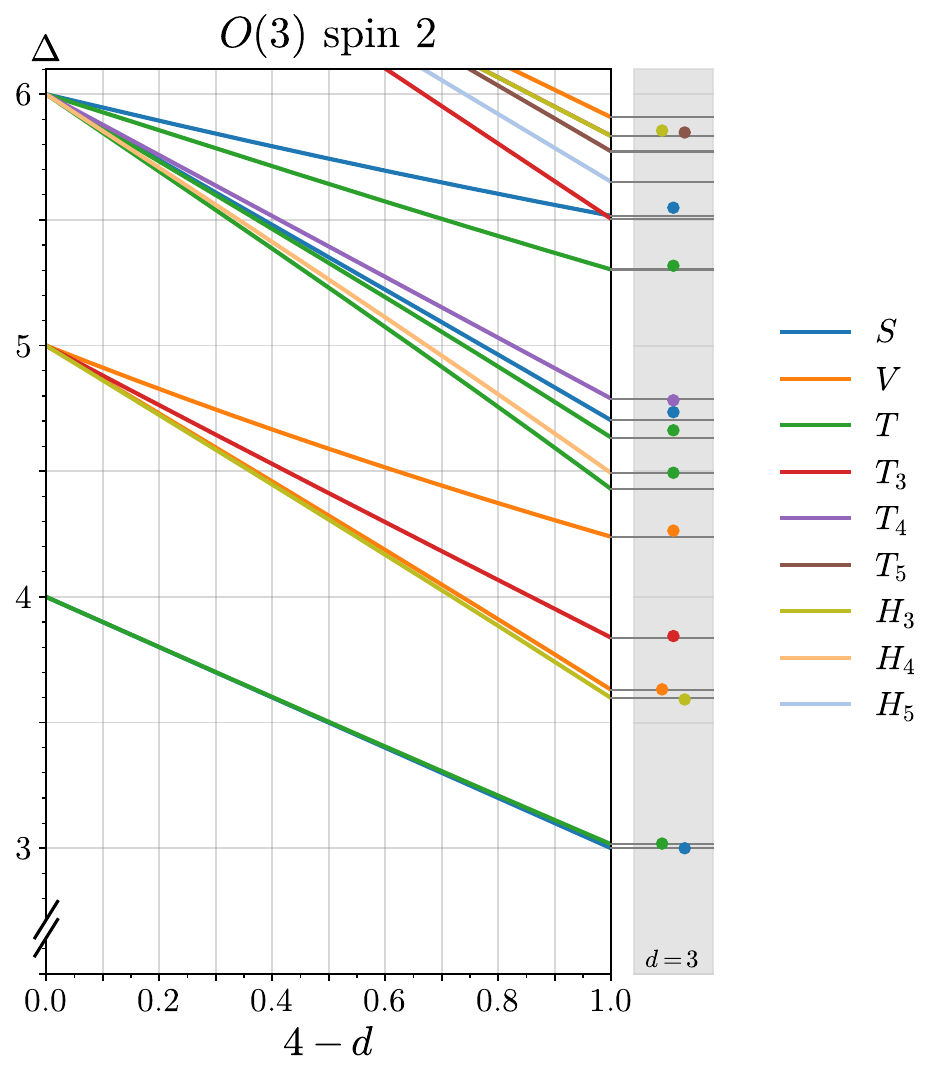}
\includegraphics[width=0.48\textwidth]{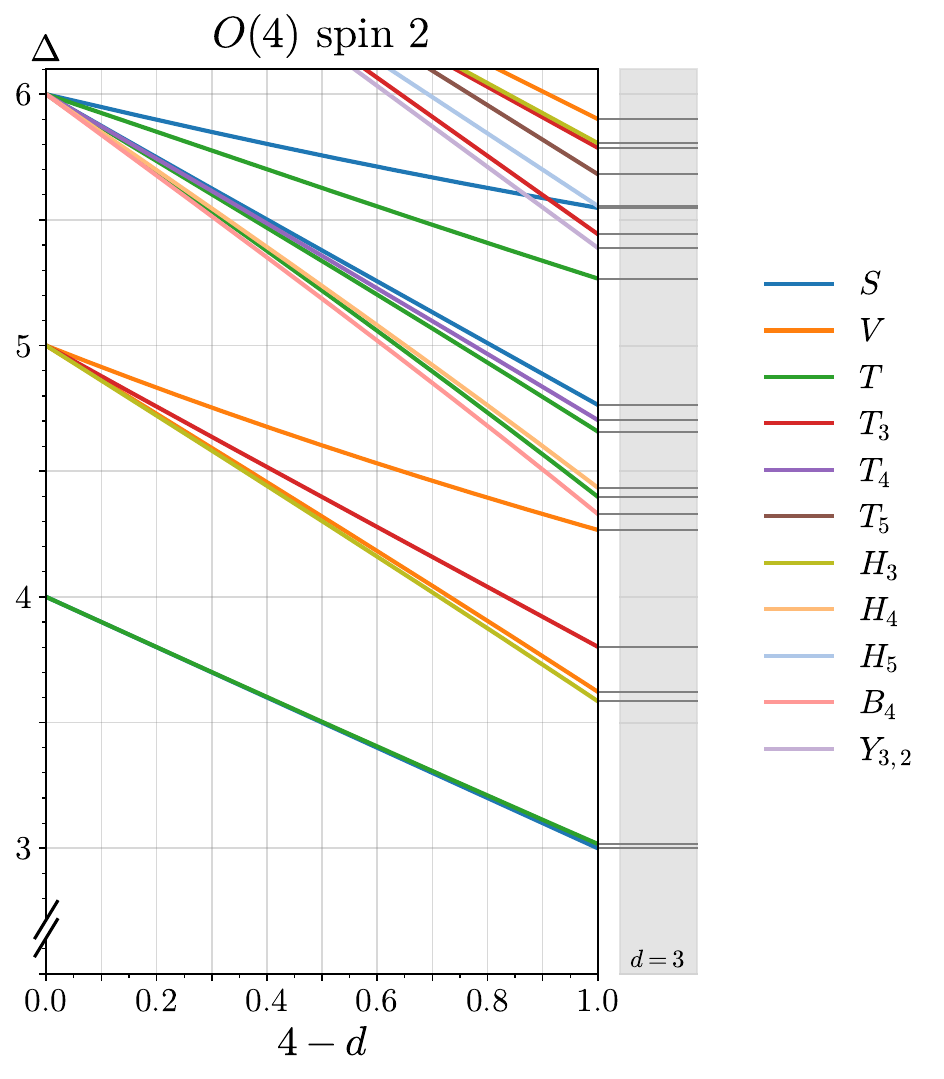}

\vspace{3mm}

    \caption{Spin-2 (traceless-symmetric) spectra for $O(n)$ symmetric theories with ${n=1,2,3,4}$. The energy levels shown on the very right are results in three dimensions: bootstrap $n=1$ \cite{Simmons-Duffin:2016wlq}, $n=2$ \cite{Liu:2020tpf}, $n=3$ \cite{Chester2020Unp} (and $T_5$ from \cite{Rong:2023owx}). We do not know of any numerical data for $O(4)$ at spin 2. 
    We plot the spin-0 and spin-1 spectra in Figures~\ref{fig:scalar-all} and \ref{fig:vectors-all}, respectively.
    }
    \label{fig:tensors-all}
\end{figure}

\paragraph{Precision spectroscopy.}
With such wealth of perturbative results, we can now perform a precision spectroscopy of the scalar, spin-1 and spin-2 sectors of the $O(n)$ CFTs for $n=1,2,3,4$. We displayed the scalars in Figure~\ref{fig:scalar-all} in the introduction, and now complement these with spins 1 and 2 in Figures~\ref{fig:vectors-all} and \ref{fig:tensors-all} here. Where available, we compare the results with non-perturbative spectra from the bootstrap and Monte Carlo methods.\footnote{For $n=3$, we use the high-precision unpublished spectrum shared with us by the authors of \cite{Chester:2020iyt}. We also point out that there is data from the fuzzy sphere regularization for a plenitude of operators; see~\cite{Han:2023lky,Dey:2025zgn}.} 
In appendix~\ref{app:ONTables} we tabulate numerical values of the Pad\'e approximants evaluated at $\eps=1$ for $n=2,3,4$.

In the figures, we display Pad\'e approximants providing a simple way to perform a resummation of our perturbative results. In particular, we use a maximal-degree Pad\'e approximant 
\begin{equation}
\label{eq:padeDef}
    \text{Pad\'e}_{[a,b]}=\frac{\sum_{k=0}^a c_k\eps^k}{1+\sum_{k=1}^bd_k\eps^k}\,,
\end{equation}
with $a\leqslant b\leqslant a+1$ and where the coefficients are fixed by matching with our perturbative series to order $O(\eps^{a+b})$. Sometimes this results in Pad\'e approximants with poles; if we encounter a pole in the range $0\leqslant\eps\leqslant 3$, we lower the total degree until no such pole occurs.\footnote{For the leading spin-2, rank-5 traceless-symmetric operator we employ $\text{Pad\'e}_{[0,2]}$ using the two-loop result we computed using the method of \cite{Kehrein:1995ia}.} This gives a set of curves representing our perturbative predictions. We then display these alongside the non-perturbative results with references given in the corresponding captions. The non-perturbative results are extracted from the bootstrap using the extremal functional method \cite{ElShowk:2012hu,El-Showk:2014dwa}, as well as from MC and fuzzy sphere \cite{Zhu:2022gjc} simulations. As can be seen from these figures, several operators enter the diagrams from above. These are operators with engineering dimension 7 for which we generically only have 1-loop results derived using the method of \cite{Kehrein:1992fn,Hogervorst:2015akt} as well as a few two-loop results using the method of \cite{Kehrein:1995ia}. These lower-order estimates may change substantially if higher orders are added, motivating further studies. 

Note that in the numerical data in $d=3$ with which we compare, there are several absent points and entire absent sectors. This is due to the unavailability of certain representations in the numerical bootstrap results to date (e.g. spin-2 $H_4$ irrep), and the choices made in which operators to include/study in the literature. We hope that our perturbative precision spectra will motivate revisiting these absences with a variety of methods.

\paragraph{Dimension-eight scalar operators.}

\begin{table}
\caption{Dimension-eight operators for which we have results at two-loop order in the $O(n)$ CFT, displaying the leading-order field content. }\label{tab:ONdata-new-at8}
\centering{ 
\renewcommand{\arraystretch}{1.25}
\begin{tabular}{|l|l|l|l|}\hline
$\O$ &  $O(n)$ reps 
\\\hline
$\phi^8$ &$S,T;T_4,T_6,T_8$ 
\\\hline
$\de^2\phi^6$ &  $S,2\times T;T_4,H_4,B_4,Y_{4,2}$ 
\\\hline
$\de^4\phi^4$ & $2\times S,2\times T,T_4,B_4$ 
\\\hline
\end{tabular}
}
\end{table}

In addition to our results up to dimension six, we have also extracted two-loop results for scalar operators at dimension eight with $O(n)$ and hypercubic couplings. In the case of $O(n)$, we performed the extraction of the 18 different operators at this level; see Table~\ref{tab:ONdata-new-at8}. The results for these operators have been included in the data file.

\subsection{New data for hypercubic-symmetric CFTs}
\label{sec:cubicRes}

Next we consider the hypercubic CFT, for which, for example, the case $n=3$ is of experimental relevance for magnets with cubic-crystal structure.\footnote{Commonly quoted to be in the Heisenberg ($O(3)$) universality class, these in fact belong to the cubic universality class since the internal rotation-breaking perturbation $\phi^4_{T_4}$ has dimension $\Delta_{\phi^4_{T_4}}<3$ \cite{Chester:2020iyt}.} This theory has the couplings given in \eqref{hypercubicLag}, which have four fixed-points: trivial ($\lambda_i=0$), decoupled Ising ($\lambda_{4,1}=0$), $O(n)$ ($\lambda_{4,2}=0$) and a non-trivial fixed-point which we call the hypercubic CFT. At this fixed-point, the critical couplings take the values
\begin{align}
    \lambda_{4,1}^*=&\frac{\eps}{n} - \frac{106 -125 n + 19 n^2)}{27 n^3} \eps^2  +O(\eps^3)\,,
\nn\\
    \lambda_{4,2}^*=&\frac{n-4}{3n}\eps
   + \frac{424 - 534 n + 93 n^2 + 17 n^3 }{81 n^3}\eps^2 +O(\eps^3)\,.
\label{cubiccoupling}
\end{align}
known to six-loop order \cite{Adzhemyan:2019gvv}. 
In the $n \rightarrow \infty$ limit, the couplings simplify to $\lambda_{4,2}^* \rightarrow \lambda^*_{\text{Ising}}$ and $\lambda_{4,1}^*\to0$, leading to the large-$n$ expansion for the cubic theory around $n$ Ising models \cite{Binder:2021vep}.

In Table~\ref{tab:Cubic-singlet} and Appendix~\ref{app:cubicTables} we tabulate all operators that exist in hypercubic theories up to engineering dimension $\Delta=6$ and Lorentz rank two. This is done for $n=3$ and $n=4$, where we remind the reader that the symmetry group of these theories is $G_{\text{hypercubic}}=S_n \ltimes (\Z_2)^n$. 
Within the tables we use boxes ($\square$) to denote derivative insertions that don't raise the spin, and  partial derivatives ($\partial$) to denote insertions that do raise the spin. Thus, for example, $\square \phi^4$ denotes a Lorentz scalar, whereas $\partial^2 \phi^4$ denotes a spin-two operator. In additional columns we also provide the scaling dimensions of the same operators but evaluated at the corresponding $O(3)$ or $O(4)$ fixed-point, and provide a final column which denotes the $O(n)$ irrep that a given hypercubic operator branches from. 

Within each table, operators are organised in terms of their Lorentz irrep and engineering dimension. For the irreps of the hypercubic group and their names as adopted in this work; see \cite[Tab.~1]{Bednyakov:2023lfj}. The numerical values for the dimension of each operator within the tables correspond to a Padé$_{[2,2]}$ approximant, or where explicitly noted a Pad\'e approximant of another degree. We chose Padé resummations due to their simplicity paired with the large amount of data we had to deal with, and because it has been observed that they were sufficiently precise in earlier work, such as \cite{Henriksson:2022gpa} and \cite{Henriksson:2025hwi}. Once sufficiently precise non-perturbative data has been produced for the hypercubic theories, it would be a very interesting question to study the effectiveness of different types of resummation.\footnote{This would probably also require more orders in perturbation theory than what we have produced here.} As of the date of writing such data does not exist. We also emphasize that number of significant digits displayed in the operator dimensions does not imply any estimate of the error, and is thus purely cosmetic.

\begin{table}
\centering
\caption{Singlet ($S$) operators in the hypercubic theory. Large expressions for one-loop anomalous dimensions have been expanded at large-$N$. We use Padé$_{[2,2]}$ approximants unless otherwise specified.
}\label{tab:Cubic-singlet}
{\small
\renewcommand{\arraystretch}{1.25}
\begin{tabular}{|c|c|l|llll|c|}
\hline
$i$ & form & $\gamma^{(1)}$ & $C_3$ & $O(3)$  & $C_4$ & $O(4)$ & $R_{O(n)}$
\\\hline
$1$ & $ \phi^2 $ &   $ \frac{2 (n-1)}{3 n} $    &    $  1.56416_{[3,3]}$   &    $ 1.56246_{[3,3]}  $   &    $ 1.61299_{[3,3]}  $   &    $  1.59925_{[3,3]} $   &  $ S $
 \\ 
$2$ & $ \phi^4 $  & $ \frac{4 (n-1)}{3 n} $    &    $ 3.01080_{[3,3]}  $   &    $ 2.99166_{[3,3]}   $   &    $  3.08113_{[3,3]} $   &    $ 2.89025_{[3,3]}  $   &  $ T_4 $
 \\ 
$3$ & $ \phi^4 $ & $ 2 $    &    $ 3.78431_{[3,3]}   $   &    $ 3.78198_{[3,3]}   $   &    $ 3.77094_{[3,3]}  $   &    $ 3.77476_{[3,3]}  $   &  $ S $
 \\ 
$4$ & $ \phi^6 $ & $ 2-\frac{2}{n}+\ldots $    &    $ 5.28692_{[2,3]}  $   &    $5.27248_{[2,3]}   $   &    $ 4.98024_{[2,3]}  $   &    $5.07115_{[2,3]}   $   &  $ T_6 $
 \\ 
$5$ & $ \square \phi^4 $ & $ \frac{4 (n-1)}{3 n} $    &    $ 5.02814  $ &  $ 5.02064  $   & $ 5.12217  $& $ 5.13792  $   &  $ S $
 \\ 
$6$ & $ \phi^6 $  & $ 5-\frac{30}{7 n}+\ldots $    &  $ 5.89107_{[2,3]}  $
   &    $ 5.93449_{[2,3]}  $   &    $ 5.94522_{[2,3]}  $   &    $ 5.75027_{[2,3]}  $   &  $ T_4 $
 \\ 
$7$ & $ \phi^6 $  & $ \frac{8}{3}+\frac{202}{21 n}+\ldots $     &  $6.55939   $  
&    $ 6.54833  $   &    $  6.49221 $   &    $  6.48220 $   &  $ S $
 \\ \hline
 1 & $ T^{\mu\nu}$ & $0$ & $3$&$3$&$3$&$3$  & $S$
 \\ 
$2$ & $ \partial^2 \phi^4 $  & $ \frac{2}{3}+\frac{74}{189 n}+\ldots $    &    $ 4.74870  $   &    $4.78720  $   &    $ 4.93486_{[2,3]}  $   &    $ 4.70737  $   &  $T_4 $
 \\ 
$3$ & $ \partial^2 \phi^4 $ & $ \frac{4}{3}-\frac{176}{27 n}+\ldots $    &    $ 4.77890  $   &    $   4.71855 $   &    $4.99619_{[2,3]}   $   &    $4.77371   $   &  $S $
 \\ 
$4$ & $ \partial^2 \phi^4 $ & $ \frac{13}{9}+\frac{260}{63 n}+\ldots $    &    $ 5.51728  $   &    $ 5.51633  $   &    $ 5.53942  $   &    $  5.54804 $   &  $ S $
 \\ \hline
\end{tabular}
}
\end{table}

To facilitate the reading of the tables, we recall the branching rules from $O(n)$ to hypercubic: 
\begin{align}
\nonumber
    &S\to S, \ V \to V,  \  T\to X\oplus Z,  \  T_3\to  Z_3 \oplus XV\oplus V
    \\
    &A\to B,  \  A_3 \to B_3,  \  H_3 \to XV\oplus VB,   \  T_4\to Z_4\oplus XX\oplus XZ\oplus X \oplus Z\oplus B\oplus S  
   \nonumber \\
   & A_4\to B_4,  \  B_4 \to BB\oplus XX\oplus XZ \nonumber \hspace{-70pt}\\
   &   H_4 \to \overline{XX}\oplus XB\oplus XZ\oplus BZ\oplus Z\oplus B 
  \nonumber  \\
    & Y_{2,1,1}\to VB_{3}\oplus XB \hspace{-70pt}  
      \nonumber\\ 
      & H_5\to V\oplus XV\oplus VB\oplus Z_3\oplus B_3\oplus \overline{XX}V\oplus \ldots \nonumber\\
      \nonumber
 &T_5\to V\oplus XV\oplus VB \oplus Z_3\oplus\ldots
    \\
    &T_6\to S\oplus X\oplus Z\oplus B\oplus \overline{XX}\oplus XZ \oplus Z_4 \oplus XX \oplus  XB \oplus BZ \oplus \ldots
    \label{eq:branching-cubic}
\end{align}
where the $\ldots$ contain representations that are not relevant for our tables (typically irreps that exist for $n \geq 5$). 
Using these and the operators listed for $O(n)$ in Table~\ref{tab:ON-operators-pres}, the operator content can be easily determined. 

For the singlet representation, we have collected our results in Table~\ref{tab:Cubic-singlet}.
We refer to Appendix~\ref{app:cubicTables} for the remaining irreps that exist at $n\geqslant 3$: $(X,Z)$ in Tab.~\ref{tab:Cubic-X};
$(B,V,XV)$ in Tab.~\ref{tab:Cubic-B}; and
$(VB, Z_3, B_3,\overline{XX})$ in Tab.~\ref{tab:JointN3}. 
The irreps that exist only at $n\geqslant 4$, namely 
$(XZ,Z_4,XX,XB,BZ,BB,VB_3,\overline{XX}V)$ are listed in Tab.~\ref{tab:Cubic-XZ}.
There are no operators with $\Delta\leqslant 6$ and $\ell\leqslant 2$ in the irreps $B_4$ and $\overline{XXX}$. We do not give any results for irreps that exist only for $n>4$.

\paragraph{Applications to the conformal bootstrap.}
We would like to emphasize that the dimensions outlined in this work have already had a direct impact on the conformal bootstrap, in part enabling the derivation of the first booststrap island for the Cubic ($n=3$) CFT \cite{Kousvos:2025ext}. The long term goal is to use our data to eventually obtain precise enough estimates for operator dimensions to at the very least rival the precision of other non-perturbative studies, such as \cite{Hasenbusch:2022zur}.

\paragraph{Broken global-symmetry current.}
One operator of particular interest is the broken global symmetry current $\mathcal J^\mu_B$ (note $A_{O(n)} \rightarrow B$, as $O(n) \rightarrow S_n \ltimes (\Z_2)^n$) , for which we now have results at five-loop order
\begin{align}
\Delta_{\mathcal J^\mu_B}&=3-\eps +\frac{(n-4)^2 \eps^2}{54 n^2}+\frac{(n-4) (109 n^3+852 n^2-4272 n+3392) \eps^3}{5832 n^4}
\nonumber\\
&\quad +\bigg(\frac{7217 n^5}{2592}+\frac{5393 n^4}{216}+\frac{21617 n^3}{108}-\frac{262274 n^2}{81}+\frac{288122
   n}{27}-\frac{355472}{27}+\frac{449440}{81 n}\nonumber\\
&\quad\qquad -4
    (n-4) (n+2) (n^3+6 n^2-32 n+28)\zeta _3\bigg)\frac{\eps^4}{{243 n^5}}
\nonumber\\&
\quad +\bigg(\frac{321511 n^5}{839808}+\frac{450427 n^4}{209952}+\frac{7598465 n^3}{104976}+\frac{5664281 n^2}{6561}-\frac{79762885
   n}{6561}+\frac{306665540}{6561}
\nonumber\\&\quad\qquad -\frac{528739864 n^2-428891264 n+133393792}{6561 n^3}-\bigg[\frac{329 n^5}{216}+\frac{422 n^4}{27}
\nonumber\\&\quad\qquad +\frac{2278 n^3}{27}-\frac{28132 n^2}{27}+672 n+\frac{187520}{27} -\frac{367488 n-189952}{27 n^2}\bigg]\zeta_3
\nonumber\\&\quad\qquad 
+(n-4) (n+2) (n^3+6 n^2-32 n+28)\zeta_4 
\nonumber\\&\quad\qquad 
-\frac{40  (n-4) (n^5+7 n^4-18 n^3-11 n^2-30 n+72)}{9 n}\zeta_5
\bigg)\frac{\eps^5}{81n^5}+O(\eps^6)
\end{align}
which agrees with the three loops available from the analytic bootstrap \cite{Dey:2016mcs}. Notice that the anomalous dimension vanishes to three loops if $n=4$.

\section{Outlook}
\label{sec:outlook}
In the present work we performed an extensive extraction of anomalous dimensions in scalar field theory. 
We renormalized the most general Lagrangian with operators up to engineering dimension six and Lorentz rank two. 
The computations are done in the minimal subtraction scheme of dimensional regularization near four dimensions, with applications both to four-dimensional EFTs, 
and to Wilson--Fisher fixed-points (CFTs) in ${d=4-\eps}$ dimensions. 
Using our five-loop results, we were able to extract the state-of-the-art perturbative spectrum for the Ising, $O(n)$ and hypercubic CFTs.
For Ising and $O(n)$, our results give a compelling picture of the spectrum of low-lying composite operators when compared with non-perturbative data -- as illustrated by the figures in this work.
In the case of the hypercubic CFT, 
for which non-perturbative data is less prevalent,
our results aim to guide further research in the conformal bootstrap.
They have have already been used to 
provide rationale for the
assumptions made in \cite{Kousvos:2025ext}.

Moreover, for theories with $O(n)$ symmetry, the wealth of extracted data, easily available in the updated data file of \cite{Henriksson:2022rnm}, can also be used to test different ways of resummation. 
Here we used the simplest type, namely Pad\'e approximants, but more elaborate techniques exist such as Borel summation. It would be interesting to inspect how such resummations improve upon our unconstrained Pad\'es in Figures~\ref{fig:Deltaphi5} and \ref{fig:Tprim}, where it seems like the continuation to known $d=2$ results pulls the non-perturbative results away from our estimates.

Beyond the specific theories considered in this work, 
our (publicly available) general results can be paired with group theory to extract results in any $\phi^4$ theory. 
Primary targets are the experimentally relevant theories summarized in Table~\ref{tab:CFTs} above, for which we outlined a general procedure to extract the scaling dimensions, but our results are equally valid for instance in tensor field theories \cite{Giombi:2017dtl,Benedetti:2020sye,Jepsen:2023pzm}. With data for many operators at several different CFTs one could also infer properties of conformal perturbation theory \cite{Cardy:1996xt} flows between them.

As another application of our results, the multiloop anomalous dimensions in the general scalar theory might yield new insights in the renormalization of the field-space geometry of EFTs~\cite{Helset:2022pde,Jenkins:2023bls,Aigner:2025xyt}. It would be interesting to extend this geometric formulation to operators with non-zero Lorentz spin. 
Rearranging our results in terms of geometric quantities has the potential to lead to new algebraic expressions for the anomalous dimensions of operators with an arbitrary number of fields; see also Appendix~\ref{sec:generalPotentialBetaFunctions}.

Besides the possible applications of our general results, 
there are many other interesting future directions to consider.
For instance, our method can 
be extended to other operators of interest in scalar theories, including those with even higher dimensions, allowing to probe various physically interesting questions. An interesting set of operators which we aim to consider in more detail are those that are odd under spacetime parity, which have larger dimensions than what we considered here.\footnote{From the point of view of the $\eps$-expansion, these have antisymmetrized Lorentz indices. Our $\ell=[1,1]$ operators correspond to parity-odd vectors (pseudovectors) in $d=3$, but also other parity-odd operators of other spin exist, which are not discussed here.} Similarly, evanescent operators in $\phi^4$ theories deserve further study, since they are responsible for non-unitarities in CFTs when the dimensionality $d$ is not an integer number \cite{Hogervorst:2015akt}.\footnote{Of course, evanescent operators in fermionic theories including 4d gauge theories are a well-known topic~\cite{Dugan:1990df}.} It would be interesting to investigate the size of the unitarity violation, and examine if there are examples lower in the spectrum than the scalar examples considered in \cite{Hogervorst:2015akt}. 
More exotic scenarios that one can consider would be Lifshitz fixed-points~\cite{Kachru:2008yh}, and Lorentz-breaking fixed-points more broadly. 
From the computational perspective, this would require to could keep e.g.~$\partial_\mu\phi^3$-type operators non-zero at the fixed-point. These operators are marginally relevant around the free theory, and hence they may lead to interesting fixed-points.\footnote{We thank A. Antunes for bringing this to our attention.}

In this work, we have emphasized the role of the basis of primary operators for structuring the computation and presenting the results. Throughout, we rely on the fact that this basis is equally suited for EFT and CFT applications. 
Here we worked with a pure scalar theory, but we would like to extend this programme towards the most general Lagrangian possible, including gauge fields~\cite{Fonseca:2025zjb,Aebischer:2025zxg,Misiak:2025xzq} as well as fermions. 
Our current method and computational setup is well-equipped for generalizations to gauged scalar EFTs~\cite{Guedes:toAppear}, while the inclusion of fermions requires additional features that consistently implement the Dirac algebra in $d$ dimensions in combination with the employed \Rs operation.
The results of this endeavor would then capture EFTs of interest, such as the SMEFT and LEFT, but also their extensions involving additional light particles.
In addition, there are also many CFT fixed-points with gauge symmetry and/or fermions, both directly in $d=4$ and in $d=3$. Results in the former category, including 4d supersymmetric CFTs and Caswell--Banks--Zaks fixed-points, would be directly accessible from results at $\eps=0$, while
continuing to $\eps=1$ would produce estimates for the latter category, including Gross--Neveu(--Yukawa) theories and 3d gauge theories.

There are also interesting Lagrangians around $d=3$ and $d=6$ dimensions that can be studied following our approach. On the EFT side, $d=3$ would describe the physics of systems at finite temperature \cite{Chala:2025cya}. On the CFT side, this would provide CFT data for novel theories which have a controlled perturbative limit in $d=3-\eps$. These include the tricritical Ising CFT \cite{Stephen1973,Henriksson:2025kws}, as well as various possible generalizations to global symmetry \cite{Hager:2002uq,Yabunaka:2017uox,BenAliZinati:2021rqc,Kapoor:2021lrr}.
On the other hand, the expansion in $6-\eps$ dimensions in the single-scalar field corresponds to the Lee--Yang CFT, recently extended to six loops~\cite{Gracey:2025rnz}, while the multicomponent case has also been considered \cite{Fei:2014xta,Gracey:2015tta,Borinsky:2021jdb}. 
More challenging on the technical side are expansions near other upper critical dimensions, e.g.~$\frac{10}3$ where $\phi^5$ becomes marginal \cite{Codello:2017epp,Katsevich:2025ojk}, and higher multicritical points \cite{Lewis:1978zz} (where $\phi^k$, $k>6$, becomes marginal). Many interesting fixed-points exist for these cases~\cite{Zinati:2019gct,Codello:2020mnt,BenAliZinati:2021rqc}, some of which may be connected to CFTs in $d=2$ in a way yet to be worked out. This could also be a way study in an $\eps$-expansion the potentially irrational 2d CFTs arising from coupling minimal models \cite{Antunes:2022vtb,Antunes:2024mfb}. It would also be useful to determine multiloop spectra in non-linear sigma models in $2+\eps$ dimensions \cite{Polyakov:1975rr,Brezin:1976qa,Hikami:1977vr,Gracey:1990sx}.\footnote{In particular, this could shed light on the status of the relation between the non-linear sigma model and the usual $O(n)$ CFT \cite{Kamal1993,Nahum:2015jya,Jones:2024ept,DeCesare:2025ukl}.}

Finally, in this paper, we have systematized the computation of multiloop anomalous dimensions of composite operators, including spinning and non-singlet ones, in dimensional regularization. This has lead to a large set of perturbative results. 
In a similar vein, it would be desirable to perform a comprehensive computation OPE coefficients perturbatively at a multiloop level. It is worthwhile to explore which aspects of our computational setup can be reused for this purpose.
On the other hand, it would be valuable to perform a systematic study of the spectrum of composite operators within other frameworks, such as the large-$n$ expansion for $O(n)$~\cite{Vasiliev:1982dc,Lang:1993ct,Broadhurst:1996ur,Gracey:1996ub,Derkachov:1997ch} and Gross--Neveu(--Yukawa) models~\cite{Gracey:1992cp,Derkachov:1993uw,Gracey:1993kc,Manashov:2017rrx,Gracey:2018fwq}.

\section*{Acknowledgements}

We are grateful for fruitful collaboration with F.~Herzog and for access to his implementation of the \Rs method, and also to A.~Stergiou for numerous helpful discussions. JRN thanks G.~Guedes for collaboration on related topics and P.~Uwer for remote access to one of the desktop computers of the phenomenology group at Humboldt University Berlin for part of this work. 
JRN is supported by the Yushan Young Scholarship 112V1039 from the Ministry of Education (MOE) of Taiwan, by the National Science and Technology Council (NSTC) grant 113-2112-M-002-038-MY3, and by the NSTC grant 114-2923-M-002-011-MY5. SRK has received funding from the European Research Council (ERC) under the European Union's Horizon 2020 research and innovation programme (grant agreement no.\ 758903), and the Marie Skłodowska-Curie Action (MSCA) High energy Intelligence (HORIZON-MSCA-2023-SE-01-101182937-HeI). JH received funding from the European Research Council (ERC) under the European Union’s Horizon 2020 research and innovation programme under grant agreements number 853507 and 949077.

\appendix

\section{Tables with data}
\label{app:Tables}

In this appendix, we display various tables of data too large to fit in the main text. 

\subsection[\texorpdfstring{$O(n)$ CFT}{O(n) CFT}]{$\boldsymbol{O(n)}$ CFT}
\label{app:ONTables}

Here we include tables with Pad\'e approximants for the scaling dimensions of primary operators for small values of $n$ in the 3d $O(n)$ CFT with $n=2,3,4$. We include all operators regardless of their perturbative field content, organized by the evaluation of the Pad\'e approximant at $n=4$, up to a certain value of $\Delta$. In Table~\ref{tab:ONscalars} we show the scalar operators with $\Delta\leqslant6.5$, in Table~\ref{tab:ONvectors} the spin-1 operators with $\Delta\leqslant5.5$, in Table~\ref{tab:ONTT} the spin-2 operators with $\Delta\leqslant5.6$, and finally in Table~\ref{tab:ONTA} the antisymmetric rank-2 operators (pseudovectors in $d=3$) with $\Delta\leqslant4.5$. The names of operators are those used in the data file of \cite{Henriksson:2022rnm}.

In preparing these tables, we have constructed Pad\'e$_{[a,b]}$ approximants with maximal order $a\leqslant b\leqslant a+1$, see \eqref{eq:padeDef}, except when these have poles for $\eps< 3$, in which case we lowered the order. Where only the $O(\eps)$ term is known, we used the linear truncation. 
We have also taken group-theoretical cancellations into account at $n=2$ and $n=3$, see e.g.~\cite{Cao:2023psi}, marking the missing operators with dashes.

\subsection{Hypercubic-symmetric CFTs}
\label{app:cubicTables}

In the main text, we provided our results for the singlet operators in the hypercubic-symmetric theory (Table~\ref{tab:Cubic-singlet}). Here we provide the remaining tables for the other operators. We use Padé$_{[2,2]}$ approximants unless otherwise specified.
Large expressions for one-loop anomalous dimensions have been expanded at large-$n$. We refer to \cite{Bednyakov:2023lfj} for the definition of the representations.

\vspace{2cm}

\begin{table}[h]
\caption{Pad\'e approximants for scalar operators in the $O(n)$ CFT. We include the operators where the approximant at $n=4$ evaluates to  $\Delta\leqslant 6.5$. Unless reference is given, the maximal order has been computed in this work. }\label{tab:ONscalars}
\centering
{\small
\begin{tabular}{|l|c|l|l|l|l|c|}
\hline
Name & $\O$ & $n=2$ & $n=3$ & $n=4$ & ord. & Pad\'e
\\\hline
 \texttt{Op[V,0,1]} & $\phi$ & .518893 & .518825 & .518178 & 8 \cite{Schnetz:2022nsc} & [4,4] \\
 \texttt{Op[T,0,1]} & $\phi^2$ & 1.23610 & 1.21081 & 1.18879 & 6 \cite{Kompaniets:2019zes} & [3,3] \\
 \texttt{Op[S,0,1]} & $\phi^2$ & 1.49766 & 1.56590 & 1.61979 & 7 \cite{Schnetz:2016fhy} & [2,2] \\
 \texttt{Op[Tm[3],0,1]} & $\phi^3$ & 2.10929 & 2.04293 & 1.98487 & 6 \cite{Bednyakov:2021ojn} & [3,3] \\
 \texttt{Op[Tm[4],0,1]} & $\phi^4$ & 3.10950 & 2.99166 & 2.89025 & 6 \cite{Bednyakov:2021ojn} & [3,3] \\
 \texttt{Op[T,0,2]} & $\phi^4$ & 3.62656 & 3.55003 & 3.48434 & 6 \cite{Bednyakov:2021ojn} & [3,3] \\
 \texttt{Op[S,0,2]} & $\phi^4$ & 3.80345 & 3.79362 & 3.78335 & 7 \cite{Schnetz:2016fhy} & [3,4] \\
 \texttt{Op[Tm[5],0,1]} & $\phi^5$ & 4.21649 & 4.03910 & 3.88513 & 6 \cite{Bednyakov:2022guj}& [3,3] \\
 \texttt{Op[B4,0,1]} &  $\square\phi^4$ & --- & --- & 4.43750 & 5 & [2,3] \\
 \texttt{Op[Tm[3],0,2]} & $\phi^5$ & 4.85021 & 4.71059 & 4.59009 & 5 & [2,3] \\
 \texttt{Op[T,0,3]} & $\square\phi^4$ & --- & 4.62792 & 4.67742 & 5 & [2,3] \\
 \texttt{Op[Tm[6],0,1]} & $\phi^6$ & 5.40689 & 5.16743 & 4.95721 & 6 \cite{Bednyakov:2022guj}  & [3,3] \\
 \texttt{Op[V,0,2]} & $\phi^5$ & 5.14493 & 5.05136 & 4.97260 & 5 & [2,3] \\
 \texttt{Op[S,0,3]} & $\square\phi^4$ & 4.85594 & 5.02573 & 5.14313 & 5 & [2,3] \\
 \texttt{Op[YT[\{3,2\}],0,1]} & $\square\phi^5$ & ---& --- & 5.50000 & 1 & [1,0] \\
 \texttt{Op[Tm[3],0,3]} &$\square\phi^5$  & --- & 5.68182 & 5.66667 & 1 & [1,0] \\
 \texttt{Op[Tm[4],0,2]} & $\phi^6$ & 6.15133 & 5.93449 & 5.75027 & 5 & [2,3] \\
 \texttt{Op[Hm[3],0,1]} & $\square\phi^5$  & --- & 5.95455 & 5.91667 & 1 & [1,0] \\
 \texttt{Op[Tm[7],0,1]} & $\phi^7$ & 6.60153 & 6.34254 & 6.08378 & 6 \cite{Bednyakov:2022guj} & [3,3] \\
 \texttt{Op[V,0,3]} & $\square\phi^5$  & 6.10000 & 6.13636 & 6.16667 & 1 & [1,0] \\
 \texttt{Op[B4,0,2]} & $\square^2\phi^4$ & --- & --- & 6.20456 & 2 & [1,1] \\
 \texttt{Op[T,0,4]} & $\phi^6$ & 6.55624 & 6.36721 & 6.20907 & 5 & [2,3] \\
 \texttt{Op[T,0,5]} & $\square^2\phi^4$ & --- & 6.25901 & 6.24382 & 2 & [1,1] \\
 \texttt{Op[S,0,5]} & $\square^2\phi^4$ & 6.28330 & 6.27224 & 6.25450 & 4 & [2,2] \\
 \texttt{Op[S,0,4]} & $\phi^6$ & 6.69438 & 6.55755 & 6.44478 & 5  & [2,3] 
\\\hline
\end{tabular}}
\end{table}

\begin{table}
\caption{Pad\'e approximants for spin-1 operators in the $O(n)$ CFT. We include the operators where the approximant at $n=4$ evaluates to  $\Delta\leqslant 5.5$. }\label{tab:ONvectors}
\centering
{\small
\begin{tabular}{|l|c|l|l|l|l|c|}
\hline
Name& $\O$ & $n=2$ & $n=3$ & $n=4$ & ord. & Pad\'e
\\\hline
 \texttt{Op[A,1,1]} & $\de\phi^2$ & 2 & 2 & 2 & & exact
\\
 \texttt{Op[Hm[3],1,1]} & $\de\phi^3$ & --- & 2.76643 & 2.73474 & 5 & [2,3] \\
 \texttt{Op[V,1,1]} & $\de\phi^3$ & 2.93795 & 3.01582 & 3.08077 & 5 & [2,3] \\
 \texttt{Op[Hm[4],1,1]} & $\de\phi^4$ & --- & 3.67644 & 3.59874 & 5 & [2,3] \\
 \texttt{Op[A,1,2]} & $\de\phi^4$ & 4.12567 & 4.09582 & 4.07219 & 5 & [2,3] \\
 \texttt{Op[Hm[5],1,1]} & $\de\phi^5$ & --- & 4.72701 & 4.59246 & 5 & [2,3] \\
 \texttt{Op[Tm[3],1,1]} & $\de\phi^5$ & 5.20749 & 5.13483 & 5.07027 & 5 & [2,3] \\
 \texttt{Op[Hm[3],1,2]} & $\de\phi^5$  & --- & 5.23304 & 5.14295 & 5 & [2,3] \\
 \texttt{Op[YT[\{2,1,1\}],1,1]} & $\de\square\phi^4$ & --- & -- & 5.22222 & 1 & [1,0] \\
 \texttt{Op[A,1,3]} & $\de\square\phi^4$ & --- & 5.27745 & 5.27778 & 1 & [1,0] \\
 \texttt{Op[V,1,2]} & $\de\phi^5$  & 5.42101 & 5.39144 & 5.36799 & 5 & [2,3] \\
 \texttt{Op[Hm[4],1,2]} & $\de\square\phi^4$ & --- & 5.48485 & 5.44444 & 1 & [1,0]
\\\hline
\end{tabular}}
\end{table}

\begin{table}
\caption{Pad\'e approximants for spin-2 operators in the $O(n)$ CFT. We include the operators where the approximant at $n=4$ evaluates to  $\Delta\leqslant 5.6$.}\label{tab:ONTT}
\centering
{\small
\begin{tabular}{|l|c|l|l|l|l|c|}
\hline
Name & $\O$ & $n=2$ & $n=3$ & $n=4$ & ord. & Pad\'e
\\\hline
 \texttt{Op[S,2,1]} & $\de^2\phi^2$& 3 & 3 & 3 & & exact
\\
 \texttt{Op[T,2,1]} & $\de^2\phi^2$ & 3.01299 & 3.01570 & 3.01715 & 5 & [2,3] \\
 \texttt{Op[Hm[3],2,1]} & $\de^2\phi^3$ & --- & 3.59756 & 3.58509 & 5 & [2,3] \\
 \texttt{Op[V,2,1]} & $\de^2\phi^3$ & 3.63461 & 3.63042 & 3.62249 & 5 & [2,3] \\
 \texttt{Op[Tm[3],2,1]} & $\de^2\phi^3$ & 3.87913 & 3.83767 & 3.80170 & 5 & [2,3] \\
 \texttt{Op[V,2,2]} & $\de^2\phi^3$ & 4.20310 & 4.23634 & 4.26667 & 5 & [2,2] \\
 \texttt{Op[B4,2,1]} & $\de^2\phi^4$  & --- & --- & 4.33014 & 5 & [2,3] \\
 \texttt{Op[T,2,2]} & $\de^2\phi^4$  & --- & 4.42893 & 4.39845 & 5 & [2,3] \\
 \texttt{Op[Hm[4],2,1]} &$\de^2\phi^4$  & --- & 4.49234 & 4.43473 & 5 & [2,3] \\
 \texttt{Op[T,2,3]} & $\de^2\phi^4$  & 4.64524 & 4.64707 & 4.66942 & 5 & [2,2] \\
 \texttt{Op[Tm[4],2,1]} & $\de^2\phi^4$  & 4.88784 & 4.78899 & 4.70379 & 5 & [2,3] \\
 \texttt{Op[S,2,2]} & $\de^2\phi^4$  & 4.61906 & 4.70246 & 4.76429 & 5 & [2,3] \\
 \texttt{Op[T,2,4]} & $\de^2\phi^4$  & 5.34496 & 5.30273 & 5.26686 & 5 & [2,3] \\
 \texttt{Op[YT[\{3,2\}],2,1]} & $\de^2\phi^5$  & --- & --- & 5.38889 & 1 & [1,0] \\
 \texttt{Op[Tm[3],2,2]} & $\de^2\phi^5$  & --- & 5.50436 & 5.44354 & 1 & [1,0] \\
 \texttt{Op[S,2,3]} & $\de^2\phi^4$  & 5.48733 & 5.51633 & 5.54804 & 5 & [2,2]\\
 \texttt{Op[Hm[5],2,1]} & $\de^2\phi^5$  &  5.76667 & 5.65152 & 5.55556 & 1 & [1,0] 
\\\hline
\end{tabular}
}
\end{table}

\begin{table}
\caption{Pad\'e approximants for antisymmetric spin-2 operators in the $O(n)$ CFT. We include the operators where the approximant at $n=4$ evaluates to  $\Delta\leqslant 4.5$.}\label{tab:ONTA}
\centering
{\small
\begin{tabular}{|l|c|l|l|l|l|c|}
\hline
Name & $\O$ & $n=2$ & $n=3$ & $n=4$ & ord. & Pad\'e
\\\hline
 \texttt{Op[A3,[1,1],1]} & $\de^2\phi^3$ & --- & 3.44636 & 3.44738 & 5 & [2,3] \\
 \texttt{Op[YT[\{2,1,1]\}],[1,1],1]} & $\de^2\phi^4$ & --- & --- & 4.25817 & 5 & [2,3] \\
 \texttt{Op[A,[1,1],1]} & $\de^2\phi^4$ & --- & 4.42699 & 4.49309 & 5 & [2,3]
\\\hline
\end{tabular}
}
\end{table}

\begin{table}
\hspace*{-1cm}
\centering
\caption{Representation $X$ and $Z$ operators in the hypercubic theory. For $X$, note the comparably large splitting of the $T_6$ operators (scalars 4 and 6). 
}\label{tab:Cubic-X}
{\small
\renewcommand{\arraystretch}{1.25}
\begin{adjustbox}{center,raise=0pt,margin*= -1cm 0 0 0}
\begin{tabular}{c|c|c|l|llll|c|}
\cline{2-9}
&$i$ & form & $\gamma^{(1)}$ & $C_3$ & $O(3)$  & $C_4$ & $O(4)$ & $R_{O(n)}$
\\\cline{2-9}
\multicolumn{2}{c}{\rule{0pt}{1ex}}\\[-5.5mm]
 \cline{2-9}
 \multirow{13}{*}{$\bm{X} \left\{ \vphantom{\rule{0pt}{4.15cm}} \right.$\hspace{-2mm}}
&$1$ & $ \phi^2 $ &   $  \frac{n-2}{3 n} $    &    $   1.20378_{[3,3]}$   &    $ 1.21081_{[3,3]}  $   &    $  1.28397_{[3,3]} $   &    $  1.18879_{[3,3]} $   &  $ T $
 \\ &
$2$ & $ \phi^4 $  & $ 1 $    &    $ 3.00296_{[3,3]}  $   &    $ 2.99166_{[3,3]}  $   &    $  2.95137_{[3,3]} $   &    $  2.89025_{[3,3]} $   &  $ T_4 $
 \\ &
$3$ & $ \phi^4 $ & $ \frac{2 (3 n-2)}{3 n} $    &    $  3.55394_{[3,3]} $   &    $ 3.55003_{[3,3]}  $   &    $  3.68480_{[3,3]} $   &    $ 3.48434_{[3,3]}  $   &  $ T $
 \\ &
$4$ & $ \phi^6 $ & $ \frac{5}{3}+\frac{2}{3 n}+\ldots $    &    $5.24670_{[2,3]}  $   &    $5.27248_{[2,3]}    $   &    $  5.25948_{[2,3]} $   &    $5.07115_{[2,3]}   $   &  $ T_6 $
 \\ &
$5$ & $ \square \phi^4 $ & $  \frac{3 n-4}{3 n} $    &    $ 4.65897  $   &    $ 4.65502   $   &    $  4.78391 $   &    $ 4.69463  $   &  $ T$
 \\ &
$6$ & $ \phi^6 $ & $  \frac{7}{3}+\frac{14}{3 n}+\ldots$    &   ---    &    ---   &    $ 4.90848_{[2,3]}  
$   &    $5.07115_{[2,3]}   $   &  $ T_6 $
 \\ &
$7$ & $ \phi^6 $  & $ \frac{8}{3}+\frac{62}{21 n}+\ldots$    &    $5.95187_{[2,3]}   $   &    $ 5.93449_{[2,3]}  $   &    $  5.78754_{[2,3]} $   &    $5.75027_{[2,3]}   $   &  $ T_4 $
 \\ &
$8$ & $ \phi^6 $  & $   5-\frac{30}{7 n}+\ldots $    &    $ 6.32679  $   &    $ 6.40469   $   &    $   6.42379$   &    $ 6.31705   $   &  $ T $
 \\ \cline{2-9} &
$ 1 $ & $ \partial^2\phi^2$ & $ 0  $ & $3.01664 $&$3.01496 $&$3.01007 $&$ 3.01648$  & $T $
 \\ &
$2$ & $ \partial^2 \phi^4 $  & $  \frac{1}{3}+\frac{22}{27 n}+\ldots $    &    $  4.46347 $   &    $ 4.45030  $   &    $4.46322   $   &    $  4.41762 $   &  $T  $
 \\ &
$3$ & $ \partial^2 \phi^4 $  & $  \frac{2}{3}+\frac{46}{189 n}+\ldots $    &    $ 4.65803  $   &    $ 4.64706  $   &    $  4.64477 $   &    $ 4.66942  $   &  $  T$
 \\ &
$4$ & $ \partial^2 \phi^4 $  & $  1-\frac{50}{27n}+\ldots$    &    $  4.80324 $   &    $  4.78720 $   &    $ 4.78596  $   &    $ 4.70737  $   &  $  T_4$   
 \\ &
$5$ & $ \partial^2 \phi^4 $  & $  \frac{13}{9}-\frac{34}{63 n}+\ldots $    &    $  5.29778  $   &    $5.28938   $   &    $5.37355_{[2,3]}  $   &    $ 5.25447  $   &  $ T $
 \\ \cline{2-9}
 \multicolumn{2}{c}{\rule{0pt}{1ex}}\\[-5.5mm]
 \cline{2-9}
 \multirow{16}{*}{$\bm{Z} \left\{ \vphantom{\rule{0pt}{5.13cm}} \right.$\hspace{-2mm}}
&$1$ & $ \phi^2 $ &   $ \frac{2}{3 n}  $    &    $   1.20704_{[3,3]}$   &    $  1.21081_{[3,3]} $   &    $ 1.11931_{[3,3]}  $   &    $   1.18879_{[3,3]}$   &  $ T $
 \\& 
$2$ & $ \phi^4 $  & $  \frac{5 n+12-\sqrt{n^2+24 n-48}}{6 n}$    &    $ 2.98797_{[3,3]}  $   &    $ 2.99166_{[3,3]}  $   &    $ 2.78167_{[3,3]}  $   &    $  2.89025_{[3,3]} $   &  $ T_4 $
 \\ &
$3$ & $ \phi^4 $ & $ \frac{5 n+12+\sqrt{n^2+24 n-48}}{6 n} $    &    $ 3.54372_{[3,3]}  $   &    $ 3.55003_{[3,3]}  $   &    $ 3.34039_{[3,3]} $   &    $3.48434_{[3,3]}  $   &  $ T $
 \\ &
$4$ & $ \phi^6 $ & $ \frac{4}{3}-\frac{2}{3 n}+\ldots  $    &    $   5.29436_{[2,3]} $   &    $5.27248_{[2,3]}    $   &    $5.04458_{[2,3]}   $   &    $5.07115_{[2,3]}    $   &  $ T_6 $
 \\ &
$5$ & $ \square \phi^4 $ & $  \frac{2}{3} $    &    $   4.66446 $   &    $ 4.65502   $   &    $  4.62555 $   &    $ 4.69463  $   &  $ T$
 \\ &
$6$ & $ \phi^6 $ & $ \frac{5}{3}+\frac{22}{3 n}+\ldots$    &      $5.25139_{[2,3]}$      &    $5.27248_{[2,3]}    $   &    $   5.23829_{[2,3]}$   &    $5.07115_{[2,3]}    $   &  $ T_6 $
 \\ &
$7$ & $ \phi^6 $ & $ 2+\frac{6}{n}-\frac{\sqrt{96}}{n^{3/2}}+\ldots  $    &   ---   &    
---  &    $  4.81069_{[2,3]} $   &    $5.07115_{[2,3]}    $   &  $ T_6 $
 \\ &
$8$ & $ \phi^6 $  & $2+\frac{6}{n}+\frac{\sqrt{96}}{n^{3/2}}+\ldots $    &    $ 5.98963_{[2,3]}$   &    $5.95187_{[2,3]}   $   &    $ 5.62653_{[2,3]}  $   &    $ 5.75027_{[2,3]}  $   &  $ T_4 $
 \\ &
$9$ & $ \phi^6 $  & $  \frac{10}{3}+\frac{2}{3 n}+\ldots  $    &    $ 6.46374  $   &    $ 6.40469   $   &    $ 6.29730  $   &    $ 6.31705   $   &  $ T $
 \\ \cline{2-9}&
$1$ & $ \partial \phi^4 $  & $ \frac{2}{3}   $    &    $ 3.70889  $   &    $  3.68387 $   &    $3.66052   $   &    $ 3.60919  $   &  $ H_4 $
 \\ \cline{2-9}&
$ 1 $ & $ \partial^2\phi^2$ & $ 0  $ & $3.01415 $&$3.01496 $&$ 3.02232$&$ 3.01648$  & $T $
 \\ &
$2$ & $ \partial^2 \phi^4 $  & $  0+\frac{58}{27 n}+\ldots $    &    $4.47814   $   &    $ 4.45030  $   &    $4.28722  $   &    $  4.41762 $   &  $T  $
 \\ &
$3$ & $ \partial^2 \phi^4 $  & $ \frac{5}{9}-\frac{8}{9 n}+\ldots $    &    $ 4.52189 $   &    $ 4.50492  $   &    $4.55369  $   &    $4.44860  $   &  $  H_4$
 \\ &
$4$ & $ \partial^2 \phi^4 $  & $\frac{2}{3}+\frac{13-\sqrt{889}}{27 n}+\ldots   $    &    $ 4.63068 $   &    $ 4.64706  $   &    $  4.68499$   &    $ 4.66942  $   &  $  T$
 \\ &
$5$ & $ \partial^2 \phi^4 $  & $ \frac{2}{3}+\frac{13+\sqrt{889}}{27 n}+\ldots $    &    $  4.79277 $   &    $  4.78720 $   &    $  4.65004 $   &    $ 4.70737  $   &  $  T_4$   
 \\ &
$6$ & $ \partial^2 \phi^4 $  & $1+\frac{2}{3 n}+\ldots  $    &    $ 5.30004  $   &    $5.28938   $   &    $ 5.17124 $   &    $ 5.25447  $   &  $ T $
 \\ \cline{2-9}
\end{tabular}
\end{adjustbox}
}
\end{table}

\begin{table}
\centering
\caption{Representation $B$, $V$ and $XV$ operators in the hypercubic theory. 
}\label{tab:Cubic-B}
{\small
\renewcommand{\arraystretch}{1.25}
\begin{adjustbox}{center,raise=0pt,margin*= -1cm 0 0 0}
\begin{tabular}{c|c|c|l|llll|c|}
\cline{2-9}
&$i$ & form & $\gamma^{(1)}$ & $C_3$ & $O(3)$  & $C_4$ & $O(4)$ & $R_{O(n)}$
\\\cline{2-9}
\multicolumn{2}{c}{\rule{0pt}{1ex}}\\[-5.5mm]
 \cline{2-9}
 \multirow{8}{*}{$\phantom{\bm{X}}\bm{B} \left\{ \vphantom{\rule{0pt}{2.5cm}} \right.$\hspace{-2mm}}
&$1$ & $ \phi^6 $ & $ \frac{5 (n+2)}{3 n}  $    &    $5.30113_{[2,3]}   $   &    $  5.27248_{[2,3]}  $   &    $5.00569_{[2,3]}   $   &    $5.07115_{[2,3]}$   &  $ T_6 $
 \\ &
$2$ & $ \phi^6 $  & $ \frac{2 (5 n+1)}{3 n} $    &    $5.91670_{[2,3]}   $   &    $  5.95187_{[2,3]}$   &    $ 5.88022_{[2,3]} $   &    $5.75027_{[2,3]}  $   &  $  T_4$
 \\ \cline{2-9}&
$ 1 $ & $ \partial \phi^2$ &$0 $ & $ 2.00092_{[2,1]}$ & $ 2$&$2.00418_{[2,3]} $&$2 $  & $ A$
 \\ &
$2$ & $ \partial \phi^4 $   & $ \frac{5 n+2-\sqrt{n^2+12 n-28}}{6 n} $    &    $ 3.70481 $   &    $  3.68387 $   &    $ 3.61676$   &    $ 3.60919  $   &  $ H_4 $
 \\ &
$3$ & $ \partial \phi^4  $    &    $  \frac{5 n+2+\sqrt{n^2+12 n-28}}{6 n} $   &    $  4.08482 $   &    $ 4.07956  $   &    $ 4.05385  $   &  $4.05888 $ &  $A $
 \\ \cline{2-9}&
$ 1 $ & $ \partial^{[1,1]}\phi^4$ & $0 $ & $ 4.45226  $&$4.44619 $&$  4.49600  $&$ 4.50576 $  & $ A$
 \\ \cline{2-9}&
$1$ & $ \partial^2 \phi^4 $  & $ \frac{11 n-\sqrt{n^2+24 n-48}}{18 n} $    &   $4.51586 $ &    $ 4.50492  $   &    $4.50151 $   &    $4.44860  $   &  $  H_4$
 \\ &
$2$ & $ \partial^2 \phi^4 $  & $\frac{11 n+\sqrt{n^2+24 n-48}}{18 n} $    &    $4.79652 $   &    $  4.78720 $   &    $4.78940  $   &    $ 4.70737  $   &  $  T_4$ 
\\\cline{2-9}
\multicolumn{2}{c}{\rule{0pt}{1ex}}\\[-5.5mm]
 \cline{2-9}
  \multirow{13}{*}{$\phantom{\bm{X}}\bm{V} \left\{ \vphantom{\rule{0pt}{4.15cm}} \right.$\hspace{-2mm}}
& $1$ & $ \phi $ &   $ 0 $    &    $  0.51837_{[3,3]} $   &    $ 0.51842_{[3,3]}  $   &    $ 0.51802_{[3,3]} $   &    $ 0.51789_{[3,3]}  $   &  $  V$
 \\ &
$2$ & $ \phi^3 $  & $ \frac{2 (n-1)}{3 n}  $    &    $   2.05409_{[3,3]}  $   &    $2.04293_{[3,3]} $   &    $ 2.08961_{[3,3]}  $   &    $ 1.98487_{[3,3]}  $   &  $ T_3 $
 \\ &
$3$ & $ \phi^5 $ & $\frac{4}{3}-\frac{4}{3 n}+\ldots   $    &    $  3.97522 $   &    $  3.97762  $   &    $ 3.73347  $   &    $ 3.84152  $   &  $ T_5 $
 \\ &
$4$ & $ \phi^5 $ & $\frac{5}{3}+\frac{10}{3 n}+\ldots   $    &    $4.05568 $   &    $  3.97762    $   &    $  4.07163 $   &    $ 3.84152  $   &  $  T_5$
 \\ &
$5$ & $ \phi^5 $ & $  2+\frac{8}{3 n}+\ldots  $    &    $ 4.59812  $   &    $ 4.71059_{[2,3]}$ &    $ 4.68930_{[2,3]}  $   &    $4.59009_{[2,3]}  $   &  $ T_3 $
 \\ &
$6$ & $ \phi^5 $ & $  \frac{10}{3}-\frac{8}{3 n}+\ldots$    &    $  5.16396 $   &    $  5.05136_{[2,3]}  $   &    $  5.01620_{[2,3]} $   &    $4.97260_{[2,3]} $   &  $ V $
 \\ \cline{2-9}&
 $1$ & $ \partial\phi^3$ & $ \frac{2 (n-1)}{3 n}   $ &    $ 3.02719  $   &    $  3.02238 $   &    $ 3.07082  $   &    $ 3.08078  $   &  $  V$
 \\ &
$2$ & $ \partial \phi^5$  & $  \frac{4}{3}-\frac{4}{3 n}+\ldots  $    &    $ 4.72580 $   &    $ 4.67814    $   &    $ 4.59226  $   &    $  4.55676 $   &  $ H_5$
 \\ &
$3$ & $ \partial \phi^5  $   & $ \frac{5}{3}+\frac{2}{3 n}+\ldots $    &    $  5.09323 $   &    $  5.08268 $   &    $ 5.10972  $   &    $ 5.02373  $   &  $ T_3$
 \\ &
$4$ & $  \partial \phi^5 $    & $  2+\frac{1}{n}+\ldots $   &    $   5.39613_{[2,3]}$   &    $  5.39144_{[2,3]} $   &    $  5.36177_{[2,3]} $   &    $  5.36799_{[2,3]} $   &  $  V$
 \\ \cline{2-9}&
$ 1  $& $ \partial^2\phi^3$ & $  0+\frac{20}{27 n}+\ldots $ & $ 3.65776$&$3.64056 $&$ 3.60358$&$ 3.63179$  & $ V$
 \\ &
$2$ & $ \partial^2 \phi^3 $  & $  \frac{5}{9}-\frac{20}{9 n}+\ldots  $    &    $ 3.83353  $   &    $3.84461   $   &    $ 3.90198  $   &    $  3.80917 $   &  $T_3  $
 \\ &
$3$ & $ \partial^2 \phi^3 $  & $ \frac{2}{3}+\frac{22}{27 n}+\ldots  $    &    $  4.23804 $   &    $4.23634   $   &    $  4.25952 $   &    $  4.26667 $   &  $  V$
\\\cline{2-9}
\multicolumn{2}{c}{\rule{0pt}{1ex}}\\[-5.5mm]
 \cline{2-9}
 \multirow{11}{*}{$\bm{XV} \left\{ \vphantom{\rule{0pt}{3.6cm}} \right.$\hspace{-2mm}}
&$1$ & $ \phi^3 $ &   $  \frac{n+2}{3 n} $    &    $  2.04313_{[3,3]} $   &    $ 2.04293_{[3,3]}  $   &    $  1.97457_{[3,3]} $   &    $  1.98487_{[3,3]} $   &  $ T_3 $
 \\ &
$2$ & $ \phi^5 $ & $ 1+\frac{4}{3 n}+\ldots $    &    $4.13122_{[2,3]} $    &    $  3.97762    $   &    $3.62914   $   &    $ 3.84152  $   &  $  T_5$
 \\ &
$3$ & $ \phi^5 $ & $ \frac{4}{3}+\frac{10}{3 n}+\ldots  $    &    $4.03161 $    &    $  3.97762    $   &    $ 4.04263_{[2,3]}  $   &    $ 3.84152  $   &  $  T_5$
 \\ &
 $4$ & $ \phi^5 $ & $ 2+\frac{4}{3 n}+\ldots$  &    $ 4.72433_{[2,3]}  $
   &    $ 4.71059_{[2,3]}$    &      $4.57158_{[2,3]}$
   &    $ 4.59009_{[2,3]}  $   &  $ T_3 $
 \\ \cline{2-9}&
$ 1$  & $ \partial\phi^3$ & $ \frac{n-1}{3 n}  $ & $2.78212 $&$2.78071 $&$2.82326 $&$2.74839 $  & $H_3 $
 \\ &
$2$ & $ \partial \phi^5$  & $   1+\frac{1}{3 n}+\ldots $    &    $ 4.70225  $   &    $ 4.67814 $   &    $  4.62253 $   &    $ 4.55676  $   &  $H_5  $
 \\ &
$3$ & $ \partial \phi^5$  & $ 1+\frac{4}{3 n}+\ldots  $    &      ---    &    
--- &    $ 4.49850  $   &    $ 4.55676  $   &  $H_5  $
 \\ &
$4$ & $ \partial \phi^5  $    &    $ \frac{4}{3}+\frac{5}{3 n}+\ldots  $   &    $ 5.11382  $   &    $ 5.08268  $   &    $ 4.99104  $   &   $  5.02373 $   &  $ T_3$
 \\ &
$5$ & $  \partial \phi^5 $    &    $2-\frac{1}{3 n}+\ldots   $   &    $  5.11359 $   &    $ 5.09317  $   &    $  5.19613 $   &   $ 5.01752  $   &  $  H_3$
 \\ \cline{2-9}&
$ 1  $ & $ \partial^2\phi^3$ & $   \frac{3 n+2-\sqrt{9 n^2-68 n+164}}{18 n}$ & $3.61665 $&$3.60745 $&$ 3.60533$&$ 3.59399$  & $H_3 $
 \\ &
$2$ & $ \partial^2 \phi^3 $  & $  \frac{3 n+2+\sqrt{9 n^2-68 n+164}}{18 n} $    &    $ 3.84307 $   &    $3.84461   $   &    $3.82098 $   &    $  3.80917 $   &  $T_3  $
 \\ \cline{2-9}
\end{tabular}
\end{adjustbox}
}
\end{table}

\begin{table}
\centering
\caption{Operators in various representations of hypercubic symmetry.}\label{tab:JointN3}
{\small
\renewcommand{\arraystretch}{1.25}
\begin{adjustbox}{center,raise=0pt,margin*= -1cm 0 0 0}
\begin{tabular}{r|c|c|l|llll|c|}
\cline{2-9}
&$i$ & form & $\gamma^{(1)}$ & $C_3$ & $O(3)$  & $C_4$ & $O(4)$ & $R_{O(n)}$
\\\cline{2-9}
\multicolumn{2}{c}{\rule{0pt}{1ex}}\\[-5.5mm]
 \cline{2-9}
 \multirow{6}{*}{$\bm{VB} \left\{ \vphantom{\rule{0pt}{2cm}} \right.$\hspace{-2mm}}
&$1$ & $ \phi^5 $ & $  \frac{3 n+8}{3 n}  $    &    $ 3.95600 $    &    $  3.97762    $   &    $ 3.75148  $   &    $ 3.84152  $   &  $  T_5$
 \\ \cline{2-9}&
$ 1 $ & $ \partial\phi^3$ & $  \frac{1}{n}  $ & $2.78406 $&$2.78071 $&$ 2.65526$&$2.74839 $  & $H_3 $
 \\ &
$ 2 $ & $ \partial\phi^5$ & $  \frac{5 n+18-\sqrt{n^2+32 n-80}}{6 n} $ & --- &---&$4.32080   $&$5.01752 $  & $ H_5$
 \\ &
$ 3 $ & $ \partial\phi^5$ & $  \frac{n+1}{n}  $ & $4.69798  $&$ 5.09317 $&$ 4.59905  $&$ 5.01752 $  & $H_5$
 \\ &
$ 4 $ & $ \partial\phi^5$ & $  \frac{5 n+18+\sqrt{n^2+32 n-80}}{6 n} $ & $ 5.25852_{[2,3]}  $ & $ 5.09317 $&$ 5.00160_{[2,3]} $&$ 5.01752 $ & $H_3$
 \\ \cline{2-9}&
$ 1  $ & $ \partial^2\phi^3$ & $ \frac49  $ & $3.60791 $&$3.60745 $&$ 3.55574$&$ 3.59399$  & $H_3 $
\\\cline{2-9}
\multicolumn{2}{c}{\rule{0pt}{1ex}}\\[-5.5mm]
 \cline{2-9}
 \multirow{6}{*}{$\bm{Z_3} \left\{ \vphantom{\rule{0pt}{2cm}} \right.$\hspace{-2mm}}
&
$1$ & $ \phi^3 $ &   $ \frac2n $    &    $  2.03072_{[2,3]} $   &    $ 2.04293_{[3,3]}  $   &    $  1.82995_{[3,3]} $   &    $   1.98487_{[3,3]}$   &  $ T_3 $
 \\ &
$2$ & $ \phi^5 $ & $ \frac{5 n+30-\sqrt{n^2+44 n-92}}{6 n}$    &     --- &   ---  &    $ 3.68837_{[2,3]}  $   &    $ 3.84152  $   &  $  T_5$
 \\ &
$3$ & $ \phi^5 $ & $\frac{5 n+30+\sqrt{n^2+44 n-92}}{6 n}$ &    $4.77087 _{[2,3]}  $   &    $ 4.71059_{[2,3]} $     &    $ 4.37167_{[2,3]} $ &    $ 4.59009 _{[2,3]} $   &  $ T_3 $
 \\ \cline{2-9}&
$1$ & $ \partial \phi^5$  & $\frac{5 n+15-\sqrt{n^2+14 n-47}}{6 n}$    &    $ 4.15531_{[2,3]} $   &    $ 4.67814    $   &    $4.36022 $   &    $  4.55676 $   &  $ H_5$
 \\ &
$2$ & $ \partial \phi^5  $   & $\frac{5 n+15+\sqrt{n^2+14 n-47}}{6 n}$    &    $  4.77087_{[2,3]} $   &    $  5.08268 $   &    $4.88432  $   &    $ 5.02373  $   &  $ T_3$
\\\cline{2-9}&
$1$ & $ \partial^2 \phi^3 $  & $ \frac{10}{9 n}  $    &    $3.85216   $   &    $3.84461   $   &    $ 3.70764$   &    $  3.80917 $   &  $T_3  $
\\\cline{2-9}
\multicolumn{2}{c}{\rule{0pt}{1ex}}\\[-5.5mm]
 \cline{2-9}
 \multirow{2}{*}{$\bm{B_3} \left\{ \vphantom{\rule{0pt}{0.65cm}} \right.$\hspace{-2mm}}
 &$1$ & $ \partial \phi^5$  & $ \frac{n+1}{n} $    &    $  4.70672 $   &    $ 4.67814    $   &    $ 4.58095 $   &    $  4.55676 $   &  $ H_5$
 \\ \cline{2-9}&
$ 1 $ & $ \partial^{[1,1]}\phi^3$ & $ 0  $ & $ 3.45055$&$ 3.45177$&$3.45708 $&$3.45225 $  & $ A_3$
\\\cline{2-9}
\multicolumn{2}{c}{\rule{0pt}{1ex}}\\[-5.5mm]
 \cline{2-9}
 \multirow{3}{*}{$\bm{\overline{XX}} \left\{ \vphantom{\rule{0pt}{1cm}} \right.$\hspace{-2mm}}
&
$1$ & $ \phi^6 $ & $  \frac{7 n+2}{3 n}  $    &    $ 5.06226   $   &    $   5.27248_{[2,3]} $   &    $  5.03299$   &    $5.07115_{[2,3]} $   &  $ T_6 $
 \\ \cline{2-9}&
$1$ & $ \partial \phi^4 $   & $  \frac{2}{3} $    &    $ 3.68242 $   &    $  3.68387 $   &    $3.70964$   &    $ 3.60919  $   &  $ H_4 $
 \\ \cline{2-9}&
$1$ & $ \partial^2 \phi^4 $  & $ \frac{n+2}{3 n} $    &   $4.51816$  &    $ 4.50492  $
&    $4.43009 $   &    $4.44860  $   &  $  H_4$
 \\ \cline{2-9}
\end{tabular}
\end{adjustbox}
}
\end{table}

\begin{table}
\centering
\caption{Operators in various representations of hypercubic symmetry.
One of the $XZ$ $\phi^6$ operators (entry $i=2$ at $\ell=0$) and the $Z_4$ ($i=2$) entry vanish identically at $n=4$.
We keep these entries to remind the reader that they exists for $n \geq 5$.
}\label{tab:Cubic-XZ}
{\small
\renewcommand{\arraystretch}{1.25}
\begin{tabular}{r|c|c|l|ll|c|}
\cline{2-7}
&$i$ & form & $\gamma^{(1)}$ & $C_4$ & $O(4)$ & $R_{O(n)}$
\\\cline{2-7}
\multicolumn{2}{c}{\rule{0pt}{1ex}}\\[-5.5mm]
 \cline{2-7}
 \multirow{9}{*}{$\bm{XZ} \left\{ \vphantom{\rule{0pt}{3cm}} \right.$\hspace{-2mm}}
&
$1$ & $ \phi^4 $  &    $ \frac{n+8}{3 n} $    &      $ 2.72527_{[3,3]}  $   &    $ 2.88576_{[3,3]}  $   &    $ T_4  $ 
 \\ &
$2$ & $ \phi^6 $  &   $ 1+\frac{10}{3 n}+\ldots $&    ---     &     ---      &    $  T_6 $
 \\ &
$3$ & $ \square \phi^4 $ & $ \frac{n+2}{3 n} $ &    $ 4.45915  $   &    $ 4.46870  $   &    $  B_4 $
 \\ &
$4$ & $ \phi^6 $  & $\frac{4}{3}+\frac{26}{3 n}+\ldots $&      $ 5.02933_{[2,3]}  $   &    $ 5.07115_{[2,3]}   $  & $T_6$
 \\ &
$5$ & $ \phi^6 $  &  $ 2+\frac{14}{3 n}+\ldots $ &     $ 5.57244_{[2,3]}  $   &    $5.75027_{[2,3]}  
$   &    $ T_4  $
\\\cline{2-7}&
$1$ & $ \partial \phi^4 $   & $ \frac{n+4}{3 n}  $    &        $3.54589_{[2,3]} $   &    $ 3.60919  $   &  $ H_4 $
 \\ \cline{2-7}&
$1$ & $ \partial^2 \phi^4 $  & $ 0+\frac{58}{27 n}+\ldots  $      &    $4.32039$   &    $4.35276 $   &  $ B_4$
 \\ &
$2$ & $ \partial^2 \phi^4 $  & $   \frac{1}{3}+\frac{22-4 \sqrt{10}}{27 n}+\ldots $   &   $4.36891 $   &    $4.44860  $   &  $  H_4$
 \\ &
$3$ & $ \partial^2 \phi^4 $  & $ \frac{1}{3}+\frac{22+4 \sqrt{10}}{27 n}+\ldots $     &    $ 4.59420 $   &    $ 4.70737  $   &  $  T_4$ 
\\\cline{2-7}
\multicolumn{2}{c}{\rule{0pt}{1ex}}\\[-5.5mm]
 \cline{2-7}
 \multirow{4}{*}{$\bm{Z_4} \left\{ \vphantom{\rule{0pt}{1.3cm}} \right.$\hspace{-2mm}}
&
$1$ & $ \phi^4 $  & $   \frac{4}{n}$    &          $ 2.62865_{[2,3]}  $   &    $  2.89025_{[3,3]} $   &  $ T_4 $
 \\ &
$2$ & $ \phi^6 $ & $  \frac{5 n+52-\sqrt{n^2+64 n-128}}{6 n} $    &    $   $    ---   &    ---  &  $ T_6 $
 \\ &
$3$ & $ \phi^6 $  & $ \frac{5 n+52+\sqrt{n^2+64 n-128}}{6 n} $    &     $5.41824_{[2,3]} $   &    $5.75027_{[2,3]}  $   &  $  T_4$
 \\ \cline{2-7}&
$1$ & $ \partial^2 \phi^4 $  & $\frac{26}{9 n}  $    &        $ 4.46503  $   &    $ 4.70737  $   &  $  T_4$ 
\\ \cline{2-7}
\multicolumn{2}{c}{\rule{0pt}{1ex}}\\[-5.5mm]
 \cline{2-7}
 \multirow{6}{*}{$\bm{XX} \left\{ \vphantom{\rule{0pt}{2cm}} \right.$\hspace{-2mm}}
&
$1$ & $ \phi^4 $  & $  \frac{2 (n+2)}{3 n} $    &     $  2.87297_{[3,3]} $   &    $  2.89025_{[3,3]}$   &  $ T_4 $
 \\ &
$2$ & $ \phi^6 $ & $  \frac{11 n+28-\sqrt{9 n^2+48 n-192}}{6 n} $    &       $4.87103_{[2,3]} $   &    $   5.07115_{[2,3]}   $   &  $ T_6 $
 \\ &
$3$ & $\square \phi^4 $  & $ \frac{2 (n-1)}{3 n}  $    &       $ 4.60121_{[2,3]}  $   &    $4.46870  $   &  $ B_4 $ 
 \\  &
$4$ & $ \phi^6 $  & $  \frac{11 n+28+\sqrt{9 n^2+48 n-192}}{6 n} $    &        $ 5.71556_{[2,3]} $   &    $ 5.75027_{[2,3]}   $   &  $  T_4$
 \\ \cline{2-7}&
$1$ & $ \partial^2 \phi^4 $  & $  \frac{9 n+4-\sqrt{9 n^2-80 n+320}}{18 n} $    &       $ 4.46009$   &    $4.35276 $   &  $ B_4$
\\&
$2$ & $ \partial^2 \phi^4 $  & $ \frac{9 n+4+\sqrt{9 n^2-80 n+320}}{18 n}$    &        $ 4.70443  $   &    $ 4.70737  $   &  $  T_4$ 
 \\ \cline{2-7}
\multicolumn{2}{c}{\rule{0pt}{1ex}}\\[-5.5mm]
 \cline{2-7}
 \multirow{4}{*}{$\bm{XB} \left\{ \vphantom{\rule{0pt}{1.33cm}} \right.$\hspace{-2mm}}&
$1$ & $ \phi^6 $ & $  \frac{2 (2 n+7)}{3 n}  $    &      $ 4.95519_{[2,3]} $   &    $5.07115_{[2,3]} $   &  $ T_6 $
\\\cline{2-7}&
$1$ & $ \partial \phi^4 $   & $  \frac{n+4}{3 n}  $    &      $ 3.53618$   &    $ 3.60919  $   &  $ H_4 $
 \\ \cline{2-7}&
$ 1 $ & $ \partial^{[1,1]}\phi^4$ & $\frac{1}{3}$ &  $4.33041$&$ 4.28020  $  & $ Y_{2,1,1}$
  \\ \cline{2-7}&
$1$ & $ \partial^2 \phi^4 $  & $  \frac{n+2}{3 n}$    &    $4.44492 $   &    $4.44860  $   &  $  H_4$
  \\ \cline{2-7}
\multicolumn{2}{c}{\rule{0pt}{1ex}}\\[-5.5mm]
 \cline{2-7}
 \multirow{3}{*}{$\bm{BZ} \left\{ \vphantom{\rule{0pt}{1cm}} \right.$\hspace{-2mm}}&
$1$ & $ \phi^6 $ & $  \frac{n+6}{n}$    &     $  4.86519_{[2,3]} $   &    $5.07115_{[2,3]} $   &  $ T_6 $
\\\cline{2-7}&
$1$ & $ \partial \phi^4 $   & $ \frac{8}{3n} $      &    $3.38273 $
   &    $ 3.60919  $   &  $ H_4 $
 \\ \cline{2-7}&
$1$ & $ \partial^2 \phi^4 $  & $ \frac{n+2}{n} $    &    $4.28473 $   &    $4.44860  $   &  $  H_4$
 \\ \cline{2-7}
\multicolumn{2}{c}{\rule{0pt}{1ex}}\\[-5.5mm]
 \cline{2-7}
 \multirow{2}{*}{$\bm{BB} \left\{ \vphantom{\rule{0pt}{0.67cm}} \right.$\hspace{-2mm}}&
$1$ & $\square \phi^4 $  & $ \frac{2}{n} $      &    $4.27758  $   &    $4.46870  $   &  $ B_4 $ 
\\\cline{2-7}&
$1$ & $ \partial^2 \phi^4 $  & $  \frac{14}{9n}$    &    $4.20540$   &    $4.35276 $   &  $ B_4$
 \\ \cline{2-7}
\multicolumn{2}{c}{\rule{0pt}{1ex}}\\[-5.5mm]
 \cline{2-7}
 \multirow{1}{*}{$\bm{VB_3}$}
 &
$ 1 $ & $ \partial^{[1,1]}\phi^4$ & $\frac{4}{3n}$ &$ 4.15510   $&$ 4.28020  $  & $ Y_{2,1,1}$
 \\ \cline{2-7}
\multicolumn{2}{c}{\rule{0pt}{1ex}}\\[-5.5mm]
 \cline{2-7}
 \multirow{1}{*}{$\bm{\overline{XX}V}$}
 &
 $1$ & $ \partial \phi^5$  & $  \frac{2 n+7}{3 n} $    &    $ 4.48312 $   &    $  4.55676 $   &  $ H_5$
 \\ \cline{2-7}
\end{tabular}
}
\end{table}

\clearpage 
\section{Scalar mixing at dimension three and four}
\label{app:scalarPotentials}

Together with our main results, which are expressed in the basis of primary operators,\footnote{Modulo the two exceptions---dimension five spin one and dimension six spin two---where we opted to retain redundant operators.} we also make the results for mixing matrices of dimension three and four Lorentz scalars in the Green's basis publicly available.%
    \footnote{See our
    \githubb repository
    under \href{https://github.com/jasperrn/EFT-RGE/tree/main/GeneralScalar/OtherCalculations}{GeneralScalar/OtherCalculations/}.} 
These allow us to illustrate, quantitatively, examples such as the mixing systems containing broken symmetry currents. Our examples also clarify when it is sufficient to retain a $\phi^k$ operator ($k=3,4,5 \ldots$) on its own ans when it is necessary to consider mixing with derivative insertions for a correct determination of the scaling dimensions. 

As we will now show, for dimension three and four, it is sufficient to only consider the diagonal elements of the mixing matrix, since there is a zero in one of the off-diagonal entries. However, for operators starting at dimension five, as we considered in the main text, one needs to take into account the mixing between $\phi^k$ type operators and operators with derivatives. These results imply in particular that general scalar potentials (Section~\ref{sec:generalPotentialBetaFunctions}) can be used to correctly generate the complete multiloop results for $\phi^k$\nobreakdash-type operators only for $k\leqslant 4$, while the one-loop results are correct for operators with any $k>4$.

\subsection{Dimension-three scalar mixing and the equations of motion}

We start by discussing the mixing matrix of $\phi^3$ and $\partial^2 \phi$, or more specifically in our case the mixing matrix of the corresponding couplings, $c_{\phi^3}$ and $c_{\partial^2\phi}$. 
We restrict to the case of one real scalar field for simplicity. The matrix is of the general form 
\begin{equation}
\label{eq:PhiCubedMixing}
    \frac{d}{d\log\mu} \begin{pmatrix}
        c_{\phi^3}\\c_{\partial^2\phi}
    \end{pmatrix} = \Gamma
    \begin{pmatrix}
        c_{\phi^3}\\c_{\partial^2\phi}
    \end{pmatrix}\,,
    \quad \text{ with } \quad
    \Gamma=\begin{pmatrix}
       \gamma_{11}(\lambda) & 0\\ \gamma_{21}(\lambda) & \gamma_{22}(\lambda)
    \end{pmatrix},
\end{equation}
where the entry $\gamma_{12}$ is zero to all orders, since insertions of $\partial^2 \phi$ (being a total derivative) do not require insertions of $\phi^3$ to be renormalized. 
One can evaluate this matrix order by order in $\lambda$. At five loops in $4-2\ep$ dimensions it evaluates to 
\begin{align}
    \gamma_{11} &= 
    -1-\ep
    +3 \lambda
    -\frac{23 \lambda ^2}{4}
   +\frac{3}{16} \left(64 \zeta _3+97\right) \lambda ^3 
   -\frac{1}{64} \left(4992 \zeta _3-1152 \zeta _4+7680 \zeta
   _5+4687\right) \lambda ^4
   \nonumber\\&\quad
    +\frac{1}{128} \left(5760 \zeta _3^2+63696 \zeta _3-18960 \zeta _4+126336 \zeta _5-43200 \zeta
   _6+169344 \zeta _7+42685\right) \lambda ^5
   \nonumber\\
   &=  \beta_\lambda/\lambda -\Delta_\phi 
   \,,
   \nonumber\\[2mm]
   %
   %
   \gamma_{21} &= -2\lambda+\frac32\lambda^2-\frac{65}8\lambda^3+\frac{1152\zeta_4-432\zeta_3+3709}{96}\lambda^4
   = -12\,\gamma_\phi/\lambda 
   \,,\nonumber\\[2mm]
   \gamma_{22} &= 
   -1+\ep+\frac{\left(432 \zeta _3-1152 \zeta _4-3709\right) \lambda ^5}{2304}+\frac{65 \lambda
   ^4}{192}-\frac{\lambda ^3}{16}+\frac{\lambda ^2}{12}\nn\\
   &= -d+2+\Delta_\phi
   \,.
\end{align}
The short forms after the second equal signs were identified \emph{a posteriori} (although expected \cite{Wegner:1974sla,Brezin:1974zr}). 
A more natural basis in which to consider this mixing problem is the one in which the EOM is explicit, i.e.\ $(\partial^2\phi+\frac{1}{3!}\lambda \phi^3,\partial^2 \phi)$. 
In this basis, the  matrix is diagonal,
\begin{equation}
\label{eq:PhiCubedMixing2}
    \bar\Gamma=\begin{pmatrix}
       \bar\gamma_{1,1}(\lambda) & 0\\ 0 & \bar\gamma_{2,2}(\lambda)
    \end{pmatrix}\,,
\end{equation}
with
\begin{align}
    \bar\gamma_{1,1}&=
    -1+\ep+\frac{\left(-432 \zeta _3+1152 \zeta _4+3709\right) \lambda ^5}{2304}-\frac{65 \lambda
   ^4}{192}+\frac{\lambda ^3}{16}-\frac{\lambda ^2}{12}
   = -\Delta_\phi
   \,,\nonumber\\[2mm]
   \bar\gamma_{2,2} &= 
   -1+\ep+\frac{\left(432 \zeta _3-1152 \zeta _4-3709\right) \lambda ^5}{2304}+\frac{65 \lambda
   ^4}{192}-\frac{\lambda ^3}{16}+\frac{\lambda ^2}{12}
   = -d+2+\Delta_\phi
   \,.
\end{align}
Relating the eigenvalues of the mixing matrix to the scaling dimensions of the corresponding operators using \eqref{3.15}, we thus find 
$\Delta_\text{EOM} = d-\Delta_\phi$ and 
$\Delta_{\partial^2\phi} = \Delta_\phi +2$.
Notice that so far we have not restricted to the fixed-point value of $\lambda$.\footnote{Hence the first eigenvalues of $\Gamma$ in \eqref{eq:PhiCubedMixing} and $\bar \Gamma$ in \eqref{eq:PhiCubedMixing2} differ with a shift proportional to the $\beta$ function. This is because away from the fixed-point, the matrices are not simply related by a similarity transform, but instead by $\bar \Gamma=\Lambda\Gamma\Lambda^{-1}+\beta\frac{d\Lambda}{d\lambda}\Lambda^{-1}$ where $\Lambda$ implements the basis change: $(\de^2\phi+\lambda\phi^3/6,\de^2\phi)=(\phi^3,\de^2\phi)\Lambda^{-1}$. Compare with \eqref{eq:basisChange} in the main text.}
At the interacting fixed-point ($\lambda = \lambda_*$), the two eigenoperators will be 
the $\text{EOM}=\partial^2\phi+\frac{1}{3!}\lambda \phi^3$ and
a level-2 descendant $\partial^2 \phi$, with scaling dimensions 
$\Delta_\text{EOM}=d-\Delta_\phi(\lambda_*)$ and
$\Delta_{\partial^2 \phi} =\Delta_\phi(\lambda_*) +2$, respectively. The result $\Delta_\text{EOM} =d -\Delta_\phi$ can also be derived to all orders by considering the scaling of correlation functions with insertions of EOM operators; see e.g.~\cite[Eq.~2.9]{Kousvos:2025ext}.
This provides a cross-check on our calculation. 

While we have worked this out explicitly in the case of a single scalar field, the fact that the mixing matrix remains of the form 

\begin{equation}
\label{eq:PhiCubedMixing0}
    \Gamma=\begin{pmatrix}
       \gamma_{11}(\lambda) & 0\\ \gamma_{21}(\lambda) & \gamma_{22}(\lambda)
    \end{pmatrix}
\end{equation}
is true for any $\lambda \phi^4$ type scalar field theory. What this means in practice, is that to extract the dimension of the primaries
it is sufficient to consider the $\gamma_{11}$
element(s) of the matrices.%
    \footnote{In the case of a single real scalar there is no $\phi^3$-type primary, but in generic theories there are; see e.g. the cases with $O(n)$ global symmetry (representation $T_3$) and cubic symmetry  ($V$, $Z_3$, and $XV$) presented in the main text.} 
This is, for example, why in \cite{Bednyakov:2021ojn} it suffices to consider the beta function of the $\phi^a \phi^b \phi^c$ coupling, and forget about mixing with the coupling of $\partial^2\phi^a$. We emphasize that this holds true as long as one is interested specifically in the eigenvalue, and not the precise eigenvector it corresponds to.

\subsection{Dimension-four scalar mixing and broken currents}
Having discussed the mixing system at dimension three and Lorentz spin zero, we now proceed to dimension four. 
Here, the anomalous dimension of all $\phi^4$-type operators can be extracted without taking the mixing with derivatives ($\phi^2\de^2$) into account. For instance, anomalous dimensions of singlet operators are eigenvalues of the derivatives of the beta function with respect to all marginal couplings $\lambda_i$,
\begin{equation}
    (\gamma^{\text{singlets}})_{ij}=\left.\frac{\de\beta_{\lambda_i}(\lambda)}{\de\lambda_j}\right|_{\lambda_*}
\end{equation}
while non-singlets can be extracted from the general $\beta$ functions $\beta_{abcd}$ by substituting $\lambda_{abcd}=\lambda^*_{abcd}+\kappa T^R_{abcd}$ where $T^R_{abcd}$ is a tensor structure of the representation $R$.

\paragraph{Single-scalar theory.}
Here we explain why $\gamma_{\phi^4}$ can be computed from the $\beta$ function in the single-scalar case, by comparing with our general treatment for any composite operator. In such a general treatment, we would consider the mixing system including  $c_{\phi^4}\phi^4$ and a redundant operator.
If the redundant operator is proportional to  the $\EOM$, this mixing has a zero off the diagonal, 
implying that $c_{\phi^4}$ is determined by a single entry in the mixing matrix.
Crucially, this basis can be obtained 
from the simple Lagrangian containing the standard kinetic and coupling terms. 
We focus on the single-scalar theory for simplicity and we drop total derivatives.

The renormalization of the dimension-four Lagrangian results in
\begin{equation}\label{A.9}
    \frac12 \partial_\mu \phi_{\mathrm b} \partial^\mu\phi_{\mathrm b} - \frac{\lambda_{\mathrm b}}{24}  (\phi^{b})^4
    = 
    \frac{1}{2} \partial_\mu (Z_\phi \phi) \partial^\mu (Z_\phi \phi) 
    - \frac{\mu^{2\ep}Z_\lambda \lambda}{24} (Z_\phi \phi)^4\,.
\end{equation}
where $Z_\phi = Z_\phi(\lambda)$ and $Z_\lambda = Z_\lambda(\lambda)$ are functions of the coupling $\lambda$. These determine the anomalous dimension of the field, $\gamma_\phi\,\phi = -\frac{d\phi}{d\log\mu}$, and the beta function, $\beta_\lambda = \frac{d\lambda}{d\log\mu}$.

The claim is now that we can extract the anomalous dimension of the operator $\phi^4$ (or in general any other $\phi^4$ type operator) directly from these expressions, without constructing a $2\times2$ anomalous dimension matrix involving also redundant operators. We want to see that
\begin{equation}
        \Delta_{\phi^4} = d+\gamma_{c_4}\,,\qquad \gamma_{c_4} = \frac{d\,\beta_\lambda}{d\lambda}\,.
\end{equation}

Let us therefore extract $\gamma_{c_4}$ from $\beta_\lambda$. Consider the redefinition $\lambda\to\lambda+c_4$ in \eqref{A.9}.
This results in 
\begin{align}
    &\frac{1}{2} \partial_\mu (\tilde Z_\phi \phi) \partial^\mu (\tilde Z_\phi \phi) 
    - \frac{\mu^{2\ep}Z_\lambda \lambda}{24} (\tilde Z_\phi \phi)^4
    - \frac{\mu^{2\ep}Z_{c_4}c_4}{24}(Z_\phi\phi)^4
    +O(c_4^2)
    \nn\\&\qquad
    = \frac{1}{2} \partial_\mu (  Z_\phi \phi) \partial^\mu ( Z_\phi \phi) 
    - \frac{\mu^{2\ep}Z_\lambda \lambda}{24} ( Z_\phi \phi)^4
    - \frac{\mu^{2\ep}Z_{c_4}c_4}{24}( Z_\phi\phi)^4
    \nn\\&\hspace{2cm}
    - \frac{c_4}{Z_\phi}
    \frac{dZ_\phi}{d\lambda}
    (Z_\phi\phi) \left(
    \partial^2(Z_\phi\phi) + \frac{\mu^{2\ep}Z_\lambda \lambda}{6} (Z_\phi\phi)^3
    \right)
    +O(c_4^2)
    \,,
    \label{A.12}
\end{align}
where we used $\tilde Z_\phi \equiv  Z_\phi(\lambda + c_4) = Z_\phi(\lambda) + c_4 \frac{d Z_\phi}{d\lambda} + O(c_4^2)$
and we defined $Z_{c_4} = \frac{d(\lambda\, Z_\lambda(\lambda))}{d\lambda}$.
In the final line, we additionally used integration by parts in the term $\de_\mu(c_4\frac{dZ_\phi}{d\lambda}Z_\phi\phi)\de^\mu(Z_\phi\phi)$ in order to factor out the EOM operator.
Indeed, as expected from an EOM operator, it can be removed using a field redefinition (working at linear order in $c_4$),
\begin{equation}
    \phi \to \phi 
    - \frac{c_4}{Z_\phi} \frac{dZ_\phi}{d\lambda} \phi\,.
\end{equation}
In this way, we arrive at the expression we would have used if treating $\phi^4$ as a separate composite operator, the scaling dimension of $\phi^4$ can be obtained from the renormalized Lagrangian
\begin{equation}\label{A.10}
    \frac12 \partial_\mu \phi^b \partial^\mu\phi^b - \frac{\lambda^b}{24}  (\phi^{b})^4
    -\frac{c_4^b}{24}(\phi^b)^4
    = 
    \frac{1}{2} \partial_\mu (Z_\phi \phi) \partial^\mu (Z_\phi \phi) 
    - \frac{\mu^{2\ep}Z_\lambda \lambda}{24} (Z_\phi \phi)^4
    - \frac{\mu^{2\ep}Z_{c_4}c_4}{24}(Z_\phi\phi)^4
    \,,
\end{equation}
where we work at leading order in $c_4$. That is, we consider only single insertions of $c_4$, such that $Z_{c_4} = Z_{c_4}(\lambda)$ is a function of $\lambda$ only. 

The mixing into the EOM operator then corresponds to $\bar \gamma_{\bar rp}(\lambda)$ in \eqref{eq:GreensMinimalBasesMain}, which is not relevant for the determination of the eigenvalues.
Said otherwise, the EOM operator does not affect the running of $c_4$, which can be determined from $Z_{c_4}$ alone, which is in turn related to $Z_\lambda$. More directly, observing that \eqref{A.12} can be obtained from \eqref{A.9} by a shift in $\lambda$, the scaling dimension of $c_4$ can be determined from
\begin{align}
    \gamma_{c_4} = \frac{d\,\beta_\lambda}{d\lambda}\,.
\end{align}

\paragraph{General Mixing Between \boldmath$\phi^4$ and \boldmath$\partial^2 \phi^2$.}
Let us now proceed to the most general mixing problem possible between $\phi^a \phi^b \phi^c \phi^d$ and operators with two fields and two derivatives. We choose the operator basis  to be
\begin{align}
    \mathcal{O}^{(4,0)}_{\phi^4} &= -\frac{1}{4!}C_{\phi^4}^{abcd} \phi^a \phi^b \phi^c \phi^d\,,\nn\\
    \mathcal{O}_\text{kin}^{(4,0)} &= -\frac{1}{4}Z^{ab}\left( 
        \phi^a \partial^2\phi^b + \partial^2\phi^a \phi^b
    \right) \,,\nn\\
    \mathcal{O}_{\phi^2_{\text{ant}}}^{(4,0)} &= 
    \frac{1}{2}C^{ab}_{\phi^2_{\text{ant}}}\left( 
        \phi^a \partial^2\phi^b - \partial^2\phi^a \phi^b
    \right)
    =
    \frac{1}{2}C^{ab}_{\phi^2_{\text{ant}}}\,\partial_\mu\left( 
        \phi^a \partial^\mu\phi^b - \partial^\mu\phi^a \phi^b
    \right)
    \,,\nn\\
    \mathcal{O}^{(4,0)}_{\phi^2_{\text{sym}}} &= 
    -\frac{1}{2}C_{\phi^2_{\text{sym}}}^{ab}
    \, \partial^2 (\phi^a \phi^b)\,,
    \label{dim4basislast}
\end{align}
which we add to the dimension-four Lagrangian $\mathcal{L}^{(4)} =\tfrac12 \partial_\mu \phi^a \partial^\mu \phi^a - \tfrac{\lambda^{abcd}}{4!}\phi^a \phi^b \phi^c \phi^d$\,. 
Here, $Z^{ab}$ is a symmetric matrix, and captures for example the kinetic term when $Z^{ab} = \delta^{ab}$, but is non-zero also for other symmetric irreps of the global symmetry in question (e.g. for $T_2$ of $O(n)$, where $Z^{ab}\subset t^a t^b$). The matrix $C^{ab}_{\phi^2_{\text{ant}}}$ is antisymmetric and $C^{ab}_{\phi^2_{\text{sym}}}$ is symmetric. This basis captures all possible operators at dimension four and at Lorentz spin zero.

Importantly, we find that the beta function of $C_{\phi^4}^{abcd}$ does not include any terms that involve $C^{ab}_{\phi^2_{\text{ant}}}$ or $C^{ab}_{\phi^2_{\text{sym}}}$. It only depends on $C_{\phi^4}^{abcd}$ and $Z^{ab}$. This mixing, of course, needs to be diagonalized in order to extract the scaling dimension of $\O_{\phi^4}^{(4,0)}$. 
Alternatively, as discussed above in the single-scalar case, this scaling dimension can be directly extracted from the $\beta$ function of $\lambda^{abcd}$. The results of this calculation can be found in our data files.

\paragraph{Broken Currents.}
Let us now consider the mixing matrix involving the divergence of global-symmetry currents $\partial_\mu J^\mu$ before and after symmetry breaking. This can be done using the results computed in the basis defined by \eqref{dim4basislast}. The mixing matrix in question, as outlined in~\cite{Kousvos:2025ext}, is central for conformal bootstrap applications. Our results hence provide another useful cross-check of those in~\cite{Kousvos:2025ext}.

For concreteness, let us start with an example in a specific theory. In particular, we discuss the symmetry breaking that occurs when flowing from $O(n)$ to the hypercubic-symmetric theory with symmetry group $C_n = S_n \ltimes (\Z_2)^n$ and how this affects the mixing matrix of the broken current. 
Consider the basis 
($B_{ab}$, $\partial_\mu J^\mu_{ab}$), where 
$B_{ab}=\phi_a \,\delta_{bcde}\phi_c \phi_d \phi_e 
    -\phi_b \,\delta_{acde}\phi_c \phi_d \phi_e $
    and $J_{ab}^\mu = \phi_a\partial^\mu \phi_b - \phi_b \partial^\mu \phi_a$.
Under $O(n)$, $B_{ab}$ is part of the rank-4 traceless-symmetric representation $T_4$, while in a hypercubic-symmetric theory, it is in the rank-2 antisymmetric representation denoted  $B$. 
Evaluating this mixing matrix, one finds
\begin{equation}
    \Gamma=\begin{pmatrix}
       \gamma_{11} & 0\\ \gamma_{21} & \gamma_{22}
    \end{pmatrix},
\end{equation}
as in the case of the $(\phi^3, \partial^2 \phi)$ mixing matrix of the previous section. In order to obtain the scaling dimensions it again suffices to evaluate $\gamma_{11}$ and $\gamma_{22}$ at the fixed-point. Notice that so far we have not committed to a choice of fixed-point ($O(n)$ versus $S_n \ltimes (\Z_2)^n$). 
For example, 
starting from the Lagrangian in \eqref{hypercubicLag}, the $O(n)$ fixed-point is obtained by
setting 
$\lambda_{4,2}=0$ and $\lambda_{4,1} =\lambda_{4,1}^{*\, O(n)}$ (see \eqref{eq:critCouplingON}), while the 
hypercubic fixed-point is obtained by 
$\lambda_{4,1} =\lambda_{4,1}^{*\, C_n}$ and 
$\lambda_{4,2} =\lambda_{4,2}^{*\, C_n}$ (see \eqref{cubiccoupling}). 
At the $O(n)$ fixed-point $\gamma_{21}=\gamma_{12}=0$ to all orders in the coupling due to $B_{ab}$ and $\partial_\mu J^\mu_{ab}$ being in different irreps, while at the $C_n$ fixed-point they are in the same irrep and hence allowed to mix with non-zero $\gamma_{21}$. However, $\gamma_{12}=0$ also in this case because $\partial_\mu J^\mu_{ab}$ is a total derivative while $B_{ab}$ is not.

The two eigenvalues of $\Gamma$ at the two fixed-points result in
\begin{align}
    O(n) &: \quad \Delta_1^{O(n)} = d\,,  \hspace{1.05cm} \Delta_2^{O(n)}=\Delta_{T_4}^{O(n)}\,,\nn\\
    C_n &: \quad \Delta^{C_n}_1 = \Delta_{B_\mu}^{C_n}+1\,,  \quad \Delta^{C_n}_2=d\,.
\end{align}
Although we explicitly derived these values at four loops, they can be argued at all loops. 
At the $O(n)$ fixed-point, $J_{ab}^\mu$ is conserved, 
\begin{align}
    O(n): \quad \partial_\mu J^\mu_{ab}=
    \phi_a \,\text{EOM}_b - \phi_b \, \text{EOM}_a\,,
\end{align}
and hence it has dimension
$\Delta_{\text{EOM}} + \Delta_{\phi} = d$.
At the hypercubic fixed-point, on the other hand, 
$J_{ab}^\mu$ is no longer conserved, and $\partial_\mu J^\mu_{ab}$ is just the descendant of $J_\mu$, with dimension $\Delta^{C_n}_{B_\mu}+1$,
where we have named the broken current $B_\mu$, since it is simply a primary in the $B$ representation of $C_n$ with Lorentz spin one. 
The composite EOM in the hypercubic-symmetric theory can be written as
\begin{align}
    C_n: \quad \phi_a \,\EOM_{b}-\phi_b \,\EOM_a =\partial_\mu J^\mu_{ab} -c \, B_{ab} \,,
\end{align}
where $c$ is a number of which the precise value which is not important here. 
This EOM again has scaling dimension $\Delta_{\text{EOM}} + \Delta_{\phi} = d$,
which agrees with the results of \cite{Bednyakov:2023lfj} at six-loop order.

Comparing back to the $(\phi^3, \partial^2 \phi)$ system, we see that the mixing matrix has the same qualitative (i.e.~triangular) form. This could have been predicted a priori, since insertions of $\partial^2 \phi$ and $\partial_\mu J^\mu_{ab}$ are renormalized once the operators $\phi$ and $J^\mu_{ab}$ are renormalized. Since both $\partial^2 \phi$ and $\partial_\mu J^\mu_{ab}$ will always be either descendants or EOMs,\footnote{Depending on if we are in the free theory for the first, and if the symmetry is broken or not for the second.} they always furnish renormalized scaling eigenoperators on their own.
This is not true however for $\phi^3$ and $B_{ab}$, where the mixing with $\partial^2 \phi$ and $\partial_\mu J^{\mu}_{ab}$ respectively needs to be taken into account to obtain finite scaling eigenoperators.

\subsection{General scalar potential}
\label{sec:generalPotentialBetaFunctions}

A compact way to package the information of anomalous dimensions of local operators is through beta functions of general scalar potentials. The three-loop scalar potential was considered in \cite{Jack:1990eb}. 
Starting using the Lagrangian density
\begin{equation}
    \mathcal L= \frac12(\de\phi)^2-V(\phi) \,,
\end{equation}
this gives a beta function (see \cite[Eq.~6.4]{Jack:1990eb})\footnote{We write $\beta_{V\to V}$ to emphasize that this expressions does not capture the mixing with operators with derivatives.}
\begin{align}
\nonumber
    \beta_{V\to V}&=(\eps-4) V(\phi)+W_{a}\hat \gamma_{ab}\phi^b+\frac12W_{ab}
    W_{ab}
    -\frac12 W_{ab}W_{acd}W_{bcd}
    \\
    &\quad +
    \frac14W_{abef}W_{bcef}W_{ac}W_{bd}-\frac3{16}W_{acde}W_{bcde}W_{af}W_{bf}+2W_{acde}W_{ab}W_{cdf}W_{bef}
   \nonumber \\
    &\quad -
    \frac14W_{cdef}W_{ab}W_{acd}W_{bef}-\frac18W_{acd}W_{bcd}W_{aef}W_{bef}+\frac{\zeta_3}2W_{abc}W_{ade}W_{bdf}W_{cef}
    \,,
    \label{eq:generalPotential}
\end{align}
where $W_{a_1\ldots a_k}=\partial_{ \phi^{a_1}}\cdots \partial_{ \phi^{a_k}}V(\phi)$ and 
\begin{equation}
    \hat\gamma_{ab}= 
    \gamma_{\phi,ab}+\Big(1-\frac\eps2\Big)\delta_{ab}=\Big(1-\frac\eps2\Big)\delta_{ab}+\frac1{12}W_{acde}W_{bcde}-\frac1{16}W_{acde}W_{bcfg}W_{defg}\,.
\end{equation}
where $\gamma_{\phi,ab}$ is the anomalous dimension of $\phi^a$ that appeared also in \eqref{eq:gammaPhiDef} in the main text. 
This formula is valid for any potential $V(\phi)$ of a multicomponent field $\phi^a$. This can now be used to derive the anomalous dimension of local operators of the form $\phi^k$ by inserting it in $V(\phi)$. However, this expression does not take mixing with derivatives into account. Hence, it only gives the correct result at the leading order in $\lambda$ (where this mixing can be ignored). From the discussion in the previous subsections, we conclude that this mixing starts from dimension five.

Consider for instance the insertion of $\phi^5$ in the single-scalar case. Using $V(\phi)=\frac\lambda{24}\phi^4-\kappa\phi^5$, 
we find
\begin{equation}
\left. -  \frac{\partial\beta_{V\to V}}{\partial \kappa}\right|_{\kappa=0} =\left[1 -\frac{3 \eps}{2}+10 \lambda-\frac{145 \lambda ^2}{4} +\lambda ^3 \left(120   \zeta_3+\frac{3455}{16}\right)\right]\phi^5\,.
\end{equation}
If there were no mixing with operators with derivatives, the coefficient of $\phi^5$ would be its anomalous dimension. 
Instead, comparing with \eqref{eq:fiveloopPhi5}, we see that only the $O(\lambda)$ term is correct. This is because the computation in the main text includes the mixing with redundant operators constructed using derivatives, thus producing the correct result. For $\phi^k$, $k=1,2,4$, however, it is easy to check against \eqref{eq:phik} that the scalar potential gives the correct result.

The same holds in the generic multi-scalar theory. The scalar potential beta function \eqref{eq:generalPotential} is thus a very compact way to capture the general 3-loop result for \emph{any} operator with field content $\phi^k$, $k=1,2,3,4$, in \emph{any} massless $\lambda\phi^4$ theory. 
In addition, it gives the correct leading-order $O(\lambda)$ anomalous dimension for \emph{any} operator of the form $\phi^k$ for any $k$, in any representation in any $\lambda\phi^4$ theory.
It would be desirable to write down higher-loop results for scalar potentials, generalizing \eqref{eq:generalPotential}. For $\phi^k$, $k\leqslant 4$, they would then package the results captured in \cite{Bednyakov:2021ojn} in a compact form. 
Likewise, in multi-critical theories with a marginal $\lambda\phi^{2p}$ interaction such scalar potentials would give the correct anomalous dimensions for $\phi^k$ operators with $k\leqslant2p$, see \cite{ODwyer:2007brp}, and likewise in more general theories with gauge fields and fermions 
\cite{Jack:1982hf,
Jack:1982sr,
Jack:1983sk,
Jack:1984vj}.

A way to incorporate operators with derivatives in a similar fashion is the work on geometrized field space \cite{Alonso:2016oah,Helset:2022tlf,Jenkins:2023bls}, which capture operators with up to two derivatives. It would be interesting to spell out the connection to the scalar potentials in more details and to work out examples at higher loop order.

\section{Example in a non-primary basis}
\label{app:example1App}

Here we consider an alternative basis to the one used in Section~\ref{s:Redex} in the main text, when considering scalar operators at dimension five.
We can, for example, consider a change of basis from the Green's basis~\eqref{eq:naive} to
\begin{equation}\label{eq:Lmin2}
    \mathcal{L}'_\text{minimal} =
        -\frac{c_1}{2} \phi^2\partial^2\phi 
        - \frac{c_2}{2} \, 
            \phi^2 \left(\partial^2\phi + 
            \frac{\lambda}{6}\phi^3 \right) .
\end{equation}
This is achieved by
\begin{equation}\label{changeofBasisToMinimal}
    \begin{split}
        b_1 &= - 10 \lambda c_2
        \,,\\
        b_2 &= c_1+c_2\,.
    \end{split}
\end{equation}
This means that the scaling dimensions in the new basis are given by
\begin{equation}
\begin{pmatrix}
    \bar \gamma_{11}& \bar \gamma_{12}\\
    \bar \gamma_{21}&\bar \gamma_{22}
\end{pmatrix}
\begin{pmatrix}
        c_1(\mu)\\c_2(\mu)
\end{pmatrix} 
    =
\left[
\begin{pmatrix}
    1/(10\lambda)&1 \\
    -1/(10\lambda)&0
\end{pmatrix}
\begin{pmatrix}
     \gamma_{11}&  \gamma_{12}\\
     \gamma_{21}& \gamma_{22}
\end{pmatrix}
\begin{pmatrix}
    0&-10\lambda \\
    1&1
\end{pmatrix}
+
\begin{pmatrix}
     -\beta_\lambda/(10\lambda^2)&0\\
     \beta_\lambda/(10\lambda^2)&0
\end{pmatrix}
\right]
\begin{pmatrix}
        c_1(\mu)\\c_2(\mu)
\end{pmatrix} ,
\end{equation} 
where
\begin{align}    
 \bar \gamma_{11} &= 
 \gamma_{22}+\frac{\gamma_{12}}{10\lambda}
 \,,\\[2mm]
 \label{3.24}
 \bar \gamma_{12}&= 
  \frac{1}{10\lambda}\left(\gamma_{12} + 10\lambda \gamma_{22} -10 \lambda \gamma_{11} -100 \lambda^2 \gamma_{21} + 10 \beta_\lambda\right)=0 
  \,,
  \\[2mm]
 \bar \gamma_{21} &=
 -\frac{\gamma_{21}}{10\lambda}
\,, \\[2mm]
 \bar \gamma_{22} &= 
 \gamma_{11}-\frac{\gamma_{12}}{10\lambda}
 \,,
\end{align}
where~\eqref{3.24} vanishes because redundant operators do not mix into non-redundant operators. We can now use this fact to write 
\begin{equation}    
\bar \gamma_{11} = 
\gamma_{22}+\frac{\gamma_{12}}{10\lambda}
 = \gamma_{11} + 10\lambda \gamma_{21} - \frac{\beta_\lambda}{\lambda}
, \qquad 
 \bar \gamma_{22} = 
 \gamma_{11}-\frac{\gamma_{12}}{10\lambda}
 = \gamma_{22} - 10 \lambda \gamma_{21} + \frac{\beta_\lambda}{\lambda}
 \,.
\end{equation}
That is, we see the relation of the anomalous dimensions to those in \eqref{eq:gammaHatPhi5example}.
In particular, the eigenvalues at the fixed-point ($\beta_\lambda=0$) are exactly the same as when computed in a different basis.

\section{Choice of operators}\label{s:Operators}

We computed multiloop anomalous dimensions in the general scalar $\phi^4$ theory, the hypercubic theory ($C_n$), the $O(n)$-model and the single scalar ($n=1$) $\phi^4$ theory.
We work with the Lagrangians
\begin{align}
    \mathcal{L}_\text{general} &= \frac12 \partial_\mu \phi^a \partial^\mu \phi^a 
    - \frac{\lambda^{abcd}}{4!}\phi^a\phi^b\phi^c\phi^d + 
    \sum_i c_i\,\mathcal{O}_i\,,\\
    \mathcal{L}_{C_\N} &= \frac12 \partial_\mu \phi^a \partial^\mu \phi^a 
    - 
    \frac{\lambda_1}{4!}
    \delta^{ab}\delta^{cd}
    \phi^a\phi^b\phi^c\phi^d - 
    \frac{\lambda_2}{4!}
    \delta^{abcd}
    \phi^a\phi^b\phi^c\phi^d+
    \sum_i c_i\,\mathcal{O}_i\,,\label{eq:cubicLagr}\\
    \mathcal{L}_{O(\N)} &= 
    \frac12 \partial_\mu \phi^a \partial^\mu \phi^a 
    - \frac{\lambda}{4!}
    \delta^{ab}\delta^{cd}
    \phi^a\phi^b\phi^c\phi^d + 
    \sum_i c_i\,\mathcal{O}_i\,,\\
    \mathcal{L}_{n=1} &= \frac12 \partial_\mu \phi\partial^\mu \phi
    - \frac{\lambda}{4!} \phi^4 + 
    \sum_i c_i\,\mathcal{O}_i\,,
\end{align}
where we sum over repeated indices and we freely raise and lower flavor indices.
In this appendix, we list the conventions for the choice of operators in the general theory.
These conventions are also listed in our \githubb repository, together with our conventions for the $O(n)$ model (scalar singlets only) and the single-scalar theory up to dimension ten.

As explained in Section~\ref{s:prim} we determine our choice of operators through the primary condition, 
$[K_\mu, \mathcal{O}] = 0$, where $K_\mu$ is the generator of special conformal transformations. We impose this condition in the free theory in $d=4$ spacetime dimensions. Except for the $O(\N)$ model at dimension eight and ten, this fixes all considered operators up to a multiplicative constant. For the general scalar theory, the primary condition takes the form of constraints on the tensors that appear in the Lagrangian.
Our choice of operator basis for the composite operators in the general scalar theory is given in Table~\ref{tabGen}, together with the constraints on the coupling constant tensors. 
The number of independent building blocks for each type of composite operator up to dimension six and spin two in the general theory is 
    \begin{align}
        H = 
        &\sum_{k=1}^6 
        \begin{pmatrix}
            n+k-1\\k
        \end{pmatrix}\phi^k
+
        \frac{n}{2}\begin{pmatrix}n+1\\3\end{pmatrix}\phi^4\square
+
    \sum_{k=2}^5(k-1)\begin{pmatrix}n+k-2\\k\end{pmatrix}
        \phi^k \partial_\mu
    \nonumber\\
    &+\begin{pmatrix}n+1\\2\end{pmatrix}\phi^2 \partial_{(\mu}\partial_{\nu)} 
    +n\begin{pmatrix}n+1\\2\end{pmatrix} \phi^3 \partial_{(\mu} \partial_{\nu)}
    +\begin{pmatrix}n+1\\2\end{pmatrix}\begin{pmatrix}n+1\\2\end{pmatrix}\phi^4 \partial_{(\mu} \partial_{\nu)}
    \nonumber\\
    &+\begin{pmatrix}n\\3\end{pmatrix} \phi^3 \partial_{[\mu} \partial_{\nu]}
    +3\begin{pmatrix}n+1\\4\end{pmatrix} \phi^4 \partial_{[\mu} \partial_{\nu]}\,,
    \end{align}
where $\binom{a}{n}$
denotes the binomial coefficient.

Beyond what can be extracted from the results in the general theory, we provide additional results in which we specified the dimension-four interactions to the hypercubic theory \eqref{eq:cubicLagr}, but we kept the higher-dimensional operators general. From these results, all irreps of scalar operators at dimension six and eight can be extracted up to five and two loops, respectively. 
The conventions at dimension six are the same as in the general theory. The operators at dimension eight are chosen to be
\begin{align}
    \mathcal{O}_{8,1}^{(8,0)} &= \frac{c_{(abcdefgh)}^{(8,0)}}{8!}\phi^a
    \phi^b\phi^c\phi^d\phi^e\phi^f\phi^g\phi^h\,,
    \nonumber\\
    \mathcal{O}_{6,1}^{(8,0)} &=
    -\frac{c^{(6,0)}_{(abcd)(ef)}}{48}
                           \phi^{a} \phi^{b} \phi^{c} \phi^{d} \partial_\mu \phi^{e} \partial^\mu \phi^{f}\,,
    \nonumber\\
    \mathcal{O}^{(8,0)}_{4,1} &= 
    c^{(8,0,4,1)}_{(abcd)} \left(
           \frac{3}{4} \, \partial_\mu \phi^{a} \partial^\mu \phi^{b} \partial_\nu \phi^{c} \partial^\nu \phi^{d}
           - \phi^{a} \partial_\mu \phi^{b} \partial_\nu \phi^{c} \partial^\mu \partial^\nu \phi^{d}
           +\frac{1}{4} \, \phi^{a} \phi^{b} \partial_\mu \partial_\nu \phi^{c} \partial^\mu \partial^\nu \phi^{d}
           \right),
    \nonumber\\
    \mathcal{O}^{(8,0)}_{4,2} &= 
    c^{(8,0,4,2)}_{(ab)(cd)} \left(
           -3 \, \partial_\mu \phi^{a} \partial^\mu \phi^{b} \partial_\nu \phi^{c} \partial^\nu \phi^{d}
           - 4 \phi^{a} \partial_\mu \phi^{b} \partial_\nu \phi^{c} \partial^\mu \partial^\nu \phi^{d}
           -\frac{1}{2} \,\phi^{a} \phi^{b} \partial_\mu \partial_\nu \phi^{c} \partial^\mu \partial^\nu \phi^{d}
                     \right) ,
                     \label{eq:dim8GeneralOperators}
\end{align}
where we indicated the symmetries of the tensors using brackets, e.g.~$c^{(6,0)}$ above is symmetric in its first four and its last two indices.
Moreover, the last tensor satisfies
\begin{align}
c_{abcd}^{(8,0,4,2)} + c_{bcad}^{(8,0,4,2)} + c_{cabd}^{(8,0,4,2)} = 0\,,
\end{align}
where we note that the constraints on $c^{(8,0,4,2)}$ are exactly the same as at dimension six; see \eqref{primdim6}. The number of independent building blocks at dimension eight is
\begin{equation}
    H = \begin{pmatrix}n+7\\8\end{pmatrix} \phi^8 + 
    \frac{3\,n}{2}\begin{pmatrix}n+3\\5\end{pmatrix} \phi^6 \square 
    +\begin{pmatrix}n+3\\4\end{pmatrix} (\phi^4\square^2)_1
    +\frac{n}{2}\begin{pmatrix}n+1\\3\end{pmatrix} (\phi^4\square^2)_2\,,
\end{equation}
where we counted the two types of $\phi^4\square^2$ operators separately.

\begin{table}[h]
\caption{Operator basis in the general scalar theory. The symmetries of the tensors under the exchange of indices are indicated by $( \ )$ (symmetric) and $[ \ ]$ (antisymmetric).
}\label{tabGen}
\centering{ 
\renewcommand{\arraystretch}{2.2}
\resizebox{\textwidth}{!}{
\begin{tabular}{|c|@{\hspace{5mm}}l@{\hskip 4mm}r@{\,}c@{\,}l|}
\hline
Dim., Spin 
& Operators & Additional && \hspace{-2.7mm}tensor relations
\\\hline
$\Delta = 3,\, \ell = 1$ 
    & $C^{[ab]}_\mu\,\phi^a \partial^\mu \phi^b $ 
    & &  &
\\\hline
$\Delta = 4,\,\ell=1$
    &  $\frac12 C^{(ab)c}_\mu\,\phi^a \phi^b \partial^\mu \phi^c$  
    & $C^{abc}_\mu$& $=$ &$-\,C^{bca}_\mu-C^{cab}_\mu$
\\\hline
 $\Delta = 4,\,\ell=2$
    &  $C^{(ab)}_{(\mu\nu)}\left(\phi^a\partial^\mu\partial^\nu \phi^b -2\,\partial^\mu\phi^a\partial^\nu\phi^b\right)  $
    & 
    &&
\\\hline
\multirow{1}{*}{$\Delta = 5,\,\ell=0$}
    & \multirow{1}{*}{$\frac1{5!} C^{(abcde)} \,\phi^a \phi^b \phi^c \phi^d \phi^e$}   
    & &&
\\\hline
    \multirow{2}{*}{\vspace{3mm}$\Delta = 5,\,\ell=1$}&  $\frac1{3!} C^{(abc)d}_\mu \,\phi^a \phi^b \phi^c \partial^\mu \phi^d$
    & $C^{abcd}$& $=$ &$-\,C^{bcda}_\mu-C^{cdab}_\mu-C^{dabc}_\mu$  
    \\[-3mm]
     &
    $C^{[ab]}_\mu \left(
      3\,\partial^\mu \phi^{a} \partial^2 \phi^{b}
      -\phi^{a} \partial^\mu \partial^2 \phi^{b}
   \right)$
    & &&
\\\hline
$\Delta = 5,\,\ell=\{1,1\}$
    &  $-C^{[abc]}_{[\mu\nu]} \, \phi^a \partial^\mu \phi^b \partial^\nu \phi^c$
    & &&
    \\
$\Delta = 5,\,\ell=2$
    &
    $2 \,C^{a(bc)}_{(\mu\nu)} \left( 
\phi^a\phi^b\partial^\mu\partial^\nu\phi^c 
    -2 \phi^a \partial^\mu\phi^b \partial^\nu \phi^c 
    \right)$\!\!\!\!\!
    & &&    
\\\hline
\multirow{2}{*}{$\Delta = 6,\,\ell=0$}
    & $\frac{1}{6!}C^{(abcdef)}\,\phi^a\phi^b\phi^c\phi^d\phi^e \phi^f$
    & &&
    \\&
    $-\frac14 C^{((ab)(cd))} \,\phi^a\phi^b\partial_\mu\phi^c\partial^\mu \phi^d$
    & $C^{abcd} $& $=$ &$ -\,C^{bcad} - C^{cabd}$
\\\hline
$\Delta = 6,\,\ell=1$
    & $\frac{1}{4!} C^{(abcd)e}_\mu\,
    \phi^a\phi^b\phi^c\phi^d \partial^\mu\phi^e$
    &  $C^{abcde}_\mu $& $=$ &$ -\,C^{...a}_\mu-C^{...b}_\mu-C^{...c}_\mu-C^{...d}_\mu$
\\\hline
$\Delta = 6,\,\ell=\{1,1\}$
    & 
    $-C^{a[bcd]}_{[\mu\nu]}\,
    \phi^a\phi^b
    \partial^\mu
    \phi^c\partial^\nu\phi^{d}$
    &$C^{abcd}_{\mu\nu} $& $=$ &$ C^{bcda}_{\mu\nu} - C^{cdab}_{\mu\nu} + C^{dabc}_{\mu\nu}$
    \\[2mm]
   \multirow{2}{*}{$\Delta = 6,\,\ell=2$}
    &
    $C^{(ab)(cd)}_{(\mu\nu)} \left(\phi^a\phi^b\phi^c\partial^\mu\partial^\nu\phi^d
    -2 \phi^a\phi^b\partial^\mu\phi^c\partial^\nu\phi^d\right)$\!\!\!\!\!\!\!\!\!\!\!\!\!\!\!
    &&&\\[-3mm]
    & $C^{(ab)}_{(\mu\nu)}\,
      \partial^\mu \partial^\nu \phi^{a} 
      \partial^2 \phi^{b}$&&&
\\\hline
\end{tabular}
}
}
\end{table}

\clearpage 
\let\oldbibliography\thebibliography
\renewcommand{\thebibliography}[1]{%
  \oldbibliography{#1}%
  \setlength{\itemsep}{0pt}%
}
\bibliographystyle{JHEP}
\bibliography{bibl}

\end{document}